\documentclass[11pt]{JHEP3}

\usepackage{epsfig}
\usepackage[latin1]{inputenc}
\usepackage{bbm,amsfonts}
\usepackage{graphicx}
\usepackage{amssymb,amsmath}
\usepackage{amsmath,amscd}
\usepackage{epsfig,multicol,bbm}
\usepackage{slashed}
\usepackage{diagrams}

\newcommand\fverb{\setbox\fverbbox=\hbox\bgroup\verb}
\newcommand\fverbdo{\egroup\medskip\noindent%
            \fbox{\unhbox\fverbbox}\ }
\newcommand\fverbit{\egroup\item[\fbox{\unhbox\fverbbox}]}
\newbox\fverbbox

\title{Quantization of Emergent Gravity}

\author{Hyun Seok Yang \\
Center for Quantum Spacetime, Sogang University, Seoul 121-741, Korea\\
E-mail:
\email{hsyang@kias.re.kr}}

\received{\today}       
\accepted{\today}       

\preprint{arXiv:1312.0580}

\abstract{Emergent gravity is based on a novel form of the equivalence principle known as the Darboux theorem or the Moser lemma in symplectic geometry stating that the electromagnetic force can always be eliminated by a local coordinate transformation as far as spacetime admits a symplectic structure, in other words, a microscopic spacetime becomes noncommutative (NC). If gravity emerges from U(1) gauge theory on NC spacetime, this picture of emergent gravity suggests a completely new quantization scheme where quantum gravity is defined by quantizing spacetime itself, leading to a dynamical NC spacetime. Therefore the quantization of emergent gravity is radically different from the conventional approach trying to quantize a phase space of metric fields. This approach for quantum gravity allows a background independent formulation where spacetime as well as matter fields is equally emergent from a universal vacuum of quantum gravity.}


\begin{document}


\section{Introduction}

This paper grew out of the author's endeavor, after fruitful interactions
with his colleagues, to clarify certain aspects of the physics
of emergent gravity \cite{hsy-ijmp09,hsy-jhep09,hsy-jpcs12} proposed by the author himself a few years ago.
The most notable feedbacks (including some confusions and fallacies) may be classified into three categories:

(I) Delusion on noncommutative (NC) spacetime,

(II) Prejudice on quantization,

(III) Globalization of emergent geometry.

Let us first defend an authentic picture from the above actions (I) and (II).
But we have to confess our opinion still does not win a good consensus.
(III) will be one of main issues addressed in this paper.

We start with the discussion why we need to change our mundane picture about gravity and spacetime
if emergent gravity picture makes sense. According to the general theory of relativity,
gravity is the dynamics of spacetime geometry where spacetime is realized as a (pseudo-)Riemannian manifold
and the gravitational field is represented by a Riemannian metric \cite{big-book,hawking-ellis}.
Therefore the dynamical field in gravity is a Riemannian metric over spacetime and
the fluctuations of metric necessarily turn a (flat) background spacetime into a dynamical structure.
Since gravity is associated to spacetime curvature, the topology of spacetime enters general relativity
through the fundamental assumption that spacetime is organized into a (pseudo-)Riemannian manifold.
A main lesson of general relativity is that spacetime itself is a dynamical entity.

The existence of gravity introduces a new physical constant, $G$.
The existence of the gravitational constant $G$, together with another physical constants $c$
and $\hbar$ originated from the special relativity and quantum mechanics, implies that spacetime
at a certain scale known as the Planck length $L_P = \sqrt{\frac{G \hbar}{c^3}} =1.6 \times
10^{-33} \mathrm{cm}$, is no longer commuting, instead spacetime coordinates obey
the commutation relation
\begin{equation}\label{nc-space}
    [y^\mu, y^\nu] = i \theta^{\mu\nu}.
\end{equation}
Note that the NC spacetime (\ref{nc-space}) is a close pedigree of quantum mechanics which
is the formulation of mechanics on NC phase space $[x^i, p_j] = i \hbar \delta^i_j$.
Once Richard Feynman said (The Character of Physical Law, 1967)
``I think it is safe to say that no one understands quantum mechanics".
So we should expect that the NC spacetime similarly brings about a radical change of physics.
Indeed a NC spacetime is much more radical and mysterious than we thought.
However we understood the NC spacetime too easily.
The delusion (I) on NC spacetime is largely rooted to our naive interpretation that
the NC spacetime (\ref{nc-space}) is an extra structure (induced by $B$ fields) defined
on a preexisting spacetime. This naive picture inevitably brings about the interpretation that
the NC spacetime (\ref{nc-space}) necessarily breaks the Lorentz symmetry.
See the introduction in \cite{lry2} for the criticism of this viewpoint.

One of the reasons why the NC spacetime (\ref{nc-space}) is so difficult to understand is that
the concept of space is doomed and the classical space should be replaced by a state in a complex
vector space $\mathcal{H}$. Since the NC spacetime (\ref{nc-space}), denoted by $\mathbb{R}_\theta^{2n}$,
is equivalent to the Heisenberg algebra of $n$-dimensional harmonic oscillator,
the Hilbert space $\mathcal{H}$ in this case is the Fock space (see eq. (\ref{fock-space})).
Since the Hilbert space $\mathcal{H}$ is a {\it complex linear} vector space, the superposition of two states
must be allowed which necessarily brings about the interference of states as in quantum mechanics.
We may easily be puzzled by a gedanken experiment that mimics the two slit experiment or
Einstein-Podolsky-Rosen experiment in quantum mechanics.
Furthermore any object $\mathcal{O}$ defined on the NC spacetime $\mathbb{R}_\theta^{2n}$ becomes
an operator acting on the Hilbert space $\mathcal{H}$.
Thus we can represent the object $\mathcal{O}$ in the Fock space $\mathcal{H}$, i.e.,
$\mathcal{O} \in \mathrm{End}(\mathcal{H})$. Since the Fock space has a countable basis,
the representation of $\mathcal{O} \in \mathrm{End}(\mathcal{H})$ is given by an $N \times N$ matrix
where $N = \mathrm{dim}(\mathcal{H})$ \cite{japan-matrix,nc-seiberg,hsy-epjc09}.
In the case at hand, $N \to \infty$.
This is the point we completely lose the concept of space.
And this is the pith of quantum gravity to define the final destination of spacetime.

To our best knowledge, quantum mechanics is the more fundamental description of nature.
Hence quantization, understood as the passage from classical physics to quantum physics,
is not a physical phenomenon. The world is already quantum and quantization is only our poor attempt
to find the quantum theoretical description starting with the classical description
which we understand better. The same philosophy should be applied to a NC spacetime.
In other words, spacetime at a microscopic scale, e.g. $L_P$, is intrinsically NC, so
spacetime at this scale should be replaced by a more fundamental quantum algebra
such as the algebra of $N \times N$ matrices denoted by $\mathcal{A}_N$.
Therefore the usual classical spacetime does not exist {\it a priori}.
Rather it must be derived from the quantum algebra like classical mechanics is derived
from quantum mechanics. But the reverse is not feasible.
In our case the quantum algebra is given by the algebra $\mathcal{A}_N$
of $N \times N$ matrices. Since $\mathcal{A}_N \cong \mathrm{End}(\mathcal{H})$,
it is isomorphic to an operator algebra $\mathcal{A}_\theta$ acting on the Hilbert space $\mathcal{H}$.
The algebra $\mathcal{A}_\theta$ is NC and generated by the Heisenberg algebra (\ref{nc-space}).
The spacetime structure derived from the NC $\star$-algebra $\mathcal{A}_\theta$ is
dubbed emergent spacetime. But this emergent spacetime is dynamical, so gravity will also be emergent
via the dynamical NC spacetime because gravity is the dynamics of spacetime geometry.
It is called emergent gravity. But it turns out \cite{hsy-ijmp09,hsy-jhep09,hsy-jpcs12}
that the dynamical NC spacetime is defined as a deformation of the NC spacetime (\ref{nc-space})
(see eq. (\ref{dynamic-nc})) and the deformation is related to U(1) gauge fields
on the NC spacetime (\ref{nc-space}). This picture will be clarified in section 4.

In this emergent gravity picture, any spacetime structure is not assumed {\it a priori} but
defined by the theory. That is, the theory of emergent gravity must be background independent.
Hence it is necessary to define a configuration in the algebra $\mathcal{A}_\theta$,
for instance, like eq. (\ref{nc-space}), to generate any kind of spacetime structure, even for flat spacetime.
A beautiful picture of emergent gravity is that the flat spacetime is emergent from
the Moyal-Heisenberg algebra (\ref{nc-space}). See section 6.2 for the verification.
Many surprising results immediately come from this dynamical origin of flat
spacetime \cite{hsy-jhep09,hsy-jpcs12}, which is absent in general relativity.
As a result, the global Lorentz symmtry, being an isometry of flat spacetime, is emergent too.
If true, the NC spacetime (\ref{nc-space}) does not break the Lorentz symmetry.
Rather it is emergent from the NC spacetime (\ref{nc-space}).
This is the picture how we correct the delusion (I).

The dynamical system is described by a Poisson manifold $(M, \theta)$ where $M$ is a
differentiable manifold whose local coordinates are denoted by $y^\mu \; \bigl(\mu = 1, \cdots, d =
\mathrm{dim}(M) \bigr)$ and the Poisson structure
\begin{equation}\label{p-bivector}
\theta = \frac{1}{2}\theta^{\mu\nu} \frac{\partial}{\partial y^\mu}
\bigwedge \frac{\partial}{\partial y^\nu} \in \Gamma(\Lambda^2 TM)
\end{equation}
is a (not necessarily nondegenerate) bivector field on $M$. Quantum mechanics is defined by quantizing
the Poisson manifold $(M, \theta)$ where $M$ is the phase space of particles with local coordinates
$y^\mu = (x^i, p_i)$. Let us call it $\hbar$-quantization.
Similarly the NC spacetime (\ref{nc-space}) is defined by quantizing
the Poisson manifold $(M, \theta)$ where $M$ is a spacetime manifold, e.g., $M = \mathbb{R}^{2n}$
and $y^\mu$'s are local spacetime coordinates. Let us call it $\theta$-quantization.
The first order deviation of the quantum or NC multiplication
from the classical one is given by the Poisson bracket of classical observables.
Thus the Poisson bracket of classical observables may be seen as a shadow of the noncommutativity
in quantum world. Then the correct statement is that spacetime always supports the spacetime Poisson
structure (\ref{p-bivector}) if a microscopic spacetime is NC.
And, if we introduce a line bundle $L \to M$ over a spacetime Poisson manifold $M$, the Poisson
manifold $(M, \theta)$ also becomes a dynamical manifold like the particle Poisson manifold.

The reason is the following. For simplicity, let us assume that the Poisson bivector (\ref{p-bivector})
is nondegenerate and define so-called a symplectic structure $B \equiv \theta^{-1} \in \Gamma(\Lambda^2 T^*M)$.
In this case the pair $(M, B)$ is called a symplectic manifold where $B$ is a nondegenerate,
closed two-form on $M$. If we consider a line bundle $L \to M$ over the symplectic manifold $(M, B)$,
the curvature $F=dA$ of the line bundle deforms the symplectic structure $B$ of the base manifold.
The resulting 2-form is given by $\mathcal{F} = B + F$ where $F=dA$ is the field strength of
dynamical U(1) gauge fields. See appendix A for the origin of this structure.
Note that the Bianchi identity $dF=0$ leads to $d\mathcal{F} = 0$
and $\mathcal{F}$ is invertible unless $\det (1+ B^{-1} F)=0$.
Then $\mathcal{F} = B + F$ is still a symplectic structure on $M$, so the dynamical gauge fields
defined on a symplectic manifold $(M, B)$ manifest themselves as a deformation of the symplectic structure.
In section 2 we show this picture also holds for a general Poisson manifold.

In consequence the dynamical spacetime Poisson manifold is modeled by a U(1) gauge theory
on $(M, \theta)$ with a fixed Poisson structure $\theta$ on a spacetime manifold $M$.
Therefore we can quantize the dynamical Poisson manifold and its quantization
leads to a dynamical NC spacetime described by a NC gauge theory. This $\theta$-quantization is
neither quantum mechanics nor quantum field theory because the underlying Poisson structure
refers to not a particle phase space but spacetime itself. Many people believe that NC gauge
theory is a classical theory because the $\hbar$-quantization does not come into play yet.
This attitude is based on a prejudice of long standing that quantization is nothing but
the $\hbar$-quantization. So a routine desk work insists on the $\hbar$-quantization of
the NC gauge theory to define a ``quantum" NC field theory. But we want to raise a question:
Is it necessary to have the ``quantum" NC field theory?

First of all, NC gauge theory is not a theory of particles but a theory of gravity,
called the emergent gravity \cite{review4}. We pointed out before that, after a matrix representation
$\mathcal{A}_N = \mathrm{End}(\mathcal{H})$ of NC spacetime, we completely lose the concept of space, 
so the concept of ``point" particles becomes ambiguous.
But the $\hbar$-quantization is just the quantization of a particle phase space
whatever it is finite-dimensional (quantum mechanics) or infinite-dimensional (quantum field theory)
because the Planck constant $\hbar$ has the physical dimension of (length) $\times$ (momentum).
Moreover the NC gauge theory describes a dynamical NC spacetime and
so formulate a theory of quantum gravity as was shown in \cite{hsy-jhep09,hsy-jpcs12} and
also in this paper. As we argued before, the dynamical NC spacetime is described by operators
acting on a Hilbert space, so the spacetime structure in emergent gravity should
be derived from the NC gauge theory. This picture leads to the concept of emergent spacetime
which would give us an insight into the origin of spacetime.
But if spacetime is emergent, everything supported on the spacetime should be emergent too
for an internal consistency of the theory. For example, quantum mechanics must be emergent
together with spacetime \cite{sladler}.\footnote{Recently Nima Arkani-Hamed advocated this viewpoint in
the Strings 2013 conference. The slides and video recording of his talk are available at http://strings2013.sogang.ac.kr.}
We will illuminate in section 6 how matter fields can be realized as
topological objects in NC $\star$-algebra $\mathcal{A}_\theta$ which correspond to stable
spacetime geometries. Recently a similar geometric model of matters was presented in \cite{ams-matter}.
To conclude, a NC gauge theory is already a quantum description because it is a quantized theory
of a spacetime Poisson manifold and it is not necessary to further consider the $\hbar$-quantization.
Rather quantum mechanics has to emerge from the NC gauge theory.
This is our objection to the prejudice (II).

If general relativity emerges from a U(1) gauge theory on a symplectic or Poisson manifold,
it is necessary to realize the equivalence principle and general covariance, the most important properties
in the theory of gravity (general relativity), from the U(1) gauge theory. How is it possible?
A remarkable aspect of emergent gravity is that there exists a novel form of the equivalence
principle even for the electromagnetic force \cite{hsy-ijmp09,hsy-jhep09}.
This assertion is based on a basic property in symplectic geometry known as the Darboux theorem
or the Moser lemma \cite{sg-book1,sg-book2} stating that the electromagnetic force can always
be eliminated by a local coordinate transformation as far as spacetime admits a symplectic structure,
in other words, a microscopic spacetime becomes NC.
The Moser lemma in symplectic geometry further implies \cite{cornalba,jur-sch,liu} that
the local coordinate transformation to a Darboux frame is equivalent to the Seiberg-Witten (SW)
map defining a spacetime field redefinition between ordinary and NC gauge fields \cite{ncft-sw}.
Therefore the equivalence principle in general relativity is realized as a noble statement \cite{lry2}
that NC gauge fields can be interpreted as the field variables defined in a locally inertial frame
and their commutative description via the SW map corresponds to the field variables in a laboratory frame represented by general curvilinear coordinates.\footnote{We think this picture may have an important implication
to black hole physics. Since we are vague so far about it, we want to quote a remark due
to Emil Martinec \cite{martinec}: ``The idea that the observables attached to different objects
do not commute in the matrix model gave a realization of the notion of {\it black hole complementarity}.
One can construct logical paradoxes if one attributes independent commuting observables
to the descriptions of events by observers who fall through the black hole horizon to probe its interior,
as well as observers who remain outside the black hole and detect a rather scrambled version
of the same information in the Hawking radiation.
If these two sets of observables do not commute, such paradoxes can be resolved."}
This beautiful statement is also true for a general Poisson manifold as will be shown in section 2.

It is possible to lift the novel form of the equivalence principle to a deformed algebra
of observables using the ``quantum" Moser lemma \cite{jsw1,jsw2}.
The quantum Moser lemma demonstrates that a star product deformed by U(1) gauge fields and
an original star product are in the same local gauge equivalence class.
See eq. (\ref{cov-star-map}). In particular two star products in the local gauge equivalence
are Morita equivalent and related by the action of a line bundle \cite{jsw-ncl,buba-me}.
In this sense we may identify the Morita equivalence of two star products
with the ``quantum" equivalence principle. This will be the subject of section 4.

The basic program of infinitesimal calculus, continuum mechanics and differential geometry is
that all the world can be reconstructed from the infinitely small.
For example, manifolds are obtained by gluing open subsets of Euclidean space where
the notion of sheaf embodies the idea of gluing their local data.
The concept of connection also plays an important role for the gluing. According to Cartan,
connection is a mathematical alias for an observer traveling in spacetime and
carrying measuring instruments. Global comparison devices are not available
owing to the restriction of the finite propagation speed.
So differential forms and vector fields on a manifold are defined locally and then glued together
to yield a global object. The gluing is possible because these objects are independent of
the choice of local coordinates. Infinitesimal spaces and the construction of global objects
from their local data are familiar in all those areas where spaces are characterized by
the algebras of functions on them. Naturally emergent gravity also faces a similar feature.
The local data for emergent gravity consist of NC gauge fields on a local Darboux chart.
They can be mapped to a Lie algebra of inner derivations on the local chart because
the NC $\star$-algebra $\mathcal{A}_\theta$ always admits a nontrivial inner automorphism.
Basically we need to glue these local data on Darboux charts to yield global vector fields
which will eventually be identified with gravitational fields, i.e., vielbeins.
This requires us to construct a global star product.
The star product is obtained by a perturbative expansion of functions in a formal deformation
parameter, e.g., the Planck's constant $\hbar$ which requires one to consider Taylor expansions
at points of $M$ \cite{kontsevich}. This suggests that a global version of the star product should
be defined in terms of deformations of the bundle of infinite jets of
functions \cite{cft,cftp,vdolg-am}. See appendix C for a brief review of jet bundles.
In section 5 we will discuss how global objects for the Poisson structure
and vector fields can be constructed from the Fedosov quantization of symplectic
and Poisson manifolds \cite{d-quant2,fedo-book}.
This will fill out the gap in our previous works which were missing the step (III).

Recent developments in string theory have revealed a remarkable and radical new picture about gravity.
For example, the AdS/CFT correspondence \cite{ads-cft1,ads-cft2,ads-cft3} shows a surprising picture
that a large $N$ gauge theory in lower dimensions defines a nonperturbative formulation of
quantum gravity in higher dimensions. In particular, the AdS/CFT duality shows a typical example of
emergent gravity and emergent space because gravity in higher dimensions is defined by a gravityless
field theory in lower dimensions. For comprehensive reviews, see, for example,
Refs. \cite{ads-cft,dho-fre,oriti,blau-theisen}. In section 6, we show that the AdS/CFT correspondence
is a particular example of emergent gravity from NC U(1) gauge fields.
Since the emergent gravity, we believe, is a significant new paradigm for quantum gravity,
it is desirable to put the emergent gravity picture on a rigorous mathematical foundation.
See also some related works \cite{richard-r,harold-p,hstein-rev,jun-r} and references therein.
We want to forward that direction in this paper although we touch only a tip of the iceberg.

The paper is organized as follows. Next three sections do not pretend to any originality.
Essential results can be found mostly in \cite{jsw1,jsw2}.
The leitmotif of these sections is to give a coherent exposition in a self-contained manner
to clarify why the symplectic structure of spacetime is arguably the essence of emergent gravity
realizing the duality between general relativity and NC U(1) gauge theory.

In section 2, we elucidate how U(1) gauge fields deform an underlying Poisson structure of spacetime.
It is shown that these deformations in terms of U(1) gauge fields can be identified via the Moser lemma
with local coordinates transformations. These coordinate transformations are represented by
Poisson U(1) gauge fields and lead to the semi-classical version of SW maps between ordinary and
NC U(1) gauge fields \cite{ncft-sw}.

In section 3, we review the Kontsevich's deformation quantization \cite{kontsevich} to understand
how to lift the results in section 2 to the case of deformed algebras.

In section 4, it is shown that dynamical NC spacetime is modeled by a NC gauge theory via
the quantum Moser lemma. It is straightforward to identify the SW map using the local covariance
map in \cite{jsw1,jsw2}. An important point is that the star product defined by a Poisson structure
deformed by U(1) gauge fields is Morita equivalent to the original undeformed one \cite{jsw-ncl,buba-me}.
This means that NC U(1) gauge theory describes their equivalent categories of modules.
We suggest that the Morita equivalence between two star products can be interpreted as the ``quantum"
equivalence principle for quantum gravity.

In section 5, we discuss how (quantum) gravity emerges from NC U(1) gauge theory.
First we identify local vector fields from NC U(1) gauge fields on a local Darboux chart.
And we consider the extension of the local data to an infinitesimal neighborhood using normal
coordinates and then present a prescription for global vector fields using the jet isomorphism
theorem \cite{epstein,ambient-metric} stating that the objects in the $\infty$-jet are represented
by the covariant tensors only. We consider a global star product using the Fedosov
quantization \cite{cft,cftp} in order to verify the prescription for global vector fields.
We also discuss a symplectic realization of Poisson manifolds \cite{d-weinstein,vaisman}
and symplectic groupoids \cite{sg-weinstein,kara-mas}.

In section 6, we show that the representation of a NC gauge theory in a Hilbert space is equivalent
to a large $N$ gauge theory which has appeared as a nonperturbative formulation of string/M theories.
We also illuminate how time emerges together with spaces from a Hamiltonian dynamical system which
is always granted by a background NC space, e.g. eq. (\ref{nc-space}), responsible for the emergent space.
We emphasize that the background is just a condensate that must be allowed for anything to develop and exist.
Finally we argue that the AdS/CFT correspondence \cite{ads-cft1,ads-cft2,ads-cft3} can be founded
on the emergent gravity from NC U(1) gauge fields.

In section 7, after a brief summary of the results obtained in this paper, we discuss possible
implications to string theory, emergent quantum mechanics and quantum entanglements building up
emergent spacetimes proposed by M. Van Raamdonk \cite{raamsdonk1,raamsdonk2}.

In appendix A, we highlight the local nature of NC U(1) gauge fields using the relation between
a Darboux transformation and SW map \cite{cornalba,jur-sch,liu} to emphasize why we need a globalization
of local vector fields obtained from them.

In appendix B, we discuss modular vector fields since emergent gravity requires unimodular Poisson
manifolds \cite{lry1}. Since Poisson manifolds can be thought of as
semiclassical limits of operator algebras, it is natural to ask whether they have modular automorphism groups,
like von Neumann algebras. It was shown \cite{cofs,omy2,mod-wein} that it is the case.
This confirms again that Poisson manifolds are intrinsically dynamical objects.

In appendix C, we give a brief exposition on jet bundles because they have been often used in this paper.
A jet bundle can be regarded as the coordinate free version of Taylor expansions, so a useful tool
for a geometrical covariant field theory though it is not widely used in physics so far.
We refer to \cite{jet-book1,jet-book2,jet-book3} for more detailed expositions.

\section{U(1) gauge theory on Poisson manifold}

In this section we recapitulate a fascinating picture that the Darboux theorem or the Moser lemma
in symplectic geometry can be interpreted as a novel form of the equivalence principle
for electromagnetic force. Fortunately all the essential details were greatly elaborated
in \cite{jsw1,jsw2} where it was shown that the local deformations of a symplectic or Poisson structure
can be transformed into a diffeomorphism symmetry using the Darboux theorem or the Moser lemma
in symplectic geometry and lead to the SW map between commutative and NC gauge fields.
Here we will review the results in \cite{jsw1,jsw2} to clarify why the symplectic structure
of spacetime leads to the novel form of the equivalence principle stating that the electromagnetic force
can be always eliminated by a local coordinate transformation.
It has been emphasized in \cite{hsy-ijmp09,hsy-jhep09} that the equivalence principle
for the electromagnetic force should be the first principle for emergent gravity.

Consider an Abelian gauge theory on a smooth real manifold $M$ that
also carries a Poisson structure (\ref{p-bivector}).
First we introduce the Schouten-Nijenhuis (SN) bracket for polyvector fields \cite{vaisman,silva-wein,dufour}.
A polyvector field of degree $k$, or $k$-vector field, on a manifold $M$ is a section of the $k$-th exterior
power $\Lambda^k TM$ of the tangent bundle and is dual to a $k$-form in $\Lambda^k T^*M$.
If $\Pi = \frac{1}{k!} \sum_{\mu_1, \cdots, \mu_k} \Pi^{\mu_1 \cdots \mu_k} \frac{\partial}{\partial y^{\mu_1}}
\bigwedge \cdots \bigwedge \frac{\partial}{\partial y^{\mu_k}}$ is a $k$-vector field, we will consider it
as a homogeneous polynomial of degree $k$ in the odd variables $\zeta_\mu \equiv \frac{\partial}{\partial y^\mu}$:
\begin{equation}\label{poly-vector}
 \Pi = \frac{1}{k!} \sum_{\mu_1, \cdots, \mu_k} \Pi^{\mu_1 \cdots \mu_k} \zeta_{\mu_1}
 \cdots \zeta_{\mu_k}.
\end{equation}
If $P$ and $Q$ are $p$- and $q$-vector fields, the SN bracket of $P$ and $Q$ is defined by \cite{kontsevich,dufour}
\begin{equation}\label{sn-bracket}
    [P,Q]_{S} = \sum_\mu \Bigl( \frac{\partial P}{\partial \zeta_\mu} \frac{\partial Q}{\partial y^\mu}
    - (-)^{(p-1)(q-1)} \frac{\partial Q}{\partial \zeta_\mu} \frac{\partial P}{\partial y^\mu} \Bigr).
\end{equation}
Clearly the bracket $[P,Q]_{S} = - (-)^{(p-1)(q-1)} [Q,P]_{S}$ defined above is a homogeneous polynomial
of degree $p+q-1$, so it is a $(p+q-1)$-vector field. The SN bracket (\ref{sn-bracket}) satisfies
a general property \cite{dufour} that, if $X$ is a vector field, then
\begin{equation}\label{lie-sn}
[X, \Pi]_S = \mathcal{L}_X \Pi
\end{equation}
for a $k$-vector field $\Pi$ where $\mathcal{L}_{X}$ is the Lie derivative with respect to the vector field $X$.
In particular, if $X$ and $Y$ are two vector fields, then the SN bracket of $X$ and $Y$ coincides
with their Lie bracket. We adopt the following differentiation rule for the odd variables
\begin{equation}\label{odd-diff}
    \frac{\partial}{\partial \zeta_\mu} \bigl(P \wedge Q) = P\frac{\partial Q}{\partial \zeta_\mu}
    + (-)^q \frac{\partial P}{\partial \zeta_\mu} Q.
\end{equation}
Then it is straightforward to verify the graded Jacobi identity for the SN bracket (\ref{sn-bracket}) \cite{dufour}:
\begin{equation}\label{sn-jacobi}
     (-)^{(p-1)(r-1)} [[P,Q]_S, R]_S + (-)^{(q-1)(p-1)} [[Q,R]_S, P]_S  + (-)^{(r-1)(q-1)} [[R,P]_S, Q]_S = 0
\end{equation}
where $P \in \Gamma(\Lambda^p TM), \; Q \in \Gamma(\Lambda^q TM), \; R \in \Gamma(\Lambda^r TM)$.
The bivector $\theta = \frac{1}{2}\theta^{\mu\nu} \zeta_\mu \zeta_\nu \in \Gamma(\Lambda^2 TM)$
is called a Poisson structure if and only if it obeys
\begin{equation}\label{poisson-str}
  [\theta, \theta]_S = \frac{1}{3} \Bigl( \theta^{\mu \lambda} \partial_\lambda \theta^{\nu\rho} +
  \theta^{\nu \lambda} \partial_\lambda \theta^{\rho \mu} + \theta^{\rho \lambda} \partial_\lambda \theta^{\mu\nu}\Bigr) \zeta_\mu \zeta_\nu \zeta_\rho = 0.
\end{equation}

Let us define the space $\mathcal{V}^\bullet(M) = \oplus_{p \geq 0} \Gamma(\Lambda^p TM)$ which forms
a graded Lie algebra under the SN bracket (\ref{sn-bracket}) if the grade of $\mathcal{V}^p (M)
\equiv \Gamma(\Lambda^p TM)$ is defined to be $p-1$.
Also introduce a differential operator $d_\theta:\mathcal{V}^p (M) \to \mathcal{V}^{p+1} (M)$ defined by
 \begin{equation}\label{p-coho}
    d_\theta \Pi \equiv - [\Pi, \theta]_S
\end{equation}
for any $p$-vector field $\Pi$ in $\mathcal{V}^p (M)$.
Using eq. (\ref{poisson-str}), one can show that the coboundary operator (\ref{p-coho}) is nilpotent, i.e.,
\begin{equation}\label{dd=0}
    d_\theta^2 \Pi = [ [\Pi, \theta]_S, \theta]_S = - \frac{1}{2} [ [\theta, \theta]_S, \Pi]_S = 0
\end{equation}
where the Jacobi identity (\ref{sn-jacobi}) was used.  Then one can define a differential complex
$(\mathcal{V}^\bullet(M), d_\theta)$ given by
\begin{equation}\label{lich-complex}
\cdots \to \mathcal{V}^{p-1} (M) \xrightarrow{d_\theta} \mathcal{V}^p (M) \xrightarrow{d_\theta}
\mathcal{V}^{p+1} (M) \to \cdots
\end{equation}
which is called the Lichnerowicz complex. The cohomology of this complex is called
the Poisson cohomology \cite{vaisman} and is defined as the quotient groups
\begin{equation}\label{poisson-coho}
    H_\theta^\bullet (M) = \mathrm{Ker}\; d_\theta/
    \mathrm{Im}\; d_\theta.
\end{equation}

Given a Poisson structure $\theta \in \mathcal{V}^2(M)$ on a manifold $M$, the Poisson bivector $\theta$
induces a natural homomorphism $\rho: T^* M \to TM$ by
\begin{equation}\label{anchor}
    A \mapsto \rho(A) = - \theta^{\mu\nu} A_\nu \frac{\partial}{\partial y^\mu}
\end{equation}
for $A = A_\mu(y) dy^\mu \in T^*_p M$. The bundle homomorphism (\ref{anchor}) is called the anchor map of $\theta$.
The anchor map (\ref{anchor}) can be written as
\begin{equation}\label{sn-anchor}
    \rho(A) \equiv A_\theta = A_\mu d_\theta y^\mu = \theta^{\mu\nu} A_\mu \zeta_\nu.
\end{equation}
Note that, if the coboundary operator $d_\theta$ acts on a smooth function $f \in C^\infty(M)$,
it generates a vector field in $\Gamma(TM)$ given by
\begin{equation}\label{ham-vec}
    d_\theta f = -[f, \theta]_S = - \theta^{\mu\nu} \partial_\nu f \partial_\mu \equiv X_f
\end{equation}
which is called the Hamiltonian vector field. Thus the operator $d_\theta$ acting on the space $C^\infty(M)$
defines the correspondence $f \mapsto X_f$ and one can introduce a bilinear map $\{-,-\}_\theta: C^\infty(M)
\times C^\infty(M) \to C^\infty(M)$, the so-called Poisson bracket, defined by
\begin{equation}\label{poisson-bracket}
    \{f, g\}_\theta \equiv (d_\theta f)(g) = X_f(g) = - X_g(f) = \langle \theta, df \wedge dg \rangle
\end{equation}
for $f, g \in C^\infty(M)$. One can also act the coboundary operator $d_\theta$ on the vector
field $A_\theta \in \Gamma(TM)$ and the corresponding bivector field is given by
\begin{equation}\label{bivec-f}
    F_\theta \equiv d_\theta A_\theta = \frac{1}{2} F_{\mu\nu}d_\theta y^\mu \wedge d_\theta y^\nu
    = \frac{1}{2} \theta^{\mu\rho} F_{\rho\sigma} \theta^{\sigma\nu}
    \zeta_\mu \zeta_\nu \in \mathcal{V}^2(M)
\end{equation}
where $F \equiv dA = \frac{1}{2}F_{\mu\nu} dy^\mu \wedge dy^\nu  \in \Gamma(\Lambda^2 T^*M)$
and the condition (\ref{poisson-str}) was used to deduce $d_\theta^2 y^\mu = 0$.
Thus one can consider the bivector $F_\theta = d_\theta A_\theta$
to be dual to the two-form $F=dA$. We will identify the one-form $A$ in eq. (\ref{anchor})
with a connection of line bundle $L \to M$ and $F=dA$ with its curvature. This identification is consistent
with the dual description in terms of polyvectors: First note that, under an infinitesimal gauge transformation
$A \mapsto A + d\lambda$, the vector field $A_\theta$ in (\ref{sn-anchor}) changes by a Hamiltonian vector
field $d_\theta \lambda$, i.e., $A_\theta \mapsto A_\theta + d_\theta \lambda$. And,
from the definition (\ref{bivec-f}), we have $d_\theta F_\theta = 0$ due to $d_\theta^2 = 0$
or $[\theta, \theta]_S = 0$, which is dual to the Bianchi identity $dF=0$ in gauge theory.

Now we perturb the Poisson structure $\theta$ by introducing a one-parameter deformation $\theta_t$
with $t \in [0,1]$ whose evolution obeys
\begin{equation}\label{t-evolution}
    \partial_t \theta_t = \mathcal{L}_{A_{\theta_t}} \theta_t
\end{equation}
with initial condition $\theta_0 = \theta$. Here $A_{\theta_t} = A_\mu d_{\theta_t} y^\mu$ is a $t$-dependent vector field defined by the anchor map (\ref{sn-anchor}) of $\theta_t$.
Note that the evolution (\ref{t-evolution}) can be written according to eq. (\ref{lie-sn}) as
\begin{equation}\label{t-evol-veca}
  \partial_t \theta_t = \mathcal{L}_{A_{\theta_t}} \theta_t = [ A_{\theta_t}, \theta_t]_S =
  - d_{\theta_t} A_{\theta_t} = - F_{\theta_t}.
\end{equation}
In terms of local coordinates, the evolution equation (\ref{t-evolution}) thus takes the form
\begin{equation}\label{t-evol-local}
   \partial_t \theta_t^{\mu\nu} = - (\theta_t F \theta_t)^{\mu\nu}, \qquad \theta_0^{\mu\nu} = \theta^{\mu\nu}.
\end{equation}
The formal solution (defined in power series of $t$) is given by \cite{jsw1,jsw2}
\begin{equation}\label{sol-t}
    \theta_t = \theta \frac{1}{1+ t F \theta}.
\end{equation}
One can see that $\theta_t$ is a Poisson structure for all $t$, i.e. $[\theta_t, \theta_t]_S=0$,
because $[\theta_t, \theta_t]_S = 0$ at $t=0$ and
\begin{equation}\label{t-jacobi}
   \partial_t [\theta_t, \theta_t]_S = 2 [ [A_{\theta_t}, \theta_t]_S, \theta_t]_S
   = [ A_{\theta_t}, [\theta_t, \theta_t]_S ]_S = \mathcal{L}_{A_{\theta_t}} [\theta_t, \theta_t]_S
\end{equation}
where we have used the Jacobi identity (\ref{sn-jacobi}).

Suppose that $\theta$ is invertible and define $B \equiv \theta^{-1} \in \Gamma(\Lambda^2 T^*M)$.
In this case the nondegenerate two-form $B$ defines a symplectic structure on $M$ because
$[\theta, \theta]_S = 0$ implies $dB=0$. Denoting $\Theta \equiv \theta_1$, the solution (\ref{sol-t})
can be written as the form
\begin{equation}\label{t-symplectic}
    \Theta = \frac{1}{B+F}
\end{equation}
and $dF=0$ due to $[\Theta, \Theta]_S = 0$. Hence the initial symplectic structure $B$ evolves
to a new symplectic structure $\mathcal{F} \equiv B + F$ according to the evolution
equation (\ref{t-evolution}) defining the one-parameter deformation. Since we have identified
the one-form $A$ in the anchor map (\ref{anchor}) with U(1) gauge fields, the fluctuations of
U(1) gauge fields can be understood as the deformation of symplectic manifold $(M,B)$.
Therefore the U(1) gauge theory, especially the Maxwell's theory of electromagnetism,
on a symplectic manifold $(M,B)$ can be understood completely in the context of symplectic geometry.

Since the $t$-evolution (\ref{t-evolution}) is generated
by the vector field $A_{\theta_t}$, it can be integrated to a flow of $A_{\theta_t}$ starting
at identity where $t$ plays a role of the affine parameter labeling points on the curve of the flow
(see Appendix A in \cite{jsw2} for the  derivation).\footnote{The derivation employs
a clever observation. Consider a $t$-dependent function $f(t)$ whose $t$-evolution is governed
by $\bigl(\partial_t + A(t) \bigr)f(t) = 0$ where $A(t)$ is a differential operator of arbitrary degree.
One can show that the function $f(t)$ satisfies the recursion relation $e^{\partial_t + A(t)}
e^{-\partial_t} f(t+1) = f(t)$. As the Baker-Campbell-Hausdorff formula implies, there are no free
$t$-derivatives acting on $f(t+1)$, so one can evaluate the recursion relation at $t=0$ and get
$e^{\partial_t + A(t)} e^{-\partial_t}|_{t=0} f(1) = f(0)$.}
The result is given by
\begin{equation}\label{t-flow}
    \rho_A^* = \exp(A_{\theta_t} + \partial_t)\exp(-\partial_t)|_{t=0}
\end{equation}
that relates the Poisson structures $\Theta = \theta_1$ and $\theta = \theta_0$, i.e.,
\begin{equation}\label{moser-flow}
   \rho_A^* \Theta = \theta.
\end{equation}
In order to calibrate the deformation in eq. (\ref{sol-t}) or (\ref{t-symplectic}) due to
local gauge fields, let us represent the exponential map $\rho_A^*$ by
\begin{equation}\label{diff-rho}
    \rho_A^* = \mathrm{id} + \mathfrak{A}_A
\end{equation}
where the differential operator $ \rho_A^*$ basically pulls back functions on $M$ to functions on $TM$.
Let $\mathfrak{P}=(C^\infty(M), \{-,-\}_\theta)$ be a Poisson algebra
on the vector space $C^\infty(M)$ equipped with the Poisson bracket (\ref{poisson-bracket}).
Then the diffeomorphism (\ref{t-flow}) defines a natural algebra homomorphism $\rho_A^*: \mathfrak{P} \to
\mathfrak{P}$ which acts on the Poisson algebra $\mathfrak{P}$ and the equivalence relation
(\ref{moser-flow}) for Poisson structures can be translated into the equivalence relation
of Poisson algebras defined by
\begin{equation}\label{equiv-poisson}
 \rho_A^*\{f, g\}_\Theta = \{\rho_A^* f, \rho_A^* g \}_\theta
\end{equation}
for $f,g \in C^\infty(M)$. The Poisson algebra $\mathfrak{P}'=(C^\infty(M), \{-,-\}_\Theta)$
now depends on the gauge field $F=dA$, so physically the map (\ref{diff-rho}) serves to incorporate
a back-reaction of gauge fields on the Poisson algebra.

The separation (\ref{diff-rho}) lucidly shows that $\mathfrak{A}_A = \rho_A^* - \mathrm{id}$ vanishes
when turning off the U(1) gauge fields, i.e. $A=0$. To examine the response of  $\mathfrak{A}_A$
under an infinitesimal gauge transformation $A \mapsto A + d\lambda$,
first recall that the vector field $A_{\theta_t}$ changes by a Hamiltonian vector field
$d_{\theta_t} \lambda = - \theta^{\mu\nu}_t \partial_\nu \lambda \partial_\mu$: $A_{\theta_t}
\mapsto A_{\theta_t} + d_{\theta_t} \lambda$. The effect of this gauge transformation on the flow (\ref{t-flow})
is given to first order in $\lambda$ by \cite{jsw2}
\begin{equation}\label{rho-gauge-tr}
  \rho_{A + d\lambda}^* = (\mathrm{id} + d_\theta \widetilde{\lambda}) \circ \rho_A^*
\end{equation}
or equivalently
\begin{equation}\label{a-gauge-tr}
 \mathfrak{A}_{A + d\lambda} = \mathfrak{A}_A + d_\theta \widetilde{\lambda}
 + \{\mathfrak{A}_A, \widetilde{\lambda} \}_\theta
\end{equation}
where
\begin{equation}\label{lambda-tilde}
    \widetilde{\lambda} (\lambda, A) = \sum_{n=0}^\infty \frac{(A_{\theta_t} + \partial_t)^n (\lambda)}{(n+1)!}|_{t=0}.
\end{equation}
The gauge transformation (\ref{rho-gauge-tr}) shows that $\rho_A^*$ is a coordinate-independent and
manifestly covariant object under a (non-Abelian) gauge transformation generated by $\widetilde{\lambda}$
in (\ref{lambda-tilde}). In this respect, $\mathfrak{A}_A$ in eq. (\ref{diff-rho}) plays the role
of a generalized gauge connection, so we call it a Poisson gauge field. We also call it
a symplectic gauge field for the case that the bivector field $\theta$ is nondegenerate.
The exponential map $\rho_A^*$ is a semi-classical version of the covariantizing map $\mathcal{D}_A$
\cite{jsw1,jsw2} which is a formal differential operator acting on a function $f \in C^\infty(M)$
such that $\mathcal{D}_A f = f + f_A$ becomes a covariant function under the NC version of
the gauge transformation (\ref{a-gauge-tr}).
The field strength of Poisson gauge fields evaluated on two functions $f,g$
is defined by \cite{jsw2}\footnote{It may be more enticing to define the field strength as $\mathcal{F}_A (f,g) = \{\rho_A^* f, \rho_A^* g \}_\theta = \rho_A^* \{f, g\}_\Theta$. A motivation for this definition
is simply to achieve a background independent object for the field
strength \cite{ncft-sw,nc-seiberg} because it may be ambiguous to discriminate dynamical fields from
a background part in the case of a generic Poisson structure.}
\begin{equation}\label{curvature-poisson}
    \mathcal{F}_A (f,g) = \{\rho_A^* f, \rho_A^* g \}_\theta - \rho_A^* \{f, g\}_\theta =
    \rho_A^* \Bigl( \{f, g\}_\Theta - \{f, g\}_\theta \Bigr)
\end{equation}
where eq. (\ref{equiv-poisson}) was used. Abstractly, the 2-cochain field strength (\ref{curvature-poisson})
can be regarded as a bidifferential operator acting on two functions and can be written in the form
\begin{equation}\label{2-cochain}
    \mathcal{F}_A = \rho_A^* \circ  \bigl( \Theta  - \theta \bigr) = \rho_A^* \circ \mathbf{F}_\theta
\end{equation}
where
\begin{equation}\label{cochain-poisson}
\mathbf{F}_\theta \equiv \frac{1}{2} \mathbf{F}_{\mu\nu} d_\theta y^\mu \wedge d_\theta y^\nu
\end{equation}
with
\begin{equation}\label{semi-field}
     \mathbf{F}_{\mu\nu} = \Bigl(\frac{1}{1+F\theta} F \Bigr)_{\mu\nu}.
\end{equation}

Now we come to an important picture about the diffeomorphism
(\ref{moser-flow}) between two different Poisson structures. The
$t$-evolution (\ref{t-evolution}) implies that all the Poisson
structures $\theta_t$ for $t \in [0,1]$ are related by coordinate transformations
generated by the flow $\rho^*_{tt'}(A)$ of $A_{\theta_t}$ such that
$\rho^*_{tt'}(A)\theta_{t'} = \theta_t$. In particular, denoting
$\rho^*_{01}(A) = \rho^*_A$, we have the relation (\ref{moser-flow}).
This constitutes an appropriate generalization of Moser's lemma from symplectic geometry
to Poisson case \cite{jsw1,jsw2}. According to the Weinstein's splitting theorem for a $d$-dimensional
Poisson manifold $(M, \theta)$,\footnote{The splitting theorem \cite{d-weinstein,vaisman,silva-wein}
states that a $d$-dimensional Poisson manifold $(M, \theta)$ is
locally equivalent to the product of $\mathbb{R}^{2n}$ equipped with
the canonical symplectic structure with $\mathbb{R}^{d-2n}$ equipped
with a Poisson structure of rank zero at a point in a local neighborhood of $M$.
That is, the Poisson manifold $(M, \theta)$ is locally isomorphic (in a
neighborhood of $P \in M$) to the direct product $S \times N$ of a
symplectic manifold $(S, \sum_{i=1}^n dq^i \wedge dp_i)$ with a
Poisson manifold $(N_P, \{-,-\}_N)$ whose Poisson tensor vanishes at $P$.
Thus the Poisson structure $\theta$ can be consistently restricted to a leaf $S$ as nondegenerate
and the leaves become $2n$-dimensional symplectic manifolds.}
one can choose coordinates $y^\mu := (q^1, \cdots, q^n, p_1, \cdots, p_n, r^1, \cdots, r^{d-2n})$
on a neighborhood centered at a point $P \in M$ such that
\begin{equation}\label{weinstein}
    \theta = \sum_{i=1}^n \frac{\partial}{\partial q^i} \bigwedge \frac{\partial}{\partial p_i} +
    \frac{1}{2} \sum_{a,b=1}^{d-2n} \varphi_{ab}(r)
    \frac{\partial}{\partial r^a} \bigwedge \frac{\partial}{\partial r^b} \qquad \mathrm{and}
    \qquad \varphi_{ab}(r)|_P = 0.
\end{equation}
We call such a coordinate system the Darboux-Weinstein frame.
In the Darboux-Weinstein frame, the Poisson bracket $\{f, g \}_\theta = \theta(df, dg)$ is given by
\begin{equation}\label{wd-poisson}
    \{f, g \}_\theta = \sum_{i=1}^n \Bigl( \frac{\partial f}{\partial q^i} \frac{\partial g}{\partial p_i}
    - \frac{\partial f}{\partial p_i} \frac{\partial g}{\partial q^i} \Bigr)
     + \sum_{a,b=1}^{d-2n} \{r^a, r^b \}_\theta
    \frac{\partial f}{\partial r^a} \frac{\partial g}{\partial r^b}
\end{equation}
where $\varphi_{ab}(r) = \{r^a, r^b \}_\theta$. Therefore the existence of the Moser flow $\rho_A^*$
implies that it is always possible to find a coordinate transformation to nullify the
local deformation of Poisson structures by U(1) gauge fields. In other words, there always exists
the Darboux-Weinstein frame where the underlying Poisson structure locally takes the form (\ref{weinstein}).
Therefore we will locally use the exponential map (\ref{t-flow}) to define the Poisson algebra
in the Darboux-Weinstein frame.

If the Poisson structure $\theta$ is nondegenerate, i.e., it defines a symplectic structure
$B \equiv \theta^{-1}$ on $M$, the equation (\ref{moser-flow}) can be written as the form
\begin{equation}\label{darboux-flow}
   \rho_A^* (B+F) = B.
\end{equation}
This is the original statement of the Moser lemma in symplectic geometry \cite{sg-book1,sg-book2}.
This rather well-known theorem in symplectic geometry suggests an important physics
\cite{hsy-ijmp09,hsy-jhep09}. The Moser lemma (\ref{darboux-flow}) implies
that the electromagnetic force $F=dA$ can always be eliminated by a local coordinate transformation
like the gravitational force in general relativity as far as an underlying space admits a symplectic structure.
Therefore there exists a novel form of the equivalence principle even for the electromagnetic force.
Moreover we observed that the equivalence principle for the electromagnetic force $F$ holds for a
general Poisson manifold $(M, \theta)$ where $\theta$ is not
necessarily nondegenerate. In the end a critical question is whether
U(1) gauge theory on a symplectic manifold $(M, B)$ or more
generally a Poisson manifold $(M, \theta)$ can be formulated as a
theory of gravity. It was positively answered in \cite{hsy-ijmp09,hsy-jhep09,yasi}
(see also reviews \cite{hsy-jpcs12,review4,lee-yang}) that Einstein gravity can emerge
from electromagnetism if spacetime admits a symplectic or Poisson structure.
We will elaborate this idea more concretely in the context of emergent gravity.

Using the bundle homomorphism $\theta: T^*M \to TM$, we defined the anchor map (\ref{anchor})
of $\theta \in \mathcal{V}^2(M)$ and introduced the correspondence $A \mapsto A_\theta$
defined by eq. (\ref{sn-anchor}) between the connection $A$ of line bundle
(regarded as a one-form in $T^*M$) and a vector field $A_\theta$.
The perturbed vector field $A_{\theta_t}$ generates a flow as an integral curve of $A_{\theta_t}$
which defines a one-parameter group of diffeomorphisms obeying the relation (\ref{moser-flow}).
The dynamical diffeomorphism (\ref{diff-rho}) then associates the U(1) gauge field
$A$ with a Poisson gauge field $\mathfrak{A}_A$ which corresponds to a semi-classical version of
the SW map \cite{ncft-sw}. But the transformation (field redefinition) from ordinary
to Poisson gauge fields has to preserve the gauge equivalence relation between
ordinary and Poisson gauge symmetries.
This condition is summarized by the following commutative diagram:
\begin{equation} \label{semi-sw-map}
\begin{diagram}
A &\rTo^{SW} & \mathfrak{A}_A\\
\dTo^{\delta_\lambda} & &\dTo_{\delta_{\widetilde{\lambda}}}\\
A+ d\lambda &\rTo^{~~~~SW~~~}
& \mathfrak{A}_A + D_A \widetilde{\lambda}
\end{diagram}
\end{equation}
where $\delta_\lambda A \equiv d\lambda$ and $\delta_{\widetilde{\lambda}} \mathfrak{A}_A \equiv D_A \widetilde{\lambda} = d_\theta \widetilde{\lambda} + \{\mathfrak{A}_A, \widetilde{\lambda} \}_\theta$.
The gauge transformation (\ref{a-gauge-tr}) shows that the Poisson gauge field $\mathfrak{A}_A$
does satisfy the equivalence relation (\ref{semi-sw-map}), i.e.,
\begin{equation}\label{sw-equiv}
  \mathfrak{A}_A + \delta_{\widetilde{\lambda}} \mathfrak{A}_A
  = \mathfrak{A}_{A + \delta_\lambda A}.
\end{equation}
It is straightforward to derive the Lie algebra structure
\begin{equation}\label{lie-poisson}
    [\delta_{\widetilde{\lambda}_1}, \delta_{\widetilde{\lambda}_2}]
    = \delta_{\{\widetilde{\lambda}_1, \widetilde{\lambda}_2\}_\theta}
\end{equation}
by applying the gauge transformation $\delta_{\widetilde{\lambda}} \mathfrak{A}_A (f) =
\{f, \widetilde{\lambda} \}_\theta + \{ \mathfrak{A}_A (f), \widetilde{\lambda} \}_\theta$ and
using the Jacobi identity for the Poisson bracket (\ref{poisson-bracket}).
It may be emphasized that the SW map $SW: A \mapsto \mathfrak{A}_A$ can be basically obtained by
finding the Moser flow (\ref{t-flow}) satisfying eq. (\ref{moser-flow}) for the Poisson case
or eq. (\ref{darboux-flow}) for the symplectic case.

Let us introduce local coordinates $y^\mu$ on a patch $U \subset M$ and consider the action of
the diffeomorphism $\rho_A$ on the coordinate function $y^\mu$. We represent the action as
\begin{equation}\label{cov-obj}
   \rho^*_A (y^\mu) \equiv x^\mu(y) = y^\mu + \mathfrak{Y}^\mu (y)
\end{equation}
which plays the role of covariant (dynamical) coordinates. According to the Weinstein's splitting theorem,
one can always choose a local coordinate chart $(\{ y^\mu \}, U)$ such that the Poisson structure $\theta$
on $U$ is given by the Darboux-Weinstein frame (\ref{weinstein}). Such a coordinate chart $(\{ y^\mu \}, U)$
on a general Poisson manifold will be called Darboux-Weinstein coordinates or simply Darboux coordinates
for the symplectic case. Therefore, for the symplectic case where the Poisson structure $\theta$ is
nondegenerate, one can assume, without loss of generality, that the initial Poisson structure
$\theta_0 = \theta$ is always constant. For the constant symplectic structure $B = \theta^{-1}$,
it is useful to represent the covariant coordinates in the form
\begin{equation}\label{cov-coord}
    x^\mu(y) = y^\mu + \theta^{\mu\nu} a_\nu (y) \in C^\infty(M)
\end{equation}
and define corresponding ``covariant momenta" by
\begin{equation}\label{cov-mom}
    C_\mu(y) \equiv B_{\mu\nu} x^\nu(y) = p_\mu +  a_\mu (y) \in C^\infty(M)
\end{equation}
with $p_\mu \equiv B_{\mu\nu} y^\nu$. In this case the field strength (\ref{curvature-poisson})
evaluated on the coordinate basis
$\{ y^\mu \}$ is given by
\begin{equation}\label{field-coord}
    \mathcal{F}_A (y^\mu, y^\nu) = \{x^\mu, x^\nu\}_\theta - \theta^{\mu\nu} = - \bigl( \theta f \theta \bigr)^{\mu\nu}
\end{equation}
or
\begin{equation}\label{field-moment}
    \mathcal{F}_A (p_\mu, p_\nu) = \{C_\mu, C_\nu\}_\theta
     + B_{\mu\nu} = f_{\mu\nu}
\end{equation}
where
\begin{equation}\label{field-comp}
 f_{\mu\nu} = \partial_\mu a_\nu - \partial_\nu a_\mu
 + \{ a_\mu, a_\nu \}_\theta.
\end{equation}
Evaluating the field strength $\mathcal{F}_A (y^\mu, y^\nu)$ with the cochain map (\ref{2-cochain})
and then identifying it with eq. (\ref{field-coord}),
one can get the relation \cite{cornalba,jur-sch,liu,hsy-mpla06,ban-yan}
\begin{equation}\label{sw-field}
     f_{\mu\nu} (y) = \Bigl(\frac{1}{1+F\theta} F \Bigr)_{\mu\nu} (x)
\end{equation}
where the coordinate transformation (\ref{cov-obj}) was assumed such that $\rho^*_A\bigl( f(y) \bigr) =
f \bigl( \rho_A (y) \bigr) = f(x)$ for a smooth function $f \in  C^\infty(M)$.
For a general Poisson structure $\theta$, the SW map (\ref{sw-field}) must be replaced by
\begin{equation}\label{gen-sw-field}
   \theta^{\mu\nu} (y) +  \Xi^{\mu\nu} (y) = \rho^*_A \Bigl( \theta^{\mu\nu} (y) +
   (\theta \mathbf{F} \theta)^{\mu\nu} (y) \Bigr)
\end{equation}
where
\begin{equation}\label{gen-sw-y}
     \Xi^{\mu\nu} (y) \equiv \{y^\mu, \mathfrak{Y}^\nu(y) \}_\theta
     - \{ y^\nu, \mathfrak{Y}^\mu(y) \}_\theta
     + \{ \mathfrak{Y}^\mu(y), \mathfrak{Y}^\nu (y) \}_\theta
\end{equation}
and $\mathbf{F}_{\mu\nu} (y)$ is given by eq. (\ref{semi-field}).
For the general case, the diffeomorphism $\rho^*_A$ in eq. (\ref{gen-sw-field}) now nontrivially
acts on the Poisson tensor $\theta^{\mu\nu} (y)$ too, so the SW map is rather complicated.

One can also find the Jacobian factor $J \equiv |\frac{\partial y}{\partial x}|$
for the coordinate transformation (\ref{cov-obj}). For the symplectic case,
it is easy to deduce from eq. (\ref{darboux-flow}) that
\begin{equation}\label{sw-jacobian}
    J(x) = \sqrt{\det(1 + F \theta)}(x).
\end{equation}
But the general Poisson case (\ref{moser-flow}) requires a careful treatment. One can first
solve eq. (\ref{moser-flow}) in a subspace where the Poisson tensor is nondegenerate and then extend
the solution (\ref{sw-jacobian}) to entire space such that it satisfies eq. (\ref{moser-flow}).
(Note that $\theta$ is placed on both sides of eq. (\ref{moser-flow}).)
Then it ends with the result (\ref{sw-jacobian}) again.
The equations (\ref{sw-field}), (\ref{gen-sw-field}) and (\ref{sw-jacobian}) consist of a semiclassical
version of the SW map \cite{ncft-sw} describing a spacetime field redefinition between ordinary and
symplectic or Poisson gauge fields in the approximation of slowly varying fields,
$\sqrt{\theta} |\frac{\partial F}{F}| \ll 1$, in the sense keeping field strengths
(without restriction on their size) but not their derivatives.

We conclude this section with a brief summary. The electromagnetic force manifests itself
as the deformation of an underlying Poisson structure and the deformation is described by
a formal solution (\ref{sol-t}) of the evolution equation (\ref{t-evolution}). But
every Poisson structures $\theta_t$ for $t \in [0,1]$ are related to the canonical Poisson
structure (\ref{weinstein}) in the Darboux-Weinstein frame by a local coordinate
transformation (\ref{t-flow}) generated by the vector field $A_{\theta_t}$.
This Darboux-Weinstein frame corresponds to a locally inertial frame in general relativity and
so constitutes a novel form of the equivalence principle for the electromagnetic force
as a viable analogue of the equivalence principle in general relativity.

\section{Deformation quantization}

Now we want to quantize the Poisson algebra $\mathfrak{P} = (C^\infty(M), \{-,-\}_\theta)$
introduced in the previous section.
The canonical quantization of the Poisson algebra $\mathfrak{P} = (C^\infty(M), \{-,-\}_\theta)$
consists of a complex Hilbert space $\mathcal{H}$ and of a quantization map $\mathcal{Q}$ to attach
to functions $f \in C^\infty(M)$ on $M$ quantum operators $\widehat{f} \in \mathcal{A}_\theta$
acting on $\mathcal{H}$. The map $\mathcal{Q}: C^\infty(M) \to \mathcal{A}_\theta$ by
$f \mapsto \mathcal{Q}(f) \equiv \widehat{f}$ should be $\mathbb{C}$-linear and an algebra homomorphism:
\begin{equation}\label{q-rule}
    f \cdot  g \mapsto \widehat{f \star g} = \widehat{f} \cdot \widehat{g}
\end{equation}
and
\begin{equation}\label{quantum-prod}
   f \star g \equiv \mathcal{Q}^{-1} \Bigl( \mathcal{Q}(f) \cdot \mathcal{Q}(g) \Bigr)
\end{equation}
for $f, g \in C^\infty(M)$ and $\widehat{f}, \widehat{g} \in \mathcal{A}_\theta$.
The Poisson bracket (\ref{poisson-bracket}) controls the failure of commutativity
\begin{equation}\label{fail-comm}
  [\widehat{f}, \widehat{g}] \sim i \{f, g \}_\theta + \mathcal{O}(\theta^2).
\end{equation}
A natural question at hand is whether such quantization is always possible for general Poisson manifolds
with a radical change in the nature of the observables. An essential step is to
construct the Hilbert space for a general Poisson manifold, which is in general highly nontrivial.
In order to postpone or rather circumvent difficult questions related to the representation theory,
we will simply choose to work within the framework of deformation quantization \cite{d-quant1,d-quant2}
which allows us to focus on the algebra itself. Later (in section 6) we will consider
a strict quantization with a Hilbert space.

M. Kontsevich proved \cite{kontsevich} that every finite-dimensional Poisson manifold $M$ admits
a canonical deformation quantization and the equivalence classes of Poisson manifolds modulo
diffeomorphisms can be naturally identified with the set of gauge equivalence classes of star products
on a smooth manifold. The existence of a star product on an arbitrary Poisson manifold follows
from the general formality theorem: The differential graded Lie algebra of Hochschild cochains defined
by polydifferential operators is quasi-isomorphic to the graded Lie algebra of polyvector fields.\footnote{A quasi-isomorphism is a morphism of (co)chain complexes such that the induced morphism of (co)homology
groups is an isomorphism.} Let $\mathcal{A}$ be an arbitrary unital associative algebra with
multiplication $\star$. The Hochschild $p$-cochains $C^p(\mathcal{A}, \mathcal{A}) \equiv \mathrm{Hom}(\mathcal{A}^{\otimes^p}, \mathcal{A})$ are the space of $p$-linear maps
$\mathcal{C}(f_1, \cdots, f_p)$ on $\mathcal{A}$ with values in $\mathcal{A}$ and
the coboundary operator $d_\star :C^p \to C^{p+1}$ is defined by \cite{kontsevich}
\begin{eqnarray}\label{hoch-d}
    (d_\star \mathcal{C}) (f_1, \cdots, f_{p+1}) &=& f_1 \star \mathcal{C} (f_2, \cdots, f_{p+1}) \nonumber \\
    && + \sum_{i=1}^p (-1)^i \mathcal{C} (f_1, \cdots, f_{i-1}, f_i \star f_{i+1}, f_{i+2}, \cdots, f_{p+1})
    \nonumber \\ && +(-)^{p+1} \mathcal{C} (f_1, \cdots, f_p) \star f_{p+1}
\end{eqnarray}
for $\mathcal{C} \in C^p(\mathcal{A}, \mathcal{A})$ and $f_1, \cdots, f_{p+1} \in \mathcal{A}$.
The Hochschild complex admits a bracket  $[-,-]_G : C^m (\mathcal{A}, \mathcal{A})
\otimes C^n (\mathcal{A}, \mathcal{A}) \to C^{m+n-1}(\mathcal{A}, \mathcal{A})$,
called the Gerstenhaber bracket or in abbreviation the $G$-bracket, and it is given by
\begin{equation}\label{g-bracket}
    [\mathcal{C}_1, \mathcal{C}_2]_G =  \mathcal{C}_1 \circ \mathcal{C}_2
    - (-)^{(m-1)(n-1)}\mathcal{C}_2 \circ \mathcal{C}_1
\end{equation}
where the composition $\circ$ for $\mathcal{C}_1 \in C^m$ and $\mathcal{C}_2 \in C^n$ is defined as
\begin{eqnarray}\label{hoch-comp}
    && (\mathcal{C}_1 \circ \mathcal{C}_2)(f_1, \cdots, f_{m+n-1}) \nonumber \\
    && = \sum_{i=1}^m (-1)^{(i-1)(n-1)}
    \mathcal{C}_1 (f_1, \cdots, f_{i-1}, \mathcal{C}_2 (f_i, \cdots, f_{i+n-1}), f_{i+n}, \cdots, f_{m+n-1}).
\end{eqnarray}
It is then straightforward to verify the graded Jacobi identity
\begin{equation}\label{gbra-jid}
 (-)^{(p_1-1)(p_3-1)} [[\mathcal{C}_1,\mathcal{C}_2]_G, \mathcal{C}_3]_G
 + (-)^{(p_1-1)(p_2-1)} [[\mathcal{C}_2,\mathcal{C}_3]_G, \mathcal{C}_1]_G
 + (-)^{(p_2-1)(p_3-1)} [[\mathcal{C}_3,\mathcal{C}_1]_G, \mathcal{C}_2]_G = 0
\end{equation}
for $\mathcal{C}_i \in C^{p_i}, \; i=1,2,3$.
Using the definition (\ref{g-bracket}) of the $G$-bracket, the action (\ref{hoch-d}) of the coboundary
operator $d_\star$ on a $p$-cochain $\mathcal{C} \in C^p$ can be compactly written as
\begin{equation}\label{hoch-dc}
  d_\star \mathcal{C} = -[\mathcal{C}, \star]_G
\end{equation}
where $\star \in C^2 = \mathrm{Hom}(\mathcal{A} \otimes \mathcal{A}, \mathcal{A})$ is the multiplication of functions
\begin{equation}\label{star-multi}
    \star(f_1, f_2) = f_1 \star f_2.
\end{equation}
Then it is easy to show that, for any $p$-cochain $\mathcal{C} \in C^p$,
\begin{equation}\label{nilp-d2}
 d_\star^2 \mathcal{C} = [\mathcal{C}, \star \circ \star]_G = \frac{1}{2}[\mathcal{C}, [\star, \star]_G]_G
\end{equation}
where we used the fact $\frac{1}{2} [\star, \star]_G = \star \circ \star \in C^3$.
Note that $[\star, \star]_G$ measures the associativity of the product $\star \in C^2$ and it
should identically vanish because the algebra $\mathcal{A}$ was assumed to be associative, i.e.,
\begin{equation}\label{associative}
\frac{1}{2}[\star, \star]_G (f,g,h) = (f \star g) \star h - f \star
(g \star h) = 0.
\end{equation}
Therefore the associativity of $\mathcal{A}$ implies that the
differential $d_\star: C^p \to C^{p+1}$ in (\ref{hoch-d}) is
nilpotent, i.e. $d_\star^2 = 0$ and the corresponding cohomology
\begin{equation}\label{h-cohomol}
    H^\bullet (\mathcal{A}, \mathcal{A}) = \mathrm{Ker}\; d_\star/
    \mathrm{Im}\; d_\star
\end{equation}
is called the Hochschild cohomology of cochain complex  $C^\bullet
(\mathcal{A}, \mathcal{A})$.

We first take $\mathcal{A} = C^\infty(M)$ to be the algebra of smooth
functions on a real manifold $M$. Then the Hochschild cochains of the
algebra $\mathcal{A} = C^\infty(M)$ are given by polydifferential
operators on $M$ denoted by $D_{\mathrm{poly}}(M)$.\footnote{An $n$-polydifferential operator in $D_{\mathrm{poly}}(M)$ is a multilinear map that acts on a tensor product of $n$ functions
and has degree $n-1$. In a local chart $\{y^\mu \}$, all elements of $D_{\mathrm{poly}}(M)$ look like
$$ f_1 \otimes \cdots \otimes f_n \mapsto \sum_{I_1, \cdots, I_n} C^{I_1 \cdots I_n} \partial_{I_1} f_1
\cdots \partial_{I_n} f_n$$ where the sum is finite and $I_k$ are multi-indices.}
There is another differential graded Lie algebra $T_{\mathrm{poly}}(M)$ we introduced in section 2.
That is the graded Lie algebra of polyvector fields on $M$
\begin{equation}\label{t-poly}
   T^k_{\mathrm{poly}}(M) = \Gamma(\Lambda^{k+1} TM), \qquad k \geq -1
\end{equation}
equipped with the SN bracket (\ref{sn-bracket}) and setting the differential $d_T \equiv 0$.
To prove the formality conjecture, Kontsevich showed \cite{kontsevich} that there exists an
$L_\infty$-quasi-isomorphism $U: T_{\mathrm{poly}}(M) \to D_{\mathrm{poly}}(M)$.
The $L_\infty$-morphism $U$ is a collection of skew-symmetric multilinear maps
$U_n: \otimes^n T_{\mathrm{poly}}(M) \to D^{m-1}_{\mathrm{poly}}(M)$ from tensor products
of $n$ $k_i$-vector fields to $m$-differential operators satisfying the formality
equation \cite{kontsevich,cf-sigma,manchon}.\footnote{The underlying complex for the $L_\infty$-morphisms
is shifted complexes. For a complex $C$, the shifted complex denoted by $C[1]$ means $C[1]^k = C^{k+1}$.}
The degree of the polydifferential operator matches with the overall degree of polyvector fields if
\begin{equation}\label{matching-condition}
m = 2-2n + \sum_{i=1}^n k_i.
\end{equation}
Since we will use the explicit formula for the formality equation later, we present it here
though it is rather complicated as well as not really inspirational:
\begin{eqnarray}\label{formality-eq}
    && \frac{1}{2}\sum_{I \sqcup J =(0,1, \cdots,n)}
    \epsilon(I,J) \widehat{Q}_2 \bigl( U_{|I|}(\alpha_{I}), U_{|J|}(\alpha_{J}) \bigr) \\
    && = \frac{1}{2} \sum_{i\neq j} \epsilon_\alpha (i,j,1, \cdots, \hat{i}, \cdots, \hat{j}, \cdots, n)
    U_{n-1} \bigl(Q_2(\alpha_i, \alpha_j), \alpha_1, \cdots, \hat{\alpha}_{i}, \cdots, \hat{\alpha}_{j},
    \cdots, \alpha_n \bigr). \nonumber
\end{eqnarray}
Here $\widehat{Q}_2(\Phi_1, \Phi_2) = (-1)^{d_1(d_2-1)} [\Phi_1, \Phi_2]_G$
where $d_i$ is the degree of the polydifferential operator $\Phi_i$, i.e.,
$d_i+1$ is the number of functions it is acting on and $Q_2(\alpha_1, \alpha_2) = (-1)^{d_1}
[\alpha_1, \alpha_2]_S$ where $d_i$ is the degree of the polyvector field $\alpha_i$.
Finally $|I|$ denotes the number of elements of multi-indices $I$ and $\epsilon(I,J)$
is an alternating sign depending on the number of transpositions of odd elements in the permutation
of $(1, \cdots, n)$ associated with the partition $(I,J)$.
As a special case, $U_0 \equiv \mu $ is defined to be the ordinary multiplication of functions:
\begin{equation}\label{formality-u0}
\mu(f_1 \otimes f_2) = f_1 f_2.
\end{equation}
It is then useful to introduce the Hochschild differential $d_H: D^{m-1}_{\mathrm{poly}}
\to D^{m}_{\mathrm{poly}}$ defined by
\begin{equation}\label{h-diff}
  d_H \Phi = - [\Phi, \mu]_G
\end{equation}
for $\Phi \in D^{m-1}_{\mathrm{poly}}$. Note that the Hochschild differential $d_H$ is a particular case
of the previous coboundary operator (\ref{hoch-d}) when the multiplication $\star$ is given by $\mu$.
It is easy to check that
\begin{equation}\label{associative-h}
  d_H^2 \Phi = \frac{1}{2} [\Phi, [\mu, \mu]_G ]_G = 0
\end{equation}
because $[\mu, \mu]_G$ measures the associativity of $C^\infty(M)$, i.e.,
\begin{equation}\label{ass-fgh}
   \frac{1}{2} [\mu, \mu]_G(f,g,h) = (fg)h-f(gh)=0.
\end{equation}
As another special case, in particular, $U_1$ is the natural map from a $k$-vector field
to a $k$-differential operator:
\begin{equation}\label{formality-u1}
    U_1(\alpha)(f_1, \cdots, f_k) = \alpha^{\mu_1 \cdots \mu_k} \partial_{\mu_1} f_1
    \cdots \partial_{\mu_k} f_k = \langle \alpha, df_1 \wedge \cdots \wedge df_k \rangle
\end{equation}
where $\alpha \in T^{k-1}_{\mathrm{poly}}(M)$ is defined by (\ref{poly-vector}).
Note that the formality condition (\ref{formality-eq}) implies that $ d_H U_1(\alpha) = - [U_1(\alpha), \mu]_G = 0$. In other words, $U_1(\alpha)$ is a derivation, i.e.,
\begin{eqnarray}\label{derivation-u1}
&& U_1(\alpha)(f_1, \cdots, f_{i-1}, f_i f_{i+1}, f_{i+2}, \cdots, f_{k+1})
= f_i U_1(\alpha)(f_1, \cdots, f_{i-1}, f_{i+1}, \cdots, f_{k+1}) \nonumber \\
&& \hspace{5cm} +  U_1(\alpha)(f_1, \cdots, f_i, f_{i+2}, \cdots, f_{k+1})  f_{i+1}
\end{eqnarray}
which is certainly satisfied by the representation (\ref{formality-u1}).

Now we will try to digest the meaning of the formality theorem in the context of
deformation quantization. Consider the following formal series \cite{cft,jsw2,behr-sykora}:
\begin{eqnarray}\label{series-star}
&& \star \equiv \sum_{n=0}^\infty \frac{\hbar^n}{n!} U_n(\theta, \cdots, \theta), \\
\label{series-vec}
&& \Phi(\alpha) \equiv \sum_{n=0}^\infty \frac{\hbar^n}{n!} U_{n+1}(\alpha, \theta, \cdots, \theta), \\
\label{series-fun}
&& \Psi(\alpha_1, \alpha_2) \equiv \sum_{n=0}^\infty \frac{\hbar^n}{n!} U_{n+2}(\alpha_1, \alpha_2,
\theta, \cdots, \theta),
\end{eqnarray}
where $\theta$ is a Poisson bivector and $\alpha, \alpha_1, \alpha_2$ are polyvector fields of some degrees
and we introduced a formal deformation parameter $\hbar$.
According to the matching condition (\ref{matching-condition}), $\star$ is a bidifferential operator
whereas $\Phi(\alpha)$ is a $k$-differential operator if $\alpha$ is a $k$-vector field and
$\Psi(\alpha_1, \alpha_2)$ is a $(k_1 + k_2 -2)$-differential operator if $\alpha_1$ and $\alpha_2$
are $k_1$- and $k_2$-vector fields, respectively. Since $\star$ is a bidifferential operator that maps
two functions to a function, this can be used to define a product
\begin{equation}\label{star-prod}
    f \star g = \sum_{n=0}^\infty \frac{\hbar^n}{n!} U_n(\theta, \cdots, \theta) (f,g)
    = fg + \hbar B_1(f,g) + \hbar^2 B_2(f,g) + \cdots,
\end{equation}
where $B_i$ are bidifferential operators.
From now on we will refer the multiplication in the Hochschild complex to the above $\star$-product.
We denote by $\mathcal{A}_\theta := C^\infty(M)[[\hbar]]$
a linear space of formal power series of the deformation parameter $\hbar$ with coefficients in $C^\infty(M)$,
which is defined by the deformation quantization map $\mathcal{Q}: C^\infty(M) \to \mathcal{A}_\theta$.

Let us represent the $\star$-product in eq. (\ref{series-star}) as
\begin{equation}\label{star-def}
    \star = \mu + \hbar \widehat{B}.
\end{equation}
Given an element $\widehat{B} \in D^{1}_{\mathrm{poly}}$, we can interpret $\mu + \hbar \widehat{B}$
as a deformation of the original product (\ref{formality-u0}).
One can apply the formality equation (\ref{formality-eq}) in each order of $\hbar^n$ to yield
\begin{equation}\label{formality-ass}
d_\star \star = - [\mu + \hbar \widehat{B}, \mu + \hbar
\widehat{B}]_G = \hbar^2 \Phi([\theta, \theta]_S).
\end{equation}
Therefore the formality theorem says \cite{kontsevich} that the formal multiplication $\star$
in eq. (\ref{star-def}) is associative, i.e. $[\star, \star]_G = 0$ if and only if
the bivector $\theta$ is a Poisson structure, i.e. $[\theta, \theta]_S=0$.
Since the original product $\mu$ is also associative, i.e. $[\mu,\mu]_G = 0$, the equation (\ref{formality-ass})
can be written as
\begin{equation}\label{mc-eq}
    d_H \widehat{B} - \frac{\hbar}{2} [\widehat{B}, \widehat{B}]_G = 0.
\end{equation}
Thus one can see that the associativity of the $\star$-product
(\ref{star-prod}) is equivalent to the Maurer-Cartan equation
(\ref{mc-eq}) for the element $\widehat{B}$ in the differential
graded Lie algebra $D_{\mathrm{poly}}(M)$. Two star products
$\widetilde{\star}$ and $\star$ are said to be equivalent if and
only if there exists a linear operator $D:
\mathcal{A}_{\widetilde{\theta}} \to \mathcal{A}_{\theta}$ of the
form\footnote{\label{star-iso}In passing from the commutative product to a NC
product, there is always an ordering problem and the explicit form
of the NC product depends on the ordering prescription. The notion
of equivalence between the products $\widetilde{\star}$ and $\star$
can be understood as an axiomatic and generalized notion of passing
from one to another ordering prescription. For example, the standard
ordered star product where one writes all momenta to the right of
coordinates is equivalent to the totally symmetrized Weyl ordered
star product. More generally, it was proven \cite{kontsevich,nt-star,bc-gutt} that the set of all star products
on a symplectic manifold $(M, \omega)$ up to the equivalence (\ref{star-equiv}) is classified by
the {\it formal} de Rham cohomology $H^2_{\mathrm{dR}}(M) \bigl[[\hbar]\bigr]$ and
there is a unique star product on $M$ for each element
$\omega + \hbar \alpha \bigl[[\hbar]\bigr]$ with $\alpha \bigl[[\hbar]\bigr] \in H^2_{\mathrm{dR}}(M) \bigl[[\hbar]\bigr]$.
In particular, there is only one equivalence class of symplectic star products on $\mathbb{R}^{2n}$.}
\begin{equation}\label{star-automorphism}
    Df = f + \sum_{n=1}^\infty \hbar^n D_n(f)
\end{equation}
such that
\begin{equation}\label{star-equiv}
    f \, \widetilde{\star} \; g = D^{-1} \bigl(Df \star Dg \bigr).
\end{equation}
The equivalence relation (\ref{star-equiv}) can be depicted by the commutativity of the diagram
\begin{equation} \label{star-diagram}
\begin{diagram}
\mathcal{A}_{\widetilde{\theta}} \times \mathcal{A}_{\widetilde{\theta}} &\rTo^{~~~~\widetilde{\star}~~~~} & \mathcal{A}_{\widetilde{\theta}} \\
\dTo^{D \times D} & &\dTo_{D}\\
\mathcal{A}_{\theta} \times \mathcal{A}_{\theta} &\rTo^{~~~~\star~~~~}
& \mathcal{A}_{\theta}
\end{diagram}
\end{equation}

When substituting the expansion (\ref{star-prod}) into the Maurer-Cartan equation (\ref{mc-eq}),
one can get the conditions for the coefficients $B_i$ in the lowest orders:
\begin{eqnarray}\label{mc-1}
    && d_H B_1 = 0, \\
    \label{mc-2}
    && d_H B_2 - \frac{1}{2}[B_1, B_1]_G = 0.
\end{eqnarray}
It may be instructive to explicitly check the above Maurer-Cartan equations by considering the following
expansion for the associativity of the $\star$-product (\ref{star-prod}):
\begin{equation}\label{star-asso}
(f \star g) \star h - f \star (g \star h) \equiv m_0 + \hbar m_1 + \hbar^2 m_2 + \cdots
\end{equation}
where
\begin{eqnarray} \label{hoch-exp1}
m_0 &=& (f g) h - f (g h) = \frac{1}{2} [\mu, \mu]_G (f, g, h),  \\
\label{hoch-exp2}
- m_1 &=& f B_1(g, h) - B_1(f g, h) + B_1(f, g h) - B_1(f, g) h = (d_H B_1) (f, g, h), \\
\label{hoch-exp3}
- m_2 &=& f B_2(g, h) - B_2(f g, h) + B_2(f, g h) - B_2(f, g) h - B_1 \bigl(B_1(f, g), h \bigr)
+ B_1 \bigl(f, B_1(g, h) \bigr) \nonumber \\
&=&  \Bigl( d_H B_2 - \frac{1}{2}[B_1, B_1]_G \Bigr) (f, g, h).
\end{eqnarray}
Since the associativity of the $\star$-product requires $m_i = 0$ for $\forall \; i=0,1,2, \cdots$,
we recover eqs. (\ref{mc-1}) and (\ref{mc-2}) at $i=1,2$, respectively.
Let us decompose the 2-cochain $B_1$ into the sum of the symmetric part and of the anti-symmetric part:
\begin{equation}\label{dec-2-cochain}
    B_1 = B_1^- + B_1^+, \qquad B_1^\pm (f, g) = \frac{1}{2} \bigl( B_1 (f, g) \pm  B_1(g, f) \bigr).
\end{equation}
One can solve the cocycle condition (\ref{mc-1}) by
\begin{equation}\label{2-cocycle}
    B_1 = B_1^H + d_H D_1
\end{equation}
where $B_1^H$ is a harmonic cochain in the second Hochschild cohomology group (\ref{h-cohomol}) (with the multiplication $\star = \mu$) and $D_1$ is an arbitrary differential operator.
Note that
\begin{equation}\label{exact-cochain}
 d_H D_1 (f,g) = fD_1(g) - D_1(fg) + D_1(f) g,
\end{equation}
so $d_H D_1 (f, g) = d_H D_1 (g, f)$. Therefore it is possible to take the symmetric part $B_1^+$
such that $B_1^+ = d_H D_1$ by choosing an appropriate differential operator $D_1$.
It is then easy to prove that $m_1=0$ in eq. (\ref{hoch-exp2})
can be solved if and only if $B_1^H = B_1^-$ and $B_1^H$ is a derivation of $C^\infty(M)$, i.e.,
\begin{equation}\label{hoch-ham}
    B_1^H (f, gh) = g B_1^H(f, h) + B_1^H(f, g) h.
\end{equation}
Thus, as we noticed in eq. (\ref{formality-u1}), the harmornic cocycle $B_1^-$ comes from a Poisson
bivector field $\theta$ on $M$:
\begin{equation}\label{hoch-theta}
    B_1^-(f,g) = \langle \theta, df \wedge dg \rangle =\{f, g\}_\theta, \qquad \theta \in \Gamma(\Lambda^2 TM).
\end{equation}
And the second Hochschild cocycle $B_1^-$ is invariant under the gauge transformation (\ref{star-equiv})
because eq. (\ref{star-equiv}) implies that
\begin{equation}\label{star-gauge}
    B_1'(f, g) = B_1(f, g) + d_H D_1 (f, g).
\end{equation}
In other words, the symmetric part $B_1^+$ in eq. (\ref{dec-2-cochain})
can be killed by the gauge transformation (\ref{star-equiv}). The
above argument in the first order approximation can be generalized
to all orders by solving (\ref{formality-ass}). Hence the formality
theorem asserts that the deformation of the commutative
multiplication $\mu$ on $C^\infty(M)$ is in bijection to deforming
the Poisson bracket on $C^\infty(M)$, so one can describe the
deformations of the Poisson algebra $\mathfrak{P} = (C^\infty(M),
\{-,-\}_\theta)$ as the elements of the Hochschild complex obeying
the formality equation (\ref{formality-eq}). For that reason we call
the $\star$-product (\ref{star-prod}) a formal quantization of the
Poisson algebra $\mathfrak{P}$. The simplest example of deformation
quantization is the Moyal star product defined by
\begin{equation}\label{moyal}
    (f \star g)(y) = e^{\frac{i}{2} \theta^{\mu\nu} \partial_\mu^x \partial_\nu^y} f(x) g(y)|_{x=y}
\end{equation}
which is a particular case of the Kontsevich's star product
(\ref{star-prod}) with a constant Poisson structure $\theta^{\mu\nu}
= \Bigl( \frac{1}{B} \Bigr)^{\mu\nu}$.

Given a Poisson structure $\theta$, i.e. $[\theta, \theta]_S=0$, the
second Hochschild cohomology (\ref{h-cohomol}) defines the
equivalence class of star multiplications if
\begin{equation}\label{star-coho}
    \widetilde{\star} = \star + d_\star \widehat{C}
\end{equation}
where $\widehat{C}$ is an arbitrary differential operator in
$D^0_{\mathrm{poly}}(M)$. Note that eq. (\ref{star-coho}) is an
infinitesimal version of eq. (\ref{star-equiv}) where $D=
\mathrm{id} + \widehat{C}$ is defined by eq. (\ref{star-automorphism}).\footnote{It may be remarked that
$D^0_{\mathrm{poly}}(M)$ is the set of differential operators, so the generators in Diff$(M)$
belong to $D^0_{\mathrm{poly}}(M)$. Therefore, if we change coordinates in star products,
we obtain a gauge equivalent star product. Cf. the Theorem 2.3 in Ref. \cite{kontsevich}.}
Two star products in the same equivalence class obeying eq. (\ref{star-equiv})
automatically satisfy the associativity condition
\begin{equation}\label{new-asso}
 [\widetilde{\star}, \widetilde{\star}]_G (f, g, h) = 2 D^{-1}
 \Bigl( (Df \star Dg ) \star Dh - Df \star (Dg \star Dh) \Bigr) = 0
\end{equation}
if the original $\star$-product is associative.

Suppose that there are two associative star products $\star'$ and $\star$. We may
consider the star product $\star'$ as a deformation of the star
product $\star$ like as eq. (\ref{star-def}), i.e.,
\begin{equation}\label{star-defor}
 \star' = \star + \hbar \widehat{B}
\end{equation}
but keeping the associativity, i.e., $[\star', \star']_G=0$. Since the initial $\star$-product is associative,
i.e., $[\star, \star]_G = 0$, the associativity condition of the new star
product $\star'$ can be written as the following Maurer-Cartan equation
\begin{equation}\label{asso-two}
    d_\star \widehat{B} - \frac{\hbar}{2} [\widehat{B}, \widehat{B}]_G = 0.
\end{equation}
Therefore, we may connect every associative products by the deformation $\widehat{B} \in D^1_{\mathrm{poly}}(M)$
which obeys the Maurer-Cartan equation (\ref{asso-two}). But they need not belong to the same equivalence class
or the same Hochschild cohomology $H^2 (\mathcal{A}, \mathcal{A})$ unless $\widehat{B} = d_\star \widehat{C}$.
In other words, the space of all associative algebras may be generated by the solutions of
the Maurer-Cartan equation (\ref{asso-two}).

We know that $\delta^\star_X \equiv \Phi(X)$ in eq. (\ref{series-vec}) is a linear differential operator
if $\alpha = X$ is a vector field in $\Gamma(TM)$. And $\widehat{f} \equiv \Phi(f)$ and $\Psi(X,Y)$
are functions in $\mathcal{A}_\theta$ if $f \in C^\infty(M)$ and $X, Y \in \Gamma(TM)$.
For a Poisson structure $\theta$, the formality condition (\ref{formality-eq}) leads
to the relations \cite{cft,cftp}
\begin{eqnarray}\label{star-der}
    && d_\star \Phi(X) = - [\Phi(X), \star]_G = \hbar \Phi(d_\theta X), \\
   \label{star-inder}
    && d_\star \widehat{f} = - [\Phi(f), \star]_G = \hbar \Phi(d_\theta f), \\
    \label{star-lie}
    &&  [\Phi(X), \Phi(Y)]_G + [\Psi(X, Y), \star]_G = \Phi([X,Y]_S) + \hbar
    \bigl(\Psi(d_\theta X, Y) + \Psi( X, d_\theta Y) \bigr),
\end{eqnarray}
where $d_\theta X = - \mathcal{L}_X \theta$ and $d_\theta f = X_f$ according
to eqs. (\ref{lie-sn}) and (\ref{ham-vec})
and $\Psi(d_\theta X, Y)$ and $\Psi( X, d_\theta Y)$ are differential operators
since $d_\theta X$ and $d_\theta Y$ are 2-vector fields.
Note that $d_\star \widehat{f} = \hbar \Phi(d_\theta f)$ is an inner derivation of the $\star$-product, i.e.,
\begin{equation} \label{star-inner}
 \bigl(d_\star \widehat{f}\bigr)(g) = (g \star \widehat{f} - \widehat{f} \star g)
    = - [\widehat{f}, g]_\star.
\end{equation}
A vector field $X$ that preserves the Poisson bracket, i.e., $d_\theta X = - \mathcal{L}_X \theta = 0$,
is called the Poisson vector field. Thus, it results from eq. (\ref{star-der}) that,
if $X$ is a Poisson vector field, then $\delta^\star_X$ is a derivation of the $\star$-product, viz.,
\begin{equation}\label{star-leibniz}
     [d_\star \Phi(X)](f, g) = -\delta^\star_X (f\star g) + f \star \delta^\star_X g
    + (\delta^\star_X f) \star g = 0.
\end{equation}
In this case, the relation (\ref{star-lie}) reads as
\begin{equation}\label{star-homo}
 [\delta^\star_X, \delta^\star_Y]_G - \delta^\star_{[X,Y]} = d_\star \Psi(X, Y)
\end{equation}
where $[X, Y]$ is the ordinary Lie bracket of vector fields. The result (\ref{star-homo}) illustrates that
the formality map $U: T_{\mathrm{poly}}(M) \to D_{\mathrm{poly}}(M)$ from the algebra of multivector fields
to the algebra of multidifferential operators fails to preserve the Lie algebra structure
but the difference between the two terms is an exact cochain in $D_{\mathrm{poly}}(M)$.
So the map $U$ induces an isomorphism between the cohomology groups of corresponding complexes.
This is exactly the role played by the $L_\infty$-morphism $U$.

Given a Poisson vector field $X$, $d_\theta X = 0$, one can solve eq. (\ref{star-der}) by
\begin{equation}\label{hoch-der}
   \Phi(X) =  \delta^\star_X + d_\star \widehat{f}
\end{equation}
where $\delta^\star_X$ is an element of the first Hochschild cohomology in eq. (\ref{h-cohomol}) and
a Hamiltonian vector field $d_\theta f$ is now mapped to the inner derivation $d_\star \widehat{f}
\equiv ad_{\widehat{f}}^\star = - [\widehat{f}, \, \cdot \,]_\star$.
Note that the first Poisson cohomology group $H^1_\theta (M)$ in
eq. (\ref{poisson-coho}) is the quotient of the space of Poisson vector
fields (i.e. vector fields $X$ such that $d_\theta X = - [X,
\theta]_S =0$) by the space of Hamiltonian vector fields (i.e. the
vector fields of the type $d_\theta f = - [f, \theta]_S = X_f$).
Therefore the first Hochschild cohomology of the deformation
quantization algebra $\mathcal{A}_\theta$ is isomorphic to the first
Poisson cohomology of the Poisson structure $\theta$.

It may be rewarding to check eqs. (\ref{star-der}) and (\ref{star-inder}) up to
next-to-leading order by considering the following expansions
\begin{eqnarray} \label{1st-def=0}
    && f \star g = fg + \hbar \theta^{\mu\nu} \partial_\mu f \partial_\nu g  + \cdots, \nonumber \\
    && \delta_X^\star = X + \hbar U_2(X, \theta) + \cdots, \\
    && \widehat{f} = f + \hbar U_2(f, \theta) + \cdots. \nonumber
\end{eqnarray}
It is straightforward to derive the result
\begin{equation}\label{exp-u2=0}
[d_\star \Phi(X)](f, g) = - \hbar \bigl(\mathcal{L}_X \theta + [U_2(X, \theta), \mu]_G \bigr)(f,g)
+ \mathcal{O} (\hbar^2).
\end{equation}
Hence one can see that $d_H U_2(X, \theta) = -[U_2(X, \theta), \mu]_G = 0$ for an arbitrary vector field  $X$
(not necessarily a Poisson vector field obeying $\mathcal{L}_X \theta = 0$) to satisfy the relation (\ref{star-der}). In other words, the differential operator $U_2(X, \theta)$ in eq. (\ref{series-vec})
is a derivation with respect to the ordinary product $\mu$.
Such a differential operator can be absorbed into the ordinary vector field $X$
in the leading term of eq. (\ref{series-vec}), which means that one can set $U_2(X, \theta)=0$.
Similarly eq. (\ref{star-inder}) requires that $\mu \bigl(U_2(f, \theta)\otimes g -
g \otimes U_2(f, \theta) \bigr) = 0$. That is, $U_2(f, \theta)$ is still a usual commutative function
at the first order in $\theta$, so it can be gauged away by redefining the commutative function $f$
which results in $U_2(f, \theta) = 0$. (See also the footnote 10 in Ref. \cite{jsw2}.)
This fact leads to an interesting consequence in emergent gravity which will be discussed later.

\section{Noncommutative gauge theory}

Suppose that there exists a line bundle $L$ on a Poisson manifold $(M, \theta)$ whose Poisson
bivector $\theta$ takes the Darboux-Weinstein frame (\ref{weinstein}).
And recall that U(1) gauge fields arise as the one-form connection $A$ of the line bundle $L$
on the Poisson manifold and they manifest themselves as a deformation of Poisson structure
as was shown in eq. (\ref{sol-t}). But we can use the exponential map (\ref{t-flow})
to define the Poisson algebra on the Darboux-Weinstein chart (\ref{weinstein}) again
even in the presence of U(1) gauge fields as was verified in eq. (\ref{equiv-poisson}).
Then Poisson gauge fields are incarnated as a coordinate representation
of the exponential map (\ref{cov-obj}) on the local chart. A NC gauge theory is basically defined by
quantizing the Poisson algebra on such Darboux-Weinstein charts (the right-hand side
of eq. (\ref{equiv-poisson})). The construction is a rather straightforward application of
the formality maps in eqs. (\ref{series-star})-(\ref{series-fun}).
We will briefly review some essential points to frame important applications to emergent gravity,
focusing only on the rank one (Abelian) case.

So far we have examined how the deformation quantization of any Poisson manifold $(M, \theta)$
can be derived from the formality theorem which stipulates the existence of
an $L_\infty$-quasi-isomorphism $U: T_{\mathrm{poly}}(M) \to D_{\mathrm{poly}}(M)$
from the differential graded Lie algebra of polyvector fields on $M$ with vanishing differential
on the SN bracket (\ref{sn-bracket}) into the differential graded Lie algebra of polydifferential
operators on $M$ with the Hochschild differential $d_\star$ and the G-bracket (\ref{g-bracket}).
Since Poisson gauge fields are defined by the exponential map (\ref{cov-obj}) and they are subject
to the Poisson bracket defined by the right-hand side of eq. (\ref{equiv-poisson}),
we can try to quantize the Poisson algebra $(M, \{-,-\}_\theta)$ to define a NC gauge theory
on the Poisson manifold $(M, \theta)$ \cite{jsw1,jsw2}.
The symplectic gauge fields and corresponding NC gauge fields will be a particular case in which
the Poisson structure is nondegenerate. Using the anchor map (\ref{anchor}),
one can always associate the U(1) gauge field $A$ with a vector field $A_\theta$ defined
by eq. (\ref{sn-anchor}). Then the formality map (\ref{series-vec}) maps the vector field $A_\theta$
to a differential operator defined by
\begin{equation} \label{formality-a}
A_\star = \sum_{n=0}^\infty \frac{\hbar^n}{n!} U_{n+1} (A_\theta, \theta, \cdots, \theta).
\end{equation}
One can also apply the formula (\ref{star-der}) to yield the bidifferential operator
$d_\star A_\star = \hbar F_\star$ corresponding to the U(1) field strength $F=dA$:
\begin{equation} \label{formality-f}
F_\star = \sum_{n=0}^\infty \frac{\hbar^{n}}{n!} U_{n+1} (F_\theta, \theta, \cdots, \theta)
\end{equation}
where $F_\theta = d_\theta A_\theta$ is given by eq. (\ref{bivec-f}).
In section 2, we have introduced the Poisson gauge fields $\mathfrak{A}_A$ in eq. (\ref{diff-rho})
through the Moser flow (\ref{t-flow}). Similarly the corresponding NC gauge fields are defined
by introducing a $t$-dependent star product
\begin{equation} \label{t-dep-star}
\star_t = \sum_{n=0}^\infty \frac{\hbar^n}{n!} U_{n} (\theta_t, \cdots, \theta_t)
\end{equation}
and a {\it quantum} evolution equation\footnote{Since the G-bracket $[A_{\star_t}, {\star_t}]_G$ is
a quantum generalization of the classical Lie bracket $[A_{\theta_t}, \theta_t]$, we may define
a quantum Lie derivative with respect to a differential operator $A_{\star_t}$ by
$\widehat{\mathcal{L}}_{A_{\star_t}} \mathfrak{S} \equiv [A_{\star_t}, \mathfrak{S}]_G$
for $\mathfrak{S} \in D_{\mathrm{poly}}(M)$.
Then the quantum evolution equaiton (\ref{q-evolution}) takes a suggestive form
$\partial_t \star_t = \hbar \widehat{\mathcal{L}}_{A_{\star_t}} \star_t$.}
\begin{equation}\label{q-evolution}
    \partial_t \star_t = - \hbar F_{\star_t}
\end{equation}
where $F_{\star_t} = d_{\star_t} A_{\star_t} = - [A_{\star_t}, {\star_t}]_G$.
First it may be instructive to consider an infinitesimal deformation
of the star product (\ref{t-dep-star}) by taking the limit $t = \epsilon \to 0$.
It is easy to show that the result is given by eq. (\ref{star-defor}) with $\widehat{B}
= - F_\star$. Therefore the bidifferential operator (\ref{formality-f}) has to
satisfy the Maurer-Cartan equation
\begin{equation}\label{mc-bidiff}
    d_\star F_\star + \frac{\hbar}{2}[F_\star, F_\star]_G = 0
\end{equation}
in order for the new star product $\star' := \star_\epsilon$ to preserve the associativity.
The evolution equation (\ref{q-evolution}) is a quantum version of the classical evolution (\ref{t-evolution})
and indeed it is derived from eq. (\ref{t-dep-star}) using eqs. (\ref{t-evol-veca}) and (\ref{formality-f}).
Since the quantum $t$-evolution is generated by the differential operator $A_{\star_t}$,
it can be integrated to a {\it quantum} Moser flow \cite{jsw1,jsw2}
\begin{equation}\label{q-moser}
    \mathcal{D}_A = \exp(\hbar A_{\star_t} + \partial_t) \exp(-\partial_t)|_{t=0}
\end{equation}
which relates two star products $\star' = \star_1$ and $\star =
\star_0$ such that $\mathcal{D}_A (\star') = \star$.

The quantum flow $\mathcal{D}_A$ plays an important role of covariantizing map $f \mapsto
\mathcal{D}_A f := f + f_A$ which maps a function $f \in \mathcal{A}_\theta$
to a covariant function $\mathcal{D}_A f \in \mathcal{A}_\theta$ that transforms under
NC gauge transformations by conjugation
\begin{equation}\label{ncg-tr}
\mathcal{D}_A f  \mapsto \Lambda \star \mathcal{D}_A f \star \Lambda^{-1}.
\end{equation}
Note that the covariance map $\mathcal{D}_A$ depends on gauge fields as eq. (\ref{q-moser}) clearly indicates.
And, as we remarked before, it is also defined on local patches of a Poisson manifold where the local
gauge field $A$ is defined. But it can be globalized by gluing local patches together using NC gauge (and coordinate) transformations between local patches \cite{jsw-ncl}. It was shown \cite{jsw1,jsw2} that
the quantum Moser flow (\ref{q-moser}) is defined as a quantization of the classical Moser flow $\rho_A^*$, i.e., $\mathcal{D}_A = D \circ \rho_A^*$. With this property,
the local covariance map $\mathcal{D}_A$ can be used to define a new star product $\star'$ via
\begin{equation}\label{cov-star-map}
     f \star' g  = \mathcal{D}^{-1}_A \bigl( \mathcal{D}_A f  \star \mathcal{D}_A g \bigr).
\end{equation}
Although the new star product $\star'$ depends on gauge fields, they appear only via the gauge invariant
field strength $F = dA$, so it can be globally defined (after a globalization \`a la \cite{cft}).
Nevertheless the equivalence relation (\ref{cov-star-map}) between two star products $\star'$ and $\star$
holds locally because it is involved with the locally defined covariance map.
Globally the star products $\star'$ and $\star$ are in general
neither gauge equivalent nor in the same cohomology class.
Instead it was shown in \cite{jsw-ncl,buba-me} that two star products are Morita equivalent
if and only if they are, modulo diffeomorphisms, related by the action of an element $F \in \mathrm{Pic}
\bigl(C^\infty(M) \bigr) \cong H^2 (M, \mathbb{Z})$, i.e., an element of equivalence classes of
a line bundle $L$ over $M$. The closed two-form $F$ representing the first Chern
class $c_1(L)$ of the line bundle $L \to M$ acts on the Poisson
structure $\theta$ as eq. (\ref{sol-t}) to generate a new Poisson structure $\theta_1 = \Theta$.
Therefore the Morita equivalent star products $\star'$ and $\star$ in the local gauge
equivalence (\ref{cov-star-map}) are related by the action of a line bundle $L$, i.e.,
\begin{equation}\label{line-star}
    L: [\star \,] \to [\star'  \,]
\end{equation}
and the invertible covariantizing map (\ref{q-moser}) can be considered as a quantum lift
of the exponential map (\ref{t-flow}).

We exactly mirror the classical case (\ref{diff-rho}) to define NC gauge fields
\begin{equation}\label{nc-gauge}
   \mathcal{D}_A = \mathrm{id} + \widehat{\mathcal{A}}_A.
\end{equation}
The NC gauge field $\widehat{\mathcal{A}}_A = \mathcal{D}_A - \mathrm{id}$ is a local 1-cochain in $D^0_{\mathrm{poly}}(M)$, i.e., a formal differential operator depending on U(1) gauge fields and
transforms under the NC gauge transformation (\ref{ncg-tr}) as
\begin{equation}\label{ncg-fa}
    f_A \mapsto f_A' = \Lambda \star  f_A  \star \Lambda^{-1} + \Lambda \star [f, \Lambda^{-1}]_\star
\end{equation}
where $f_A \equiv \widehat{\mathcal{A}}_A(f)$. The star product
(\ref{t-dep-star}) describes a formal associative deformation of the
Poisson structure $\theta_t$ defined by eq. (\ref{sol-t}).
The star product $\star'$ is a formal deformation of the original star product $\star$
in the sense of eq. (\ref {star-defor}) if an element $\widehat{B} = \frac{1}{\hbar} (\star' - \star)
\in D^1_{\mathrm{poly}}(M)$ obeys the Maurer-Cartan equation (\ref{asso-two}).
Note that in terms of NC gauge fields in eq. (\ref{nc-gauge}), the gauge equivalence (\ref{cov-star-map})
can be written as the form
\begin{equation}\label{gequiv-f}
    \mathcal{D}_A \circ (\star' - \star) = d_\star \widehat{\mathcal{A}}_A +
    \widehat{\mathcal{A}}_A \star \widehat{\mathcal{A}}_A \equiv \mathbb{F}_A.
\end{equation}
Therefore $\star'$ is gauge equivalent to $\star$ provided $\widehat{B} = \frac{1}{\hbar} \mathcal{D}^{-1}_A
\circ \mathbb{F}_A \in D^1_{\mathrm{poly}}(M)$. Then the Morita equivalence between the formal
deformations of Poisson structures $\theta_1 = \Theta$ and $\theta_0 = \theta$ means that
the Maurer-Cartan element $\widehat{B}$ describes an orbit of the action (\ref{line-star})
of the formal diffeomorphism $\mathcal{D}_A$.

All quantities in deformation quantization are defined by formal power series of
a deformation parameter, e.g., typically $\hbar$. Hence it is necessary to keep track
of $\hbar$ to control the expansion of the power series. We are applying this expansion
to the formal series in eqs. (\ref{series-star})-(\ref{series-fun}) by simply taking the replacement
$\theta \to \hbar \theta$ in the formality maps.
Then the classical (or commutative) limit corresponds to the limit $\hbar \to 0$.
In this classical limit, the covariance map $\mathcal{D}_A = D \circ \rho_A^*$ in eq. (\ref{q-moser})
as well as the equivalence map $D$ in eq. (\ref{star-automorphism})
reduce to the identity operators, so the Moser flow $\rho_A^*$ in eq. (\ref{t-flow}) must also become
an identity operator in the limit. But it is not obvious to see how the exponential map (\ref{t-flow})
reduces to the identity operator in the limit $\hbar \to 0$.
In order to cure this situation, it may be necessary to introduce a formal vector field $\mathbf{X} \in \Gamma(TM)\bigl[[\hbar]\bigr]$ and a formal Poisson tensor
$\vartheta \in \Gamma(\Lambda^2 TM)\bigl[[\hbar]\bigr]$:
\begin{eqnarray} \label{formal-x}
\mathbf{X} &=& \hbar \widetilde{X}_1 + \hbar^2 \widetilde{X}_2 + \cdots := \hbar X,  \\
\label{formal-t}
\vartheta &=& \hbar \widetilde{\theta}_1 + \hbar^2 \widetilde{\theta}_2 + \cdots := \hbar \theta.
\end{eqnarray}
This can be simply implemented by replacing $\theta$ by $\hbar \theta$
and, accordingly, $A_\theta \to \hbar A_\theta$, for all formulas in section 2.
Important changes, for example, are given by
\begin{eqnarray} \label{h-change1}
&& (\ref{t-evol-veca}) \quad \to \quad \partial_t \theta_t = - \hbar F_{\theta_t}, \\
\label{h-change2}
&& (\ref{sol-t}) \quad \to \quad \theta_t = \theta \frac{1}{1+ t \hbar F\theta}, \\
\label{h-change3}
&& (\ref{t-flow}) \quad \to \quad
    \rho_A^* = \exp(\hbar A_{\theta_t} + \partial_t)\exp(-\partial_t)|_{t=0}.
\end{eqnarray}
Now the Moser flow (\ref{h-change3}) shows the desired behavior, $\rho_A^* \to \mathrm{id}$,
in the classical limit $\hbar \to 0$.\footnote{If we implemented the $\hbar$-expansions
(\ref{formal-x}) and (\ref{formal-t}) in the formality maps (\ref{formality-a})
and (\ref{formality-f}) (by the replacements $A_\theta \to \hbar A_{\theta}$ and
$F_\theta \to \hbar^2 F_{\theta}$) from the outset,
the formulas (\ref{q-evolution}) and (\ref{q-moser}) did not get the $\hbar$-factor.
But the differential operator $A_{\star_t}$ would instead start from an $\mathcal{O}(\hbar)$-term, 
so the classical limit of $\mathcal{D}_A$ in eq. (\ref{q-moser}) will be same.}
In particular, the new formula (\ref{h-change2}) gives us the formal relation
\begin{equation}\label{new-formal}
    \Theta = \theta \frac{1}{1+ \hbar F\theta}
\end{equation}
between Poisson structures $\theta_1 = \Theta$ and $\theta_0 = \theta$,
which suggests an attractive picture.\footnote{One may understand the formal power series (\ref{formal-t})
as the deformation in terms of $U(1)$ gauge fields which is akin to eq. (\ref{new-formal}), i.e.,
$\theta = \widetilde{\theta}_1 \frac{1}{1+ \hbar \widetilde{f} \widetilde{\theta}_1}$ where
$\widetilde{f} = d \widetilde{a}$. Substituting this expression into eq. (\ref{new-formal}) yields the result
$\Theta = \widetilde{\theta}_1 \frac{1}{1+ \hbar (F+ \widetilde{f})\widetilde{\theta}_1}$.
Hence one may attribute the quantum $\hbar$-corrections in eq. (\ref{new-formal}) to the deformation
due to $U(1)$ gauge fields by a simple shift $F \to F + \widetilde{f}$.}
In the classical limit, there is a primitive Poisson structure
$\theta$ and other Poisson structures can be obtained by deforming the primitive Poisson structure
$\theta$ in terms of line bundle $L \to M$ such that the curvature $F=dA$ acts on the Poisson structure.
Then the formality map $U: \Theta \mapsto \star'$ in eq. (\ref{t-dep-star})
gives us the star product $\star'$ locally equivalent to the star product $\star$ of the original
Poisson structure $\theta = \rho_A^* (\Theta)$ which realizes the action (\ref{line-star})
of the line bundle $L$. Therefore it is not necessary to consider the set of all Poisson structures
in the classical limit. Instead, without loss of generality, we can assume the primitive Poisson
structure $\theta$ to be in the Darboux-Weinstein frame (\ref{weinstein}) and then deform it by turning
on U(1) gauge fields to generate the set of all possible Poisson structures with a fixed first order $\widetilde{\theta}_1$. This approach provides a more unified description
of formal deformations of a given Poisson structure $\widetilde{\theta}_1$.

The equivalence classes for the deformation quantization of a Poisson manifold $M$ are characterized
by geometric properties of the underlying manifold $M$. For example, for a symplectic manifold $M$,
they are parameterized by the second de Rham cohomology space $H^2(M, \mathbb{R})$.
See the footnote \ref{star-iso}. Now we are considering a line bundle $L \to M$
on a fixed Poisson manifold $(M, \theta)$.
Since the line bundle $L$ acts on the Poisson structure as eq. (\ref{new-formal}), one can define
NC U(1) gauge theory as the deformation quantization of the (dynamical) Poisson manifold $(M, \Theta)$.
The equivalence relation (\ref{cov-star-map}) then implies that the deformation quantization $\star'$
of the dynamical Poisson manifold $(M, \Theta)$ is Morita-equivalent to the deformation quantization
$\star$ of the base manifold $(M, \theta)$ and the NC U(1) gauge theory describes
their equivalent categories of modules. Thus, given two star products $\star'$ and $\star$
in the Morita equivalent class, we can always associate a NC U(1) gauge theory to realize
the action (\ref{line-star}). For example, the deformation quantization of any symplectic manifold
is described by a NC U(1) gauge theory where the system can be described
by the Moyal star product (\ref{moyal}). But this description is local in nature because
it is linked to a specific choice of coordinates known as the Darboux coordinates.

The topology of a line bundle $L$ on commutative $\mathbb{R}^{2n}$ is trivial and this fact
leads to the conclusion in the footnote \ref{star-iso}. But it becomes nontrivial on a symplectic
or NC $\mathbb{R}^{2n}$ because the NC space admits the existence of NC U(1)
instantons \cite{ns-inst,nek-cmp03}. Furthermore it was shown in \cite{lry2,lry3} that the singularities
of U(1) instantons on commutative $\mathbb{R}^{4}$ are blown up to noncontractible two cycles
after turning on the noncommutativity of the space $\mathbb{R}^{4}$, so incorporating the backreaction
of NC U(1) instantons brings about a topology change of $\mathbb{R}^{4}$ in the context
of emergent gravity. As a result, the first Chern class $c_1(L)$ of the line bundle $L$ supported
on those two cycles is nonzero. Therefore it may be necessary to also include singular line bundles of
U(1) instantons in the classification of star products on $\mathbb{R}^{2n}$ because
the singularity of U(1) instantons can be resolved by NC deformations.\footnote{We thank
Stefan Waldmann for helpful discussions related to this aspect.}
That is, the Morita equivalence of star products may be more subtle due to the topology change of
an underlying manifold triggered by NC U(1) instantons.
This means that there exists a nontrivial class of star products if the line bundle $L \to M = \mathbb{R}^{2n}$
describes (generalized) U(1) instantons (for instance, Hermitian U(1) instantons
obeying the Donaldson-Uhlenbeck-Yau equations in six dimensions) whose singularities at a finite number
of points are blown up to two cycles.
In the end the classification of star products can be interpreted as the (Morita) equivalence classes
of NC U(1) gauge theories which are parameterized by the first Chern class $c_1(L)$ of the line bundle $L$
replacing the second de Rham cohomology space $H^2(M, \mathbb{R})$ for Poisson manifolds.

Let us recapitulate how we got the NC gauge field (\ref{nc-gauge}). We started with a U(1) gauge theory
defined on a Poisson manifold $(M, \theta)$. The U(1) gauge field $A$ is mapped via
the anchor map (\ref{anchor}) to a general vector field $A_\theta$ with the bivector field $F_\theta
= d_\theta A_\theta$ in $T_{\mathrm{poly}} (M)$. The vector field $A_\theta$ is then mapped
via the formality map (\ref{formality-a}) to an arbitrary differential operator
$A_\star$ in $D_{\mathrm{poly}} (M)$ with the field strength $F_\star = \frac{1}{\hbar} d_\star A_\star$.
The whole mapping is depicted by the diagram:
\begin{equation} \label{quantum-anchor}
\begin{diagram}
A &\rTo^{~~~~\rho~~~~} & A_\theta &\rTo^{~~~~U~~~~} & A_\star \\
\dTo^{d} & &\dTo^{d_\theta} &  & \dTo_{d_\star} \\
F &\rTo^{~~~~\rho~~~~} & F_\theta &\rTo^{~~~~U~~~~} & F_\star
\end{diagram}
\end{equation}
The chain of maps in eq. (\ref{quantum-anchor}) is in general defined only locally because closed two-forms
are only locally exact, i.e., $F=dA$. In this correspondence the role of U(1) gauge fields can be understood
as a deformation of the Poisson structure $\theta$ of a base manifold $(M, \theta)$ which can be described by
the evolution equation (\ref{h-change1}) in $T_{\mathrm{poly}} (M)$.
According to the formality map (\ref{t-dep-star}), the classical evolution equation in $T_{\mathrm{poly}} (M)$
can be lifted to the quantum evolution equation (\ref{q-evolution}) in $D_{\mathrm{poly}} (M)$.
The NC U(1) gauge field $\widehat{\mathcal{A}}_A$ in eq. (\ref{nc-gauge}) is defined through the integral
curve (\ref{q-moser}) of the quantum evolution equation.
Since $A$ is a U(1) gauge field whose gauge transformation is given by $A + d\lambda$,
one can also consider the corresponding mapping (\ref{quantum-anchor}) after the gauge transformation:
\begin{equation}\label{quantum-gt-anchor}
A + d \lambda \xrightarrow{~\rho~} A_\theta + d_\theta \lambda \xrightarrow{U}
A_\star + \frac{1}{\hbar} d_\star \widehat{\lambda}
\end{equation}
where the relation (\ref{star-inder}) was used. Thus one can consider a quantum Moser flow
generated by the gauge-transformed differential operator $A_\star + \frac{1}{\hbar}
d_\star \widehat{\lambda}$ \cite{jsw2}:\footnote{In order to derive (\ref{gauge-q-moser}), it may be
necessary to use the Leibniz rule of the Hochschild differential (\ref{hoch-dc}),
$d_\star (\mathcal{C}_1 \circ \mathcal{C}_2) = (-1)^{n-1}
(d_\star \mathcal{C}_1) \circ \mathcal{C}_2 + \mathcal{C}_1 \circ (d_\star \mathcal{C}_2)$,
for $\mathcal{C}_1 \in C^m$ and $\mathcal{C}_2 \in C^n$ and $[\hbar A_{\star_t} + \partial_t,
d_{\star_t} \widehat{\lambda}]_G = d_{\star_t} [\bigl(\hbar A_{\star_t} + \partial_t \bigr)
(\widehat{\lambda})]$ where $[\partial_t, d_{\star_t} \widehat{\lambda}]_G (f)
= \partial_t \bigl([f, \widehat{\lambda}]_{\star_t} \bigr)$.}
\begin{eqnarray} \label{gauge-q-moser}
\mathcal{D}_{A+d\lambda} &=& \exp \bigl(\hbar A_{\star_t} + d_{\star_t} \widehat{\lambda}
+ \partial_t \bigr) \exp(-\partial_t)|_{t=0} \nonumber \\
&=& \bigl(\mathrm{id} + d_{\star} \widehat{\Lambda} \bigr) \circ \mathcal{D}_A +
  \mathcal{O} (\widehat{\lambda}^2)
\end{eqnarray}
where $\widehat{\Lambda}(\lambda, A)$ is a quantum version of the gauge parameter
(\ref{lambda-tilde}) and is given by
\begin{equation}\label{q-gauge-para}
   \widehat{\Lambda}(\lambda, A) = \sum_{n=0}^\infty \frac{(\hbar A_{\star_t} + \partial_t)^n
   (\widehat{\lambda})}{(n+1)!}|_{t=0}.
\end{equation}
In terms of NC gauge fields, the NC gauge transformation (\ref{gauge-q-moser}) reads as
\begin{equation}\label{q-sw-map}
   \widehat{ \mathcal{A}}_{A + d\lambda} = \widehat{\mathcal{A}}_A +
    \bigl( d_\star \widehat{\Lambda} + [\widehat{\mathcal{A}}_A, \widehat{\Lambda}]_\star \bigr)
\end{equation}
or
\begin{equation}\label{q-sw-map1}
   \widehat{ \mathcal{A}}_{A + \delta_\lambda A} = \widehat{\mathcal{A}}_A +
    \widehat{\delta}_{\widehat{\Lambda}} \widehat{\mathcal{A}}_A,
\end{equation}
where $\delta_\lambda A =  d\lambda$ and $\widehat{\delta}_{\widehat{\Lambda}} \widehat{\mathcal{A}}_A
\equiv \widehat{\mathcal{D}}_\star {\widehat{\Lambda}} = d_\star \widehat{\Lambda}
+ [\widehat{\mathcal{A}}_A, \widehat{\Lambda}]_\star$.
Note that the right-hand side of eq. (\ref{q-sw-map}) is an infinitesimal version of the finite NC gauge
transformation (\ref{ncg-fa}) with $\Xi \approx 1 - \widehat{\Lambda}$.
The gauge equivalence relation (\ref{q-sw-map1}) constitutes the SW map from commutative U(1) gauge
fields to NC U(1) gauge fields, which is a quantum version of (semi-)classical SW map (\ref{sw-equiv}).
The SW map can be depicted by the commutativity of the following diagram:
\begin{equation} \label{sw-map}
\begin{diagram}
A &\rTo^{SW} & \widehat{\mathcal{A}}_A \\
\dTo^{\delta_\lambda} & &\dTo_{\widehat{\delta}_{\widehat{\Lambda}}}\\
 A + \delta_\lambda A &\rTo^{~~~SW~~~}
& \widehat{\mathcal{A}}_A + \widehat{\delta}_{\widehat{\Lambda}} \widehat{\mathcal{A}}_A
\end{diagram}
\end{equation}
It is straightforward to show that the NC gauge transformations in eq. (\ref{q-sw-map1}) form
a Lie algebra under the $\star$-product, i.e.,
\begin{equation}\label{star-lie-gtr}
 \bigl[\widehat{\delta}_{\widehat{\Lambda}_1}, \widehat{\delta}_{\widehat{\Lambda}_2} \bigr]
 = \widehat{\delta}_{{[\widehat{\Lambda}_1, \widehat{\Lambda}_2]_\star}}
\end{equation}
which of course recovers (\ref{lie-poisson}) in the (semi-)classical limit.

In order to define the field strength of NC gauge fields, it is useful to introduce
the Chevalley-Eilenberg (CE) complex $(C^\bullet(\mathcal{A}, \mathcal{A}), d_{CE} \equiv \widehat{d})$
and a cup product $\circledast$. The CE complex is defined as follows \cite{cheeil}.
A $p$-cochain $C^p = \mathrm{Hom} (\mathcal{A}^{\wedge^p}, \mathcal{A})$ is the space of
$p$-linear skew-symmetric maps with values in an associative algebra $\mathcal{A}$
and the differential $\widehat{d}: C^p \to C^{p+1}$ is defined on homogeneous elements by
\begin{eqnarray}\label{ce-diff}
   2\hbar (\widehat{d} \mathcal{C})(f_1, \cdots, f_{p+1}) &=& \sum^{p+1}_{i=1} (-1)^{i+1}
    \bigl[f_i, \mathcal{C}(f_1, \cdots, \widehat{f}_i, \cdots, f_{p+1}) \bigr]_\star \nonumber \\
    && + \sum_{i<j} (-1)^{i+j}  \mathcal{C}\bigl( [f_i, f_j]_\star, f_1, \cdots, \widehat{f}_i, \cdots,
    \widehat{f}_j, \cdots, f_{p+1} \bigr)
\end{eqnarray}
where the hat stands for an omitted symbol. It is straightforward to prove that $\widehat{d}^2=0$,
using the Jacobi identity of the associative $\star$-algebra $\mathcal{A}$. It may be instructive
to explicitly check that $(\widehat{d}^2 \mathcal{C})(f_1, f_2, f_3) = 0$.
The cup product $\circledast: C^m \times C^n \to C^{m+n}$ is defined by
\begin{eqnarray}\label{cup-prod}
\hbar (\mathcal{C}_1 \circledast \mathcal{C}_2) (f_1, \cdots, f_{m+n})
&=& \frac{1}{(m+n)!}  \sum_{\sigma \in \mathfrak{S}_{m+n}}
    \epsilon(\sigma) \mathcal{C}_1 (f_{\sigma_1}, \cdots, f_{\sigma_m}) \star
    \mathcal{C}_2 (f_{\sigma_{m+1}}, \cdots, f_{\sigma_{m+n}}) \nonumber \\
&\equiv & \hbar (\mathcal{C}_1 \circledast \mathcal{C}_2) (f_1 \wedge \cdots \wedge f_{m+n})
\end{eqnarray}
where we introduced the symmetrization map defined by
\begin{equation}\label{symm-map}
    (f_1, \cdots, f_k) \mapsto (f_1 \wedge \cdots \wedge f_k) = \frac{1}{k!} \sum_{\sigma \in \mathfrak{S}_k}
    \epsilon(\sigma) (f_{\sigma_1}, \cdots, f_{\sigma_k})
\end{equation}
and $\epsilon(\sigma)$ is an alternating sign depending on the number of transpositions in the permutation
$\sigma \in \mathfrak{S}_k$ of $k$ elements.

We define the field strength of NC gauge fields evaluated on two functions $f, g$ by lifting
the case (\ref{curvature-poisson}) of Poisson gauge fields to the NC version:
\begin{eqnarray}\label{nc-field-fg}
    \widehat{\mathfrak{F}}_A (f, g) &=& \frac{1}{2\hbar} \Bigl([\mathcal{D}_A(f), \mathcal{D}_A(g)]_\star
    - \mathcal{D}_A \bigl([f, g]_\star \bigr)  \Bigr) =
    \Bigl(\mathcal{D}_A \circledast \mathcal{D}_A - \frac{1}{\hbar} \mathcal{D}_A \circ \star \Bigr)
    (f \wedge g) \nonumber \\
    &=& \widehat{\mathfrak{F}}_A (f \wedge g)
\end{eqnarray}
where $\widehat{\mathfrak{F}}_A \equiv \widehat{d}\widehat{\mathcal{A}}_A
+ \widehat{\mathcal{A}}_A \wedge \widehat{\mathcal{A}}_A$ and $(\mathcal{D}_A \circ \star) (f \wedge g)
= \frac{1}{2}\mathcal{D}_A ( [f, g]_\star)$.\footnote{\label{p-convention}We combined the definition
of the cup product $\circledast: C^1 \times C^1 \to C^{2}$ and the skew-symmetric property of $f \wedge g$
to define the wedge product $(\widehat{\mathcal{A}}_A \circledast \widehat{\mathcal{A}}_A ) (f \wedge g)
:= (\widehat{\mathcal{A}}_A \wedge \widehat{\mathcal{A}}_A)(f \wedge g)$.
The $\hbar$ factor in eqs. (\ref{ce-diff}) and (\ref{cup-prod}) and the factor 2 in eq. (\ref{ce-diff})
are originated from the fact that $[f, g]_\star = 2 \hbar \{f, g \}_\theta + \mathcal{O}(\hbar^3)$.
Furthermore it may be necessary to replace the formal deformation
parameter $\hbar$ to $\frac{i\hbar}{2}$ to match with the physicist's convention such as eq. (\ref{moyal}).}
Note that the CE-differential $\widehat{d}$ in (\ref{ce-diff}) is an antisymmetric version of
the Hochschild differential $d_\star$ in (\ref{hoch-d}), e.g.,
$\hbar (\widehat{d} \widehat{\mathcal{A}}_A) (f,g)
= (d_\star \widehat{\mathcal{A}}_A) (f \wedge g)$ and $2 \hbar (\widehat{d} \,
\widehat{\mathfrak{F}}_A) (f, g, h) = 3 (d_\star \widehat{\mathfrak{F}}_A) (f \wedge g \wedge h)$.
It leads to the relation that the field strength (\ref{nc-field-fg}) is the antisymmetrized version
of the Hochschild field strength (\ref{gequiv-f}), i.e.,
\begin{equation}\label{ncf-hfs}
    \widehat{\mathfrak{F}}_A (f, g) = \frac{1}{\hbar}\mathbb{F}_A (f \wedge g).
\end{equation}
Combining the field strength (\ref{nc-field-fg})
with the equivalence relation (\ref{cov-star-map}) which reads as $\widehat{\mathfrak{F}}_A (f, g)
= \frac{1}{2\hbar} \mathcal{D}_A([f, g]_{\star'} -  [f, g]_\star)
= \frac{1}{\hbar} \bigl(\mathcal{D}_A \circ (\star'-\star) \bigr)
(f \wedge g)$, the NC field strength $\widehat{\mathfrak{F}}_A$ can compactly be written as
\begin{equation}\label{nc-com-f}
  \widehat{\mathfrak{F}}_A = \frac{1}{\hbar} \mathcal{D}_A \circ (\star' - \star).
\end{equation}
Using the formality map (\ref{t-dep-star}) and $\theta_1 = \Theta = \theta + \mathbf{F}_\theta$ where
$\mathbf{F}_\theta$ is given by eq. (\ref{cochain-poisson}), one can calculate the right-hand side of
eq. (\ref{nc-com-f}) and the result is given by
\begin{equation}\label{sw-gen-f}
  \widehat{\mathfrak{F}}_A = \mathcal{D}_A \circ \Bigl( \Phi(\mathbf{F}_\theta)
  + \frac{\hbar}{2} \Psi(\mathbf{F}_\theta, \mathbf{F}_\theta) + \cdots \Bigr).
\end{equation}
Note that both sides of eq. (\ref{sw-gen-f}) are bidifferential operators. The map (\ref{sw-gen-f})
between ordinary and NC gauge fields constitutes the exact SW map for NC U(1) gauge theory \cite{ncft-sw}
and represents a quantum version of the (semi-)classical SW map (\ref{sw-field}).
By comparing eq. (\ref{nc-com-f}) with eq. (\ref{star-defor}) and identifying $\widehat{B} =
\mathcal{D}_A^{-1} \circ \widehat{\mathfrak{F}}_A$, one can also see that
$\mathcal{D}_A^{-1} \circ \widehat{\mathfrak{F}}_A$ is a solution of the Maurer-Cartan
equation (\ref{asso-two}) \cite{jsw2}. Finally the nilpotent property of the CE-differential,
i.e. $\widehat{d}^2 =0$, leads to the integrability condition that the NC field strength
$\widehat{\mathfrak{F}}_A$ in eq. (\ref{nc-field-fg}) obeys the Bianchi identity, i.e.,
\begin{equation}\label{nc-bianchi}
 \widehat{d}\widehat{\mathfrak{F}}_A + \widehat{\mathcal{A}}_A \wedge \widehat{\mathfrak{F}}_A -
    \widehat{\mathfrak{F}}_A \wedge \widehat{\mathcal{A}}_A = 0.
\end{equation}

It is worthwhile to notice that the CE-complex $(C^\bullet(\mathcal{A}, \mathcal{A}), \widehat{d})$
in the classical limit $\hbar \to 0$ reduces to the Lichnerowicz complex (\ref{lich-complex})
for the Poisson cohomology. For example, one may check this property with the NC field strength (\ref{nc-field-fg}).
First notice that the field strength (\ref{curvature-poisson}) of Poisson gauge fields can be written as
\begin{equation}\label{poisson-f-limit}
    \mathcal{F}_A (f, g) = \langle d_\theta \mathfrak{A}_A + \mathfrak{A}_A \wedge \mathfrak{A}_A,
    df \wedge dg \rangle
\end{equation}
where
\begin{equation}\label{cup-limit}
    \{ \mathfrak{A}_A(f), \mathfrak{A}_A(g) \}_\theta \equiv
    \langle \mathfrak{A}_A \wedge \mathfrak{A}_A, df \wedge dg \rangle
\end{equation}
is a commutative analogue of the cup product (\ref{cup-prod}). And look at the classical limit of
the CE-differential (\ref{ce-diff}) which reads as
\begin{eqnarray}\label{ce-limit}
 (\widehat{d} \widehat{\mathcal{A}}_A)(f,g)  &\overset{\hbar \to 0}{=}&
 \{f, \mathfrak{A}_A(g) \}_\theta -\{ g, \mathfrak{A}_A(f) \}_\theta
 - \mathfrak{A}_A \bigl(\{f, g\}_\theta \bigr) \nonumber \\
 &=&  \langle d_\theta \mathfrak{A}_A, df \wedge dg \rangle
\end{eqnarray}
with the definition $\widehat{\mathcal{A}}_A(f)|_{\hbar \to 0} \equiv \mathfrak{A}_A(f)$.
Combining all together, one can get the relation
\begin{equation}\label{bif-limit}
    \widehat{\mathfrak{F}}_A (f,g)|_{\hbar \to 0} = \langle \widehat{\mathfrak{F}}_A|_{\hbar \to 0},
    df \wedge dg \rangle = \mathcal{F}_A (f, g)
\end{equation}
where
\begin{equation}\label{ncf-limit}
    \widehat{\mathfrak{F}}_A = \widehat{d}\widehat{\mathcal{A}}_A
+ \widehat{\mathcal{A}}_A \wedge \widehat{\mathcal{A}}_A \overset{\hbar \to 0}{=}
d_\theta \mathfrak{A}_A + \mathfrak{A}_A \wedge \mathfrak{A}_A.
\end{equation}
In this sense the CE-complex defined by eqs. (\ref{ce-diff}) and (\ref{cup-prod}) may be regarded as
a (deformation) quantization of the Lichnerowicz complex (\ref{lich-complex}) for the Poisson cohomology.

\section{Emergent gravity from U(1) gauge fields}

\subsection{Interlude for local life}

Let $M$ be a $d$-dimensional Riemannian manifold whose metric is given by
\begin{equation}\label{r-metric}
    ds^2 = g_{\mu\nu}(x) dx^\mu \otimes dx^\nu, \qquad \mu, \nu = 1, \cdots, d.
\end{equation}
In general relativity, the metric (\ref{r-metric}) of the Riemannian manifold $M$ is also defined
by introducing at each spacetime point in $M$ a local reference frame of tangent bundle $TM$ in the form
of $d$ linearly independent vectors (so-called vielbeins) $E_a = E_a^\mu \partial_\mu \in \Gamma(TM)$.
The frame basis $\{E_a: a = 1, \cdots, d \}$ defines dual vectors $E^a = E^a_\mu dx^\mu
\in \Gamma(T^* M)$ by a natural pairing $\langle E^a, E_b \rangle = \delta^a_b$.
This pairing leads to the relation $E^a_\mu E^\mu_b = \delta^a_b$. In terms of the basis in
$\Gamma(TM)$ or $\Gamma(T^* M)$, the metric (\ref{r-metric}) can be written as
\begin{equation}\label{e-metric}
    ds^2 = \delta_{ab} E^a \otimes E^b = \delta_{ab} E_\mu^a (x) E_\nu^b(x) dx^\mu \otimes dx^\nu
\end{equation}
or
\begin{equation}\label{ie-metric}
    \Bigl(\frac{\partial}{\partial s} \Bigr)^2
    = \delta^{ab} E_a \otimes E_b = \delta^{ab} E_a^\mu (x) E_b^\nu (x) \partial_\mu \otimes \partial_\nu.
\end{equation}
We remark that, mathematically, a vector field $X$ on a smooth manifold $M$ is a derivation of the algebra
$\mathcal{A} = C^\infty(M)$.

As was observed before, given a Poisson structure $\theta \in \Gamma(\Lambda^2 TM)$ on a manifold $M$,
there is a natural homomorphism $\rho: T^* M \to TM$ induced by the Poisson bivector $\theta$.
Since any smooth vector field generates a one-parameter family of deformations described
by eq. (\ref{t-evolution}), one can perturb the Poisson structure $(M, \theta)$ by a line bundle $L \to M$.
Then the deformed Poisson bivector is given by eq. (\ref{sol-t}) where $F=dA$ is the curvature
of the line bundle $L$. Conversely, one can find a one-parameter family of diffeomorphisms
generated by vector fields $A_{\theta_t}= A_\mu (y) d_{\theta_t} y^\mu$ such that all the Poisson structures
$\theta_t$ for $t \in [0,1]$ are related by coordinate transformations defined by the exponential
map $\rho^*_{tt'}$, i.e., $\rho^*_{tt'} \theta_{t'} = \theta_t$.
In particular, denoting $\rho^*_{01} = \rho_A^*$, we have the relation (\ref{moser-flow}).
Thus we conceive a novel form of the equivalence principle because the electromagnetic force is to the deformation of a Poisson manifold what the gravitational force is to the deformation of a Riemannian manifold.
As the equivalence principle in general relativity beautifully explains why
the gravitational force manifests itself as a spacetime geometry, one may ask whether
a similar geometrization of the electromagnetic force can arise from the novel form of
the equivalence principle based on the Darboux theorem or the Moser lemma in symplectic or Poisson geometry.
For example, one may wonder whether the frame basis $E_a \in \Gamma(TM)$ in the metric (\ref{ie-metric})
can arise from U(1) gauge fields. This poses a profound question about the possibility that a $d$-dimensional Riemannian manifold $(M, g)$ can be defined by U(1) gauge theory on a Poisson manifold $(M, \theta)$.
We will address this issue in the context of emergent gravity but let us proceed with a general discussion.

So far we have mostly been stuck to a coordinate independent description.
But the physics will be of benefit to an explicit description by introducing a local coordinate system.
We will choose a local coordinate system $\{y^\mu \}$ on a local patch $U_i \subset M$ such that
$\{y^\mu \}_{U_i}$ are in the Darboux-Weinstein frame (\ref{weinstein}).
And consider the action of the formal diffeomorphism $\mathcal{D}_A$ in eq. (\ref{nc-gauge}) on
the local coordinates $\{y^\mu \}$ on $U_i \subset M$ and define the so-called covariant (dynamical)
coordinates by
\begin{equation}\label{d-coord}
X^\mu \equiv  \mathcal{D}_A (y^\mu) = y^\mu + \widehat{Y}^\mu(y)
\end{equation}
where $\widehat{Y}^\mu(y) \equiv \widehat{\mathcal{A}}_A (y^\mu)$.
From now on we will adopt the physicist's convention by replacing $\hbar$ by $\frac{i\hbar}{2}$.
(See the footnote \ref{p-convention}.) And we will often set $\hbar = 1$ whenever it is not necessary
to keep in with it. The Darboux-Weinstein coordinates obey the following commutation relation\footnote{\label{minimal-theta}Although
NC gauge theory construction in Refs. \cite{jsw1,jsw2} can be applied to a general Poisson structure $\theta$,
we will assume the unperturbed Poisson structure $\theta$ to be of the form in the Darboux-Weinstein
frame (\ref{weinstein}) keeping in mind the remark below (\ref{new-formal}). This minimal choice will
be of great benefit to the construction of a Hilbert space of NC algebra $\mathcal{A}_\theta$ for some cases,
e.g., the Heisenberg-Moyal algebra with constant $\theta^{\mu\nu}$ and Lie algebras
with a linear Poisson structure $\theta^{\mu\nu} = {f^{\mu\nu}}_\lambda y^\lambda$.}
\begin{equation}\label{nc-space5}
    [y^\mu, y^\nu]_\star = i \theta^{\mu\nu}(y)
\end{equation}
where higher order terms except the $\mathcal{O}(\hbar)$-term identically vanish.
When the bivector $\theta \equiv B^{-1}$ is an invertible symplectic structure,
$\theta^{\mu\nu}$ in eq. (\ref{nc-space5}) become a nondegenerate constant $d \times d$ matrix with $d$ = even.
In this case, as in eq. (\ref{cov-mom}), it is convenient to designate covariant ``momentum"
variables\footnote{\label{symp-realize}The most famous example of (quantized) Poisson algebra
is the SO(3) algebra of angular momenta $L_i = {\varepsilon_{ij}}^k x^j p_k$ in quantum mechanics.
In this case or the Poisson algebra case in general, there is no distinction (polarization)
between coordinates and momenta. Moreover this example illustrates that the generators $L_i$ of
Poisson algebra arise from the composite operators of $x^i$ and $p_i$ obeying the symplectic
algebra $[x^i, p_j] = i \hbar \delta^i_j$. It is known \cite{sg-weinstein} that for a general Poisson
manifold there exists globally an essentially unique symplectic realization of the Poisson manifold
which possesses a local groupoid structure compatible with the symplectic structure.
Later we will think of the possibility for the quantization of a Poisson manifold.}
\begin{equation}\label{c-momemtum}
 \widehat{C}_\mu (y) \equiv  \mathcal{D}_A (p_\mu) = p_\mu + \widehat{A}_\mu(y)
\end{equation}
where $p_\mu = B_{\mu\nu} y^\nu$ and $\widehat{A}_\mu(y) \equiv B_{\mu\nu} \widehat{Y}^\nu(y)$
define NC U(1) gauge fields used to formulate a NC gauge theory.
It is then straightforward to calculate the NC field strength (\ref{nc-field-fg})
in the Darboux-Weinstein frame:
\begin{eqnarray}\label{nc-fs}
    \widehat{F}_{\mu\nu}(y) &\equiv& \widehat{\mathfrak{F}}_A (p_\mu, p_\nu) \nonumber \\
    &=& \partial_\mu \widehat{A}_\nu(y) - \partial_\nu \widehat{A}_\mu(y)
    -i [\widehat{A}_\mu(y), \widehat{A}_\nu(y)]_\star.
\end{eqnarray}
Note that NC gauge fields can also be viewed as sections of a deformed vector bundle $L\bigl[[\hbar]\bigr]
\to M$ or naturally interpreted as connections over the NC algebra $\mathcal{A}_\theta$.

Here we intend to view the vielbeins $E_a \in \Gamma(TM)$ on a Riemannian manifold $M$
as a derivation of the algebra $\mathcal{A} = C^\infty(M)$.
And we want to understand, if any, whether the vielbeins $E_a \in \Gamma(TM)$ in the Riemannian
metric (\ref{ie-metric}) can arise from a derivation of NC $\star$-algebra $\mathcal{A}_\theta =
C^\infty(M) \bigl[[\hbar]\bigr]$ in commutative limit $|\theta| \to 0$.
Thereby let us again look at the chain of maps:
\begin{equation}\label{chain-anchor}
A \xrightarrow{~\rho~} A_\theta \xrightarrow{~U~} A_\star.
\end{equation}
Given U(1) gauge fields viewed as a one-form $A$ on a Poisson manifold $(M, \theta)$,
the first step is to associate a vector field $A_\theta \in \Gamma(TM)$ using the anchor map (\ref{anchor})
(though in general not injective except for the symplectic case).
And the next step is then to apply the formality
map $U: T_{\mathrm{poly}} (M) \to D_{\mathrm{poly}} (M)$ to get a differential operator $A_\star$
acting on the NC algebra $\mathcal{A}_\theta$. Hence it is natural to consider whether it is possible
to take the differential operator $A_\star$ as a candidate of the derivation for the metric fields.
But it turns out that it is not a proper choice for the following two reasons.
First note that combining eqs. (\ref{star-der}) and (\ref{star-leibniz}) for $X = A_\theta$ leads to the relation
\begin{equation}\label{non-der}
- \delta^\star_X (f\star g) + f \star \delta^\star_X g
    + (\delta^\star_X f) \star g = \hbar [\Phi(d_\theta A_\theta)](f, g).
\end{equation}
This result immediately implies that the differential operator $\delta_X^\star = \Phi(X) = A_\star$
for $X = A_\theta$ is not a derivation of $\star$-algebra $\mathcal{A}_\theta$
unless $F_\theta = d_\theta A_\theta = 0$.
The bivector field $F_\theta = d_\theta A_\theta$ is dual to the U(1) field strength $F = dA$
which is usually non-vanishing in NC gauge theory. Another reason is that we want to define
the derivation algebra relevant to NC gauge fields, but NC gauge fields are
defined via the covariantizing map (\ref{q-moser}) instead of $A_\star$ itself.
It should be also remarked for later arguments that the NC gauge fields in eq. (\ref{d-coord})
or (\ref{c-momemtum}) are locally defined because the covariantizing map $\mathcal{D}_A$ depends
on gauge fields and their coordinate representation (\ref{d-coord}) or (\ref{c-momemtum})
is also defined in the Darboux-Weinstein frame (\ref{weinstein}) on a local patch $U_i \subset M$.
In appendix A we will illuminate this local nature of NC gauge fields by showing that
they are basically defined by local coordinate transformations into a Darboux frame
which corresponds to the SW map between commutative and NC gauge fields \cite{ncft-sw}.
It is also worthwhile to remark that the Kontsevich's star product is defined on an open subset
of $\mathbb{R}^d$ whose Poisson bivector is given by the Darboux-Weinstein frame (\ref{weinstein}), 
so it should be used locally on a general Poisson manifold.
But the local expressions can be glued together to obtain a global star product \cite{cft}.
We will thus allow all compatible coordinate systems by an {\it atlas}
on $M$ as a family of local Darboux charts $\{(U_i, \varphi_i): i \in I \}$
where $\varphi_i: U_i \to \mathbb{R}^d$ and assume for the moment that they are consistently
glued together as in \cite{jsw-ncl} to define a global star product {\it \`a la} \cite{cft}.

Hence we may first define a derivation algebra $\mathfrak{X}_i := \Gamma(TU_i)\bigl[[\hbar]\bigr]$
(or in general $\mathfrak{X}_i := \mathcal{V}^\bullet (U_i)\bigl[[\hbar]\bigr]$)
on a local Darboux chart $(U_i, \varphi_i)$ as a subalgebra of $D_{\mathrm{poly}} (M)$ and
then try to glue the algebras $\{\mathfrak{X}_i: i \in I \}$ altogether to yield a globally defined algebra $\mathfrak{X} = \bigcup_{i \in I} \mathfrak{X}_i$ of derivations.
On a local Darboux chart $(U_i, \varphi_i)$, we have NC gauge fields in eq. (\ref{d-coord}) or
(\ref{c-momemtum}) and consider them as elements in the NC algebra $\mathcal{A}_\theta$.
Then there is a natural map from the NC algebra $\mathcal{A}_\theta$
to the Lie algebra $\mathfrak{X}_i$ defined by the inner derivation (\ref{star-inder}).
To be specific, we define an adjoint action of the NC algebra $\mathcal{A}_\theta$ as
\begin{equation}\label{adj-nc}
    \mathrm{ad}_{\widehat{f}}: \widehat{g} \mapsto -i [\widehat{f}, \widehat{g}]_\star
\end{equation}
for $\widehat{f}, \widehat{g} \in \mathcal{A}_\theta$. Obviously the adjoint action (\ref{adj-nc})
satisfies the Leibniz rule, i.e.,
\begin{equation}\label{leibniz}
    \mathrm{ad}_{\widehat{f}}(\widehat{g} \star \widehat{h}) =
    (\mathrm{ad}_{\widehat{f}} \widehat{g}) \star \widehat{h} + \widehat{g} \star (\mathrm{ad}_{\widehat{f}} \widehat{h})
\end{equation}
for $\widehat{g}, \widehat{h} \in \mathcal{A}_\theta$, so it defines a derivation of $\mathcal{A}_\theta$.
Using the Kontsevich's formula (\ref{star-prod}), one can expand the commutator in eq. (\ref{adj-nc}) to get
an explicit form of the polydifferential operator $X_{\widehat{f}}^\star \equiv \mathrm{ad}_{\widehat{f}}
\in \mathfrak{X}_i$ given by
\begin{equation}\label{poly-vec}
 X_{\widehat{f}}^\star = X_f + \sum_{n=2}^\infty \xi_f^{\mu_1 \cdots \mu_n}(y) \partial_{\mu_1}
 \cdots \partial_{\mu_n}
\end{equation}
where $X_f$ is an ordinary Hamiltonian vector field defined by $X_f (g) = \{f, g \}_\theta$.
An explicit formula for $X_{\widehat{f}}^\star$ up to second order can be found in \cite{behr-sykora}.
The Jacobi identity for the $\star$-commutator $[\widehat{f}, \widehat{g}]_\star$ leads to the result
that the polydifferential operator (\ref{poly-vec}) on $\mathcal{A}_\theta$ satisfies the deformed Lie algebra
\begin{equation}\label{def-lie}
    [X_{\widehat{f}}^\star ,  X_{\widehat{g}}^\star] =  X_{-i[\widehat{f}, \widehat{g}]_\star}^\star.
\end{equation}
It should be noted that the polydifferential operator (\ref{poly-vec}) recovers the usual vector field
at leading order. Thus it is obvious that the left-hand side of eq. (\ref{def-lie})
is a deformation of the ordinary Lie bracket of vector fields.

For simplicity, let us first consider the symplectic case.
On a local Darboux chart $U_i$, we have the set of dynamical gauge fields given by eq. (\ref{c-momemtum}).
Hence, according to the adjoint map $\mathcal{A}_\theta \to \mathfrak{X}_i$ in eq. (\ref{adj-nc}),
one can derive generalized vector fields for the set
$\{ \widehat{C}_a (y) \in \mathcal{A}_\theta: a=1, \cdots, d=2n \}$ on $U_i \subset M$
and they are given by\footnote{It may be convenient to distinguish local vector fields from
global ones which will be considered later. For this purpose we use small letters to denote
local vector fields and large letters to indicate global vector fields introduced later.}
\begin{equation}\label{gen-vec}
    v_a^\star \equiv X_{\widehat{C}_a}^\star \in \mathfrak{X}_i : v_a^\star(\widehat{f})
    = -i [\widehat{C}_a (y), \widehat{f}]_\star
\end{equation}
for any $\widehat{f} \in \mathcal{A}_\theta$.
Using the $\star$-commutator relations
\begin{eqnarray} \label{comm-rel-f}
&& -i[\widehat{C}_a (y), \widehat{C}_b (y)]_\star = -B_{ab} + \widehat{F}_{ab}(y) \in \mathcal{A}_\theta, \\
\label{comm-rel-df}
&&  -[\widehat{C}_a (y), [\widehat{C}_b (y), \widehat{C}_c (y)]_\star ]_\star
= \widehat{D}_a \widehat{F}_{bc}(y) \in \mathcal{A}_\theta,
\end{eqnarray}
one can apply the Lie algebra homomorphism (\ref{def-lie}) to the above gauge fields to yield
the differential operators given by
\begin{eqnarray} \label{gen-vec-f}
&& X_{\widehat{F}_{ab}}^\star = [v^\star_a , v^\star_b] \in \mathfrak{X}_i, \\
\label{gen-vec-df}
&&  X_{\widehat{D}_a \widehat{F}_{bc}}^\star =
[v^\star_a, [v^\star_b, v^\star_c] ] \in \mathfrak{X}_i.
\end{eqnarray}
Then one can use the above relations to transform the equations of NC gauge fields
in $\mathcal{A}_\theta$ into the (geometric) equations of generalized vector fields in $\mathfrak{X}_i$.
For example, NC U(1) gauge fields in $d=4$ dimensions admit (anti-)self-dual connections,
the so-called NC U(1) instatons, obeying the self-duality equations
\begin{equation}\label{nc-sde}
 \widehat{F}_{ab}(y) = \pm \frac{1}{2} {\varepsilon_{ab}}^{cd} \widehat{F}_{cd}(y).
\end{equation}
According to the map (\ref{gen-vec-f}), the NC U(1) instantons can thus be understood as some
(geometric) objects described by the self-duality equations
\begin{equation}\label{sde-gec}
 [v^\star_a , v^\star_b]   = \pm \frac{1}{2} {\varepsilon_{ab}}^{cd} [v^\star_c , v^\star_d ].
\end{equation}
Similarly, according to the map (\ref{gen-vec-df}), the equations of motion as well as the Bianchi identity
for general NC U(1) gauge fields are transformed into the following differential equations:
\begin{eqnarray} \label{eom-riem}
&&  \widehat{D}^a \widehat{F}_{ab} = 0 \qquad \Leftrightarrow
\qquad  [{v^\star}^a, [v^\star_a, v^\star_b] ] = 0, \\
\label{bianchi-riem}
&& \frac{1}{3!} \delta_{abc}^{def} \widehat{D}_d \widehat{F}_{ef} = 0 \qquad \Leftrightarrow
\qquad \frac{1}{3!} \delta_{abc}^{def} [v^\star_d, [v^\star_e, v^\star_f] ] = 0.
\end{eqnarray}

In order to identify geometric objects described by the differential equations (\ref{eom-riem}) and
(\ref{bianchi-riem}), it is necessary first to know the relation between the (inverse) vielbeins
$E_a \in \Gamma(TM)$ and the generalized vector fields $v_a^\star \in \mathfrak{X}_i$.
Note that the vector fields (vielbeins) in the gravitational metric (\ref{ie-metric}) are globally defined.
Therefore in order to identify a gravitational metric from locally defined NC gauge fields
we need to consider a global version of the Lie algebra of derivations. For this purpose, we can use
NC U(1) gauge transformations as well as coordinate transformations to glue the locally defined derivations
on overlapping regions of an open covering $ M = \bigcup_{i \in I} U_i$.
First it will be instructive to understand a corresponding situation in general relativity.
On relying on the equivalence principle in general relativity,
which is mathematically tantamount to the simple statement that every manifold is locally flat,
at every point $x'$ of spacetime one can choose a set of coordinates $\xi^a$ that are {\it locally inertial}
at $x'$. The metric components in any general non-inertial coordinate system are then given by
\begin{equation}\label{li-metric}
    \widetilde{g}^{\mu\nu}(x) = \delta^{ab} e^\mu_a (x) e^\nu_b(x)
\end{equation}
where
\begin{equation}\label{li-vielbein}
 e^\mu_a (x) = \frac{\partial x^\mu}{\partial \xi^a} (x)|_{x=x'}.
\end{equation}
Therefore the equivalence principle always guarantees the existence of $d$ linearly independent
flat coordinates $(\xi^1, \cdots, \xi^d)$ such that the metric locally becomes flat, i.e.,
\begin{equation}\label{lic-metric}
    \widetilde{g}^{\mu\nu}(x) \frac{\partial \xi^a}{\partial x^\mu}\frac{\partial \xi^b}{\partial x^\nu}
    = \delta^{ab}.
\end{equation}
Note that we have intentionally distinguished the locally defined
inertial frame (\ref{li-vielbein}) from the globally defined basis
$E_a = E_a^\mu(x) \frac{\partial}{\partial x^\mu}$ in
eq. (\ref{e-metric}) which in general cannot be written as the form
(\ref{li-vielbein}) unless spacetime is flat. But the basis $E_a$
can be restricted to an infinitesimal neighborhood $U_{x'}$ centered
at $x'$ so that it can be represented in the locally inertial frame
(\ref{li-vielbein}), i.e., $E^\mu_a |_{U_{x'}} = e_a^\mu$. As the
metric $\widetilde{g}^{\mu\nu}(x)$ varies smoothly with $x$, there is no
obstacle to find a more general basis by allowing the $d \times d$
matrix $e_a^\mu(x)$ to vary smoothly with $x$. This should be the case because every manifold
can be constructed by gluing together Euclidean domains. This suggests a
simple recipe to get a globally defined frame $E_a$ from a locally
defined coordinate basis $e_a$ on $U_{x'}$:
\begin{equation}\label{global-recipe}
  e_a |_{U_{x'}}  \to E_a.
\end{equation}
However the replacement (\ref{global-recipe}) should be compatible with the orthonormality of the bases:
\begin{equation}\label{orthonormal}
    e_a \cdot e_b = \delta_{ab} \qquad \Leftrightarrow \qquad E_a \cdot E_b = \delta_{ab}
\end{equation}
which means that $\widetilde{g}_{\mu\nu} (x) |_{U_{x'}}  \to g_{\mu\nu} (x) = E^a_\mu(x) E^a_\nu (x)$.
With this replacement the metric (\ref{li-metric}) in a locally inertial frame can be extended to
an entire region with the metric (\ref{ie-metric}) because the frames $E_a$ are now coordinate independent
and so globally defined.

We need a similar globalization for locally defined vector fields $v_a^\star \in \mathfrak{X}_i$.
Deferring a detailed exegesis later, let us take a simple recipe analogous to
the replacement (\ref{global-recipe}). That is, we will implicitly assume that local Darboux charts
$(U_i, \varphi_i)$ and derivation algebras $\mathfrak{X}_i$ defined over there are consistently
glued together by coordinate transformations and NC U(1) gauge transformations
on overlapping regions \cite{jsw-ncl,cft}.
We will denote by $V^\star_a \in \mathfrak{X}$ the global version of generalized vector fields
obtained through the gluing of local data:
\begin{eqnarray}\label{global-vector}
  v^\star_a |_{U_i} &=& \widetilde{v}_a + \sum_{n=2}^\infty \xi_a^{\mu_1 \cdots \mu_n}(y)
  \partial_{\mu_1} \cdots
  \partial_{\mu_n} \in \mathfrak{X}_i \nonumber \\
&&  \to V^\star_a = V_a + \sum_{n=2}^\infty \Xi_a^{\mu_1 \cdots \mu_n}(x) \partial_{\mu_1} \cdots
  \partial_{\mu_n} \in \mathfrak{X}.
\end{eqnarray}
Note that the vector fields $v^\star_a |_{U_i} \in \mathfrak{X}_i$ are defined by the inner
derivation (\ref{gen-vec}) in a local Darboux frame and the local vector fields $\widetilde{v}_a
= \widetilde{v}_a^\mu (y) \partial_\mu \in \Gamma(TU_i)$ are divergence-free, i.e.,
$\partial_\mu \widetilde{v}_a^\mu (y) = 0$. This means that the vector fields $\widetilde{v}_a$ are volume-preserving, i.e., $\mathcal{L}_{\widetilde{v}_a} \nu_D = 0$
for the flat volume form $\nu_D = \frac{B^n}{n! \mathrm{Pf}B} = d^{2n} y$ in a Darboux frame.
As a parallel analogue of (\ref{orthonormal}) in general relativity, the replacement (\ref{global-vector})
similarly needs to keep the volume-preserving condition for global vector fields
$V_a = V_a^\mu (x) \partial_\mu \in \Gamma(TM)$ such that
\begin{equation}\label{v-preserving}
\mathcal{L}_{\widetilde{v}_a} \nu_D = 0     \qquad \Leftrightarrow \qquad \mathcal{L}_{V_a} \nu = 0
\end{equation}
for some volume form $\nu$. A Poisson manifold with the above property is called unimodular and
any symplectic manifold is unimodular. We give a brief exposition in appendix B for modular vector fields,
Poisson homology and their deformation quantization. Suppose that the volume form is given
by $\nu = \lambda^2 V^1 \wedge \cdots \wedge V^d$ where $V^a = V_\mu^a(x)dx^\mu \in \Gamma(T^* M)$ are
globally defined covectors, i.e., $\langle V^a, V_b \rangle =\delta^a_b$. Then, by definition, we get
\begin{equation}\label{vol-lambda}
    \lambda^2 = \nu(V_1, \cdots, V_d).
\end{equation}
One can see that the right-hand side of eq. (\ref{v-preserving}), when restricted to a local Darboux chart,
reduces to the left-hand side, as it should be.
It must be emphasized that the above globalization will be compatible with the derivation
structure (\ref{leibniz}) as well as the Lie algebra structure (\ref{def-lie}) because the differential
operators $V^\star_a \in \mathfrak{X}$ are realized as an inner derivation of global star product,
as will be shown later. Nevertheless it turns out that the global vector fields $V_a \in \Gamma(TM)$
will reproduce a general volume-preserving vector fields. Since we eventually want to achieve
a background independent formulation of NC gauge theory in terms of the algebra of (large $N$) matrices,
this property is actually desirable because any derivation of the matrix algebra is well-known to be inner.

One can choose the conformal factor $\lambda$ such that the orthonormal vectors $E_a$
preserve the volume form $\widetilde{\nu} = \lambda^{3-d} \nu_g$ where $\nu_g = E^1 \wedge \cdots \wedge E^d
= \sqrt{\det g_{\mu\nu}} d^{d} x$ is the Riemannian volume form \cite{lry1,lee-yang}. This means that
the gauge theory vectors $V_a = V_a^\mu (x) \partial_\mu \in \Gamma(TM)$ are related to the basis
of orthonormal tangent vectors $E_a = E_a^\mu (x) \partial_\mu$ by
\begin{equation}\label{vec-2rel}
    V_a = \lambda E_a
\end{equation}
and their covectors in $\Gamma(T^* M)$ are related by
\begin{equation}\label{covec-2rel}
    E^a = \lambda V^a.
\end{equation}
This can be checked as follows:\footnote{\label{foot-volpc}The standard formula for the covariant
divergence $\nabla \cdot V$ of a vector field $V$ is given by $\mathcal{L}_V \nu_g = (\nabla \cdot V) \nu_g$.
Therefore we get $\mathcal{L}_{V_a} \nu = \bigl( \nabla \cdot V_a + (2-d)V_a \log \lambda \bigr) \nu$.
Unfortunately there was a stupid mistake for the divergence formula in the footnote 19 of \cite{hsy-jhep09}.
But fortunately this remained a harmless slip and did not affect any results.}
\begin{equation}\label{der-vec2}
     \mathcal{L}_{V_a} \nu = \mathcal{L}_{\lambda E_a} (\lambda^{2-d} \nu_g )
     = \mathcal{L}_{E_a} (\lambda^{3-d} \nu_g ) = \mathcal{L}_{E_a} \widetilde{\nu} = 0.
\end{equation}
Using the relation (\ref{covec-2rel}), one can completely determine the Riemannian metric (\ref{e-metric})
in terms of the global vector fields defined by NC gauge fields via eqs. (\ref{gen-vec})
and (\ref{global-vector}) and it takes the form
\begin{equation}\label{emergent-metric}
    ds^2 = E^a \otimes E^a = \lambda^2 V^a \otimes V^a = \lambda^2 V_\mu^a V^a_\nu dx^\mu \otimes dx^\nu.
\end{equation}

Now we are ready to translate the equations of motion (\ref{eom-riem}) for NC gauge fields together with
the Bianchi identity (\ref{bianchi-riem}) into some geometric equations related to Riemann curvature tensors
determined by the metric (\ref{emergent-metric}) at leading order.
To see what they are, recall that, in terms of covariant derivatives, the torsion $T$ and the curvature $R$
are given by well-known formulae
\begin{eqnarray} \label{torsion-t}
T(X,Y) &=& \nabla_X Y - \nabla_Y X - [X, Y], \\
\label{curvature-r}
R(X,Y)Z  &=& [\nabla_X, \nabla_Y] Z - \nabla_{[X, Y]} Z,
\end{eqnarray}
where $X,Y, Z$ are vector fields on $M$. Because $T$ and $R$ are multilinear differential operators,
one can easily deduce the following relations
\begin{eqnarray} \label{torsion-s2}
T(V_a, V_b) &=& \lambda^2 T(E_a,E_b), \\
\label{curvature-s3}
R(V_a, V_b)V_c  &=& \lambda^3 R(E_a, E_b)E_c.
\end{eqnarray}
Since we want to recover the general relativity from the emergent gravity approach, we will impose
the torsion free condition, $T(E_a,E_b) = 0$. Then it is straightforward,
by using eqs. (\ref{torsion-t}) and (\ref{curvature-r}) and repeatedly converting
$\nabla_a V_b - \nabla_b V_a$ into $[V_a, V_b]$, to derive the identity below
\begin{equation}\label{bianchi-gg}
 R(E_a, E_b)E_c + \mathrm{cyclic} (a \to b \to c) =  \lambda^{-3} \Bigl(
[V_a, [V_b, V_c]] + \mathrm{cyclic} (a \to b \to c) \Bigr).
\end{equation}
Therefore we arrived at a pleasing result that the Bianchi identity (\ref{bianchi-riem}) for NC gauge fields
reduces at leading order to the first Bianchi identity for Riemann curvature tensors, i.e.,
\begin{equation}\label{eg-bianchi}
\widehat{D}_a \widehat{F}_{bc} + \mathrm{cyclic} (a \to b \to c) = 0 \quad \Leftrightarrow
\quad \Bigl( R_{abcd} + \mathrm{cyclic} (a \to b \to c) \Bigr) + \mathcal{O}(\theta^2) = 0.
\end{equation}
We will discuss later how the classical general relativity is corrected due to the NC structure of spacetime.

The transformation for the equations of motion (\ref{eom-riem}) into gravitational equations requires
more algebras. But it is natural to expect the Einstein equations from NC gauge fields
at leading order\footnote{A frugal way to derive eq. (\ref{eg-eom}) is to first calculate the Ricci
tensor $R_{ab}$ in terms of vector fields $V_a$ \cite{hsy-jhep09} and then add appropriate terms
by inspection on both sides of $[V^a, [V_a,V_b]] = 0$ so that the left-hand side together with
the added terms yields $R_{ab}$.}
\begin{equation}\label{eg-eom}
\widehat{D}^a \widehat{F}_{ab} = 0 \qquad \Leftrightarrow
\qquad  R_{ab} - 8 \pi G (T_{ab} - \frac{1}{2} \delta_{ab} T) + \mathcal{O}(\theta^2) = 0.
\end{equation}
Thus the upshot of the analysis is to determine the form of energy-momentum tensor $T_{ab}$.
It was determined in \cite{hsy-jhep09} only in lower dimensions $d \leq 4$. Since we do not know
the result in higher dimensions, let us focus on the four dimensions.
First note that the Einstein gravity arises at the first order of NC gauge fields, i.e.,
$R_{ab} \sim \mathcal{O}(\theta)$ and the parameters, $G_{YM}$ and $\theta$, defining the NC gauge theory
are related to the gravitational constant $G$ by
\begin{equation}\label{g-ncym}
    \frac{G \hbar^2}{c^2} \sim G^2_{YM} |\theta|
\end{equation}
where $|\theta| \equiv (\mathrm{Pf} \theta)^{\frac{1}{n}}$. Therefore the Einstein equations (\ref{eg-eom})
imply that $T_{ab} \sim \mathcal{O}(1)$. We know that NC U(1) gauge theory reduces to the ordinary
Maxwell theory at $\mathcal{O}(1)$, so $T_{ab}$ in eq. (\ref{eg-eom}) has to contain the Maxwell
energy-momentum tensor. Indeed the detailed analysis reveals some surprise \cite{hsy-jhep09,hsy-jpcs12}.
In addition to the Maxwell energy-momentum tensor, it also contains an exotic energy-momentum tensor
which is absent in Einstein gravity. The reason is as follows. Define the structure equation of
vector fields $V_a \in \Gamma(TM)$ as
\begin{equation}\label{st-eq}
    [V_a, V_b] = - {g_{ab}}^c V_c
\end{equation}
and take the canonical decomposition
\begin{equation}\label{dec-st}
  g_{abc} = g_c^{(+)i} \eta^i_{ab} + g_c^{(-)i} \overline{\eta}^i_{ab}
\end{equation}
according to the Lie algebra splitting $so(4) = su(2)_L \oplus su(2)_R$.
It turns out \cite{hsy-jhep09} that the energy-momentum tensor $T_{ab}$ consists of purely interaction terms
between self-dual and anti-self-dual parts in eq. (\ref{dec-st}). However the energy-momentum tensor $T_{ab}$
has a nonvanishing trace although it originates from the mixed sectors, i.e.,  $(su(2)_L, su(2)_R)$
and $(su(2)_R, su(2)_L)$. Normally the trace for the mixed sectors vanishes because the Ricci scalar
in general relativity belongs to $(su(2)_L, su(2)_L)$ and $(su(2)_R, su(2)_R)$ sectors \cite{ahs-1978,oh-yang,loy-jhep}. (The traceless Ricci tensor and the Ricci scalar were denoted by $B$ and $s$, respectively, in eq. (4.29) in \cite{loy-jhep}.) Nevertheless, the Ricci scalar deduced from eq. (\ref{eg-eom})
becomes nonzero since it is determined by a mixed tensor $g_{abc}$ in Eq. (\ref{dec-st}) which
is not simply an antisymmetric second-rank tensor.
In addition, in a long-wavelength limit where the scalar modes are dominant, it reduces to
\begin{equation}\label{dark-energy}
    T_{\mu\nu} \approx - \frac{R}{32 \pi G} g_{\mu\nu}
\end{equation}
where $R$ is the Ricci scalar for the metric tensor (\ref{emergent-metric}).
Hence, one can see that the mystic energy (\ref{dark-energy}) cannot be realized in Einstein gravity.
Moreover eq. (\ref{dark-energy}) implies that the mystic energy behaves like the dark energy with $w=-1$
after the Wick rotation into the Lorentzian signature \cite{hsy-jpcs12}.
Actually it copies all the properties of dark energy, so it was suggested
in Refs. \cite{hsy-jhep09,hsy-jpcs12} as a possible candidate of dark energy.

\subsection{Equivalence principle and Riemann normal coordinates}

First let us understand how to realize the globalization (\ref{global-recipe}) of vector fields
from local data in an inertial frame. The underlying idea is that local invariants of a metric
in Riemannian geometry are quantities expressible in local coordinates in terms of the metric and
its derivatives and they have an invariance property under changes of coordinates.
It is a fundamental result in Riemannian geometry that such invariants can be written in terms of
the curvature tensor of the metric and its covariant derivatives.
Hence the full Taylor expansion of the metric can be recovered from the iterated covariant
derivatives of curvature tensors. As a consequence, any local invariant of Riemannian
metrics has a universal expression in terms of the curvature tensor and its covariant derivatives.
This is known as the jet isomorphism theorem \cite{epstein,ambient-metric} stating that
the space of infinite order jets of metrics modulo coordinate changes is isomorphic to a space of
curvature tensors and their covariant derivatives modulo the orthogonal group.
(One may view the jet bundle as a coordinate free version of Taylor expansions.
See appendix C for a brief exposition of jet bundles.)

The Taylor expansion of a metric at a point $p \in M$ can be more explicit by considering
a coordinate system which is locally flat at that point on a curved manifold.
The coordinates in an open disk centered at the origin are normal coordinates \cite{epstein,besse}
arising from an orthonormal basis at the origin if and only if for each point $p$ in the disk,
\begin{equation}\label{rnc}
    g_{ab}(p) \xi^b  = \delta_{ab} \xi^b.
\end{equation}
As one can see from eq. (\ref{rnc}), geodesic normal coordinates are determined up to the orthogonal
group $O(d)$, i.e., different normal coordinates are related by an element of $O(d)$.
The basic idea behind the so-called
Riemann normal coordinates (RNCs) is to use the geodesics through a given point to define the coordinates
for nearby points. They have an appealing feature that the geodesic equations
\begin{equation}\label{geodesic}
    \frac{d^2 x^\mu}{dt^2} + {\Gamma^\mu}_{\rho\sigma} \frac{d x^\rho}{dt}
    \frac{d x^\sigma}{dt} = 0
\end{equation}
passing through the point have the same form as the equations in the Cartesian coordinate system
in Euclidean geometry because the Levi-Civita connections ${\Gamma^\mu}_{\rho\sigma}$ vanish at that point.
It may be useful to recall \cite{kobnom} that the projection onto any integral curve of
a standard horizontal vector field of the bundle $F(M)$ of linear frames over $M$ is a geodesic
and, conversely, every geodesic is obtained in this way.

We may construct the normal coordinates around each point $p$ of $M$ using the exponential map
$\exp_p: T_p M \to M$. Basically the normal coordinates are the coordinates of the tangent space
at $p \in M$ pulled back to the base manifold.
Recall that the exponential map $\exp_p: T_p M \to M$ is defined by
$\exp_p (v):= \gamma_v (1)$ where $\gamma_v: [0, 1] \to M$ is a geodesic curve
for which $\gamma_v (0) = p$ and $\dot{\gamma}_v (0) = v \in T_p M$. Thus, for any $p \in M$,
there exists a neighborhood $\mathcal{U}$ of $0$ in $T_p M$ and a neighborhood $U$ of $p$ in $M$
so that $\exp_p: \mathcal{U} \to U$ is a (local) diffeomorphism. By the construction,
for every $q \in U$, there exists a unique geodesic which joins $p$ to $q$ and lies entirely in $U$.
Given an orthonormal frame $\{ e_a \}_{a=1}^d$ of $T_p M$, the linear isomorphism $\xi: \mathbb{R}^d \to T_p M$
by $(\xi^1, \cdots, \xi^d) \mapsto \xi^a e_a$ defines a coordinate system in $\mathcal{U}$ in a natural manner.
Therefore the map
\begin{equation}\label{gd-coord}
\exp_p \circ \, \xi : \xi^{-1}(\mathcal{U}) \to U
\end{equation}
is a local chart for $M$ around $p$ and its inverse defines the normal coordinate system in $U$.
The normal coordinates on $U$ are then given by $\exp_p^{-1}(q) = y^\mu (q) e_\mu$
or equivalently
\begin{equation}\label{normal-q}
  y^\mu = l^\mu \circ \exp_p^{-1}
\end{equation}
where $(l^1, \cdots, l^d)$ is the dual basis of $e_a$.
In terms of the normal coordinates, the vielbeins $\widetilde{e}^a = \widetilde{e}^a_\mu (y) dy^\mu$
(objects in the $\infty$-jet will be denoted with the tilde) are given by \cite{closed-german}
\begin{eqnarray}\label{rnc-vielbein}
    \widetilde{e}^a_{\mu} &=& \delta^a_{\mu} - \frac{1}{6} R_{a \rho \mu \sigma} y^\rho y^\sigma
    - \frac{1}{12} \nabla_\lambda R_{a \rho \mu \sigma} y^\rho y^\sigma y^\lambda \nonumber \\
    && - \Bigl(\frac{1}{40} \nabla_\rho \nabla_\sigma R_{a \lambda \mu \nu}
    - \frac{1}{120} R_{a \nu \kappa \lambda} R_{\mu \rho \kappa \sigma} \Bigr)
    y^\nu y^\lambda y^\rho y^\sigma + \mathcal{O} (y^5).
\end{eqnarray}
Then the metric $\widetilde{g}_{\mu\nu} = \widetilde{e}^a_\mu \widetilde{e}^a_\nu$ in the $\infty$-jet
is given by \cite{closed-german}
\begin{eqnarray}\label{rnc-metric}
    \widetilde{g}_{\mu\nu} &=& \delta_{\mu\nu} - \frac{1}{3} R_{\mu \rho \nu \sigma} y^\rho y^\sigma
    - \frac{1}{6} \nabla_\lambda R_{\mu \rho \nu \sigma} y^\rho y^\sigma y^\lambda \nonumber \\
    && - \Bigl(\frac{1}{20} \nabla_\rho \nabla_\sigma R_{\mu \alpha \nu \beta}
    - \frac{2}{45} R_{\mu \rho \lambda \sigma} R_{\nu \alpha \lambda \beta} \Bigr)
    y^\rho y^\sigma y^\alpha y^\beta + \mathcal{O} (y^5),
\end{eqnarray}
and so
\begin{eqnarray}\label{rnc-det}
 \det \widetilde{g}_{\mu \nu} &=& 1 - \frac{1}{3} R_{\mu \nu} y^\mu y^\nu
    - \frac{1}{6} \nabla_\lambda R_{\mu \nu} y^\mu y^\nu y^\lambda \nonumber \\
    && - \Bigl(\frac{1}{20} \nabla_\rho \nabla_\sigma R_{\mu \nu}
    + \frac{1}{90} R_{\mu \lambda \nu \kappa} R_{\rho \lambda \sigma \kappa}
    - \frac{1}{18} R_{\mu\nu} R_{\rho\sigma} \Bigr) y^\mu y^\nu y^\rho y^\sigma + \mathcal{O} (y^5).
\end{eqnarray}
These formulas exhibit how the curvature and its derivatives locally affect the metric and
volume form $\nu_g = \sqrt{\det \widetilde{g}_{\mu\nu}} d^d y$.
A closed formula for the vielbein as well as the metric in the RNC expansion
is now available due to a remarkable paper \cite{closed-german}
which demonstrates the jet isomorphism theorem \cite{epstein,ambient-metric}.

Therefore we may compare local invariants at a point $p \in M$ determined by the objects
in the $\infty$-jet such as $\widetilde{e}^a_\mu$ and $\widetilde{g}_{\mu\nu}$ with those
determined by the global quantities such as $E^a_\mu$ and $g_{\mu\nu}$ (if they are known).
If they coincide each other up to any arbitrary order, we can identify these two quantities, i.e.,
\begin{equation}\label{id-two}
 \widetilde{e}^a_\mu \cong E^a_\mu  \qquad \mathrm{and} \qquad
 \widetilde{g}_{\mu\nu} \cong g_{\mu\nu}.
\end{equation}
The identification (\ref{id-two}) makes sense if the geodesic Taylor expansion in a patch around $p$
converges, so the patch has to be taken sufficiently small in a strongly curved region.
The above prescription means that sections in the $\infty$-jet
bundle $\mathcal{E}$ (the infinite jet prolongations of the frame bundle and its symmetric tensor
product--see appendix C) belong to the same equivalence class and thus it holds everywhere
because the objects in the $\infty$-jet are represented by the covariant tensors only (or the natural tensors
in the terminology of \cite{epstein}) that respect an invariance property under changes of coordinates.
Hence we can implement the identification (\ref{id-two}) to define a prescription
for the globalization (\ref{global-recipe}).

\subsection{Fedosov manifolds and global deformation quantization}

Note that the NC gauge theory was defined by quantizing the Poisson algebra $\mathfrak{P}
= (C^\infty(M), \{-,-\}_\theta)$ of Poisson gauge fields on local Darboux-Weinstein charts.
And we defined the inner derivation (\ref{gen-vec}) from local NC gauge fields in $\mathcal{A}_\theta$.
Thus it is necessary to glue together local objects defined on Darboux charts to yield global objects.
We will employ a similar prescription as eq. (\ref{id-two}) for the globalization (\ref{global-vector})
using the Fedosov's approach of deformation quantization. Let us first consider a symplectic
manifold $(M, \omega)$ and $\{(U_i, \varphi_i): i \in I \}$ an atlas on $M$.
By $\omega$ we mean the symplectic 2-form. We introduce a connection $\partial^S$ on
the symplectic manifold $(M, \omega)$ which preserves the symplectic form $\omega$ \cite{d-quant1}.
The Christoffel symbols ${\Gamma^\lambda}_{\mu\nu}$ are defined as usual
by $\partial^S_{\partial_\mu} \partial_\nu = {\Gamma^\lambda}_{\mu\nu} \partial_\lambda$.
The curvature tensor of a symplectic connection is also defined by the usual formula (\ref{curvature-r})
and is given in the holonomic basis by
\begin{equation}\label{symp-curv}
    {\mathfrak{R}^\mu}_{\nu\rho\sigma} = \partial_\rho {\Gamma^\mu}_{\sigma\nu}
    - \partial_\sigma {\Gamma^\mu}_{\rho\nu} + {\Gamma^\mu}_{\rho\lambda} {\Gamma^\lambda}_{\sigma\nu}
-  {\Gamma^\mu}_{\sigma\lambda} {\Gamma^\lambda}_{\rho\nu}.
\end{equation}
Here we use the Fraktur letter for the symplectic curvature tensor to avoid a confusion with
the Riemann curvature tensor in the previous sections. The symplectic connection $\Gamma = ({\Gamma^\lambda}_{\mu\nu})$ on $M$ is a torsion free connection locally satisfying the condition
\begin{equation}\label{symp-conn}
    \partial^S \omega = 0.
\end{equation}
In any Darboux coordinates on a local chart $(U_i, \varphi_i)$ where $\omega_{\mu\nu}$ are constants, (\ref{symp-conn}) reduces to
\begin{equation}\label{symp-conn-comp}
     \omega_{\mu\rho} {\Gamma^\rho}_{\lambda\nu}
    - \omega_{\nu\rho} {\Gamma^\rho}_{\lambda \mu} = \Gamma_{\mu \lambda \nu} - \Gamma_{\nu \lambda \mu} = 0
\end{equation}
where $\Gamma_{\mu \nu \lambda} \equiv \omega_{\mu\rho} {\Gamma^\rho}_{\nu\lambda}$.
The connection (\ref{symp-conn}) is thus a symplectic analogue of the Levi-Civita connection
in Riemannian geometry. Combining the torsion free (i.e., symmetric) condition, i.e. $\Gamma_{\lambda\mu\nu}
= \Gamma_{\lambda\nu\mu}$, the symmetric connections preserving $\omega$ are exactly the connections with
the Christoffel symbols $\Gamma_{\lambda\mu\nu}$ which are completely symmetric with respect
to all indices $\mu, \nu, \lambda$. Such a symplectic connection exits on any symplectic
manifold.\footnote{\label{foot-sympconn}It
is a well-known fact (e.g., see Remark 1.4 in \cite{fedo-man}) that a symmetric connection preserving $\omega$
exists if and only if $\omega$ is closed. If $\omega$ is a symplectic 2-form, then locally we can take
the trivial connection in Darboux coordinates. Globally we can glue symmetric connections preserving $\omega$
using a partition of unity. We will basically use this fact for constructing global vector fields on
a symplectic manifold.}
In particular, the triple $(M, \omega, \Gamma)$ is called a Fedosov manifold and the deformation quantization
on a symplectic manifold $(M, \omega)$ is defined by the data $(M, \omega, \Gamma)$.
Note that every K\"ahler manifold is a Fedosov manifold. Indeed Fedosov manifolds constitute
a natural generalization of K\"ahler manifolds. However we will not refer to the existence of
any Riemannian metric when considering a Fedosov manifold since we want to derive the former from
the latter according to the spirit of emergent gravity.

For a symplectic manifold $(M, \omega)$, each tangent space $T_p M$ at $p \in M$ is a symplectic
vector space and $(TM = \bigcup_{p \in M} T_p M, \omega)$ becomes a symplectic vector bundle over $M$.
Given a point $p \in U \subset M$ we can construct the exponential map $\exp_p: \mathcal{U} \to U$
defined by the symplectic connection $\partial^S$ where $\mathcal{U}$ is a small neighborhood of 0
in $T_p M$. Let $x(t)$ be a curve in $U$ satisfing the geodesic equation (\ref{geodesic})
defined in local coordinates $(x^1, \cdots, x^d)$ such that $x(0) = p \in U, \; \dot{x}(0)
= v \in \mathcal{U}$ and $\exp_p (v) = x(1)$ where the Christoffel symbols ${\Gamma^\lambda}_{\mu\nu}$
are now defined by eq. (\ref{symp-conn}).\footnote{Given a connection $\nabla$, the covariant derivative of
a tensor $\mathbf{T}$ along a curve $\lambda(t)$ is defined by $\frac{D \mathbf{T}}{\partial t}
\equiv \nabla_{\frac{\partial}{\partial t}} \mathbf{T}$ and the tensor $\mathbf{T}$ is said to be
parallelly transported along $\lambda$ if $\frac{D \mathbf{T}}{\partial t} = 0$.
And the curve $\lambda(t)$ is said to be a geodesic curve if $\frac{D}{\partial t}
\Bigl(\frac{\partial}{\partial t}\Bigr)_\lambda = 0$ for an affine parameter $t$.
If we choose local coordinates so that $\lambda(t)$ has the coordinates $x^\mu(t)$
and so $\Bigl(\frac{\partial}{\partial t}\Bigr)_\lambda = \frac{d x^\mu(t)}{d t} \partial_\mu$,
we get the geodesic equation (\ref{geodesic}) for the connection $\nabla$ which might be
either the Levi-Civita connection or a symplectic connection.}
Using the exponential map, we can construct the normal coordinate system on $U$ defined by
$\exp_p^{-1} (q) = y^\mu(q) e_\mu$ in the same way as (\ref{normal-q}).
In other words, if $v = y^1 \frac{\partial}{\partial x^1} + \cdots + y^d \frac{\partial}{\partial x^d}
\in T_p M$, then $(y^1, \cdots, y^d)$ are the normal coordinates of $\exp_p v$.
In this case the geodesic equation (\ref{geodesic}) along the curve $(x^1, \cdots, x^d) =
t (y^1, \cdots, y^d)$ enforces
\begin{equation}\label{symp-normal}
    \Gamma_{\mu\nu\lambda}(x) y^\nu y^\lambda = 0
\end{equation}
and, taking the limit as $t \to 0$, eq. (\ref{symp-normal}) in turn implies
\begin{equation}\label{normal-cond}
    \Gamma_{\mu\nu\lambda}(0)  = 0.
\end{equation}

Let us take the Taylor expansion of $\omega_{\mu\nu}$ in terms of these normal coordinates
(the tilde denotes an $\infty$-jet object):
\begin{eqnarray}\label{to-expansion}
 \widetilde{\omega}_{\mu\nu}(y) &=& B_{\mu\nu} + \sum^\infty_{n=2} \frac{1}{n!}
 \omega_{\mu\nu, \lambda_1 \cdots \lambda_n}(0) y^{\lambda_1} \cdots y^{\lambda_n} \nonumber \\
  &\equiv &  B_{\mu\nu} + F^x_{\mu\nu} (y)
\end{eqnarray}
where $B_{\mu\nu} = \widetilde{\omega}_{\mu\nu}(0)$ are constant values of the symplectic two-form at $p \in M$ which is assumed to be in the Darboux frame. An important point is that different normal coordinates
with the same origin differ by a linear transformation, so the expansion coefficients
in eq. (\ref{to-expansion}) define tensors, called affine normal tensors, on $M$ \cite{fedo-man}.
Note that the $\omega$-preserving condition (\ref{symp-conn}) reduces to $\partial_\lambda \omega_{\mu\nu} = \Gamma_{\mu\lambda\nu} - \Gamma_{\nu\lambda\mu}$, 
so the condition (\ref{normal-cond}) was already imposed in the expansion (\ref{to-expansion})
(where $y$-dependent terms start from $\mathcal{O}(y^2))$.
It may be noted that the expansion (\ref{to-expansion}) can be formally written as the exponential map
\begin{equation}\label{fedo-expmap}
 \widetilde{\omega}_{\mu\nu}(y) = \Bigl( \exp_p(v) \omega(0) \Bigr)_{\mu\nu}.
\end{equation}
There exists an analogue of the jet isomorphism theorem for a Fedosov manifold (see, for example,
Theorem 5.11 in \cite{fedo-man} and also \cite{tamarkin}) stating that any local invariant
of a Fedosov manifold is a function of $\omega_{\mu\nu}$ and a finite number of covariant derivatives
of its curvature tensor $\mathfrak{R}_{\mu\nu\rho\sigma} = \omega_{\mu\lambda} {\mathfrak{R}^\lambda}_{\nu\rho\sigma}$ which does not depend on the choice of
local coordinates.\footnote{As is well-known,
a symplectic manifold does not admit local invariants such as curvature due to the famous Darboux theorem.
The dimension is the only local invariant of symplectic manifolds up to symplectomorphisms.
But, by introducing the concept of the symplectic connection, it is now possible to construct
curvature tensors and their covariant derivatives. So the symplectic connection corresponds
to the operation to make a Darboux chart be infinitesimally small. This ``infinitesimal" approach
will bring about another benefit to bypass the need of gluing together star-products defined
on (large) Darboux charts.} Note that, for a Fedosov manifold, the curvature tensor (\ref{symp-curv})
has the following symmetry property \cite{fedo-man}
\begin{equation}\label{fedo-curvature}
 \mathfrak{R}_{\mu\nu\rho\sigma} = - \mathfrak{R}_{\mu\nu\sigma\rho}, \qquad
 \mathfrak{R}_{\mu\nu\rho\sigma} = \mathfrak{R}_{\nu\mu\rho\sigma}
\end{equation}
that is slightly different from Riemannian manifolds. Using several identities for the curvature tensors
(e.g., Proposition 5.2, Lemma 5.14 and Theorem 5.18 in \cite{fedo-man}), it can be shown that
the Taylor expansion (\ref{to-expansion}) starts as follows:
\begin{eqnarray}\label{exp-expansion}
 \widetilde{\omega}_{\mu\nu}(y) &=& B_{\mu\nu} + \frac{1}{6} \mathfrak{R}_{\rho\sigma\mu\nu} y^\rho y^\sigma
 - \frac{1}{12} \bigl(\partial^S_\lambda \mathfrak{R}_{\rho\mu\nu\sigma}
 - \partial^S_\lambda \mathfrak{R}_{\rho\nu\mu\sigma} \bigr) y^\rho y^\sigma y^\lambda + \mathcal{O}(y^4).
\end{eqnarray}
It is easy to invert the above result to yield the corresponding Poisson bivector $\widetilde{\theta}^{\mu\nu}(y) =  \bigl (\widetilde{\omega}^{-1} \bigr)^{\mu\nu}(y)$:
\begin{eqnarray}\label{poisson-expansion}
 \widetilde{\theta}^{\mu\nu}(y) &=& \theta^{\mu\nu} - \frac{1}{6} \mathfrak{R}_{\rho\sigma\alpha\beta}
 \theta^{\mu\alpha} \theta^{\beta\nu} y^\rho y^\sigma \nonumber \\
&&  + \frac{1}{12} \bigl(\partial^S_\lambda \mathfrak{R}_{\rho\alpha\beta\sigma}
 - \partial^S_\lambda \mathfrak{R}_{\rho\beta\alpha\sigma} \bigr) \theta^{\mu\alpha} \theta^{\beta\nu}
 y^\rho y^\sigma y^\lambda + \mathcal{O}(y^4).
\end{eqnarray}
Using the first Bianchi identity $\mathfrak{R}_{\mu(\nu\rho\sigma)} = 0$,\footnote{We use the bracket notation
for symmetrization and antisymmetrization over tensor indices:
$X_{(\mu\nu\rho)} \equiv X_{\mu\nu\rho} + X_{\nu\rho\mu} + X_{\rho\mu\nu}$ and $X^{[\mu\nu]}
\equiv X^{\mu\nu} - X^{\nu\mu}$.} the second term in eq. (\ref{poisson-expansion})
can be rewritten as $\mathfrak{R}_{\rho\sigma\alpha\beta} \theta^{\mu\alpha} \theta^{\beta\nu}
y^\rho y^\sigma = {\mathfrak{R}^{[\mu}}_{\rho\lambda\sigma} \theta^{\nu]\lambda} y^\rho y^\sigma$, 
so recovers eq. (3.9) in Ref. \cite{cftp}.
Similarly we can consider the Taylor expansion of a vector field $v_a = v_a^\mu(x) \partial_\mu \in \Gamma(TM)$ in terms of normal coordinates. The leading order terms are given by\footnote{Note that $\widetilde{v}_a
\in \Gamma(TM)$ in eq. (\ref{global-vector}) are (locally) Hamiltonian vector fields, i.e., $\widetilde{v}_a
= d_\theta C_a = \partial_a + \cdots$. Since they will be identified with eq. (\ref{vec-expansion})
by definition, the vector fields $\widetilde{v}_a$ describe the mutation from the flat basis,
i.e., $v^\mu_a = \delta^\mu_a$.}
\begin{eqnarray}\label{vec-expansion}
 \widetilde{v}_a^{\mu}(y) &=& v^{\mu}_a  + (\partial^S_\nu v^{\mu}_a) y^\nu
 + \frac{1}{2} \Bigl( \partial^S_\rho \partial^S_\sigma v^{\mu}_a
 + \frac{1}{3} {\mathfrak{R}^\mu}_{\rho\lambda\sigma} v^\lambda_a \Bigr) y^\rho y^\sigma \nonumber \\
 && + \frac{1}{6} \Bigl( \partial^S_\nu \partial^S_\rho \partial^S_\sigma v^{\mu}_a
+ {\mathfrak{R}^\mu}_{\nu\lambda\rho} \partial^S_\sigma v^\lambda_a
+ \frac{1}{6} \partial^S_\nu {\mathfrak{R}^\mu}_{\rho\lambda\sigma} v^\lambda_a \Bigr)
 y^\nu y^\rho y^\sigma + \mathcal{O}(y^4).
\end{eqnarray}
In order to derive the above result, we used the identities (4.9) and (4.10) in Ref. \cite{fedo-man} and
the relation $\partial_\nu {\Gamma^\mu}_{\rho\sigma} = \frac{1}{3} \bigl( {\mathfrak{R}^\mu}_{\rho\nu\sigma}
+ {\mathfrak{R}^\mu}_{\sigma \nu \rho} \bigr)$ in the geodesic coordinates obeying (\ref{normal-cond})
which is also true in Riemannian geometry.

Consider a Fedosov manifold $(M, \Omega, \partial^S)$ where $\Omega = \frac{1}{2}
\Omega_{\mu\nu} (x) dx^\mu \wedge dx^\nu \in \Gamma(\Lambda^2 T^* M)$ is a globally defined symplectic
two-form and $\partial^S$ is a symplectic connection, i.e., $\partial^S \Omega = 0$.
Also introduce a complete set of global vector fields $V_a = V_a^\mu (x) \partial_\mu \in \Gamma (TM),
\; a=1, \cdots, d$. Since a Fedosov manifold $(M, \Omega, \partial^S)$ has the connection $\partial^S$,
it is possible to construct local invariants of the Fedosov manifold,
e.g. curvature tensors and their covariant derivatives. It was shown in \cite{fedo-man} (see, in particular, Theorem 5.11) and \cite{tamarkin} that any local invariant of a Fedosov manifold is an appropriate function of
the components of $\Omega$ and of the covariant derivatives of the curvature tensor.
The above Taylor expansions in terms of normal coordinates exhibit
such local invariants at lowest orders. Therefore we can calculate local invariants at a point $p \in M$
determined by the symplectic two-form $\Omega$ and global vector fields $V_a$ and compare them with
those determined by $\widetilde{\omega}_{\mu\nu}(x;y)$ and $\widetilde{v}_a^\mu (x;y)$
on a geodesic extension of Darboux section. But there is an ambiguity coming from the symplectic connection.
Unlike the Riemannian connection, the symplectic connection is not unique. Any two symplectic connections
differ by a completely symmetric tensor $S_{\mu\nu\lambda}$. See, for example, section 2.5 in \cite{fedo-book}.
So we may impose an additional condition requiring $S_{\mu\nu\lambda} = \Gamma_{(\mu\nu\lambda)} = 0$.
Then the symplectic connection is uniquely determined by $\Omega_{\mu\nu}$ as
\begin{equation}\label{symp-csymbol}
    \Gamma_{\lambda\mu\nu} = \frac{1}{3}\bigl( \partial_\mu \Omega_{\lambda\nu} + \partial_\nu \Omega_{\lambda\mu} \bigr).
\end{equation}
Note that this choice is compatible with the geodesic condition (\ref{normal-cond})
because $\Gamma_{\lambda\mu\nu}$ are completely symmetric in Darboux coordinates.
So we can consistently implement the following identification
\begin{equation}\label{id-symp}
 \widetilde{\omega}_{\mu\nu} \cong \Omega_{\mu\nu}  \qquad \mathrm{and} \qquad
 \widetilde{v}_a^{\mu} \cong V_a^{\mu}
\end{equation}
if their local invariants at $p \in M$ coincide each other up to any arbitrary order.
It should be globally well-defined because the Taylor expansion is independent of the choice
of local coordinates. And the prescription (\ref{id-symp}) simply means the passage from local
to global objects by gluing together the local data on the left-hand side.
This prescription for the globalization constitutes a symplectic counterpart
of the Riemannian case (\ref{id-two}).

A standard mathematical device for patching the local information together to obtain a global theory is
to use the notion of formal geometry \cite{cft,cftp}. Formal geometry provides a convenient language
to describe the global behavior of objects defined locally in terms of coordinates.
Now we will explain how the above prescription (\ref{id-symp}) can be obtained by introducing formal local
coordinates defined by a smooth map $\phi: \mathcal{U} \to M$ from a neighborhood $\mathcal{U}$ of
the zero section of $TM$ to $M$. The smooth map, $(x, y \in \mathcal{U}_x) \mapsto
\phi_x (y)$, is called a generalized exponential map if $\phi_x(0) = x$ and $d_y \phi_x (0)
= \mathrm{id}, \; \forall x \in M$. Here we shall look at the exponential map for a torsion free
but not necessarily symplectic connection. If $f$ is a smooth function on $M$, we can define
the pullback $\phi^* f := f \circ \phi \in C^\infty(\mathcal{U})$ which satisfies
$d(\phi^* f)= df \circ d\phi$. Since we are interested in the Taylor expansion of $\phi_x^* f(y)$
at $y = 0 \in \mathcal{U}$ which will be denoted
by $f_\phi(x;y)$,\footnote{\label{symm-cot}In \cite{bcm-jhep},
it was denoted by $T\phi_x^* f \in \widehat{S} T_x^* M$ where $T$ means the Taylor expansion
in the $y \in \mathcal{U}_x$-variables around $y=0$ and $\widehat{S}$ denotes the
formal completion of the symmetric algebra. The bundle $\widehat{S} T^* M$ of formally completed
symmetric algebra of the cotangent bundle $T^*M$ is defined as a jet bundle whose sections are
given by eq. (\ref{jet-section}). Instead we will denote this bundle by $\mathcal{E}$ according to \cite{cftp}.}
we define an equivalence relation for two generalized exponential maps, $ \phi \sim \psi $,
if all partial derivatives of $\phi_x$ and $\psi_x$ at $y=0$ coincide.
A formal exponential map is an equivalence class of such maps.
Choosing local coordinates $\{x^\mu \}$ on the base and $\{y^\mu \}$ on the fiber,
we can write such a formal exponential map generated by a tangent vector
$v = y^1 \frac{\partial}{\partial x^1} + \cdots + y^d \frac{\partial}{\partial x^d} \in T_p M$
as a formal power series
\begin{equation}\label{reformal}
    \bigl(\exp_p (v) \phi \bigr)^\mu := \phi^\mu_x(y) =
    x^\mu + \sum_{n=1}^\infty \frac{1}{n!} \phi^\mu_{x,\lambda_1 \lambda_2
    \cdots \lambda_n} y^{\lambda_1} y^{\lambda_2} \cdots y^{\lambda_n}
\end{equation}
that depends smoothly on $x \in M$. The coefficients in the exponential map (\ref{reformal})
can be determined using the geodesic flow of a torsion free connection defined by
\begin{equation}\label{g-eq-exp}
    \ddot{\Phi}^\mu_x + {\Gamma^\mu}_{\rho\sigma} (\Phi_x) \dot{\Phi}_x^\rho \dot{\Phi}_x^\sigma = 0
\end{equation}
where $\Phi_x(t, y), \; t \in [0,1]$, is a formal curve with initial conditions $\Phi_x(0, y) = x$
and $\dot{\Phi}_x(0, y) = y$. It is easy to show that the required formal exponential
map $\phi_x(y) = \Phi_x(1, y)$ is given in local coordinates by
\begin{equation}\label{formal-emap}
   \phi^\mu_x(y) = x^\mu + y^\mu - \frac{1}{2} {\Gamma^\mu}_{\rho\sigma} (x) y^{\rho} y^{\sigma}
   - \frac{1}{3!} \Bigl( \partial_\lambda {\Gamma^\mu}_{\rho\sigma} (x)
   - 2 {\Gamma^\mu}_{\nu\lambda}(x) {\Gamma^\nu}_{\rho\sigma} (x) \Bigr)
      y^{\rho} y^{\sigma} y^{\lambda} + \cdots.
\end{equation}
By putting $\phi^\mu_x(y)$ at the origin, i.e., $x=0$, we note the similarity with the exponential
map (\ref{cov-obj}) determined by the Moser flow (\ref{t-evolution}).
We will see later that they are related to each other.

It is now straightforward to consider the Taylor expansion of the pullback $f_\phi(x;y) =
f \bigl(\phi_x(y) \bigr)$ of a smooth function $f \in C^\infty(M)$ via the formal exponential
map $\phi: \mathcal{U} \to M$. We write
\begin{equation}\label{formal-fmap}
 f_\phi(x;y) = f(x) + \sum_{n=1}^\infty \frac{1}{n!} f^{(n)}_{\phi,\lambda_1 \lambda_2
    \cdots \lambda_n} (x) y^{\lambda_1} y^{\lambda_2} \cdots y^{\lambda_n}
\end{equation}
where the coefficient $f_\phi^{(n)}$ is a covariant symmetric tensor
of rank $n$ and smoothly depends on $x \in M$. It turns out \cite{cftp,acgutt} that $f_\phi$ is
a particular example of a section of the jet bundle $\mathcal{E} \to M$ (where $\mathcal{E}$ is
the bundle $F(M) \times_{GL(d, \mathbb{R})} \mathbb{R}[[y^1, \cdots, y^d]]$ associated
to the frame bundle $F(M)$ on $M$) with the fiber $\mathbb{R}[[y^1, \cdots, y^d]]$
(i.e., formal power series in $y$ with real coefficients) and transition functions induced
from the transition functions of $TM$. In general any section of $\mathcal{E}$ is of the form
\begin{equation}\label{jet-section}
    \sigma(x;y) = \sum_{n=0}^\infty \frac{1}{n!} a^{(n)}_{\lambda_1 \lambda_2
    \cdots \lambda_n} (x) y^{\lambda_1} y^{\lambda_2} \cdots y^{\lambda_n}
\end{equation}
where $a^{(n)}_{\lambda_1 \lambda_2 \cdots \lambda_n}$ define covariant tensors on $M$.
In this way the variables $y^\lambda$ may be thought of as formal coordinates on the fibers of
the tangent bundle $TM$. Also recall \cite{cftp} that a section $\sigma$ of the jet
bundle $\mathcal{E}$ is the pullback of a function, i.e. $\sigma = f_\phi$ if and only if
\begin{equation}\label{flat-jsection}
    D^{(0)}_X \sigma = 0, \qquad \forall X \in \Gamma(TM)
\end{equation}
where $D^{(0)}_X$ is the differential operator given by
\begin{equation}\label{diff-jop}
    D^{(0)}_X = X - X^\mu(x) \frac{\partial \phi_x^\nu}{\partial x^\mu}
    \Bigl[\Bigl(\frac{\partial \phi_x}{\partial y} \Bigr)^{-1} \Bigr]^\lambda_\nu
    \frac{\partial}{\partial y^\lambda} =: X + \widetilde{X}.
\end{equation}
It is easy, using the expansion (\ref{formal-emap}), to yield the inverse of the Jacobian matrix
$\bigl( \frac{\partial \phi_x^\mu}{\partial y^\lambda} \bigr)$ which is given by
\begin{equation}\label{inv-texp}
 \Bigl[\Bigl(\frac{\partial \phi_x}{\partial y} \Bigr)^{-1} \Bigr]^\lambda_\nu = \delta_\nu^\lambda
 + {\Gamma^\lambda}_{\nu\rho} y^\rho + \Bigl( \frac{1}{2} \partial_{\nu} {\Gamma^\lambda}_{\rho\sigma}
 - \frac{1}{3} {\mathfrak{R}^\lambda}_{\rho\nu\sigma} \Bigr) y^\rho y^\sigma + \cdots.
\end{equation}
The property (\ref{flat-jsection}) is simply a result of the chain rule
for the section $\sigma = f \circ \phi$. By observing that
\begin{equation}\label{g-flow}
 D^{(0)}_X \sigma (x;y) = \frac{d}{dt}|_{t=0}  \sigma \Bigl( x(t); \phi_{x(t)}^{-1}
 \bigl(\phi_x(y) \bigr) \Bigr)
\end{equation}
for any curve $t \mapsto x(t) \in M$ such that $x(0) = x$ and $\dot{x}(0) = X \in T_x M$,
it can be proven \cite{acgutt} that $[D^{(0)}_X, D^{(0)}_Y] = D^{(0)}_{[X,Y]}$. See also appendix C.
Its immediate consequence is that the covariant derivative $D^{(0)}=dx^\mu D^{(0)}_\mu: \Gamma(\mathcal{E})
\to \Omega^1(\mathcal{E}, M)$ defines a flat connection, i.e., $\bigl(D^{(0)} \bigr)^2 = 0$.
This is also called the Grothendieck connection.

Let us write the flat connection $D^{(0)}=dx^\mu D^{(0)}_\mu$ as the form
\begin{equation}\label{flat-conn}
    D^{(0)}_\mu = \frac{\partial}{\partial x^\mu} - R^\lambda_\mu (x;y) \frac{\partial}{\partial y^\lambda}
\end{equation}
where
\begin{equation}\label{r-conn}
  R^\lambda_\mu (x;y) \equiv \frac{\partial \phi_x^\nu}{\partial x^\mu}
    \Bigl[\Bigl(\frac{\partial \phi_x}{\partial y} \Bigr)^{-1} \Bigr]^\lambda_\nu
\end{equation}
is a formal power series in $y$ which begins with $\delta^\lambda_\mu$ and whose coefficients
are smooth in $x$. By these properties it immediately follows \cite{cftp,acgutt} that a section of
the jet bundle $\mathcal{E}$ is the Taylor expansion of a globally defined function if and only if
it is $D^{(0)}$-closed. Obviously $D^{(0)}$ is a derivation of the usual product of
sections of $\mathcal{E}$, i.e. $D^{(0)}(\sigma \tau)= \bigl( D^{(0)} \sigma \bigr) \tau
+ \sigma D^{(0)} \tau$ for $\sigma, \tau \in \Gamma(\mathcal{E})$. Thus the algebra of global functions
on $M$ can be identified with the subalgebra of $D^{(0)}$-closed sections.
A differential form with values in $\mathcal{E}$ is a section of the bundle $\mathcal{E} \otimes
\Lambda^m T^* M$, which can be expressed locally as
\begin{equation}\label{jet-mform}
    \Sigma(x;y) = \sum_{n=0}^\infty \frac{1}{m! n!} a^{(m,n)}_{\rho_1\rho_2 \cdots \rho_m \lambda_1 \lambda_2
    \cdots \lambda_n} y^{\lambda_1} y^{\lambda_2} \cdots y^{\lambda_n}
    dx^{\rho_1} \wedge dx^{\rho_2} \wedge \cdots \wedge dx^{\rho_m}.
\end{equation}
It is useful to define the total degree of a form on $M$ taking values in sections of $\mathcal{E}$ as
the sum of the form degree and the degree in $y$ and then to decompose the Grothendieck connection (\ref{flat-conn}) in the following way
\begin{equation}\label{dec-grothendieck}
    D^{(0)} = - \delta + d^S + \texttt{A}
\end{equation}
where
\begin{equation}\label{deg-0}
    \delta \equiv dx^\mu \frac{\partial}{\partial y^\mu}
\end{equation}
is the zero-degree part and
\begin{equation}\label{deg-1}
  d^S \equiv dx^\mu \Bigl( \frac{\partial}{\partial x^\mu} -  {\Gamma^\lambda}_{\mu\nu} y^\nu \frac{\partial}{\partial y^\lambda} \Bigr)
\end{equation}
is the degree-one part and finally
\begin{equation}\label{deg-high}
     \texttt{A} \equiv dx^\mu A_\mu^\lambda(x;y) \frac{\partial}{\partial y^\lambda}
     = dx^\mu \Bigl(-\frac{1}{3} {\mathfrak{R}^\lambda}_{\rho\mu\sigma} y^\rho y^\sigma
     + \mathcal{O}(y^3) \Bigr) \frac{\partial}{\partial y^\lambda}
\end{equation}
is at least of second degree in $y$. The requirement of the vanishing of the curvature
$\bigl(D^{(0)} \bigr)^2 \equiv \Upsilon$ yields the condition
\begin{equation}\label{curvature-free}
 0 = \Upsilon = - {\mathfrak{R}^\mu}_\nu y^\nu \frac{\partial}{\partial y^\mu}
 + \digamma - \delta \texttt{A}
\end{equation}
with ${\mathfrak{R}^\mu}_\nu = \frac{1}{2} {\mathfrak{R}^\mu}_{\nu\rho\sigma} dx^\rho \wedge dx^\sigma$
and $\digamma = d^S \texttt{A} +\texttt{A}^2$.

Define the ``inverse" operator of $\delta$ by
\begin{equation}\label{inverse-delta}
    \delta^{-1} \Sigma_{(n)} = \frac{1}{m+n} \iota_v \Sigma_{(n)}
\end{equation}
when $m+n>0$ and $\delta^{-1} \Sigma_{(n)} = 0$ when $m+n=0$, where $\Sigma_{(n)}$
is a monomial with degree $m+n$ in eq. (\ref{jet-mform}) and $v = y^\mu \frac{\partial}{\partial x^\mu}
\in T_p M$. Then there is a Hodge-decomposition \cite{d-quant2,fedo-book} that any form $\Sigma \in \Gamma(\mathcal{E}) \otimes \Lambda^* M$ has the representation
\begin{equation}\label{hodge-dec}
   \Sigma = \delta \delta^{-1} \Sigma + \delta^{-1}\delta  \Sigma + a^{(0,0)}
\end{equation}
where $a^{(0,0)}$ is a function on $M$ (independent of $y$) in eq. (\ref{jet-mform}).
Since $\delta^2=0$, the decomposition (\ref{hodge-dec}) means that the cohomology of $\delta$
consists of zero forms constant in $y$. Note that, at least at leading order,
\begin{equation}\label{zerodel-inv}
  \delta^{-1} \texttt{A} = 0
\end{equation}
but it can be proven that it is generally true. Since $\mathrm{deg}\; (\texttt{A}) \geq 2$ and
so $\texttt{A}^{(0,0)} =0$, the Hodge decomposition (\ref{hodge-dec}) together with
eqs. (\ref{curvature-free}) and (\ref{zerodel-inv}) leads to the relation
\begin{equation}\label{hodge-sol}
   \texttt{A} = - \delta^{-1} \mathfrak{R} + \delta^{-1} \digamma
\end{equation}
where $\mathfrak{R} \equiv {\mathfrak{R}^\mu}_\nu y^\nu \frac{\partial}{\partial y^\mu}$.
By cohomological perturbation theory, it is not difficult to prove \cite{d-quant2,fedo-book} that
the cohomology of $D^{(0)}$ for the bundle $\mathcal{E} \otimes \Lambda^* M$ is almost trivial
and concentrated in degree 0, i.e., functions $a^{(0,0)}(x)$ in eq. (\ref{jet-mform}).
This fact will be important later for the global version of deformation
quantization for Poisson manifolds as well as our construction of global vector fields.

The above Taylor expansion can be generalized to a polyvector field $\Xi \in \mathcal{V}^k(M)$
using the exponential map $\phi_x$ again. Consider the push-forward $(\phi_x)^{-1}_* \Xi$
of a $k$-vector field $\Xi$ defined on $M$. Its Taylor expansion denoted by $\Xi_\phi$ becomes a
formal multivector field in $y$ for any $x \in M$. For example, if $X$ is a vector field on $M$,
then we get the coefficients
\begin{equation}\label{formal-vector}
    X_\phi^\mu(x; y) = X^\lambda \bigl(\phi_x(y) \bigr)
    \Bigl[\Bigl(\frac{\partial \phi_x}{\partial y} \Bigr)^{-1} \Bigr]_\lambda^\mu.
\end{equation}
The result is exactly the same as eq. (\ref{vec-expansion}) with the replacement $v_a^\mu \to X^\mu$.
Similarly, for a Poisson bivector $\Pi \in \mathcal{V}^2(M)$, the corresponding Taylor
expansion is given by
\begin{equation}\label{f-poisson-exp}
 \Pi_\phi^{\mu\nu}(x;y) = \Pi^{\mu\nu}(x) + \partial^S_\lambda \Pi^{\mu\nu}(x) y^\lambda
 + \frac{1}{2} \Bigl(\partial^S_\rho \partial^S_\sigma \Pi^{\mu\nu}(x) - \frac{1}{3} {\mathfrak{R}^{[\mu}}_{\rho\lambda\sigma} \Pi^{\nu] \lambda} (x) \Bigr) y^\rho y^\sigma + \cdots.
\end{equation}
Note that the above result coincides with eq. (\ref{poisson-expansion}) when $\Pi^{\mu\nu}$ are constants
in a Darboux frame obeying eq. (\ref{normal-cond}). Recall that
the tangent bundle of a manifold is an example of a vector bundle.
Thus, given a Poisson manifold $(M, \Pi)$, the Poisson structure on $M$ induces a Poisson structure
on each fiber of the tangent bundle $TM$, so each tangent space $T_x M$ for any $x \in M$
can be considered as an affine space with the fiberwise Poisson structure. In this way,
the tangent bundle $TM$ becomes a Poisson manifold with the fiberwise Poisson bracket.
In particular, for a symplectic manifold $(M, \Omega)$, the tangent space $T_x M$ becomes
a symplectic vector space equipped with a constant symplectic structure.
Since $\Pi_\phi = (\phi_x)^{-1}_* \Pi \in \mathcal{V}^2 (M)$ and the push-forward is a Poisson map,
if $\Pi$ is Poisson, so is $\Pi_\phi$.
Therefore we will regard the bivector $\Pi_\phi$ as an induced Poisson
structure on $\mathcal{U}_x \subset TM$.
Choosing local coordinates $(x, y)$ for $\mathcal{U}_x$, its expression is locally given by
\begin{equation}\label{local-poi}
\Pi_\phi  = \frac{1}{2} \Pi_\phi^{\mu\nu}(x;y) \frac{\partial}{\partial y^\mu}
\bigwedge \frac{\partial}{\partial y^\nu}.
\end{equation}
The local fiber coordinates $\{ y^\mu \}_{\mathcal{U}_x}$ by construction will be given by
the Darboux-Weinstein coordinates on $\mathcal{U}_x$ such that the coefficients $\Pi_\phi^{\mu\nu}(x;y)$
take the simplest form (\ref{wd-poisson}), so they become constants (independent of $y$
but $x$-dependent) for a symplectic vector space.

We introduce the Poisson bracket on sections of $\mathcal{E}$ by
\begin{equation}\label{epoi-bracket}
    \{ \sigma, \tau \}_{\Pi_\phi} (x;y) = \Pi_\phi^{\mu\nu}(x;y)
    \frac{\partial \sigma(x;y)}{\partial y^\mu}\frac{\partial \tau(x;y)}{\partial y^\nu}
\end{equation}
for $\tau, \sigma \in \Gamma(\mathcal{E})$. Since the set of flat sections obeying eq. (\ref{flat-jsection}),
denoted by $\ker D^{(0)}$, forms a subalgebra, we can restrict the Poisson bracket (\ref{epoi-bracket})
to $\ker D^{(0)}$. Using the one-to-one correspondence between $C^\infty(M)$
and $\ker D^{(0)}$, we identify $\sigma = f_\phi = f \circ \phi$ and $\tau = g_\phi = g \circ \phi$.
Then it is straightforward, using the chain rule \cite{cftp}
\begin{equation}\label{schain-rule}
\frac{\partial f_\phi}{\partial x^\lambda} =
    \frac{\partial f}{\partial x^\mu}\frac{\partial \phi^\mu_x}{\partial x^\lambda}, \qquad
    \frac{\partial f_\phi}{\partial y^\lambda} =
    \frac{\partial f}{\partial x^\mu}\frac{\partial \phi^\mu_x}{\partial y^\lambda},
\end{equation}
to show that
\begin{equation}\label{iso-poisson}
 \{ f_\phi, g_\phi \}_{\Pi_\phi} (x;y) = \{f, g \}_\Pi \bigl( \phi_x(y) \bigr)
\end{equation}
where the right-hand side is the Poisson bracket of global functions on $M$.
Similarly, using eq. (\ref{formal-vector}), one can show that
\begin{equation}\label{iso-vector}
    X_\phi (f_\phi) = \bigl( Xf \bigr)_\phi
\end{equation}
for a global vector field $X \in \Gamma(TM)$ and its push-forward $X_\phi := (\phi_x)^{-1}_* X$.
Observe that by assumption $d_y \phi_x (0) = \mathrm{id}$ we can recover the global objects
$X \in \Gamma(TM)$ and $\Pi \in \mathcal{V}^2(M)$ from $X_\phi$ and $\Pi_\phi$, respectively,
by evaluating their components at $y=0$ and replacing formally each $\frac{\partial}{ \partial y^\mu}$
by $\frac{\partial}{ \partial x^\mu}$ as one can explicitly see from eqs. (\ref{formal-vector}) and
(\ref{f-poisson-exp}). In general, if $\Xi_\phi$ and $\Omega_\phi$ are in the image of $T\phi^*$
(see the footnote \ref{symm-cot}) of a polyvector field $\Xi \in \mathcal{V}^k(M)$ and
a $k$-form $\Omega \in \Gamma(\Lambda^k T^* M)$, respectively, i.e.,
\begin{equation}\label{image-fexp}
\Xi_\phi  =  (\phi_x)^{-1}_* \Xi, \qquad \Omega_\phi  =  (\phi_x)^* \Omega,
\end{equation}
it is enough to evaluate the components of $\Xi_\phi$ and $\Omega_\phi$ at $y=0$ and to replace
formally each $dy^\mu$ by $dx^\mu$ and each $\frac{\partial}{ \partial y^\mu}$
by $\frac{\partial}{ \partial x^\mu}$ in order to recover the global objects
$\Xi$ and $\Omega$ \cite{bcm-jhep}. More explicitly, if $\Xi_\phi(x;y) = \Xi^{\mu_1
\cdots \mu_k} (x;y) \frac{\partial}{\partial y^{\mu_1}} \bigwedge \cdots \bigwedge
\frac{\partial}{\partial y^{\mu_k}}$ is equal to $T\phi^* \Xi = T \phi^{-1}_* \Xi$, then, in local coordinates,
\begin{equation}\label{global-xi}
    \Xi(x) = \Xi^{\mu_1 \cdots \mu_k} (x;0) \frac{\partial}{\partial x^{\mu_1}} \bigwedge \cdots \bigwedge
\frac{\partial}{\partial x^{\mu_k}},
\end{equation}
and, if $\Omega_\phi(x;y) = \Omega_{\mu_1 \cdots \mu_k} (x;y) dy^{\mu_1} \wedge \cdots \wedge
dy^{\mu_k}$ is equal to $T\phi^* \Omega$, then
\begin{equation}\label{global-omega}
    \Omega(x) = \Omega_{\mu_1 \cdots \mu_k} (x;0) dx^{\mu_1} \wedge \cdots \wedge dx^{\mu_k}.
\end{equation}

Let us summarize how we deal with global objects. Let us focus on the symplectic case for simplicity.
At the outset we prepare the system of multivector fields $\{\Xi \in \mathcal{V}^k(M): k=0,1, \cdots, d\}$
at a point $p \in M$ whose local coordinates are $x$. Then we develop the system along a geodesic curve
described by eq. (\ref{g-eq-exp}). We extend the system such that it goes into a Darboux frame (\ref{weinstein})
at the end point of the geodesic flow whose coordinates are $\phi_x(y)$.
We denote the system in the Darboux frame as $\{\Xi_\phi = (\phi_x)^{-1}_* \Xi \in \mathcal{V}_\phi^k (\mathcal{U}) : k=0,1, \cdots, d\}$. In particular, we can construct the Grothendieck connection
for an infinite jet bundle $\mathcal{E}$ of functions using the formal coordinates $\phi_x(y)$.
This connection allows us to identify smooth functions on $M$ with flat (or integrable) sections
of the jet bundle $\mathcal{E}$. But the Darboux frame is malleable because the torsion free connection
will vanish there, so it can be further extended using normal coordinates as we discussed before.
Indeed this extension corresponds to the situation that the system is initially prepared in
the Darboux frame, so the exponential map is given by $\phi_x(y) = x + y$.\footnote{\label{foot-27}Note
that the exponential map in eq. (\ref{fedo-expmap}) can be identified with $\phi_x(y)$, i.e.,
$\exp_p(v) =\phi_x(y) = x + y$ where the point $p$ was taken to be the origin, $x=0$.}
That is the reason why we get a parallel result with the normal coordinate system.
In this case we attribute the infinitesimal development on a Darboux chart to U(1) gauge fields
as applied in eq. (\ref{to-expansion}). Hence we have the relation $(\phi_x)^{-1}_* \widetilde{\theta}
= \phi_x^* \widetilde{\theta} = \theta$ since $\phi_x: \mathcal{U}_x \to M$ is a diffeomorphism, 
so we can identify the exponential map $\phi_x^*$ with the Moser flow (\ref{t-flow}).
Eventually we will quantize a symplectic manifold $(M, \Omega)$ in the Darboux
frame where the Poisson bivector $\Pi_\phi \in \mathcal{V}_\phi^2$ takes the simplest form.
See the footnote \ref{minimal-theta} for the advantage of this frame.

The previous identification (\ref{id-symp}) can now be well founded on this global approach.
Consider two local Darboux charts $(U_1, \phi_1)$ and $(U_2, \phi_2)$
such that $U_1 \cap U_2 \neq \emptyset$ and $\phi_i: \mathcal{U}_i \to M$ is a formal exponential
map given by (\ref{formal-emap}) on $U_i \subset M$ for $i=1,2$.
On each chart the Poisson structure is defined from the global Poisson structure
$\Pi \equiv \Omega^{-1} \in \mathcal{V}^2(M)$ with its own exponential map:
$\Pi_{\phi_1} = (\phi_{1x})^{-1}_* \Pi$ and $\Pi_{\phi_2} = (\phi_{2x})^{-1}_* \Pi$.
On an overlap $U_{21} = U_1 \cap U_2$, they are definitely related to each other by
\begin{equation}\label{overlap-poisson}
 \Pi_{\phi_2} =   \bigl( \phi_{21x} \bigr)^{-1}_* \Pi_{\phi_1}
\end{equation}
where $\phi_{21x} \equiv \phi_{2x} \circ \phi^{-1}_{1x}: \mathcal{U}_x \to U_{21}$.
Note that the exponential map $\phi_{21}$ is a diffeomorphism between nearby Darboux charts, 
so it can be generated by a normal coordinate system. As a result, two Poisson structures must 
be related to each other according to eq. (\ref{poisson-expansion}), 
so the exponential map $\phi_{21}$ will be of the form (\ref{t-flow}).
This gluing procedure was described in \cite{jsw-ncl}. Similarly, for the exponential maps
obeying eq. (\ref{iso-vector}) on each Darboux chart, we have $X_{\phi_1} (f_{\phi_1})
= \bigl( Xf \bigr)_{\phi_1}$ and $X_{\phi_2} (f_{\phi_2}) = \bigl( Xf \bigr)_{\phi_2}$.
Thus, on the intersection $U_{21} = U_1 \cap U_2$, the gluing condition for vector fields
on local charts is given by
\begin{equation}\label{overlap-vector}
 X_{\phi_2} =   \bigl( \phi_{21x} \bigr)^{-1}_* X_{\phi_1}.
\end{equation}
Note that, if $X_\phi$ is a Hamiltonian vector field on $\mathcal{U}_x$, i.e. $X_\phi := X_{f_\phi}$
for any global function $f$ on $M$, the relation (\ref{iso-vector}) reduces to
\begin{equation}\label{hamiso-vector}
    X_{f_\phi} (g_\phi) = \bigl( X_f g \bigr)_\phi
\end{equation}
which precisely means eq. (\ref{iso-poisson}). Thus we see that a global Hamiltonian vector field
$X_f \in \Gamma(TM)$ is mapped via the formal exponential map to a Hamiltonian vector field $X_\sigma$
of a flat section $\sigma = f_\phi$ on the jet bundle $\mathcal{E}$.
Finally, we can apply the rule (\ref{global-xi}) and (\ref{global-omega}) to identify
the global objects in eq. (\ref{id-symp}):
\begin{equation}\label{global-id}
    \Omega_{\mu\nu} (x) = \bigl( \Pi_\phi^{-1} \bigr)_{\mu\nu} (x;y=0), \qquad
    V^\mu_a (x) = \bigl( V_a \bigr)^\mu_\phi (x; y=0)
\end{equation}
where the set of Hamiltonian vector fields obeys the relation (\ref{hamiso-vector}), i.e.,
$(V_a)_\phi (f_\phi) = \bigl( V_a (f) \bigr)_\phi $.

Now next step is to quantize the symplectic manifold $(M, \Omega)$. Since the emergent gravity
claims that a symplectic manifold $(M, \Omega)$ gives rise to a Riemannian manifold $(M, g)$,
according to the emergent gravity picture, we understand the quantization of the dynamical symplectic
manifold $(M, \Omega)$ as the quantization of the corresponding (emergent) Riemannian manifold $(M, g)$.
Thus the emergent gravity picture suggests a completely new quantization scheme of Riemannian manifolds
where quantum gravity is defined by a dynamical NC spacetime.
For example, U(1) gauge theory on a symplectic manifold can be identified with a Fedosov manifold
which includes any K\"ahler manifolds, so the NC gauge theory corresponds to
the quantization of the Fedosov manifold which should contain ``quantized K\"ahler manifolds".
In order to quantize the manifold $M$, it is important to note that we have an isomorphism of Poisson algebras
\begin{equation}\label{iso-poiss}
    \iota: C^\infty(M) \to \ker D^{(0)}
\end{equation}
from the algebra of smooth functions on $M$ onto the algebra of horizontal sections of $\mathcal{E}$.
Therefore one may try to quantize $(\mathcal{E}, D^{(0)})$ and to identify a subalgebra of the quantized
algebra $\widehat{\mathcal{E}} := \mathcal{E} \bigl[[\hbar]\bigr]$ with the vector space $C^\infty (M) \bigl[[\hbar]\bigr]$ in such a way that the induced multiplication on $\mathcal{A}_\theta
= C^\infty (M) \bigl[[\hbar]\bigr]$ gives a deformation quantization of $M$.
For this program to work, we need the quantum Grothendieck
connection $\mathfrak{D} := \sum_{n=0}^\infty \hbar^n D^{(n)}$ to be a derivation (so that
the space of flat sections of $\widehat{\mathcal{E}}$ becomes a subalgebra) and to be flat
(so that there is no obstruction to the integrability of the horizontal distribution defining
the quantum connection). It is straightforward to quantize the space of sections of
the jet bundle $\mathcal{E}$ which is subject to the Poisson bracket (\ref{epoi-bracket})
with the $y^\mu$ as quantized variables and $x^\mu$ as parameters.
Using the Kontsevich's formality map \cite{cft,cftp}
\begin{equation}\label{kw-formality}
\widehat{\star} \equiv \sum_{n=0}^\infty \frac{\hbar^n}{n!} U_n (\Pi_\phi, \cdots, \Pi_\phi),
\end{equation}
we define the star product for sections $\sigma, \tau \in \Gamma(\mathcal{E})$ by
\begin{equation}\label{sec-star}
\widehat{\star} (\sigma \otimes \tau) = \sigma \, \widehat{\star} \; \tau.
\end{equation}
The star product (\ref{kw-formality}) is basically the Moyal star-product on $T_x M$ and
the pair $(\Gamma(\mathcal{E})\bigl[[\hbar]\bigr], \widehat{\star})$ is known as the Weyl algebra
and will be denoted by $W_x$. The algebras $W_x, \; x \in M$, can be smoothly patched and
we get a fiber bundle $W = \cup_{x \in M} W_x$ on $M$, called the Weyl algebra bundle over $M$.
Hence the Weyl algebra bundle may be thought of as a ``quantum tangent bundle".

Although the covariant derivative (\ref{diff-jop}) is a derivation of the usual product of sections
of $\mathcal{E}$, it is not a derivation of $\widehat{\star}$. So we introduce a quantum covariant
derivative in the direction of $X \in T_x M$ defined by
\begin{equation}\label{q-covx}
    \widetilde{\mathfrak{D}}_X = X + A(\widetilde{X}) = X^\mu (x) \widetilde{\mathfrak{D}}_\mu
\end{equation}
where $\widetilde{X} =  - X^\mu (x) R^\lambda_\mu(x;y) \frac{\partial}{\partial y^\lambda}$ is
given by eq. (\ref{diff-jop}) and the formality map (\ref{series-vec}) for the quantum connection
is defined by
\begin{equation}\label{der-vec-phi}
  A(\widetilde{X}) = \sum_{n=0}^\infty \frac{\hbar^n}{n!} U_{n+1}
  (\widetilde{X}, \Pi_\phi, \cdots, \Pi_\phi).
\end{equation}
Since $A(\widetilde{X}) = \widetilde{X} + \cdots$ and $U_{2} (\widetilde{X}, \Pi_\phi) = 0$
(see eq. (\ref{exp-u2=0}) and comments below),\footnote{\label{foot-x}Furthermore, $U_{n}
  (\xi, \Pi_\phi, \cdots, \Pi_\phi) = 0$ for $n \geq 2$ if $\xi$ is a linear vector field \cite{acgutt}.
  Thus $\widetilde{X}$ in eq. (\ref{q-cov-exp}) can be replaced by $\widetilde{\texttt{X}}
  = \iota_{X} (- \delta + \texttt{A})$.}
\begin{equation}\label{q-cov-exp}
 \widetilde{\mathfrak{D}}_X = D^{(0)}_X + \sum_{n=2}^\infty \frac{\hbar^n}{n!} U_{n+1}
  (\widetilde{X}, \Pi_\phi, \cdots, \Pi_\phi).
\end{equation}
The formality identity (\ref{star-der}) applied to the quantum covariant derivative (\ref{q-covx})
implies the crucial statement that $\widetilde{\mathfrak{D}}$ is a derivation of $\widehat{\star}$
(see Proposition 4.2 in \cite{cftp} for the proof), i.e.,
\begin{equation}\label{der-starhat}
    \widetilde{\mathfrak{D}} (\sigma \, \widehat{\star} \; \tau) = (\widetilde{\mathfrak{D}} \sigma) \, \widehat{\star} \; \tau
    + \sigma \, \widehat{\star} \; (\widetilde{\mathfrak{D}} \tau)
\end{equation}
for $\sigma, \tau \in \Gamma (W)$.

But the quantum connection $\widetilde{\mathfrak{D}} = dx^\mu \widetilde{\mathfrak{D}}_\mu$ is not flat
in general but it has a curvature given by \cite{cftp}
\begin{equation}\label{weyl-curvature}
 \widetilde{\mathfrak{D}}^2 \sigma  = [F^M, \sigma]_{\widehat{\star}}
\end{equation}
where $F^M$ is a 2-form on $M$ with values in the section of $W$ defined using the formality
identity (\ref{star-lie}) as
\begin{equation}\label{weyl-curv-def}
F^M (X, Y) = \Psi(\widetilde{X}, \widetilde{Y}) = \sum_{n=0}^\infty \frac{\hbar^n}{n!} U_{n+2}
  (\widetilde{X}, \widetilde{Y}, \Pi_\phi, \cdots, \Pi_\phi).
\end{equation}
Hence we need to modify $\widetilde{\mathfrak{D}}$ by adding more ``quantum corrections" to have a flat
quantum connection so that the new covariant derivative
\begin{equation}\label{new-qconn}
 \mathfrak{D} = \widetilde{\mathfrak{D}}  + [ \gamma, \cdot \;]_{\widehat{\star}}
\end{equation}
is again a derivation where $\gamma$ is a one-form on $M$ with values in the section of $W$.
Note that $\mathfrak{D}$ is certainly a derivation of $\widehat{\star}$ because the adjoint action $\mathrm{ad}(\gamma)$ in eq. (\ref{new-qconn}) is automatically a derivation of $\widehat{\star}$ and
so we only need to find the one-form $\gamma$ such that the quantum corrected connection $\mathfrak{D}$
is flat, i.e., $\mathfrak{D}^2 = 0$. The flatness condition $\mathfrak{D}^2 = 0$ can be stated
as the form
\begin{equation}\label{flat-fedocurvature}
  G^M \equiv  F^M  + \widetilde{\mathfrak{D}}\gamma + \gamma \widehat{\star} \gamma = \omega
\end{equation}
where $\omega$ is a central element, i.e., $[\omega, \sigma]_{\widehat{\star}} = 0,
\; \forall \sigma \in \Gamma(W)$ and the wedge product between forms was implicitly assumed.
From the definition (\ref{weyl-curvature}), it is obvious that the Bianchi identity,
$\widetilde{\mathfrak{D}} F^M = 0$,
is guaranteed \cite{cftp}. This in turn leads to the Bianchi identity $\mathfrak{D} G^M = 0$.
When we write $\widetilde{\mathfrak{D}}$ in eq. (\ref{q-cov-exp}) as the form $\widetilde{\mathfrak{D}} =  D^{(0)}
+ A'(\widetilde{\texttt{X}})$ (see the footnote \ref{foot-x} for the definition of $\widetilde{\texttt{X}}$) where $A'(\widetilde{\texttt{X}})$ at most starts from $\mathcal{O}(\hbar^2)$,
the curvature $F^M$ is given by
\begin{equation}\label{fedo-curv}
    F^M = D^{(0)} A'(\widetilde{\texttt{X}}) +
    A'(\widetilde{\texttt{X}}) \widehat{\star} A'(\widetilde{\texttt{X}})
\end{equation}
which starts from $\mathcal{O}(\hbar^2)$ too. The connection $\widetilde{\mathfrak{D}}$ with
the above properties is called the Fedosov connection \cite{d-quant2,fedo-book}.

If we are able to find the one-form $\gamma$ so that $G^M = \omega$, then $\mathfrak{D}$-closed
sections, denoted by $\ker \mathfrak{D}$, will form a nontrivial subalgebra of $W$.
A basic observation to determine $\gamma$ from the equation (\ref{flat-fedocurvature}) is that
the $D^{(0)}$-cohomology is trivial \cite{d-quant2,fedo-book}.
The procedure is similar to that to yield eq. (\ref{hodge-sol}).
For this purpose it is convenient to split the Fedosov connection as
the form $\widetilde{\mathfrak{D}} = - \delta + \widetilde{\mathfrak{D}}'$ and then write the equation (\ref{flat-fedocurvature}) as the following form
\begin{equation}\label{fedo-eq}
  \delta \gamma = F^M  - \omega + \widetilde{\mathfrak{D}}' \gamma + \gamma \widehat{\star} \gamma.
\end{equation}
In particular it is enough to choose $\gamma$ so that it starts from $\mathcal{O}(\hbar^2)$
because $F^M$ at most starts from $\mathcal{O}(\hbar^2)$.
Then the one-form $\gamma$ is uniquely determined by the lowest term $F^M-\omega$
using the iteration method with the filtration defined by the grading $\mathrm{deg} (y) =1$
and $\mathrm{deg} (\hbar) =2$. Consequently we have a quantum version of the classical
isomorphism (\ref{iso-poiss}) stating that there is a module isomorphism between $\mathcal{A}_\theta
= C^\infty (M) \bigl[[\hbar]\bigr]$ and $\ker \mathfrak{D}$ such that the star product in
$\mathcal{A}_\theta$ inherits from the star product $\widehat{\star}$ in $\ker \mathfrak{D}$.
More precisely there exists a quantization map $\rho: \Gamma(W) \to \Gamma(W)$ so that
the formal series $\rho = \mathrm{id} + \sum_{n=1}^\infty \hbar^n \rho_n$ obeys the relation
\begin{equation}\label{iso-fedo}
 \mathfrak{D} \rho(\sigma) = \rho \bigl( D^{(0)} \sigma \bigl)
\end{equation}
for every $\sigma \in \Gamma(W)$. The quantization map $\rho$ can be uniquely determined by
solving eq. (\ref{iso-fedo}) using the same iteration method as eq. (\ref{fedo-eq}).
In particular it is easy to show that $\rho_1 = 0$. Therefore the image under $\rho$ of the space of
$D^{(0)}$-flat sections of $W$ is the subalgebra of $\mathfrak{D}$-flat sections of $W$.
Since $\ker D^{(0)}$ is isomorphic to the space of formal series of functions on $M$,
we can finally define a global star product on $M$ by \cite{cft,cftp}
\begin{equation}\label{global-star}
    f \star g = \Bigl[ \rho^{-1} \Bigl( \rho(f_\phi) \widehat{\star} \rho(g_\phi) \Bigr) \Bigr]_{y=0}.
\end{equation}
This constitutes the quantum version of the first part for the globalization (\ref{global-id})
where we regard the left-hand side of eq. (\ref{global-star}) as the star product of
the global Poisson structure $\Pi = \Omega^{-1}$. It recovers the classical result
when $\hbar \to 0$.

Now it is straightforward to prescribe the quantum version of the second part for the globalization (\ref{global-id}). Let $\{\widehat{\mathfrak{C}}_a \in \ker \mathfrak{D}: a=1, \cdots, d\}$
be the set of global $\mathfrak{D}$-flat sections of $W$ which may be obtained from the space
$\ker D^{(0)}$ with the quantization map $\rho$ obeying the relation (\ref{iso-fedo})
or gluing the local covariant momentum variables (\ref{c-momemtum}) {\it \'a la} \cite{jsw-ncl}.
Then the adjoint action
\begin{equation}\label{gmap-derivation}
    \mathrm{ad} (\widehat{\mathfrak{C}}_a) \equiv -i [\widehat{\mathfrak{C}}_a, \; \cdot \;]_{\widehat{\star}}
\end{equation}
defines a derivation of $\ker \mathfrak{D}$ since $\mathrm{ad} (\widehat{\mathfrak{C}}_a)$
definitely satisfies the derivation property and preserves the space $\ker \mathfrak{D}$, i.e.,
$\mathfrak{D} \bigl(\mathrm{ad} (\widehat{\mathfrak{C}}_a)(\sigma) \bigr) = -i[\mathfrak{D}\widehat{\mathfrak{C}}_a, \sigma]_{\widehat{\star}} -i[\widehat{\mathfrak{C}}_a, \mathfrak{D}\sigma]_{\widehat{\star}} = 0$ for any $\sigma \in \ker \mathfrak{D}$.
Actually more is true; it is enough for $\mathfrak{D}\widehat{\mathfrak{C}}_a := \psi_a$ to be central
closed one-forms on $M$, i.e., $[\psi_a, \sigma]_{\widehat{\star}} = 0, \; \forall \sigma
\in \ker \mathfrak{D}$ and $d\psi_a = \mathfrak{D}\psi_a = \mathfrak{D}^2\widehat{\mathfrak{C}}_a = 0$.
(See Appendix A in \cite{pxu} for a succinct summary of derivation algebras from the Fedosov
quantization approach.) The global vector fields $V_a^\star$ can then be obtained by applying
the rule (\ref{global-xi}) to $\mathrm{ad} (\widehat{\mathfrak{C}}_a)$ \cite{cftp}.
Explicitly they are given by
\begin{equation}\label{gmap-did}
    V^\star_a \cong \mathrm{ad} (\widehat{\mathfrak{C}}_a)|_{y=0;
    \frac{\partial}{\partial y^\mu} \to \frac{\partial}{\partial x^\mu}}.
\end{equation}
Together with the globally defined star product (\ref{global-star}), we realize the prescription (\ref{global-vector}) for the globalization of the vector fields $V^\star_a \in \mathfrak{X}$.

In order to complete the globalization, it is also necessary to understand how to
lift the volume preserving condition (\ref{v-preserving}) to quantum vector fields
$V^\star_a \in \mathfrak{X}$. To understand this issue, we need to look at the modular class of
Poisson manifolds in the first Poisson cohomology space of the manifold \cite{mod-wein},
i.e., the equivalence class of Poisson vector fields modulo Hamiltonian vector fields
which is an infinitesimal Poisson automorphism.
We devote appendix B to a brief review of modular vector fields and Poisson homology.
Recall that, on an orientable manifold, there exists a volume form invariant under
all Hamiltonian vector fields if and only if there exists a modular vector field which vanishes.
Such a Poisson manifold is called unimodular \cite{mod-wein}.
It turns out (see appendix B) that it is possible to define a trace as a NC version
of integration and to lift the modular vector fields up to a quantized Poisson manifold if a Poisson
manifold was originally unimodular. Thus it is necessary to restrict to unimodular Poisson manifolds
in order to realize the volume preserving condition (\ref{v-preserving}) even in the quantum level.
It is in no way a restriction to symplectic manifolds since any symplectic manifold
is unimodular \cite{cofs,omy2}.
In fact, the Liouville volume form is invariant under all Hamiltonian vector fields.
However, up to our best knowledge, it is not well understood yet how to define a trace for a general
(non-unimodular) Poisson manifold. Henceforth we will consider only unimodular Poisson manifolds.

Let us write the NC vector field (\ref{global-vector}) as $V^\star_a = V_a + \Xi_a$ where $V_a$
represents a global Hamiltonian vector field defined by eq. (\ref{global-id}) while $\Xi_a$ is
a polydifferential operator comprising derivative corrections due to the NC structure of spacetime.
Definitely the NC vector field $V^\star_a$ represents a NC deformation of the usual vector field $V_a$.
As we observed before, Einstein gravity arises from the vector fields $V_a$ at leading order.
Therefore it should be obvious that the polydifferential operator $\Xi_a$ will generate the derivative
corrections of Einstein gravity. It is thus remained to determine the precise form of the
derivative corrections (which may be a challenging problem). Nevertheless we expect the NC emergent
gravity may be very similar to the NC gravity in Refs. \cite{nc-gravity1,nc-gravity2}
as was conjectured in \cite{hsy-ijmp09} since the NC gravity is also based on a NC deformation
of the diffeomorphism symmetry. But it should be remarked that the emergent gravity does not allow
a coupling of cosmological constant like $\int d^d x \sqrt{g} \Lambda$ which is of prime importance
to resolve the cosmological constant problem \cite{hsy-jpcs12}.
We will not in this paper calculate the corrections except to note that the leading NC corrections
will identically vanish. We showed in eq. (\ref{exp-u2=0}) (see also the footnote 10 in \cite{jsw2})
that any arbitrary vector field $X$
is stable at least up to the first order of NC deformations, i.e., $U_2 (X, \theta) = 0$.
This statement is similarly true for a smooth function $f \in C^\infty(M)$, i.e., $U_2 (f, \theta) = 0$.
Indeed it must be a generic property for the deformation quantization because it comes
only from the associativity of an underlying algebra.
This fact implies that the NC corrections of the vector field $V_a$ start at most from the second order,
i.e., $\Xi_a^{\mu\nu} (x) \partial_\mu \partial_\nu = 0$.\footnote{It must be remarked
that, although we pretend the vector field $V_a$ is independent of $\theta$,
$V_a$ is actually $\mathcal{O}(\theta)$ since it is a Hamiltonian vector field defined
by the Poisson bracket with the Poisson gauge fields (\ref{cov-mom}).
(It may be noted that the Poisson gauge fields in eq. (\ref{cov-mom}) are given
by $C_\mu (y) = B_{\mu\nu} x^\nu (y)$ which cancels out $\theta^{\mu\nu}$ in the Poisson bracket.)
Therefore, precisely speaking, the $\mathcal{O}(\theta^2)$ correction identically vanishes and
the nontrivial NC corrections start from $\mathcal{O}(\theta^3)$.}
As a result, the general relativity, which is emergent from the leading order of NC gauge fields,
receives no next-to-leading order corrections. In other words, the emergent gravity from NC gauge fields
predicts an intriguing result that Einstein gravity corresponds to a (local) minimum of moduli space of
Poisson (or Riemannian) structure deformations and is stable up to first order against quantum deformations
due to the NC structure of spacetime.

\subsection{Generalization to Poisson manifolds}

So far we have mostly kept symplectic manifolds in mind for the construction of global vector fields
and their quantization. Now we will think of the generalization to Poisson manifolds.
In the context of Poisson geometry $(M, \Pi)$, generally speaking,
one cannot find a ``Poisson connection"  $\nabla^P$ such that $\nabla^P \Pi = 0$
since parallel transport preserves the rank of the Poisson tensor, 
so the Poisson manifold must be regular in order for such a connection to exist.
But it turns out \cite{cft,cftp,vdolg-am,acgutt} that it is enough to consider only a torsion
free linear connection to formulate a star product on any Poisson manifold.
Thus one can work only with an affine torsion free connection and construct the formal exponential
map (\ref{reformal}) for the torsion free connection. Actually the previous formalism except the earlier discussion referring to the Fedosov manifold can also be applied to any Poisson manifold with impunity.
So let us recapitulate the essential steps for the deformation quantization of Poisson manifolds we have discussed starting from the paragraph containing (\ref{reformal}).
Given a torsion free connection on a Poisson manifold $(M, \Pi)$, one builds an identification
of the commutative algebra $C^\infty (M)$ of smooth functions on $M$ with the algebra
of flat sections of the jet bundle $\mathcal{E} \to M$, for the Grothendieck connection $D^{(0)}$.
And then one quantizes this situation to yield a quantum jet bundle $\widehat{\mathcal{E}} \to M$.
A deformed algebra structure on $\Gamma(\widehat{\mathcal{E}})$ is obtained through fiberwise quantization
of the jet bundle using Kontsevich star product on $\mathbb{R}^d$, and a deformed flat connection $\mathfrak{D}$ which is a derivation of this deformed algebra structure is constructed ``\`a la Fedosov".
Again one constructs an identification between the formal series of functions on $M$ and
the algebra of flat sections of this quantized bundle of algebras. Finally this identification defines
the star product on $M$. This quantization procedure can also be implemented to a general Poisson
manifold \cite{cft,cftp}. Hence the globalization in eq. (\ref{global-id}) can be simply generalized
to the Poisson case if the first one is replaced by
\begin{equation}\label{global-pid}
    \Pi_{\mu\nu} (x) = \bigl( \Pi_\phi \bigr)_{\mu\nu} (x;y=0).
\end{equation}
However we also need to lift the volume preserving condition (\ref{v-preserving}) to a quantum
Poisson algebra. Thus it is necessary to restrict Poisson manifolds to unimodular ones
as we discussed before.

But, as we discussed in the footnote \ref{symp-realize}, there is an interesting realization of
Poisson manifolds in terms of symplectic realizations. A symplectic realization
of a Poisson manifold $(M, \Pi)$ is a Poisson map
\begin{equation}\label{poisson-map}
    \varphi: (S, \Omega) \to (M, \Pi)
\end{equation}
from a symplectic manifold $(S, \Omega)$ to $(M, \Pi)$.
More precisely there is a collection of functions of the canonical variables $(q^1, \cdots, q^n,
p_1, \cdots, p_n)$ which is a subalgebra under the canonical Poisson bracket and
generated by a finite number of independent functions $\varphi_1, \cdots, \varphi_r$.
This means that $\mathbb{R}^r$ has a Poisson structure induced from the canonical symplectic
structure $\mathbb{R}^{2n}$ in the sense that $\Phi=(\varphi_1, \cdots, \varphi_r):\mathbb{R}^{2n}
\to \mathbb{R}^r$ is a Poisson map. For the SO(3) algebra of angular momenta in the footnote \ref{symp-realize},  for example, we have the Poisson map $\varphi_i := L_i = \varepsilon_{ijk} x^j p_k$
from the symplectic manifold $(S, \Omega)= (\mathbb{R}^6, \sum_{i=1}^3 dx^i \wedge dp_i)$
to the Poisson manifold $(M, \Pi) = (\mathbb{R}^3, \frac{1}{2} \varepsilon^{ijk} \varphi_k
\frac{\partial}{\partial \varphi_i} \wedge \frac{\partial}{\partial \varphi_j})$.
The symplectic realization is a very natural object in Poisson geometry, in particular,
from the point of view of the quantization theory of Poisson manifolds.
For Poisson manifolds which are the classical analogue of associative algebras,
symplectic realizations play a similar role as representations do for associative algebras.
The symplectic realization of a Poisson manifold was first introduced by Lie who proved that
such a realization always exists locally for any Poisson manifold of constant rank.
After almost a century, Weinstein proved \cite{d-weinstein} the local existence theorem of
symplectic realizations for general Poisson manifolds and later found \cite{sg-weinstein,kara-mas} that
there exists globally a unique symplectic realization which possesses a local groupoid structure
(though there is in general an obstruction for the existence of a global groupoid structure)
compatible with the symplectic structure. There is also a direct global proof for the existence
of symplectic realizations of arbitrary Poisson manifolds \cite{srsg4}.

In mathematics literatures, the Poisson map $\varphi$ is assumed to be a surjective submersion.
But we will not assume it because there are some examples in quantum mechanics with $\dim M \geq \dim S$.
We will illustrate such kind of symplectic realizations known as the boson realization of Lie algebras
or the Schwinger representation \cite{georgi}.
Suppose that the Poisson structure of $2n$-dimensional NC space (\ref{nc-space}) is given by $\theta^{2i-1, 2i} = \zeta^i > 0, \; i=1, \cdots, n$,
otherwise $\theta^{\mu\nu} = 0$. Using this canonical pairing (polarization) of symplectic structure,
define $n$-dimensional annihilation and creation operators as
\begin{equation}\label{ndim-ho}
    a_i = \frac{y^{2i-1} + i y^{2i}}{\sqrt{2\zeta^i}}, \qquad
    a_i^\dagger = \frac{y^{2i-1} - i y^{2i}}{\sqrt{2\zeta^i}}.
\end{equation}
Then the Moyal-Heisenberg algebra (\ref{nc-space}) reduces to the Heisenberg algebra $\mathcal{A}$
of $n$-dimensional harmonic oscillator, i.e.,
\begin{equation}\label{h-algebra}
    [a_i, a_j^\dagger]= \delta_{ij}.
\end{equation}
The Schwinger representation of Lie algebra generators for $G = SU(n)$ is defined by
\begin{equation}\label{schwinger-rep}
    Q^I = a_i^\dagger T^I_{ij} a_j, \qquad I = 1, \cdots, r,
\end{equation}
where $r = \dim G = n^2-1$ and $T^I$'s are constant $n \times n$ Hermitian matrices satisfying
the $su(n)$ Lie algebra $[T^I , T^J ] = if^{IJK} T^K$.
It is easy to verify that the operators $Q^I$ obey the commutation relation of $su(n)$ Lie algebra
\begin{equation}\label{q-lie-alg}
[Q^I , Q^J ] = if^{IJK} Q^K.
\end{equation}
In this example, we have the Poisson map $\varphi^I : = Q^I$ from the symplectic manifold
$(S, \Omega) = (\mathbb{R}^{2n}, \frac{1}{2} \theta^{-1}_{\mu\nu} dy^\mu \wedge dy^\nu)$
to an $r$-dimensional Poisson manifold $(M, \Pi) = (\mathbb{R}^{r}, \frac{1}{2} f^{IJK} \varphi^K
\frac{\partial}{\partial \varphi^I} \wedge \frac{\partial}{\partial \varphi^J})$.
In the case $n \geq 3$, $\dim M > \dim S$, so it is possible to have a generalized Poisson map
which is not necessarily a surjective submersion.

The above Schwinger representation can be generalized to any semi-simple Lie algebras.
Suppose that $\mathcal{A}$ is the Heisenberg algebra defined by eq. (\ref{h-algebra}) and
$\mathfrak{g}$ is an arbitrary Lie algebra of $G$ and $(\rho, V)$ is an $n$-dimensional
representation of $\mathfrak{g}$ where $V$ is the representation space and $\rho$ is the homomorphic
mapping from $\mathfrak{g}$ to End$(V)$. Then the Schwinger representation of $\mathfrak{g}$ is
given by a Poisson map $\varphi$ defined by
\begin{equation}\label{poisson-schwinger}
    X \mapsto \varphi(X) = a_i^\dagger (\rho X)_{ij} a_j, \qquad \forall X \in \mathfrak{g}.
\end{equation}
It is easy to check that the Poisson map is a Lie algebra homomorphism from $\mathfrak{g}$ to
$\mathcal{A}$, i.e.,
\begin{equation}\label{lie-homo-hs}
    [\varphi(X), \varphi(Y)] = \varphi\bigl( [X,Y] \bigr), \qquad \forall X, Y \in \mathfrak{g}.
\end{equation}
The Poisson map (\ref{poisson-schwinger}) provides a symplectic realization (which is not necessarily
a surjective submersion) to all semi-simple Lie group families, including the five exceptional groups.
Note that we already quantized Poisson manifolds via the symplectic realization (\ref{schwinger-rep})
or (\ref{poisson-schwinger}) although the complete classification of irreducible representations
for the quantized Poisson algebras (\ref{q-lie-alg}) and (\ref{lie-homo-hs}) still remains.
Fortunately the 20th century had been completed the latter problem at least
for semi-simple Lie algebras \cite{lie-rep1,lie-rep2}.
Later we will discuss why the symplectic realization of a Poisson manifold supplies a great benefit
for the quantization of Poisson manifolds.

There is another important realization of Poisson manifolds by the so-called {\it symplectic groupoids}
which was initiated with the works \cite{sg-weinstein,kara-mas}.
This can be seen as a generalization of the famous Lie's third theorem in the theory of Lie groups
in the sense that the correspondence between symplectic groupoids and Poisson manifolds is a natural
extension of the one between Lie groups and Lie algebras. We refer to Chapters 8 and 9 in \cite{vaisman}
for a nice exposition of symplectic realizations and symplectic groupoids.
The symplectic groupoid is a natural object in Poisson geometry for the following reason.
We know that a Poisson algebra $\mathfrak{P} = (C^\infty(M), \{-,-\}_\theta)$ is a Lie algebra on the
vector space $C^\infty(M)$ with respect to the Poisson bracket $\{-,-\}_\theta$ on a manifold $M$.
As every finite-dimensional Lie algebra $\mathfrak{g}$ over $\mathbb{R}$ is associated
to a Lie group $G$, a natural question is then whether there is a Lie group integrating this
Poisson-Lie algebra. We may pose the issue with the following basic construction \cite{d-weinstein}.
Given a finite dimensional real Lie algebra $\mathfrak{g}$, its dual space $\mathfrak{g}^*$
carries a Poisson structure, the Kirillov-Kostant structure. Let $G$ be any Lie group
whose Lie algebra is $\mathfrak{g}$, and let $T^*G$ be its cotangent bundle with its canonical
symplectic structure. Then $\mathfrak{g}^*$ may be embedded as the cotangent space
at the identity, a Lagrangian submanifold of $T^*G$. Thus the Lie group $G$ leads to a symplectic
realization of $\mathfrak{g}^*$ by the cotangent bundle $T^*G$ with the symplectic form
$\Omega = d \Lambda$, where $\Lambda$ is the Liouville one-form of $T^*G$.

For a general Poisson manifold $M$, the program is to embed $M$ as a Lagrangian submanifold of
a symplectic groupoid $\mathfrak{G}$ in such a way that $\mathfrak{G}$ integrates the cotangent bundle
$T^* M$ of the Poisson manifold $M$. For a given Poisson manifold $M$, the Poisson bracket on
its functions extends to a Lie bracket among all differential one-forms, which is the contravariant
analogue of the Lie bracket on vector fields. For exact one-forms, it can be defined
by $[df, dg] = d\{f, g\}_\theta$. For arbitrary one-forms, the so-called Koszul bracket
is generalized to the formula \cite{vaisman}
\begin{equation}\label{gelfand-dorfman}
    [\xi, \eta] = \mathcal{L}_{\xi_\theta} \eta - \mathcal{L}_{\eta_\theta} \xi
    - d\bigl(\theta(\xi, \eta) \bigr),
\end{equation}
where $\xi_\theta = \rho(\xi)$ and $\eta_\theta = \rho(\eta)$ are the anchor map (\ref{anchor})
for the one-forms $\xi$ and $\eta$, respectively. This bracket of differential forms satisfies
the following two important properties
\begin{eqnarray}\label{bra-2}
  && [\xi,  f \eta] = f [\xi, \eta] + (\xi_\theta f) \eta, \nonumber \\
  && [\rho(\xi), \rho(\eta)] = \rho([\xi, \eta]),
\end{eqnarray}
where $f \in C^\infty(M)$. All these properties described by the triple $(T^* M, [-,-], \rho)$
make the cotangent bundle of a Poisson manifold $M$ a special case of a more general object
in differential geometry, called a Lie algebroid.
A Lie algebroid is a straightforward generalization of a Lie algebra and
the Lie algebroid of a symplectic groupoid $\mathfrak{G}$ is canonically isomorphic to $T^*M$.
Therefore the Lie groupoid integrating a Poisson manifold has a natural symplectic structure.

To define a general Lie algebroid, one can simply replace $T^* M$ by a general vector bundle $E$.
A Lie algebroid $\mathfrak{L}$ is then a triple $(E, [-, -], \rho)$ consisting of a vector bundle $E$
over a manifold $M$, together with a Lie algebra structure $[-, -]$ on the vector space $\Gamma(E)$
of the smooth global sections of $E$, and the anchor map of vector bundles $\rho: E \to TM$.
The above properties (\ref{bra-2}) are generalized in an obvious way \cite{vaisman} to
\begin{eqnarray}\label{liebra-2}
  && [X, f Y] = f[X, Y] + \bigl(\rho(X)f \bigr) Y, \nonumber \\
  && [\rho(X), \rho(Y)] = \rho([X, Y]),
\end{eqnarray}
for all $X, Y \in \Gamma(E)$ and $f \in C^\infty(M)$.
Here $\rho(X)f$ is the derivative of $f$ along the vector field $\rho(X)$.
The anchor map $\rho$ defines a Lie algebra homomorphism from the Lie algebra of global sections of $E$,
with Lie bracket $[-, -]$, into the Lie algebra of vector fields on $M$.
Hence Lie algebroids can be thought as ``infinite dimensional Lie algebras of geometric type", or
``generalized tangent bundles". To every Lie groupoid there is an associated Lie algebroid.
But the converse is not true because there are obstructions to the integrations of Lie algebroids
to Lie groupoids. For the case where $E = T^*M$ and $M$ is a Poisson manifold, for example,
there are topological obstructions encoded in what are called the monodromy groups \cite{srsg5}.

The notion of symplectic groupoid provides a framework for studying the collection of all
symplectic realizations of a given Poisson manifold.
A Poisson manifold $M$ is called integrable if a symplectic groupoid $\mathfrak{G}$ exists
such that its infinitesimal version corresponds to a given Lie algebroid $(T^* M, [-,-], \rho)$.
And it turns out that symplectic realizations contain a lot of information
about integrability. For instance, it was shown in \cite{srsg3} that a Poisson manifold is
globally integrable if and only if it admits a complete symplectic realization.
It is also interesting to note that Morita equivalent Poisson manifolds have equivalent categories
of complete symplectic realizations \cite{pxu-cmp}.
It was proven in \cite{srsg1} that the reduced phase space of
the Poisson sigma model under certain boundary conditions has a natural groupoid structure,
assuming that it is a smooth manifold and the symplectic groupoid integrating a given Poisson
manifold is explicitly constructed for the integrable case.
Furthermore it was shown in \cite{srsg1} that the perturbative quantization of this model yields
the Kontsevich star product formula. A formal version of the integration of Poisson manifolds
by symplectic groupoids was also given in \cite{srsg2}. Therefore the symplectic realization
of Poisson manifolds in terms of symplectic groupoids provides an efficient route of quantization of
Poisson manifolds \cite{awe-pxu} so that the deformation or geometric quantization of a symplectic
groupoid $\mathfrak{G}$ descends to the quantization of a Poisson manifold $M$ though it is nontrivial
to quantize $\mathfrak{G}$ in such a way that the quantization descends to a quantization of $M$.

\subsection{Towards a global geometry}

Now we will apply our globalization in eq. (\ref{global-id}) to the formulation of a global geometry
in emergent gravity. First let us consider a symplectic structure $\Omega = \Pi^{-1}$
on open subsets of $M = \mathbb{R}^{2n}$ (see section 7 in \cite{cft-pre}).
In this case we can take the formal exponential map (\ref{formal-emap}) given by
\begin{equation}\label{f-exp-sr2n}
    \phi_x^\mu(y) = x^\mu + y^\mu.
\end{equation}
Then the flat connection (\ref{flat-conn}) takes the form \cite{cft-pre}
\begin{equation}\label{flat-2n}
    D^{(0)} = dx^\mu \Bigl(\frac{\partial}{\partial x^\mu} - \frac{\partial}{\partial y^\mu} \Bigr)
\end{equation}
and thus the algebra of flat sections of the jet bundle $\mathcal{E}$ is generated
by the set of smooth functions on $M$ of the form
\begin{equation}\label{kerd-2n}
    \ker  D^{(0)} = \{ (C_a)_\phi (x;y) = C_a (x+y): C_a \in C^\infty (M), \; d = 1, \cdots, 2n \}.
\end{equation}
Using the relations (\ref{iso-poisson}) and (\ref{iso-vector}), it is easy to find the global vector
fields defined by eq. (\ref{global-id}) and one yields the result
\begin{equation}\label{vec-2n}
    V_a = \Pi^{\mu\nu} (x) \frac{\partial C_a(x)}{\partial x^\mu} \frac{\partial}{\partial x^\nu}.
\end{equation}

On an open subset of $M = \mathbb{R}^{2n}$, we can identify the exponential
map $\phi_x^*$ with the Moser flow (\ref{t-flow}) as we discussed before
(see the argument below the footnote \ref{foot-27}) and represent the symplectic
form $\Omega = \Pi^{-1}$ as the form (\ref{to-expansion}), i.e., $\Omega = B + F$
where $B$ is a constant asymptotic value in the Darboux frame such that $F \to 0$ at $|x| \to \infty$.
In this case, the exponential map $\phi_x^\mu (y)$ can be identified with covariant coordinates
on the symplectic vector space $T_x M$ defined by eq. (\ref{cov-obj}), i.e.,
$\phi_x^\mu (y) = \rho^*_A (y^\mu)$ and so $(C_a)_\phi (x;y) = B_{a\mu} (x+y)^\mu$.\footnote{Note that
$x$ simply refers to local coordinates of a base point $p \in M$ of the tangent bundle $T_p M$, 
so we may put the tangent space $T_p M$ at the origin, $x=0$.
But it is convenient to consider the tangent space $T_p M$ at an arbitrary base point $p \in M$ for
the construction of global vector fields (\ref{vec-2n}).}
In the end the vector fields (\ref{vec-2n}) are given by
\begin{equation}\label{vec-2nsec}
    V_a = B_{a\mu} \Pi^{\mu\nu} (x) \frac{\partial}{\partial x^\nu} \in \Gamma(TM).
\end{equation}
The dual one-forms $V^a = V^a_\mu (x) dx^\mu$ are defined by the natural pairing $\langle V^a, V_b \rangle = \delta^a_b$, so they are given by
\begin{equation}\label{dual1-2n}
    V^a = dx^\mu \Omega_{\mu\nu}\theta^{\nu a} = dx^\mu \Bigl(\delta_{\mu a}
    + (F\theta)_{\mu a} (x) \Bigr) \in \Gamma(T^*M).
\end{equation}

Given the vector fields (\ref{vec-2nsec}), we can solve the volume preserving condition (\ref{v-preserving})
which is equivalent to finding a volume form $\nu = \lambda^2 V^1 \wedge \cdots \wedge V^{2n}$
such that the modular vector field in (\ref{mod-lvec}) identically vanishes, i.e.,
\begin{equation}\label{mvf-2n}
    \phi_\nu = - X_{\log \upsilon} - \partial_\mu \Pi^{\mu\nu} \frac{\partial}{\partial x^\nu} = 0,
\end{equation}
where $\upsilon(x) = \lambda^2 \det V_\mu^a$.
The condition (\ref{mvf-2n}) can be written as
\begin{eqnarray} \label{vpc-2n}
  0 &=& \partial_\mu \Pi^{\mu\nu} + \Pi^{\mu\nu} \partial_\mu \log \upsilon \nonumber \\
   &=& \Pi^{\mu\nu} \bigl( \partial_\mu \log \upsilon - \Pi^{\rho\sigma} \partial_\sigma
   \Omega_{\mu\rho} \bigr) \nonumber \\
   &=& \Pi^{\mu\nu} \bigl( \partial_\mu \log \upsilon - \frac{1}{2} \Pi^{\rho\sigma}
   \partial_\mu \Omega_{\sigma\rho} \bigr) \nonumber \\
   &=& \Pi^{\mu\nu} \bigl( \partial_\mu \log \upsilon - \frac{1}{2}
   \partial_\mu \log \det V_\mu^a \bigr) \nonumber \\
   &=& \Pi^{\mu\nu}  \partial_\mu  \log \bigl( \lambda^2 \sqrt{\det V_\mu^a} \bigr)
\end{eqnarray}
where we used the Bianchi identity $\partial_{[\sigma} \Omega_{\mu\rho]} = 0$ in the third step and
the formula $\partial_\mu  \log \det A = \mathrm{Tr} A^{-1} \partial_\mu A$ for a matrix $A$
in the fourth step. Therefore we get
\begin{equation}\label{lambda-v2n}
 \lambda^2 (x) = \frac{1}{\sqrt{\det V_\mu^a}} \qquad \mathrm{or} \qquad
 \upsilon(x) = \sqrt{\det V_\mu^a}.
\end{equation}
The above result can be understood as follows. On an open subset of $M = \mathbb{R}^{2n}$,
the invariant volume form is certainly given by $\nu_\phi = d^{2n} y = \frac{B^n}{n! \mathrm{Pf}B}$.
Using the formula (\ref{image-fexp}) for the volume form, we have the relation $\nu_\phi
= (\phi_x)^* \nu$. Since $(\phi_x)^* \Omega = B$ as we argued above (see also eq. (\ref{darboux-flow})),
we get the volume form $\nu = \frac{\Omega^n}{n! \mathrm{Pf}B}$ which can be written as
\begin{equation} \label{vol-2n}
  \nu = \frac{\mathrm{Pf} \Omega}{\mathrm{Pf} B} d^{2n} x  = \frac{\mathrm{Pf} B}{\mathrm{Pf} \Omega}
 V^1 \wedge \cdots \wedge V^{2n}  = \frac{1}{\sqrt{\det V_\mu^a}} V^1 \wedge \cdots \wedge V^{2n}
\end{equation}
where we used the result (\ref{dual1-2n}). This is consistent with the result (\ref{lambda-v2n}).

Finally the emergent metric (\ref{emergent-metric}) determined by the one-form basis (\ref{dual1-2n})
is given by
\begin{eqnarray}\label{emetric-2n}
    g_{\mu\nu} (x) &=& \frac{V^a_\mu(x) V^a_\nu(x)}{\sqrt{\det V_\mu^a}} \nonumber \\
    &=& \frac{\Bigl(\delta_{\mu a} + (F\theta)_{\mu a} (x) \Bigr)\Bigl(\delta_{a \nu}
    + (\theta F)_{a \nu} (x) \Bigr)}{\sqrt{\det \bigl(1 + (F \theta)(x) \bigr)}}.
\end{eqnarray}
This form of the metric was also appeared in different contexts in \cite{ban-yan}
(see eq. (4.15) which can be identified with $g_{\mu\nu} = e^{-\phi} \mathrm{g}_{\mu\nu}$)
and in \cite{hstein-rev} (see eq. (50) which coincides with eq. (\ref{emetric-2n}) in four dimensions).
Note that $\sqrt{\det g_{\mu\nu}} = \bigl(\det (1 + F \theta) \bigr)^{1-\frac{n}{2}}$
and so $\sqrt{\det g_{\mu\nu}} = 1$ in four dimensions $(n=2)$.
One can expand the metric (\ref{emetric-2n}) in powers of $F$ or $\theta$ which leads to
the expansion
\begin{eqnarray}\label{emetric-2nep}
    g_{\mu\nu} (x) &=& \delta_{\mu \nu} + (F\theta)_{\mu\nu} + (\theta F)_{\mu\nu}
    + (F\theta \theta F)_{\mu\nu} - \frac{1}{2} \delta_{\mu \nu} \Bigl( \mathrm{Tr} F\theta
    - \frac{1}{2}\mathrm{Tr} (F\theta)^2 - \frac{1}{4} \bigl(\mathrm{Tr} F\theta \bigr)^2 \Bigr)
    \nonumber \\
    && - \frac{1}{2} \bigl( F\theta + \theta F \bigr)_{\mu\nu} \mathrm{Tr} (F\theta) + \cdots.
\end{eqnarray}
The linear order metric of the above expanded form looks like the gravitational metric derived from
the SW map (see eq. (50) in \cite{lry2}) except the trace term $- \frac{1}{2} \delta_{\mu \nu}
\mathrm{Tr} F\theta$. Interestingly, for symplectic U(1) instantons, all the trace terms in $\mathcal{O}(F^m)$
coming from the determinant of the denominator in eq. (\ref{emetric-2n}) are canceled
by the diagonal components of next higher order terms in $\mathcal{O}(F^{m+1})$.
For this cancelation, it is crucial to use the identity $\sqrt{\det (1 + F \theta )} =
1 + \frac{1}{4} \mathrm{Tr} F\theta$ (that is eq. (20) in \cite{yas-prl} derived from
the instanton equation (18)). To be more specific, the metric (\ref{emetric-2n}) can be simplified
as $g_{\mu\nu} (x) = \bigl(\delta_{\mu \nu} + (F\theta + \theta F)_{\mu\nu}
    + (F\theta \theta F)_{\mu\nu} \bigr) \bigl(1 + \frac{1}{4} \mathrm{Tr} F\theta \bigr)^{-1}$.
Using the identity again, one can show that the trace term $\delta_{\mu \nu} \bigl(-\frac{1}{4} \bigr)^{m}
(\mathrm{Tr} F\theta)^m$ for $m \geq 1$ is canceled by the diagonal components of
$\bigl(-\frac{1}{4} \bigr)^{m-1} \Bigl((F\theta \theta F)_{\mu\nu} - \frac{1}{4}
\bigl( F\theta + \theta F \bigr)_{\mu\nu} \mathrm{Tr} (F\theta) \Bigr) (\mathrm{Tr} F\theta)^{m-1}$
in $\mathcal{O}(F^{m+1})$. After this cancelation, the metric (\ref{emetric-2n}) can be written as the form
\begin{equation}\label{emetric-perp}
g_{\mu\nu} (x) = \delta_{\mu \nu} + (F\theta + \theta F)^{\parallel}_{\mu\nu}
+ \frac{(F\theta + \theta F)^\perp_{\mu\nu} + (F\theta \theta F)^\perp_{\mu\nu}}
{1 + \frac{1}{4} \mathrm{Tr} F\theta}
\end{equation}
where $\parallel$ and $^\perp$ denote the diagonal and off-diagonal parts, respectively.
Thus the four-dimensional gravitational metric for symplectic U(1) instantons
is approximated up to linear order by
\begin{equation}\label{emetric-u1}
    g_{\mu\nu} (x) \approx \delta_{\mu \nu} + (F\theta)_{\mu\nu} + (\theta F)_{\mu\nu}
\end{equation}
which is symmetric for arbitrary U(1) field strengths. This metric is precisely the form identified by
the SW map \cite{lry2,sty-plb}. This analysis suggests that
higher order terms, $\mathcal{O}(F^2)$, in the metric (\ref{emetric-perp}) and its inverse metric
must be regarded as derivative corrections of general relativity that certainly exist
as we discussed at the end of section 5.3. Moreover only tensor (spin 2) modes generate the higher order
corrections. This is a desirable property since there is no scalar graviton in general relativity
thanks to general covariance.

For a general symplectic manifold other than $\mathbb{R}^{2n}$, the Grothendieck connection $D^{(0)}$
becomes more complicated whose general form is given by eq. (\ref{flat-conn}), so it is nontrivial
to find the set, $\ker  D^{(0)}$, of flat sections. (It is nontrivial even in two dimensions.
See \cite{kishimoto} for the deformation quantization of two-dimensional constant curvature spaces.)
In this case the global vector fields $V_a$ will also take a complicated form than eq. (\ref{vec-2n}).
And we also need to solve the volume preserving condition (\ref{v-preserving}) given by
\begin{eqnarray} \label{gvpc-2n}
0 = \mathcal{L}_{V_a} \nu &=& \bigl( \partial_\mu V^\mu_a
+ V^\mu_a \partial_\mu \log \upsilon \bigr) \nu \nonumber \\
&=& \bigl( \nabla \cdot V_a + 2(1-n) V_a \log \lambda \bigr) \nu
\end{eqnarray}
where $\upsilon(x) = \lambda^2 \det V_\mu^a$. There is some interesting class of metrics, e.g.,
the Gibbons-Hawking metric \cite{gh-inst} and the real heaven \cite{real-heaven},
which obeys $\partial_\mu V^\mu_a = 0$ \cite{hsy-epl09}. In this case the condition (\ref{gvpc-2n})
reduces to
\begin{equation}\label{lambda-ghrh}
\upsilon(x) = 1  \qquad \mathrm{or} \qquad \lambda^2 (x) = \frac{1}{\det V_\mu^a}, 
\end{equation}
so the emergent metric (\ref{emergent-metric}) is given by
\begin{equation}\label{ghrh-metric}
    g_{\mu\nu} (x) = \frac{V^a_\mu(x) V^a_\nu(x)}{\det V_\mu^a}.
\end{equation}
(There is a confusing flip of indices, $(\mu, \nu) \leftrightarrow (a, b)$, between eq. (\ref{ghrh-metric})
and eq. (35) in \cite{hsy-epl09}. We hope readers are not bothered by it.)
For a general class of symplectic manifolds, it may be necessary to fully solve the volume preserving
condition (\ref{gvpc-2n}). In general it is not easy to find a solution such that the first term, $\partial_\mu V^\mu_a$, cancels out the second term, $V^\mu_a \partial_\mu \log \upsilon$, because
they occasionally have a quite different functional form. One way to circumvent this difficulty is to attach
a nonlocal dipole-like object to vector fields $V_a$ similar to an open Wilson line such that the
inverse vielbeins $E_a = \lambda^{-1} V_a$ still become local by compensating the nonlocal object
in $V_a$ by $\lambda$. It was argued in \cite{lry1} that such kind of a nonlocal object is necessary
for the LeBrun metric \cite{lebrun}, that is the most general scalar-flat K\"ahler metric
with a U(1) isometry and contains the Gibbons-Hawking metric, the real heaven as well as
the multi-blown up Burns metric which is a scalar-flat K\"ahler metric on $\mathbb{C}^2$
with $n$ points blown up.

\begin{figure}
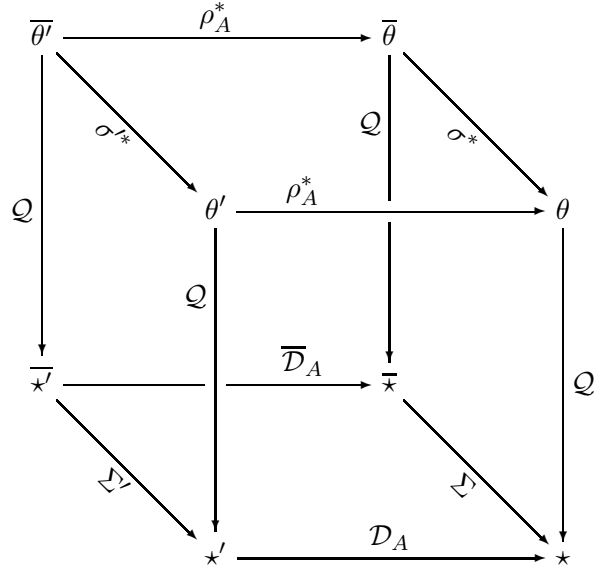

\label{moser-map}
\begin{diagram}
\overline{\theta'} & & \rTo^{\overline{\rho_A^*}} & & \overline{\theta} & & \\
& \rdTo_{\sigma'^*} & & & \vLine^{\mathcal{Q}} & \rdTo_{\sigma^*} & \\
\dTo^{\mathcal{Q}} & & \theta' & \rTo^{\rho_A^*} & \HonV & & \theta \\
& & \dTo^{\mathcal{Q}} & & \dTo & & \\
\overline{\star'} & \hLine & \VonH & \rTo^{\overline{\mathcal{D}}_A} & \overline{\star}
& & \dTo_{\mathcal{Q}} \\
& \rdTo_{\Sigma'} & & & & \rdTo_\Sigma & \\
& & \star' & & \rTo^{\mathcal{D}_A} & & \star \\
\end{diagram}
\caption{Deformation quantization and covariance \cite{jsw2}}
\end{figure}

It may be worthwhile to point out some (surmisable) aspect of emergent Riemannian metrics.
As in general relativity, the explicit form of a global metric depends on the choice of coordinates.
For example, the usual spherical coordinate representation of Eguchi-Hanson
metric \cite{egu-han1,egu-han2} is equivalent to the two-center Gibbons-Hawking metric \cite{gh-inst}
by a coordinate transformation \cite{pra-equi} though their bare appearance looks very different.
A similar feature also arises in emergent gravity. It was recently shown \cite{lry2,sty-plb} that
the Cartesian coordinate representation of Eguchi-Hanson metric takes the form (\ref{emetric-u1}), 
so it belongs to the class (\ref{emetric-2n}). However the Gibbons-Hawking representation of the
Eguchi-Hanson metric takes the form (\ref{ghrh-metric}) as was shown in \cite{joyce,osaka1,osaka2,hsy-epl09}.
This implies that there exists a coordinate transformation relating two types of metrics if they describe
the same manifold. And this coordinate transformation needs to be globally defined because it has to relate
one globally defined metric to the other global metric. Actually the general covariance for the choice
of coordinates is reflected in deformation quantization \cite{kontsevich} and
clearly discussed in section 3.3 in \cite{jsw2} in the context of NC gauge theory.
The covariance in deformation quantization can be summarized with Fig. 1 which illustrates how
the semi-classical and quantum constructions are affected by a change of coordinates $\sigma^*$.
The vertical arrow in Fig. 1 indicates a quantization map defined by eq. (\ref{quantum-prod}).
The deformation quantization of $\overline{\theta}$ and $\overline{\theta'}$ in new coordinates leads
to star products $\overline{\star}$ and $\overline{\star'}$ that are related to the star products $\star$
and $\star'$ in old coordinates by the equivalence maps $\Sigma$ and $\Sigma'$, respectively.
It might be emphasized that the coordinate transformation $\sigma^*$ is globally defined
as we remarked above whereas the coordinate transformation in the Moser flow $\rho_A^*$ is
locally defined as was verified by eq. (\ref{darboux-flow}). Hence it is naturally expected that their
quantization maps $\Sigma \;(\Sigma')$ and $\mathcal{D}_A \; (\overline{\mathcal{D}}_A)$ will also
keep the property. The relation between local covariance maps $\mathcal{D}_A$ and $\overline{\mathcal{D}}_A$
in the old and new coordinates can be deduced from the commutativity of the diagram of the horizontal plane
in Fig. 1 \cite{jsw2}:
\begin{eqnarray} \label{dcov-diagram}
 && \overline{\mathcal{D}}_A   = \Sigma^{-1} \circ \mathcal{D}_A \circ \Sigma', \\
 \label{acov-diagram}
 && \overline{\widehat{\mathcal{A}}}_A = \Sigma^{-1} \circ (\Sigma' - \Sigma) +
 \Sigma^{-1} \circ \widehat{\mathcal{A}}_A \circ \Sigma'.
\end{eqnarray}

It is well-known that there are three kinds of geometries (see, for example, Chapter 12 in \cite{book-dasilva}):

A. Symplectic geometry $(M, \Omega)$: geometry of a closed, nondegenerate, skew-symmetric bilinear form $\Omega$.

B. Riemannian geometry $(M, g)$: geometry of a positive-definite symmetric bilinear map $g$.

C. Complex geometry $(M, J)$: geometry of an integrable linear map $J$ with $J^2 = -1$.

Each category has a generalization to a more general geometry. For example,
in the category A, we can relax the closedness condition, then we get an almost symplectic manifold,
i.e., manifold $M$ with a nondegenerate, not necessarily closed, exterior 2-form $\Omega$.
Or we can relax the nondegenerate condition. In this case we get a Poisson geometry $(M, \Pi)$ where
the closedness condition is replaced by a generalized version (\ref{poisson-str}), i.e., $[\Pi, \Pi]_S = 0$.

The generalization of the category B is historical and still going on.
The most famous (and rational) attempt is the Cartan geometry based on the Cartan connection generalizing
the Levi-Civita connection.

The category C also admits a generalization to an almost complex manifold $(M, J)$ which is
a smooth manifold $M$ having an endomorphism $J: TM \to TM$ of the real tangent bundle which
satisfies $J^2 = -1$ but is not necessarily integrable. The integrability of an almost complex
structure $J$ is measured by the Nijenhuis tensor defined by
\begin{equation}\label{nijenhuis-j}
    N_J(X, Y) = [X, Y] + J[JX, Y] + J[X, JY] - [JX, JY]
\end{equation}
for any two vector fields $X, Y$. The Newlander-Nirenberg theorem \cite{newnir} states that
an almost complex structure $J$ is integrable if and only if $N_J = 0$.

The emergent gravity is to aim for the derivation of a Riemannian geometry from the symplectic geometry
or more generally the Poisson geometry.
The question is what kind of Riemannian geometry arises from a given symplectic geometry.
An onset is to think of a role of complex geometry for the emergent gravity.
Though there is a close similarity between the symplectic geometry and the complex geometry,
there do not exist any definite inclusion relations between them.
For example, there exists a symplectic manifold with no complex structure:
A circle bundle over circle bundle over a 2-torus \cite{comnosym}. And there is a complex manifold that
admits no symplectic structure: the Hopf surface $\mathbb{S}^3 \times \mathbb{S}^1$ \cite{book-dasilva}.
Thus it may be necessary to start with either a Poisson geometry or an almost symplectic geometry
in order to derive a general Riemannian geometry. For instance, it can be shown \cite{lry2} that
the K\"ahler condition is equivalent to the Bianchi identity, $d \Omega = 0$,
for a U(1) field strength $\Omega$, so it is required to consider the almost symplectic or Poisson
structure to construct the four-dimensional Euclidean Schwarzschild black-hole geometry \cite{hawking-ebh}
since it is not a K\"ahler manifold (though it is a Ricci-flat manifold).
Unfortunately it is very nontrivial to derive the emergent Riemannian geometry
from a general almost symplectic or Poisson geometry. Therefore let us focus on the symplectic geometry.
See \cite{yasi} for emergent Riemannian geometries with a constant curvature derived from the Poisson geometry.

Given a symplectic manifold $(M, \Omega)$, we can introduce a symplectic connection (see the footnote \ref{foot-sympconn}) obeying eq. (\ref{symp-conn}) which reads as
\begin{equation}\label{symp-conn-vec}
    Z \bigl( \Omega(X, Y) \bigr) = \Omega(\partial^S_Z X, Y) + \Omega (X, \partial_Z^S Y)
\end{equation}
for any vector fields $X, Y, Z$. In this case, if we can find a compatible complex structure $J$
such that $g_J(X,Y) \equiv \Omega (X, JY)$ for any $X, Y \in \Gamma(TM)$ is a positive-definite
symmetric bilinear map, we can identify $g_J(X,Y)$ with a Riemannian metric (\ref{r-metric}).
Indeed, if we take $X=\partial_\mu, Y = J \partial_\nu, Z = \partial_\lambda$,
the above condition (\ref{symp-conn-vec}) can be written as
\begin{equation}\label{lc-symp}
    \partial_\lambda g_{\mu\nu} = {\Gamma^\rho}_{\lambda \mu} g_{\rho\nu}
    + {\Gamma^\rho}_{\lambda \nu} g_{\rho\mu} + \Omega \bigl(\partial_\mu,
    (\partial_\lambda^S J) \partial_\nu \bigr).
\end{equation}
Therefore, if we assume that $\partial_\lambda^S J = 0$, the Christoffel symbol ${\Gamma^\rho}_{\mu \nu}$
in eq. (\ref{lc-symp}) should be regarded as the Levi-Civita connection, $\nabla^{LC}$,
in general relativity because (\ref{lc-symp}) in that case is precisely the metric compatibility
condition, $\nabla^{LC} g = 0$. Thus, the condition, $\partial_\lambda^S J = 0$, can be interpreted
as the one that parallel translation should preserve the almost complex structure, i.e.,
$\nabla^{LC} J = 0$ or equivalently the metric is K\"ahler \cite{dgms-km}.
Consequently, in order to define an emergent geometry from a symplectic
manifold $(M, \Omega)$, the problem at hand is whether it is possible to find the compatible complex
structure $J$ within the context of symplectic geometry such that $g_J(X,Y) \equiv \Omega (X, JY)$
becomes a positive-definite Riemannian metric. Note that to any symplectic manifold $(M, \Omega)$
one can always find almost complex structures $J$ that are compatible with $\Omega$
(see section 12.3 in \cite{book-dasilva}).

Actually this compatible complex structure may be inherited from the symplectic vector space
$(T M, \Omega_\phi)$ where $\Omega_\phi(x; y) = \frac{1}{2} \Omega_{\mu\nu} (x;y) dy^\mu \wedge dy^\nu$
is the symplectic structure dual to the Poisson bivector (\ref{local-poi}).
Since the global vector fields $(V_a)_\phi \in \Gamma(TM)$ and their dual covectors
$(V^a)_\phi \in \Gamma(T^* M)$ form a complete basis for $TM$ and $T^* M$, respectively,
we use them to represent
\begin{eqnarray}\label{symp-jtan}
 && \Omega_{\phi} (x;y) = - \frac{\lambda^2}{2} I_{ab} V^a(x;y) \wedge V^b(x;y), \\
 \label{pois-jtan}
 && \Pi_{\phi} (x;y) = \frac{1}{2\lambda^2} I^{ab} V_a(x;y) \wedge V_b(x;y),
\end{eqnarray}
where the constant symplectic matrix $I_{ab}$ is given by
\begin{equation}\label{symp-matrix}
    I_{2i-1, 2j} = \delta_{ij} = - I_{2j, 2i-1}, \qquad i,j = 1, \cdots, n.
\end{equation}
Here we supposed to choose the smooth function $\lambda(x;y)$ so that $\Omega_{\phi} (x;y)$ becomes
a closed two-form. It turns out that $\lambda = \lambda(x;0)$ can be identified with eq. (\ref{vol-lambda})
for K\"ahler manifolds. The complex structure $J_\phi = (\phi_x)_*^{-1} J =  (\phi_x)^* J$ on $T_p M$,
$p \in M$, is a smooth tensor field of type $(1,1)$ and is locally in a coordinate chart given by
\begin{equation}\label{complex-jtan}
 J_\phi(x; y) = J^\mu_\nu(x;y) \frac{\partial}{\partial y^\mu} \otimes dy^\nu
\end{equation}
where smooth functions $J^\mu_\nu(x;y)$ on $T_p M$ satisfy a matrix relation
\begin{equation}\label{comp-j}
     J_\mu^\lambda(x;y) J_{\lambda}^{\nu} (x;y) = - \delta_\mu^\nu.
\end{equation}
We can borrow the complex structure (\ref{complex-jtan}) from the symplectic structure (\ref{symp-jtan})
as follows:
\begin{equation}\label{comp-symp}
 J_\phi(x; y) = I_{ab} V_a(x;y) \otimes V^b(x;y).
\end{equation}
It is then easy to check that $g_\phi (V_a, V_b) = \Omega_\phi (V_a, JV_b) = \lambda^2 \delta_{ab}$.
If we define the metric $g_{\mu\nu}(x) = (g_\phi)_{\mu\nu} (x;0)$ according to the rule used
for eqs. (\ref{global-xi}) and (\ref{global-omega}), we finally get the emergent metric (\ref{emergent-metric})
given by
\begin{equation}\label{emetric-symp}
    g_{\mu\nu} (x) = \lambda^2 V^a_\mu(x) V^b_\nu(x) \delta_{ab}.
\end{equation}
Depending on the choice of $\lambda^2$, the above metric reproduces either eq. (\ref{emetric-2n}) or
(\ref{ghrh-metric}). But it is interesting to note that the complex structure (\ref{comp-symp})
is immune from the different choice of $\lambda$ because it is the tensor field of type $(1,1)$.

In order for the metric (\ref{emetric-symp}) to be K\"ahler, there are several constraints.
First we have the volume preserving condition (\ref{gvpc-2n}) which can be represented as the form
$g_{bab} = V_a \log \lambda^2$ \cite{lry1} using the structure equation (\ref{st-eq}). Of course
it is in principle solved if the complete set of vector fields $V_a$ and the invariant volume form
$\nu = \lambda^2 V^1 \wedge \cdots \wedge V^{2n}$ are determined. Essential constraints are given
by the integrability of the almost complex structure (\ref{comp-symp}), $N_J(V_a, V_b) = 0$, and
the closedness condition of the symplectic two-form (\ref{symp-jtan}),  $d\Omega_{\phi} (x;0) = 0$.
In this case, $\lambda^2$ in eq. (\ref{vol-lambda}) should be identified with that in $\Omega_{\phi}$
because the compatibility condition, $g(X,Y) = \Omega(X, JY)$, must be obeyed for K\"ahler metrics.
But they do not have to be identified for non-K\"ahler metrics since the compatibility condition
is not necessarily satisfied. Let us summarize all these constraints:
\begin{eqnarray}
\label{kahler-cons1}
\mathcal{L}_{V_a} \nu = 0 \quad &\Leftrightarrow&  \quad  g_{bab} = V_a \log \lambda^2, \\
\label{kahler-cons2}
d\Omega_{\phi} (x;0) = 0  \quad  &\Leftrightarrow&  \quad  I_{(ab} V_{c)} \log \lambda^2
= - I_{d(a} g_{bc)}^{~~d}, \\
\label{kahler-cons3}
N_J(V_a, V_b) = 0  \quad  &\Leftrightarrow&  \quad   (I \lrcorner \, g)_{ab}^{~~c}
= (I^3 \lrcorner \, g)_{ab}^{~~c},
\end{eqnarray}
where the contraction symbols mean that $(I \lrcorner \, g)_{ab}^{~~c} = I_{ad}  g_{db}^{~~c}
+ I_{bd}  g_{ad}^{~~c} + I_{cd}  g_{ab}^{~~d}$ and $(I^3 \lrcorner \, g)_{ab}^{~~c} =
I_{ad} I_{be} I_{cf}  g_{de}^{~~f}$. To derive (\ref{kahler-cons2}), we used
the Maurer-Cartan equation, $dV^a = \frac{1}{2} g_{bc}^{~~a} V^b \wedge V^c$,
dual to the structure equation (\ref{st-eq}). If we introduce the structure equation for orthonormal
frames defined by
\begin{equation}\label{st-eq-onf}
    [E_a, E_b] = -f_{ab}^{~~c} E_c,
\end{equation}
the structure functions are related each other by \cite{hsy-jhep09}
\begin{equation}\label{sf-ve}
    g_{ab}^{~~c} = \lambda \bigl( f_{ab}^{~~c} - E_a \log \lambda \delta_{bc}
    + E_b \log \lambda \delta_{ac} \bigr).
\end{equation}
In terms of the structure functions $f_{ab}^{~~c}$ in eq. (\ref{st-eq-onf}), the above three constraints
take more simpler forms given by
\begin{eqnarray}
\label{kahler-fcons1}
\mathcal{L}_{E_a} \widetilde{\nu} = 0 \quad &\Leftrightarrow&  \quad  f_{bab} = (3-2n) E_a \log \lambda, \\
\label{kahler-fcons2}
d\Omega_{\phi} (x;0) = 0  \quad  &\Leftrightarrow&  \quad I_{d(a} f_{bc)}^{~~d} = 0, \\
\label{kahler-fcons3}
N_J(E_a, E_b) = 0  \quad  &\Leftrightarrow&  \quad   (I \lrcorner \, f)_{ab}^{~~c}
= (I^3 \lrcorner \, f)_{ab}^{~~c}.
\end{eqnarray}
The first condition is consistent with the fact (\ref{der-vec2}) that the vector fields $E_a$ preserve
the volume form $\widetilde{\nu} = \lambda^{3-2n} \nu_g$. The third condition should be expected
since the complex structure (\ref{comp-symp}) is invariant under the Weyl
scaling $V_a \to E_a = \lambda^{-1} V_a$.

By definition a K\"ahler manifold is a symplectic manifold $(M, \Omega)$ equipped with an
integrable compatible almost complex structure. The symplectic form $\Omega$ is then called
a K\"ahler form. That is, any K\"ahler manifold must be both symplectic and complex.
We also know any symplectic or complex manifold admits almost complex structures.
Then one might ask: If $M$ is both symplectic and complex, is it necessarily K\"ahler?
It is known \cite{gromov-book} that any noncompact almost complex manifold has a compatible
symplectic structure. But the situation is very different for closed symplectic manifolds.
The answer for them is strikingly no. The Kodaira-Thurston example \cite{book-dasilva} demonstrates
such a case which is given by a manifold $M = \mathbb{R}^4/\Gamma$ where $\Gamma$ is a discrete
group generated by symplectomorphisms acting on $\mathbb{R}^4$. The manifold $M$ is a flat torus
bundle over a torus and is both symplectic and complex. But it admits no K\"ahler structure.
Thus a question is to what extent the emergent geometry obeying eqs. (\ref{kahler-cons1})-(\ref{kahler-cons3})
describes an almost complex manifold. Since the almost complex structure (\ref{comp-symp})
was already chosen to be compatible with $\Omega$, the above argument implies that
non-K\"ahler manifolds may not solve both eqs. (\ref{kahler-cons1}) and (\ref{kahler-cons2})
simultaneously.\footnote{\label{non-kahler}In this case, it may be necessary to take a more general
form for the global symplectic structure (\ref{symp-jtan}) in such a way that
$\widetilde{\Omega}_{\phi} (x;0) = - \frac{\widetilde{\lambda}^2}{2} I_{ab} V^a(x) \wedge V^b(x)$
becomes a closed two-form. In addition it has to be required that the symplectic form
$\widetilde{\Omega}_{\phi}$ is compatible with the other two conditions which will determine
$\widetilde{\lambda} = \widetilde{\lambda}(\lambda)$. But the compatibility condition will be failed
in the sense that $g_\phi (V_a, V_b) = \frac{\lambda^2}{\widetilde{\lambda}^2}
\widetilde{\Omega}_\phi (V_a, JV_b)$.}
There is a useful object to detect an obstruction for a symplectic manifold $M$ to be K\"ahler
which is the Massey products in $H^\bullet (M, \mathbb{R})$ (for the definition of the Massey products,
see, for example, \cite{caval-massey}).
Symplectic manifolds can have non-trivial rational Massey triple products,
but all the Massey triple products on closed K\"ahler manifolds are zero \cite{dgms-km}.
Therefore a symplectic manifold with nontrivial Massey products must be non-K\"ahler.
In this regard, it is also worthwhile to recall an important theorem \cite{dgms-km} stating
that all compact K\"ahler manifold $M$ is formal and, if $M$ is formal, all its Massey products vanish.
The formality of a space $M$ means that its real homotopy type of $M$ is completely defined by
the real cohomology ring $H^\bullet (M, \mathbb{R})$.
Hence the existence of nontrivial Massey products indicates that there
is more to the rational homotopy type of the manifold than can be seen from the
cohomology algebra. Based on these observations, we conjecture that it is necessary to generalize
global objects in eq. (\ref{global-id}) to ``locally" Hamiltonian vector fields to describe
non-K\"ahler manifolds emergent from a generic symplectic geometry.
But we are not adept in adding any remark for a global Riemannian geometry
emergent from a general Poisson algebra.

We conclude this section with the calculation of the K\"ahler constraints
eqs. (\ref{kahler-cons1})-(\ref{kahler-cons3}) in four dimensions.
In four dimensions, the constant symplectic matrix (\ref{symp-matrix}) is equal to the third
self-dual 't Hooft matrix, i.e., $I_{ab} = \eta^3_{ab}$ and we utilize the canonical
splitting (\ref{dec-st}). The result is given by
\begin{eqnarray}\label{kahler-cons41}
    && (\ref{kahler-cons1}) \to \qquad \eta^i_{ab} g^{(+)i}_b + \overline{\eta}^i_{ab} g^{(-)i}_b
    = - V_a \log \lambda^2, \\
    \label{kahler-cons42}
    && (\ref{kahler-cons2}) \to \qquad \eta^i_{(ab} \eta^3_{c)d} g^{(+)i}_d
    + \overline{\eta}^i_{(ab} \eta^3_{c)d} g^{(-)i}_d = \eta^3_{(ab} V_{c)} \log \lambda^2, \\
    \label{kahler-cons43}
    && (\ref{kahler-cons3}) \to \qquad \eta^i_{ab} \eta^3_{cd} g^{(+)i}_d
    - \eta^3_{ab} \eta^3_{cd} g^{(+)3}_d = - \varepsilon^{3ij} \eta^i_{ab} g^{(+)j}_c,
\end{eqnarray}
where $i,j =1,2,3$. Contracting $\eta^3_{ab}$ with eq. (\ref{kahler-cons42}) and
using eq. (\ref{kahler-cons41}) leads to the result, $g^{(+)3}_a = 0, \; \forall a$.
After solving eq. (\ref{kahler-cons43}) in a similar way,
we yield the K\"ahler constraints in four dimensions:
\begin{equation}\label{kahler-4}
g^{(+)1}_a =  \eta^3_{ab}  g^{(+)2}_b, \qquad g^{(+)3}_a = 0.
\end{equation}
An interesting point is that the anti-self-dual parts, $g^{(-)i}_a$, are not constrained when
$I_{ab} = \eta^3_{ab}$. But, if we chose a different complex structure, e.g.,
$I_{ab} = \overline{\eta}^3_{ab}$, the situation would be flipped so that instead
the self-dual parts, $g^{(+)i}_a$, are not constrained.

\section{Noncommutative geometry and quantum gravity}

The quantization of a classical system has proved to be a delicate as well as difficult problem.
(See, for example, \cite{gotay-etal} for some obstructions and difficulties in quantization theory.)
There are two main approaches to the quantization of a general symplectic or Poisson manifold.
The formal deformation quantization \cite{d-quant1,d-quant2,fedo-book} gives rise to
a NC deformation of the algebra of smooth functions, whereas the emphasis of geometric
quantization \cite{geoq-book1,geoq-book2,vaisman} is on the construction of a Hilbert space,
the ``space of states". The deformation quantization embodies quantum dynamics as much as
possible in terms of deformed algebra structures without using the customary representations
in Hilbert spaces. However a physically reasonable concept for the states is necessary in order
to understand the spectral structure of the observable algebra, which is missing in deformation
quantization. Moreover many classical concepts such as the concept of spaces and points
still remain even after (formal) quantization. Therefore the deformation quantization should not
be regarded as a complete quantization \cite{gukov-witten} but rather an intermediate stage of
an ultimate quantization. See also \cite{waldmann-drep} for various aspects of representation theory
in deformation quantization and the concepts of states as positive functionals and the GNS construction.

\subsection{Quantum geometry and matrix models}

We will use deformation quantization to find a canonical method to quantize a general symplectic
or Poisson manifold. First recall the remark below eq. (\ref{new-formal}) that a general Poisson manifold
$(M, \Theta)$ can be constructed by the deformation in terms of a line bundle $L \to M$ on a primitive
Poisson manifold $(M, \theta)$. Without loss of generality, we can assume the primitive Poisson
structure $\theta$ to be in the Darboux-Weinstein frame (\ref{weinstein}).
It was shown \cite{jsw1,jsw2} that the deformation of a Poisson manifold $(M, \theta)$ in terms of
a line bundle $L \to M$ can be formulated by a NC gauge theory using the local covariance map
$\mathcal{D}_A$ obeying the property (\ref{cov-star-map}). In the end the quantization of a general
Poisson manifold is modeled by a NC gauge theory whose star product is defined by the (formal)
deformation quantization of the primitive Poisson manifold $(M, \theta)$. In order to construct
a natural Hilbert space $\mathcal{H}$, the space of states, on which the deformed algebra acts,
we will consider the representation theory of the quantized Poisson
algebra $(\mathcal{A}_\theta, \star)$ which is Morita-equivalent to the dynamical quantum Poisson
algebra $(\mathcal{A}_\Theta, \star')$ as was shown in Refs. \cite{jsw-ncl,buba-me}.
We discussed this aspect before in section 4.

Of course the representation theory of the quantized Poisson algebra $(\mathcal{A}_\theta, \star)$
is in general nontrivial except as the case of symplectic manifolds for which the primitive Poisson
structure $\theta$ is reduced to the canonical one, familiar in quantum mechanics,
in the Darboux-Weinstein frame (\ref{weinstein}). In this case the quantization of the primitive
Poisson manifold $(M, \theta)$ gives rise to the Moyal-Heisenberg algebra (\ref{nc-space}) and
we will regard it as a vacuum algebra. In the presence of a line bundle $L \to M$, the vacuum
coordinates $y^\mu$ should be promoted to the covariant dynamical coordinates defined by eq. (\ref{d-coord})
and a general NC space
\begin{equation}\label{dynamic-nc}
    [X^\mu, X^\nu]_\star = i \theta^{\mu\nu} - i(\theta \widehat{F} \theta)^{\mu\nu} := i \Theta^{\mu\nu}
\end{equation}
generated by the dynamical coordinates is regarded as a (large) deformation of the vacuum
NC space (\ref{nc-space}) due to NC gauge fields. As we discussed before, the vacuum algebra (\ref{nc-space})
is equivalent to the Heisenberg algebra (\ref{h-algebra}) of $n$-dimensional harmonic oscillator.
Hence the underlying Hilbert space on which the deformed algebra (\ref{dynamic-nc}) acts is given by
the representation space of the Heisenberg algebra (\ref{h-algebra}). It is the Fock space defined by
\begin{equation}\label{fock-space}
    \mathcal{H} = \{|\vec{n}\rangle \equiv |n_1, \cdots, n_n \rangle | \; n_i \in \mathbb{Z}_{\geq 0},
    \; i=1, \cdots, n \},
\end{equation}
which is orthonormal, i.e., $\langle \vec{n}|\vec{m} \rangle = \delta_{\vec{n},\vec{m}}$ and
complete, i.e., $\sum_{\vec{n} = 0}^{\infty} | \vec{n}\rangle \langle \vec{n}| = \mathbf{1}_{\mathcal{H}}$,
as is well-known from quantum mechanics. Since the Fock basis (\ref{fock-space}) is a countable basis,
it is convenient to introduce a one-dimensional basis using the ``Cantor diagonal method" to put the
$n$-dimensional non-negative integer lattice in $\mathcal{H}$ into one-to-one correspondence
with the infinite set of natural numbers (i.e., 1-dimensional positive integer lattice) \cite{hsy-epjc09}:
\begin{equation}\label{cantor}
\mathbb{Z}^n_{\geq 0} \leftrightarrow \mathbb{Z}_{> 0}: |\vec{n}\rangle \leftrightarrow |n \rangle, \; n = 1, \cdots, N \to \infty.
\end{equation}
In this one-dimensional basis, the completeness relation of the Fock space (\ref{fock-space}) is now
given by $\sum_{n = 1}^{\infty} | n \rangle \langle n| = \mathbf{1}_{\mathcal{H}}$.

Consider two arbitrary dynamical fields $\widehat{C}_1(y)$ and $\widehat{C}_2(y)$ on the NC
space (\ref{nc-space}) which are elements of the deformed algebra $\mathcal{A}_\theta$.
In quantum mechanics physical observables are considered as operators acting on a Hilbert space.
Similarly the dynamical variables on NC space $\mathbb{R}_{\theta}^{2n}$ can be regarded as operators
acting on the Hilbert space (\ref{fock-space}). Thus we can represent the operators acting
on the Fock space (\ref{fock-space}) as $N \times N$ matrices in $\mathrm{End}(\mathcal{H})
\equiv \mathcal{A}_N$ where $N = \mathrm{dim}(\mathcal{H}) \to \infty$:
\begin{eqnarray}\label{matrix-rep}
     && \widehat{C}_1(y) = \sum_{n,m=1}^\infty | n \rangle \langle n| \widehat{C}_1 (y) | m \rangle \langle m|
      := \sum_{n,m=1}^\infty (\Phi_1)_{nm} | n \rangle \langle m|, \nonumber \\
     && \widehat{C}_2(y) = \sum_{n,m=1}^\infty | n \rangle \langle n| \widehat{C}_2 (y) | m \rangle \langle m|
      := \sum_{n,m=1}^\infty (\Phi_2)_{nm} | n \rangle \langle m|,
\end{eqnarray}
where $\Phi_1$ and $\Phi_2$ are $N \times N$ matrices in $\mathcal{A}_N = \mathrm{End}(\mathcal{H})$.
Then we get a natural composition rule
\begin{equation}\label{matrix-comp}
 (\widehat{C}_1 \star \widehat{C}_2) (y) = \sum_{n,l,m=1}^\infty | n \rangle \langle n|
 \widehat{C}_1 (y) | l \rangle \langle l| \widehat{C}_2(y) | m \rangle \langle m|
      = \sum_{n,l,m=1}^\infty (\Phi_1)_{nl}(\Phi_2)_{lm} | n \rangle \langle m|.
\end{equation}
The above composition rule implies that the ordering in the NC algebra $\mathcal{A}_\theta$
is perfectly compatible with the ordering in the matrix algebra $\mathcal{A}_N$.
Thus we can straightforwardly translate multiplications of NC fields in $\mathcal{A}_\theta$
into those of matrices in $\mathcal{A}_N$ using the matrix representation (\ref{matrix-rep})
without any ordering ambiguity. Furthermore, since symplectic manifolds are always unimodular as
we discussed in appendix B, we can define a trace on the deformed algebra $\mathcal{A}_\theta$ as
the integral (\ref{star-trace}) over $M=\mathbb{R}^{2n}$. Using the map (\ref{matrix-rep}) between the NC $\star$-algebra $\mathcal{A}_\theta$ and the matrix algebra $\mathcal{A}_N$,
the trace over $\mathcal{A}_\theta$ can be transformed into the trace over $\mathcal{A}_N$, i.e.,
\begin{equation}\label{matrix-trace}
    \int_M \frac{d^{2n} y}{(2 \pi )^n |\mathrm{Pf}\theta|} = \mathrm{Tr}_{\mathcal{H}} = \mathrm{Tr}_N.
\end{equation}

Let us apply the matrix representation (\ref{matrix-rep}) to a NC gauge theory. For this purpose,
consider a $d=(m+2n)$-dimensional NC U(1) gauge theory on $\mathbb{R}^m \times \mathbb{R}^{2n}_{\theta}$
whose coordinates are $X^M = (x^\mu, y^a), \; M = 0, 1, \cdots, d-1, \; \mu = 0, 1, \cdots, m-1,
\; a=1, \cdots, 2n$ where $\mathbb{R}^m \ni x^\mu$ is an $m$-dimensional either Minkowski or Euclidean
spacetime and $\mathbb{R}^{2n}_{\theta} \ni y^a$ is a $2n$-dimensional NC space obeying the commutation
relation
\begin{equation}\label{extra-nc2n}
    [y^a, y^b]_\star = i \theta^{ab}.
\end{equation}
The $d$-dimensional U(1) connections are similarly split as
\begin{equation}\label{D-conn}
    D_M (X) = \partial_M - iA_M (x,y) = (D_\mu, D_a)(x,y)
\end{equation}
where
\begin{eqnarray} \label{d-u1conn}
&& D_\mu (x,y) = \partial_\mu - iA_\mu(x,y), \\
\label{2n-u1conn}
&& D_a (x,y) = -i\bigl( B_{ab}y^b + A_a(x,y) \bigr) \equiv -iC_a(x,y).
\end{eqnarray}
Here we used the fact (\ref{adj-nc}) that $\partial_a = \mathrm{ad}_{p_a} = -i[B_{ab}y^b, -]_\star$ and
we omitted the hat symbol to indicate NC gauge fields for notational simplicity.
Using the matrix representation (\ref{matrix-rep}) defined by $\mathcal{A}_\theta \to \mathcal{A}_N:
D_M (x,y) \mapsto (D_\mu, - i\Phi_a)(x)$, the $d$-dimensional NC U(1) gauge theory is exactly mapped to
the $m$-dimensional U($N \to \infty$) Yang-Mills theory \cite{japan-matrix,nc-seiberg,hsy-epjc09,hsy-jhep09}:
\begin{eqnarray} \label{equiv-ncu1}
 S &=& - \frac{1}{4 G_{YM}^2} \int d^d X (F_{MN} - B_{MN})^2 \\
 \label{equiv-u1un}
   &=& - \frac{1}{g_{YM}^2} \int d^m x \mathrm{Tr}_N \Bigl( \frac{1}{4} F_{\mu\nu}F^{\mu\nu}
   + \frac{1}{2} D_\mu \Phi_a D^\mu \Phi^a - \frac{1}{4}[\Phi_a, \Phi_b]^2 \Bigr)
\end{eqnarray}
where $F_{MN} - B_{MN} = i [D_M, D_N]_\star, \; G_{YM}^2 = (2 \pi )^n |\mathrm{Pf}\theta| g_{YM}^2$
and $B_{MN} = \left(
                  \begin{array}{cc}
                    0 & 0 \\
                  0 & B_{ab} \\
                  \end{array}
                \right)$. Note that, according to the matrix representation (\ref{matrix-rep}),
the NC U(1) gauge symmetry in $d$ dimensions is also exactly mapped to the ordinary U($N \to \infty$)
gauge symmetry in $m$ dimensions, i.e.,
\begin{eqnarray}\label{gmap-u1un}
   D'_M (X)  &=& U(X) \star  D_M (X) \star U(X)^\dagger, \quad U(X) \in U(1)_\star \nonumber \\
   && \leftrightarrow \quad (D_\mu, \Phi_a)'(x) = U(x) (D_\mu, \Phi_a)(x) U(x)^\dagger,
   \quad U(x) \in U(N \to \infty).
\end{eqnarray}

According to our construction, the above large $N$ gauge theory can be regarded as a (strict) quantization
of a $d$-dimensional manifold $\mathcal{M}$ along the directions of Poisson structure $\theta = B^{-1}$
which is extended only along the $2n$-dimensional subspace. A remarkable point is that the resulting
matrix models or large $N$ gauge theories described by the action (\ref{equiv-u1un}) arise as a
nonperturbative formulation of string/M theories. For instance, we get the IKKT matrix model
for $m=0$ \cite{ikkt}, the BFSS matrix quantum mechanics for $m=1$ \cite{bfss} and
the matrix string theory for $m=2$ \cite{mst}. The most interesting case arises for $m=4$ and $n=3$
which suggests an engrossing connection that the 10-dimensional NC U(1) gauge theory
on $\mathbb{R}^{3,1} \times \mathbb{R}^{6}_{\theta}$ is equivalent to the bosonic action of
4-dimensional $\mathcal{N} = 4$ supersymmetric U(N) Yang-Mills theory, which is the large $N$
gauge theory of the AdS/CFT duality \cite{ads-cft1,ads-cft2,ads-cft3}. According to the large $N$ duality
or gauge/gravity duality, the large $N$ matrix model (\ref{equiv-u1un}) is dual to a higher dimensional
gravity or string theory. Hence it should not be surprising that the $d$-dimensional NC U(1) gauge theory
should describe a theory of gravity (or a string theory) in $d$ dimensions.\footnote{We may emphasize that
the equivalence between the $d$-dimensional NC U(1) gauge theory (\ref{equiv-ncu1}) and
$m$-dimensional U($N \to \infty$) Yang-Mill theory (\ref{equiv-u1un}) is a mathematical identity
and has been known long ago, for example, in \cite{japan-matrix,nc-seiberg}.
Nevertheless the possibility that gravity can emerge
from NC U(1) gauge fields has been largely ignored until recently. But the emergent gravity picture
based on NC U(1) gauge theory debunks that this coincidence did not arise by some fortuity, 
so we want to quote an epigram due to John H. Schwarz \cite{highly}: ``Take coincidences seriously".}
In other words, the emergent gravity from NC gauge fields is actually the manifestation of
the gauge/gravity duality or large $N$ duality in string/M theories.
Therefore the emergent gravity from NC gauge fields opens a lucid avenue to understand the gauge/gravity
duality such as the AdS/CFT correspondence. While the large $N$ duality is still a conjectural duality
and its understanding is far from being complete to identify an underlying first principle for the duality,
we are reasonably understanding the first principle for the emergent gravity from NC U(1) gauge fields
and we know how to derive gravitational variables from gauge theory quantities.
Later we will show that the 4-dimensional $\mathcal{N} = 4$ supersymmetric U(N) Yang-Mills theory
is equivalent to the 10-dimensional $\mathcal{N} = 1$ supersymmetric NC U(1) gauge theory
on $\mathbb{R}^{3,1} \times \mathbb{R}^{6}_{\theta}$ if we consider the Moyal-Heisenberg vacuum
(\ref{extra-nc2n}) which is a consistent solution of the former -- the $\mathcal{N} = 4$ super Yang-Mills theory.

We showed above that the $m$-dimensional U($N \to \infty$) Yang-Mills theory
is equivalent to the $d=(m+2n)$-dimensional NC U(1) gauge theory on $\mathbb{R}^{m} \times
\mathbb{R}^{2n}_{\theta}$. Thus we can apply the emergent gravity picture in section 5 to
the $d=(m+2n)$-dimensional NC U(1) gauge theory to derive a $d$-dimensional Einstein gravity
which is certainly expected to be dual to the $m$-dimensional U($N \to \infty$) Yang-Mills theory.
We think this trinity relation between large $N$ gauge theories, NC U(1) gauge theories
and gravitational theories in various dimensions will shed light on the gauge/gravity duality
or large $N$ duality. For this reason, let us focus
on the commutative limit of the NC gauge theory in (\ref{equiv-ncu1}) which corresponds to a planar
limit ($N \to \infty$) of large $N$ gauge theory. Suppose that the global Poisson structure (see eqs. (\ref{f-poisson-exp}) and (\ref{local-poi}) where $y^\mu$'s are fiber coordinates and are not
related to $y^a$ in eq. (\ref{extra-nc2n})) is given by $\Pi = \frac{1}{2} \Pi^{ab}(x,y)
\frac{\partial}{\partial y^a} \bigwedge \frac{\partial}{\partial y^b} \in \Gamma(\wedge^2 TM_{2n})$
obtained by gluing together local Poisson structures $(U_i \subset
M_{2n}, \Theta)$ on open subsets in $2n$-dimensional symplectic
manifold $M_{2n} = \bigcup_i U_i$. Let $\mathcal{M}$ be an emergent
$d$-dimensional manifold which locally looks like $\mathcal{M} \approx \mathbb{R}^{m} \times M_{2n}$, 
so may be regarded as a regular Poisson manifold. We can follow the procedure in section 5 to
derive $d$-dimensional global vector fields $V_A = (V_\mu, V_a) \in
\Gamma(T\mathcal{M})$ using the map (\ref{global-id}) from the $d$-dimensional NC U(1) connections
(\ref{D-conn}). For example, on open subsets of $\mathbb{R}^{m}
\times \mathbb{R}^{2n}$ (where life becomes simple as was illustrated in eq. (\ref{vec-2n})),
they are given by \cite{hsy-jhep09,hsy-epjc09}
\begin{equation}\label{d2n-vector}
V_A (f) = \left\{
                  \begin{array}{ll}
                  \partial_\mu f(x,y) + \{A_\mu(x,y), f(x,y) \}_\Pi,
& \quad \hbox{$A = \mu$;} \\
                   \{C_a(x,y), f(x,y) \}_\Pi, & \quad \hbox{$A=a$}
                   \end{array}
                          \right.
\end{equation}
for any $f \in C^\infty(\mathcal{M})$ where
\begin{equation}\label{d2n-vec}
    V_\mu (X) = \partial_\mu + A_\mu^a (X) \frac{\partial}{\partial y^a},
    \qquad V_a (X) = C_a^b (X) \frac{\partial}{\partial y^b}.
\end{equation}
Then the dual covectors $V^A = V^A_M (X) dX^M  \in \Gamma(T^* \mathcal{M})$ are given by
\begin{equation}\label{d2n-covec}
    V^A (X) = \Bigl(dx^\mu, V^a_b(X) \bigl( dy^b - A_\mu^b(X) dx^\mu \bigr) \Bigr),
\end{equation}
where $C_a^c V_c^b = \delta^b_a$. The vector fields $V_A$ are volume
preserving as before, i.e. $\mathcal{L}_{V_A} \nu = 0$, with respect
to the volume form
\begin{equation}\label{d2n-volume}
    \nu = \lambda^2 V^1 \wedge \cdots \wedge V^d = d^m x \wedge \nu_{2n}
\end{equation}
where $\nu_{2n} = \lambda^2 V^1 \wedge \cdots \wedge V^{2n}$. Therefore the $d$-dimensional
Lorentzian metric on $\mathcal{M}$ emergent from the NC U(1) gauge fields or large $N$ matrices
is given by
\begin{eqnarray} \label{d2n-metric}
  ds^2 &=& \eta_{AB} E^A \otimes E^B = \lambda^2 \eta_{AB} V^A \otimes V^B \nonumber \\
   &=&  \lambda^2 \Bigl(\eta_{\mu\nu} dx^\mu dx^\nu + \delta_{ab}V_c^a V_d^b \bigl( dy^c - \mathbf{A}^c \bigr)
   \bigl( dy^d - \mathbf{A}^d \bigr) \Bigr)
\end{eqnarray}
where $\mathbf{A}^a = A^a_\mu dx^\mu$ and
\begin{equation}\label{d2n-lambda}
    \lambda^2 = \nu(V_1, \cdots, V_d).
\end{equation}

The $d$-dimensional emergent gravity described by the metric (\ref{d2n-metric}) is completely determined by
the configuration of $d$-dimensional symplectic U(1) gauge fields $A_M(x,y)$ or the $m$-dimensional
gauge-Higgs system $(A_\mu, \Phi_a)(x)$ in U(N) gauge theory. In other words, the equations of motion and
the Bianchi identity for dynamical gauge fields in the NC U(1) gauge theory or U(N) gauge theory can
be mapped to the corresponding equations for the $d$-dimensional Lorentzian metric (\ref{d2n-metric})
in a similar way as eqs. (\ref{eg-bianchi}) and (\ref{eg-eom}).
As expected, it will be difficult to complete the mission for general gauge fields
and indeed we do not yet know the precise form of Einstein equations determined by symplectic or large $N$
gauge fields except lower dimensions $d \leq 4$ \cite{hsy-jhep09}. Hence it may be instructive to consider
a more simpler system. For this purpose, we may introduce linear algebraic conditions of $d$-dimensional
field strengths $F_{AB}$ as a higher-dimensional analogue of four-dimensional self-duality equations
such that the Yang-Mills equations of motion for the action (\ref{equiv-ncu1}) follow automatically.
These are of the following type \cite{highsde1,highsde2,highsde3}:
\begin{equation}\label{highsde}
    \frac{1}{2} T_{ABCD} F_{CD} = \zeta F_{AB}
\end{equation}
with a constant 4-form tensor $T_{ABCD}$. The relation (\ref{highsde}) clearly implies via the Bianchi
identity, $D_{[A} F_{BC]} = 0$, that the equations of motion, $D^{A} F_{AB} = 0$, are satisfied
provided $\zeta$ is nonzero. Keeping in with the action (\ref{equiv-u1un}), a particularly interesting choice
for the tensor $T_{ABCD}$ will be the case; $T_{abcd} \neq 0$, otherwise $T_{ABCD} = 0$.
In this case, nontrivial gauge fields are mapped to adjoint Higgs fields $\Phi_a (x)$
in U(N) Yang-Mills theory that obey the commutation relation
\begin{equation}\label{higgs-comm}
    -i[\Phi_a (x), \Phi_b (x)] = -i [C_a (x,y), C_b (x,y)]_\star = -B_{ab} + F_{ab} (x,y).
\end{equation}
For instance, the important examples in four $(n=2)$ and six $(n=3)$ dimensions are given by
\begin{eqnarray} \label{high4}
  && n=2: \; T_{abcd} = \varepsilon_{abcd},  \qquad \zeta = \pm 1, \\
  \label{high6}
  && n=3: \; T_{abcd} = \frac{1}{2} \varepsilon_{abcdef} I_{ef},  \qquad \zeta = - 1,
\end{eqnarray}
where $I_{ab}$ is the constant symplectic matrix (\ref{symp-matrix}) in six dimensions.
In the case (\ref{high4}), we recover the self-duality equation (\ref{nc-sde}) for NC U(1) instantons.
In the 6-dimensional case (\ref{high6}), we have the so-called Hermitian Yang-Mills equations given by
\begin{eqnarray}\label{hym-eq}
    && F_{ab} = - \frac{1}{4} \varepsilon_{abcdef} F_{cd} I_{ef}, \nonumber \\
    && I^{ab} F_{ab} = 0.
\end{eqnarray}
Actually the second equation needs not be imposed separately because it can be derived from the first
one by using the identity $\frac{1}{8} \varepsilon_{abcdef} I_{cd} I_{ef} = I_{ab}$.
The above Hermitian Yang-Mills equations can be understood
as follows. For $d > 4$, the 4-form tensor $T_{ABCD}$ cannot be invariant under $G = SO(d)$ rotations
and the equation (\ref{highsde}) breaks the rotational symmetry to a subgroup $H \subset SO(d)$.
In the 6-dimensional case, the 4-form tensor (\ref{high6}) breaks the rotational symmetry
$G = SO(6) = SU(4)/\mathbb{Z}_2$ to a subgroup $H=U(3) \subset SO(6)$. Then we can decompose
the 15-dimensional vector space of 2-forms $\Lambda^2 T^* M$ under the unbroken symmetry group $H$
into three subspaces \cite{yanyun}:
\begin{equation}\label{2form-dec}
    \Lambda^2 T^* M = \Lambda_1^2 \oplus \Lambda^2_6 \oplus \Lambda^2_8
\end{equation}
where $\Lambda_1^2, \; \Lambda^2_6, \;\Lambda^2_8$ are one-dimensional (singlet), six-dimensional and
eight-dimensional vector spaces  taking values in $U(1) \subset U(3),\; G/H =\mathbb{C}\mathbb{P}^3$,
and $SU(3) \subset U(3)$, respectively.
The Hermitian Yang-Mills equations (\ref{hym-eq}) project the vector space $\Lambda^2 T^* M$ into
the eight-dimensional subspace $\Lambda^2_8$ which preserves the $SU(3)$ rotational symmetry \cite{yanyun}.

Using the map (\ref{global-vector}) or (\ref{global-id}) (whose simplest version is given by
eq. (\ref{d2n-vector})), we can identify the emergent metric (\ref{d2n-metric}) for gauge fields obeying
the self-duality equations, (\ref{nc-sde}) and (\ref{hym-eq}), in four and six dimensions, respectively.
It was shown \cite{hsy-ijmp09,hsy-jhep09,hsy-epjc09,hsy-epl09} that the classical limit of 4-dimensional
NC U(1) instantons, called symplectic U(1) instantons, is equivalent to gravitational instantons which
are Ricci-flat, K\"ahler manifolds and so Calabi-Yau 2-folds $M_4= CY_2$. If we consider 6-dimensional
NC Hermitian U(1) instantons defined by (\ref{hym-eq}), the first condition is translated into
the K\"ahler condition of a six-dimensional manifold $M_6$ and the second condition demands a Ricci-flat
condition on $M_6$. In the end, the classical limit of 6-dimensional NC Hermitian U(1) instantons
will be mapped to 6-dimensional Ricci-flat and K\"ahler manifolds, namely,
Calabi-Yau 3-folds $M_6= CY_3$ \cite{yanyun-unp}. Remember that NC U(1) gauge fields in extra dimensions
(i.e., along the space $M_{2n}$) originally come from the adjoint scalar fields in U(N) gauge theory and
obey the commutation relation (\ref{higgs-comm}). According to our scheme, we thus expect that
NC (Hermitian) U(1) instantons correspond to the quantization of symplectic U(1) instantons
in four and six dimensions and so equivalently ``quantized" Calabi-Yau manifolds.
Thus they consist of some topological objects made out of large $N$ matrices $\Phi_a \in \mathcal{A}_N$
in the Hilbert space (\ref{fock-space}). It was claimed in \cite{hsy-jhep09,hsy-jpcs12} that the
topological objects take values in the K-theory $K(\mathcal{A}_\theta)$ for NC $\star$-algebra
$\mathcal{A}_\theta$. Via the Atiyah-Bott-Shapiro isomorphism \cite{abs-magic} that
relates complex and real Clifford algebras to K-theory, combined with the trinity relation \cite{loy-jhep}
between NC U(1) instantons, $SU(n)$ Yang-Mills instantons and Calabi-Yau $n$-folds,
it was conjectured there that the topological objects made out of large $N$ matrices
$\Phi_a \in \mathcal{A}_N$ should be realized as leptons and quarks
in the fundamental representation of the holonomy group $SU(n)$ of Calabi-Yau $n$-folds.
Recently a similar geometric model of matters was advocated in \cite{ams-matter,geom2}.
Later we will further discuss this geometric model of matters (or emergent quantum mechanics).

We remark closely related approaches for the quantization of symplectic (or Poisson) manifolds.
Bressler and Soibelman \cite{b-soibelman} studied some relationship between mirror symmetry and
deformation quantization and suggested that the A-model is related to deformation quantization
in the sense that there is a category of holonomic modules (that are the modules with smallest possible
characteristic varieties) over the quantized algebra of smooth functions on a symplectic manifold
and it becomes equivalent (at least locally) to the Fukaya category of
the same symplectic manifold.\footnote{The homological mirror symmetry \cite{hms} states that
the derived category of coherent sheaves on a K\"ahler manifold should be isomorphic to the Fukaya
category of a mirror symplectic manifold. The Fukaya category is described by the Lagrangian
submanifolds of a given symplectic manifold as its objects and the Floer homology groups
as their morphisms.}
Kapustin \cite{kapustin} argued that for a certain class of symplectic manifolds the category of A-branes
is equivalent to a NC deformation of the category of B-branes on the same manifold, so
A-branes can also be described in terms of modules over a NC algebra. He also observed that
generalized complex manifolds are in some sense a semi-classical approximation to NC complex manifolds
with $B$-fields. In particular he showed that the equivalence arises from the SW transformation that
relates gauge theories on commutative and NC spaces. Later this suggestion has been extended and
made more precise in \cite{vpestun,gualtieri}, partly using the framework of generalized complex geometry.
Gukov and Witten \cite{gukov-witten} formulated the problem of quantizing a symplectic manifold $(M, \omega)$
in terms of the A-model of a complexification of $M$ where the Hilbert space obtained by the quantization
of $(M, \omega)$ is the space of strings connecting an ordinary Lagrangian A-brane and a space-filling
coisotropic A-brane. Recently Kay \cite{mkay} showed how affine and projective special K\"ahler manifolds
(arising as moduli spaces of vector multiplets in 4-dimensional $\mathcal{N}=2$ supersymmetric gauge theories) emerge from the structure of Fedosov quantization of symplectic manifolds.

\subsection{Emergent time}

So far we have kept silent to a notorious issue in quantum gravity known as the ``emergent time" \cite{isham}.
We have considered only the emergence of spaces from NC $\star$-algebra $\mathcal{A}_\theta$.
But the special relativity unifies space and time into a single entity -- spacetime.
Furthermore the general relativity dictates that space and time should be subject to the general
covariance and they must be coalesced into the form of Minkowski spacetime in a locally inertial frame.
Hence, if we want to realize the (quantum) general relativity from a NC $\star$-algebra $\mathcal{A}_\theta$,
it is desirable to put space and time on an equal footing. If a space is emergent, so should time. But
the concept of time is more stringent since it is difficult to give up the causality and
unitarity.\footnote{Therefore we believe that a naive introduction of NC time, e.g., $[t, x] = i\theta$,
will be problematic because it is impossible to keep the locality in time with the NC time and
so to protect the causality and unitarity. Moreover we know that the time variable in a conservative
dynamical system behaves like a completely classical variable. But it is difficult to recover this classical
nature of time from the NC time because the commutation relation $[t, x] = i\theta$
leads to the spacetime uncertainty relation, so the finite size squeezing of space
gives rise to a finite extension of time uncertainty even in the commutative limit.}
So we want to define the emergent time together with the emergent space. The craving picture is that
time is entangled with spaces to unfold into spacetime and to take the shape
of Lorentz covariance. Essentially our leitmotif is to understand what is time.
We will see soon that quantum mechanics, a close pedigree of NC space, gives us a decisive lesson
for this question.

Before addressing the issue of emergent time, it will be important to identify where the flat space,
i.e. the space with the metric $g_{\mu\nu} = \delta_{\mu\nu}$, comes from.
In order to trace out the origin of flat space, let us look at the emergent
metric (\ref{emergent-metric}). Definitely the emergent metric becomes flat when $V^a
= \delta^a_\mu dx^\mu \in \Gamma(T^*M)$ or equivalently
$V_a = \delta_a^\mu \frac{\partial}{\partial x^\mu} \in \Gamma(TM)$ in which $\lambda^2 = 1$.
Then the definition (\ref{gen-vec}) of vector fields immediately implies that the (flat) vector field
$V^{(0)}_\mu = \partial_\mu$ is coming from the (vacuum) gauge field given by
$\widehat{C}^{(0)}_a = p_a = B_{a\mu} y^\mu$, turning off all fluctuations.
Note that the vielbein $E^{(0)}_\mu = V^{(0)}_\mu = \partial_\mu$ in this case can be extended
to entire space, so we do not need the globalization (\ref{global-vector}).
We now get into the most beautiful and remarkable point of emergent gravity that is the underlying
key point to resolve the cosmological constant problem \cite{hsy-jhep09,hsy-jpcs12,lee-yang}:
The vacuum algebra (\ref{extra-nc2n}) is responsible for the generation of flat spacetime that
is not an empty space unlike general relativity. Instead the flat spacetime is emergent from a uniform
condensation of gauge fields in vacuum. Here we have embraced time too because we will eventually
describe the evolution of spacetime geometry in terms of derivations of an underlying NC algebra
generated by the vacuum algebra (\ref{extra-nc2n}).

In quantum mechanics, the time evolution of a dynamical system is defined as an inner automorphism
of NC algebra $\mathcal{A}_\hbar$ generated by the NC phase space
\begin{equation}\label{nc-phase}
[x^i, p_j] = i \hbar \delta^i_j.
\end{equation}
It is worthwhile to realize that the mathematical structure of emergent gravity is basically
the same as quantum mechanics. The former is based on the NC space (\ref{extra-nc2n}) while the
latter is based on the NC phase space (\ref{nc-phase}).
Another fundamental fact for the concept of emergent time is that any Poisson manifold $(M, \Pi)$
always admits a dynamical Hamiltonian system on $M$ where the Poisson structure $\Pi$ is
a bivector in $\Gamma(\Lambda^2 TM)$ and the dynamics of the system is described by the Hamiltonian
vector field $X_f = \Pi(df)$ for any energy function $f \in C^\infty(M)$ of an underlying Poisson
algebra \cite{sg-book1,sg-book2}. Since the concept of emergent time has been explored in \cite{hsy-jhep09,hsy-jpcs12,lee-yang} along this viewpoint, let us here consider this issue from different  perspective.

For this reason, let us look at the $d=(m+2n)$-dimensional emergent metric (\ref{d2n-metric}).
According to the gauge/gravity duality, we regard the $d$-dimensional emergent spacetime
described by the metric (\ref{d2n-metric}) as a bulk geometry $\mathcal{M}$ dual to
the $m$-dimensional large $N$ gauge theory (\ref{equiv-u1un}).
However we have to note that the $m$-dimensional commutative spacetime $\mathbb{R}^{m-1,1}$ was not
emergent but preexisted from the beginning. Of course this spacetime also becomes dynamical when
the gauge fields $A_\mu(x)$ are nontrivial fluctuations. But the original background spacetime
$\mathbb{R}^{m-1,1}$ was preexisting\footnote{It is interesting to notice that this part of geometry is
described by $\partial_\mu$ in the covariant derivative (\ref{d-u1conn}).} unlike the entirely emergent
space $M_{2n}$. Initially the emergent space $M_{2n}$ was not existent in the large $N$ gauge
theory (\ref{equiv-u1un}). This space is only emergent as a result of the vacuum condensate described
by $\Phi_a^{(0)} = B_{ab} y^b$ where $y^a$'s satisfy the commutation relation (\ref{extra-nc2n}).
Note that the configuration of vacuum gauge fields $\Phi_a^{(0)}$ is a consistent solution of
U(N) Yang-Mills theory (\ref{equiv-u1un}) and is achieved by turning off all fluctuations,
i.e., $A_\mu = A_a = 0$.
It might be emphasized that the vacuum expectation value $\langle \Phi_a \rangle_\mathrm{vac}
= B_{ab} y^b$ of adjoint scalar fields does not break the Lorentz symmetry $SO(m-1, 1)$ as in the Higgs
mechanism $\langle \phi \rangle_\mathrm{vac} = v$ because $\langle \Phi_a \rangle_\mathrm{vac}$
are $SO(m-1, 1)$ scalars. Even it should not be interpreted as the breaking of Lorentz
symmetry $SO(2n)$ in extra dimensions since the extra space $\mathbb{R}^{2n}$ is newly emergent
from the vacuum condensate \cite{hsy-jpcs12}.
For this vacuum solution, the $d$-dimensional metric (\ref{d2n-metric})
precisely reduces to $\mathbb{R}^{d-1,1} = \mathbb{R}^{m-1,1} \times \mathbb{R}^{2n}$ where
$\mathbb{R}^{2n}$ is the emergent space triggered by the Moyal-Heisenberg algebra (\ref{extra-nc2n}).
The enticing point for us is that the NC space (\ref{extra-nc2n}) plays a similar role in doing
the NC phase space (\ref{nc-phase}) in quantum mechanics. To be precise, we can introduce a
Hamiltonian system, i.e., Heisenberg equations, describing the evolution of spacetime geometry
using the NC algebra (\ref{extra-nc2n}) in the exactly same way as quantum mechanics.

To illuminate this aspect, let us reconsider the action (\ref{equiv-u1un})
for the case $m=1$ \cite{hsy-jpcs12}:
\begin{eqnarray} \label{mqm-action}
 S &=& - \frac{1}{g_{YM}^2} \int dt \mathrm{Tr}_N \Bigl( \frac{1}{2} (D_0 \Phi_a)^2
 - \frac{1}{4}[\Phi_a, \Phi_b]^2 \Bigr) \nonumber \\
   &=& - \frac{1}{4 G_{YM}^2} \int d^d X (F_{MN} - B_{MN})^2
\end{eqnarray}
where we derived the $d=(2n+1)$-dimensional NC gauge theory using the fact that the Moyal-Heisenberg
algebra (\ref{extra-nc2n}) is a solution of the matrix quantum mechanics.
The $(2n+1)$-dimensional Lorentzian metric emergent from the matrix quantum mechanics (\ref{mqm-action})
is simply given by the metric (\ref{d2n-metric}) for the case of $m=1$:
\begin{equation} \label{mqm-metric}
  ds^2 = \lambda^2 \Bigl(-dt^2 + \delta_{ab}V_c^a V_d^b \bigl( dy^c - \mathbf{A}^c \bigr)
   \bigl( dy^d - \mathbf{A}^d \bigr) \Bigr)
\end{equation}
where $\mathbf{A}^a = A^a_0 dt$ and $\lambda^2$ is determined by an invariant volume
form $\nu = dt \wedge \nu_{2n}$.
The above metric is generated by vector fields $V_A = (V_0, V_a)(t, y)
\in \Gamma \bigl(T(\mathbb{R} \times M_{2n}) \bigr)$ which are, for example,
on open subsets of $\mathbb{R} \times \mathbb{R}^{2n}$, given by eq. (\ref{d2n-vector}):
\begin{eqnarray}\label{time-vector}
&& V_0 (f) = \frac{\partial}{\partial t} f(t,y) + \{A_0, f \}_\Pi (t,y), \\
\label{space2n-vector}
&& V_a (f) = \{C_a, f \}_\Pi(t,y),
\end{eqnarray}
for any smooth function  $f \in C^\infty(\mathbb{R} \times \mathbb{R}^{2n})$.
If all fluctuations are turned off, we can see that the emergent geometry (\ref{mqm-metric}) reduces
to flat Minkowski spacetime $\mathbb{R}^{d-1,1}$ and the global Lorentz symmetry $SO(d-1,1)$ is
emergent too as an isometry of the vacuum geometry  $\mathbb{R}^{d-1,1}$. Note that,
if we identity $A_0(t,y) := - H(t, y)$ with a Hamiltonian $H(t, y)$ of a dynamical system whose
phase space is characterized by the global Poisson structure $\Pi = \Pi^{ab}(t, y) \partial_a \wedge \partial_b$,
the first equation (\ref{time-vector}) for a temporal vector field $V_0$ is precisely
the Hamilton's equation of the dynamical system.
It is obvious that the dynamical system in our case is a spacetime geometry described by
the Lorentzian metric (\ref{mqm-metric}) and the quantization of the dynamical system
should be described by the action (\ref{mqm-action}). In this sense, the matrix quantum mechanics,
known as the BFSS matrix model \cite{bfss}, should describe a quantum geometry of space and time.

It should be noted that the time evolution (\ref{time-vector}) for a general time-dependent
system is not completely generated by an inner automorphism since the first term is not
an inner but an outer derivation. But it is well-known \cite{sg-book2} that the time evolution
of a time-dependent system can be defined by the inner automorphism of
an extended phase space whose extended Poisson bivector is given by
\begin{equation}\label{ext-poisson}
    \widetilde{\Pi} = \Pi + \frac{\partial}{\partial t} \bigwedge \frac{\partial}{\partial H}.
\end{equation}
Then one can see that the temporal vector field (\ref{time-vector}) is equal to
the generalized Hamiltonian vector field defined by
\begin{equation}\label{ext-hamvec}
    V_0 = \widetilde{X}_H = - \widetilde{\Pi} (dH) = \Pi(dA_0) + \frac{\partial}{\partial t}.
\end{equation}
However the spatial vector fields (\ref{space2n-vector}) remain intact because of the relation
$\widetilde{V}_a = \widetilde{\Pi}(dC_a) = \Pi(dC_a) = V_a$. Here we remark that the extended Poisson
structure (\ref{ext-poisson}) raises a serious issue if the time variable might also be quantized, i.e.,
time also becomes an operator obeying the commutation relation $[t, H] = i$,
for a general time-dependent system. We will not dwell into this issue since it is a challenging open
question even in quantum mechanics. See \cite{time-qm} for a comprehensive, up-to-date review of this
and related topics. This retreat admits that we do not have any clear understanding on
the issue of emergent time for a general ``time-dependent" geometry.
For the moment, we want to evade this perverse quantization issue of time by simply tolerating that
the evolution of a spacetime geometry in nonequilibrium (we intentionally elude
the tautology with the ``evolution" by saying ``in nonequilibrium" instead of ``time-dependent")
is generated by both inner and outer automorphisms.

Our argument so far implies that the BFSS matrix model (\ref{mqm-action}) can be interpreted
as a Hamiltonian system of IKKT matrix model whose action is given by \cite{ikkt}
\begin{equation} \label{ikkt-action}
 S = - \frac{1}{4g_{YM}^2} \mathrm{Tr}_N [\Phi_a, \Phi_b]^2.
\end{equation}
The above matrix model is a 0-dimensional theory, so it does not assume any kind of spacetime
structures from the beginning. The theory is defined only with a bunch of $N \times N$ matrices (as objects)
which are subject to the following algebraic relations (as morphisms):
\begin{eqnarray}\label{matrix-jacobi}
    && [\Phi_a, [\Phi_b, \Phi_c]] + \mathrm{cyclic}(a \to  b \to c) = 0, \\
    \label{matrix-eom}
    && [\Phi^a, [\Phi_a, \Phi_b]] = 0.
\end{eqnarray}
Physical solutions consist of all possible matrix configurations obeying the above matrix morphisms
up to U(N) gauge transformations.
We adopt a traditional picture so that general matrix configurations are constructed by
considering all possible deformations over a vacuum solution, especially, the most primitive vacuum.
Hence the prime step is to find the primitive vacuum on which all fluctuations are supported.
In particular, we are interested in large $N$ limit, typically, $N \to \infty$.
In this limit, a most natural primitive vacuum is given by the the Moyal-Heisenberg
algebra (\ref{extra-nc2n}), i.e.,
\begin{equation}\label{prim-vacuum}
    \Phi_a^{(0)} \equiv \langle \Phi_a \rangle_{\mathrm{vac}} = B_{ab} y^b \in \mathcal{A}_N
\end{equation}
where $B_{ab} = (\theta^{-1})_{ab}$. We now consider all possible deformations of the vacuum
(\ref{prim-vacuum}) and parameterize them as
\begin{equation}\label{all-fluct}
 \Phi_a (y) =  B_{ab} y^b + A_a (y) \in \mathcal{A}_N.
\end{equation}
We notice that $-i \Phi_a (y)$ becomes a covariant derivative, $D_a (y)
= \partial_a - i A_a (y)$, because the matrix model (\ref{ikkt-action}) contains only
the adjoint operation between matrices under which we can identify $p_a \equiv B_{ab} y^b$ with
$\mathrm{ad}_{p_a} = -i [B_{ab} y^b, -] = \partial_a$. Moreover, using the map (\ref{matrix-rep})
between the matrix algebra $\mathcal{A}_N$ and NC gauge fields in $\mathcal{A}_\theta$,
we can realize a NC field theory representation of the matrix model (\ref{ikkt-action}).
In particular, the adjoint scalar fields (\ref{all-fluct}) are mapped to NC U(1) gauge fields
(\ref{2n-u1conn}):
\begin{equation}\label{ncmatrix-map}
\mathcal{A}_N \to \mathcal{A}_\theta:  \Phi_a \mapsto C_a (y) = iD_a(y).
\end{equation}
Thus we can represent the matrix action (\ref{ikkt-action}) using  NC U(1) gauge fields and
the resulting action is given by \cite{japan-matrix,nc-seiberg}
\begin{equation}\label{mm-ncft}
S = \frac{1}{4 G_{YM}^2} \int d^{2n} y  (F_{ab} - B_{ab})^2.
\end{equation}

Recall that the NC U(1) gauge fields $A_a(y)$ in eq. (\ref{all-fluct}) were introduced as fluctuations
around the vacuum (\ref{prim-vacuum}) which supports an intrinsic symplectic or Poisson structure
represented by the Heisenberg algebra (\ref{extra-nc2n}). Therefore the deformations of the vacuum
(\ref{prim-vacuum}) in terms of NC U(1) gauge fields must be regarded as a dynamical system.
The corresponding Heisenberg equation for an observable $f \in \mathcal{A}_\theta$ is defined by
\begin{equation}\label{q-heisenberg}
    \frac{df(y)}{dt}  = - i [A_0(y), f(y)]_\star,
\end{equation}
that is precisely an analogue of quantum mechanics defined by the symplectic structure (\ref{nc-phase}).
Here we implicitly assumed that the dynamical mechanism we have considered is in the conservative process.
For general time-dependent fluctuations, the above Heisenberg equation has to be replaced by
\begin{equation}\label{td-heisenberg}
    \frac{df(t,y)}{dt}  = \frac{\partial f(t,y)}{\partial t}- i [A_0(t,y), f(t, y)]_\star.
\end{equation}
Its commutative limit will recover the Hamilton's equation (\ref{time-vector}) that is organized into
the temporal vector field, i.e., $\frac{d}{dt} := V_0$. It should be remarked that, if gravity is emergent
from a more fundamental theory, for an internal consistency of the theory, spacetime as well as gravity
should be simultaneously emergent from some fundamental degrees of freedom in the theory.
We observed that the emergent gravity from NC gauge fields is indeed the case.
Consequently, the emergent (quantum) gravity derived from the NC algebra (\ref{extra-nc2n}) provides
a natural concept of emergent time via the Hamiltonian system of spacetime geometry
though the time-dependent case is still elusive.

Finally we want to point out that the above picture of emergent time is consistent with that
in general relativity. In the Hamiltonian formulation of general relativity,
in particular, in the ADM formalism \cite{big-book,sg-book2}, the Hamiltonian $H$ is a constraint
rather than a dynamical variable. We claim that this should be the case too in NC gauge theory because
the $\Lambda$-symmetry in NC gauge theory is equivalent to diffeomorphism symmetry as we showed in appendix A.
Unfortunately this diffeomorphism symmetry is not manifest in the action (\ref{mqm-action})
because it has been represented in a particular vacuum state, e.g., eq. (\ref{extra-nc2n}).
However we observed in section 4 that the (dynamical) diffeomorphism symmetry in NC gauge theory is
realized as the (local) gauge equivalence (\ref{cov-star-map}) between star products or Morita equivalence
(\ref{line-star}) in representation theory (i.e., the ring-theoretic equivalence of bimodules).
In general relativity this choice of a particular vacuum corresponds to a particular background manifold
whose metric is $\overline{g}$. In this case, the diffeomorphism symmetry is reduced to
a Killing symmetry, $\mathcal{L}_X \overline{g} = 0$, of the background metric $\overline{g}$.
Precisely the corresponding situation also arises in NC gauge theory. In a particular vacuum characterized
by a specific symplectic 2-form $\omega$, e.g. $\omega = B$, the $\Lambda$-symmetry is (spontaneously) broken
to the symplectomorphism, $\mathcal{L}_X B = 0$,
which is equivalent to NC U(1) gauge symmetry \cite{hsy-ijmp09}. To be specific,
the NC or large $N$ gauge theory (\ref{mqm-action}) respects the NC U(1) or U($N \to \infty$)
gauge symmetry (\ref{gmap-u1un}). Thus the temporal gauge field $A_0$ becomes a Lagrange multiplier
rather than a dynamical variable. The local gauge transformations will be generated by the first class
constraints which leave the physical states invariant like as general relativity \cite{dirac-cq}.
Of course, one should not expect that the temporal gauge field $A_0$ is directly related
to the Hamiltonian in general relativity since we do not take into account the full diffeomorphism
symmetry in NC gauge theory. Nonetheless we want to put forward that the structure of gauge symmetries and
constraints is compatible each other in two theories, so the concept of emergent time congruous
with general relativity will ensue too.

\subsection{Matrix representation of Poisson manifolds}

In this subsection we will briefly discuss the matrix representation of quantized Poisson manifolds
in a Hilbert space on which deformed Poisson algebra acts. We will focus on its physical correspondence
rather than a mathematical scrutiny.

As Kontsevich proved, every finite dimensional Poisson manifold admits a deformation quantization.
But its representation in a Hilbert space on which quantized Poisson algebra acts is in general
a challenging open problem. Fortunately the most important examples of Poisson manifolds in physics
occur in semi-simple Lie groups and their representation theory had been mathematically completed
in the 20th century. The Hilbert space for an irreducible representation of a compact
Lie algebra is a ``finite" dimensional (complex) vector space unlike the Moyal-Heisenberg
algebra (\ref{nc-space}) whose irreducible representation is infinite dimensional.
We may also apply the Schwinger representation (\ref{poisson-schwinger}) of Lie algebras.
Indeed we will see later that, when we try to realize matter fields such as leptons and quarks
and their non-Abelian interactions, i.e., weak and strong forces in the context of emergent geometry,
this symplectic realization of quantum Poisson algebras becomes more relevant \cite{hsy-jhep09,hsy-jpcs12}.
For a general Poisson manifold, first we can employ the symplectic realization (\ref{poisson-map})
of it and then quantize an ambient symplectic manifold or a symplectic groupoid as before, i.e.,
represent a corresponding NC algebra $\mathcal{A}_\theta$ in the Hilbert space (\ref{fock-space})
which can also be modeled by either a NC gauge theory or a large $N$ gauge theory.
Finally, guided by the Poisson map from the ambient symplectic manifold to the original Poisson manifold,
we can try to find an irreducible representation $(\rho, V)$ of a quantized Poisson algebra $\mathcal{A}_{\mathfrak{P}} \subset \mathcal{A}_\theta$ where $V \subset \mathcal{H}$ is
an $n$-dimensional representation space and $\rho$ is a Lie algebra homomorphism
from $\mathcal{A}_{\mathfrak{P}}$ to $\mathrm{End}(V)$.

We will consider two situations which incorporate a Poisson manifold.
The first situation is that a Poisson manifold directly arises as a vacuum solution of a
NC gauge theory or a large $N$ gauge theory. In some cases the relevant action needs to contain
a mass deformation. One can show that the IKKT matrix model (\ref{ikkt-action}) cannot admit
a compact vacuum such as the Lie algebra (\ref{q-lie-alg}). This must be true too for
the action (\ref{equiv-u1un}) because the latter can be obtained by applying
the ``matrix T-duality" \cite{m-tduality} to the former. It was shown \cite{yasi} that
the mass deformation is actually required to realize constant curvature spacetimes such as
$d$-dimensional sphere, de Sitter and anti-de Sitter spaces. However there seems to be a novel
realization of constant curvature spacetimes as was recently verified in \cite{lry3} with explicit examples.
This realization is involved with a large topology change concomitant with the change of
the compactness of spacetime geometry. This novel mechanism for compactifications is realized
as follows. Consider the Moyal-Heisenberg vacuum (\ref{extra-nc2n}) as a vacuum solution and
then incorporate generic U(1) gauge fields whose field strength does not necessarily vanish
at asymptotic infinity. For instance, the gauge fields $A_a(y)$ in eq. (\ref{all-fluct}) can be arranged
to breed further vacuum condensates, $\langle F_{ab}(y) \rangle_{\mathrm{vac}} \neq 0$,
which are superposed on the original background field $B_{ab}$ for
the Moyal-Heisenberg vacuum (\ref{extra-nc2n}). The analysis in \cite{lry3} shows that
the additional condensate triggered by the U(1) field strength $F_{ab}(y)$ leads to the topology
change of spacetime geometry from a noncompact space to a compact space.
We may envisage a generalization so that the extra vacuum condensates in
$\langle F_{ab}(y) \rangle_{\mathrm{vac}} \neq 0$ occur only in a subspace of $\mathbb{R}^{2n}$,
i.e., $\mathrm{rank} F_{|y| \to \infty} \leq \mathrm{rank} B$.
In this case the compactification of spacetime geometry will arise only in the subspace.
If so, it may be possible to realize Poisson manifolds by turning on general U(1) gauge
fields with a nontrivial asymptotic behavior. This mechanism may be called ``dynamical symplectic
realizations". We think that the dynamical symplectic realization is physically more enticing than
the mass deformation because the mass deformed matrix model does not reproduce the usual massless
U(1) gauge theory in a commutative limit, so it is phenomenologically unviable.

The second situation is largely motivated by the speculation in \cite{hsy-jhep09, hsy-jpcs12}
realizing matter fields such as leptons and quarks in terms of NC U(1) instantons along extra dimensions.
A similar geometric model of matters was recently appeared in \cite{ams-matter,geom2} where matters such as
electron, proton, neutorn and neutrino are realized in terms of four-manifolds such as Taub-NUT, Atiyah-Hitchin,
$\mathbb{C}\mathbb{P}^2$ and $\mathbb{S}^4$. Note that all these four-manifolds in our case
arise from NC U(1) gauge fields \cite{hsy-epl09,lry3}. We start with the relation (\ref{higgs-comm}).
It demonstrates that the adjoint scalar fields in U($N \to \infty$) gauge theory over
the Moyal-Heisenberg vacuum (\ref{extra-nc2n}) are mapped to higher dimensional NC U(1) gauge fields.
We are interested in a time-independent ``stable" solution in U($N \to \infty$) gauge theory.
To construct such a stable solution, consider a NC gauge field configuration described by the
generalized self-duality equation (\ref{highsde}) where $T_{abcd}$ are only nonvanishing structure constants
and others identically vanish. For example, in four $(n=2)$ and six $(n=3)$ dimensions,
they are given by eq. (\ref{high4}) and eq. (\ref{high6}), respectively. In these cases,
the NC gauge field configurations describe NC U(1) instantons in four and six dimensions obeying
(\ref{nc-sde}) and (\ref{hym-eq}), respectively. But these solutions partially break the Lorentz
symmetry $SO(2n)$ which is the isometry of $\mathbb{R}^{2n}$ emergent from the vacuum gauge fields
$\Phi_a^{(0)} = B_{ab} y^b$. In four dimensions, on one hand, $T_{abcd}$ in eq. (\ref{high4}) does not break
the Lorentz symmetry $SO(4) = SU(2)_L \times SU(2)_R/\mathbb{Z}_2$. But the self-duality equation
(\ref{nc-sde}) breaks it into $SU(2)_L$ or $SU(2)_R$ depending on the self-duality. On the other hand,
in six dimensions, $T_{abcd}$ in (\ref{high6}) breaks the Lorentz symmetry $SO(6) = SU(4)/\mathbb{Z}_2$
into $U(3) \subset SO(6)$ because $I_{ab}$ was inherited from the background K\"ahler
form $B_{ab} = |\theta|^{-1} I_{ab}$. The Hermitian U(1) instantons obeying (\ref{hym-eq})
further break $U(3)$ into $SU(3)$.

The NC U(1) instantons in four or six dimensions (though it can be generalized to higher dimensions,
we want to mainly focus on these two cases for simplicity and they seem to be mostly relevant to physics)
will be realized as four or six dimensional submanifolds in $d$-dimensional spacetime described by
the metric (\ref{d2n-metric}).\footnote{\label{dp-brane}We may simplify the situation by assuming
that NC U(1) gauge fields $A_a(x,y)$ on emergent space $\mathbb{R}^{2n}$ depend only on NC coordinates $y^a$,
i.e. $A_a(y)$. Thus NC U(1) instantons in this case are extended along $\mathbb{R}^{p,1}$ with $p=m-1$
whose thickness is set by $\zeta = \sqrt{|\theta|}$. See Fig. \ref{d3}.
Therefore we may identify the NC U(1) instantons with
$Dp$-branes extended along $\mathbb{R}^{p,1}$ but their internal structures depend on their substances, i.e.,
NC U(1) instantons or equivalently Calabi-Yau $n$-folds with different dimensionality.}
As we argued in section 6.1, they are Calabi-Yau $n$-folds. And the unbroken Lorentz symmetry of
NC U(1) instantons, e.g., $SU(2)$ or $SU(3)$, precisely coincides with the holonomy group of
Calabi-Yau $n$-folds. From a theoretical perspective, when there is a symmetry breaking,
order parameters arise as is well-known in condensed matter systems such as superconducting or
ferromagnetic materials. An example of the order parameter is the net magnetization in a ferromagnetic
system, whose direction is spontaneously chosen when the system cooled below the Curie temperature.
A similar phenomenon should happen in NC U(1) instantons or Calabi-Yau $n$-folds. In our case they
are either $SU(2)$ or $SU(3)$ variables depending on solutions in extra dimensions.
To be specific, let us consider NC U(1) gauge fields on $\mathbb{R}^{p,1}$ with $p=m-1$
that appear in the covariant derivative (\ref{d-u1conn}). As we argued above, the NC coordinates
$y^a \in \mathcal{A}_\theta$ should arrange themselves in the form of $SU(2)$ or $SU(3)$ variables
due to the internal structure of $\mathbb{R}^{p,1}$ originated from NC U(1) instantons or
Calabi-Yau $n$-folds (see the footnote \ref{dp-brane}). Certainly they are given by eq. (\ref{schwinger-rep}),
which can be regarded as low-energy order parameters (or collective modes) in the vicinity of
the solution of eq. (\ref{highsde}). For this reason, let us expand NC U(1) gauge fields on $\mathbb{R}^{p,1}$
in terms of the order parameters \cite{hsy-jhep09,hsy-jpcs12}:
\begin{equation}\label{u1exp-order}
    A_{\mu}(x,y) = A_\mu(x) + A_\mu^I(x) Q^I + A_\mu^{IJ}(x) Q^I Q^J + \cdots,
\end{equation}
where we assumed that each term in (\ref{u1exp-order}) belongs to an irreducible representation of
$\rho = \mathrm{End}(V)$ and $V = L^2(\mathbb{C}^n)$. Remarkably SU($n$) gauge fields $A_\mu^I(x)$
as well as ordinary U(1) gauge fields $A_\mu (x)$ arise as low lying excitations on $\mathbb{R}^{p,1}$
of NC U(1) gauge fields when there exists a nontrivial solution obeying eq. (\ref{highsde}) in extra dimensions.

As usual, the Poisson algebra appears as symmetry generators which are composite operators,
namely a symplectic realization (\ref{poisson-schwinger}), rather than fundamental variables
in emergent spacetime. Furthermore it was conjectured \cite{hsy-jhep09,hsy-jpcs12} for the cases
of $m=4$ and $m=6$ that the representation of four or six dimensional NC U(1) instantons
in a (subspace of) Hilbert space $\mathcal{H}$, e.g. eq. (\ref{fock-space}), has an incarnation
in terms of chiral fermions in four-dimensional spacetime $\mathbb{R}^{3,1}$. The curious conjecture
was motivated by a very mysterious (at least for us) connection between homotopy groups, K-theory and
Clifford modules \cite{abs-magic,k-book1,k-book2}.\footnote{\label{deep-connection}It
is amusing to note that the Clifford algebra from a modern viewpoint can be thought of as a quantization
of the exterior algebra, in the same way that the Weyl algebra is a quantization of the symmetric algebra.
In this correspondence the ``volume operator" $\gamma_{d+1}=\gamma_1 \cdots \gamma_d$ in the Clifford algebra corresponds to the Hodge-dual operator $*$ in the exterior algebra.
Note also that any physical force is represented by a 2-form in the exterior algebra taking values
in a classical Lie algebra and 2-forms are in one-to-one correspondence with Lorentz symmetry
generators $J^{\mu\nu} = \frac{1}{4} [\gamma^\mu, \gamma^\nu]$ in the Clifford algebra whose
irreducible representations are spinors, in particular, chiral fermions in even dimensions.
Hence the most engrossing connection is that the chiral fermions in the Clifford algebra correspond
to self-dual instantons in the exterior algebra. Useful references for this point of view, for example,
are Wikipedia (http://en.wikipedia.org/wiki/Clifford${}_{-}$algebra) and Ref. \cite{meinren}.}
An underlying reasoning is the following \cite{hsy-jpcs12}: NC U(1) instantons made out of
time-independent adjoint scalar fields (\ref{higgs-comm}) in U($N \to \infty$) Yang-Mills theory
can be regarded as a homotopy map
\begin{equation}\label{homotopy-map}
    \Phi_a: \mathbb{S}^3 \to GL(N, \mathbb{C})
\end{equation}
from $\mathbb{S}^3$ to the group of nondegenerate complex $N \times N$ matrices.
Thus the topological class of (perturbatively) stable solutions can be characterized by the homotopy
group $\pi_3 \bigl( GL(N, \mathbb{C}) \bigr)$.\footnote{Any Lie group deformation retracts onto
a maximal compact subgroup by the Iwasawa decomposition. In particular, we have homotopy equivalences
$GL(N, \mathbb{C}) \cong U(N), \; GL(N, \mathbb{R}) \cong O(N)$.} As is well-known, in the stable
regime where $N > 3/2$, the homotopy group of $GL(N, \mathbb{C})$ or $U(N)$ defines a generalized
cohomology theory, known as the K-theory $K(X)$. In our case where $X = \mathbb{R}^{3,1}$,
this group with compact support is given by
\begin{equation}\label{stk-group}
    K(\mathbb{R}^{3,1}) = \pi_3 \bigl( U(N) \bigr) = \mathbb{Z}.
\end{equation}
We now come to the connection with K-theory, via the celebrated Atiyah-Bott-Shapiro
isomorphism \cite{abs-magic} that relates complex and real Clifford algebras to K-theory.
It turns out \cite{hsy-jpcs12} that the chiral fermions representing (or emergent from)
the K-theory state (\ref{stk-group}) are in the fundamental representation of gauge group $SU(2)$
or $SU(3)$ that is coming from the unbroken Lorentz symmetry in extra dimensions or the holonomy group
of Calabi-Yau $n$-folds. Through the minimal coupling of the (coarse-grained) fermion with $SU(2)$
or $SU(3)$ gauge fields in eq. (\ref{u1exp-order}), it was claimed in \cite{hsy-jpcs12} that
four (six)-dimensional NC U(1) instantons or Calabi-Yau 2 (3)-folds give rise to leptons (quarks).
This phenomenon is very reminiscent of low-energy phenomenology via Calabi-Yau compactifications
in string theory since a Calabi-Yau manifold serves as an internal geometry whose shapes and
topology determine a detailed structure of the multiplets for elementary particles and gauge
fields through the compactification. But it might be remarked that the Calabi-Yau manifolds in our case
are non-compact and we do not yet know how to construct compact Calabi-Yau manifolds
although we discussed a possible dynamical compactification mechanism earlier in this subsection.

\subsection{Noncommutative field theory representation of AdS/CFT correspondence}

Consolidating all the results obtained so far, here we want to argue that the AdS/CFT correspondence \cite{ads-cft1,ads-cft2,ads-cft3} is a particular case of emergent gravity from NC U(1) gauge fields.
But we will address only some essential features and any extensive progress along this approach
will be reported elsewhere. The AdS/CFT correspondence implies that a wide variety of quantum field
theories provide a nonperturbative realization of quantum gravity. In the AdS/CFT duality,
the dynamical variables are large $N$ matrices, 
so gravitational physics at a fundamental level is described by NC operators.
A field theory of gravity like Einstein's general relativity defined in higher dimensions is a purely
low-energy or large-distance approximation to some large $N$ gauge theory in lower dimensions
where the relevant observables are approximately commutative. Conventional geometry and general relativity
arise as collective phenomena, akin to fluid dynamics arising out of molecular dynamics.
A key point to the AdS/CFT correspondence is that the dynamical variables belong to the $\mathcal{N}=4$
vector multiplet in the adjoint representation of U(N), so they are all $N \times N$
matrices. In particular, classical geometries or a supergravity limit appears in the planar limit
$N \to \infty$. This is a motive why we have to stare again the equivalence between higher-dimensional
NC U(1) gauge theory (\ref{equiv-ncu1}) and lower-dimensional U($N \to \infty$) Yang-Mills
theory (\ref{equiv-u1un}).

Keeping this picture in mind, let us consider four-dimensional $\mathcal{N} = 4$ supersymmetric Yang-Mills
theory with gauge group U(N). The $\mathcal{N} = 4$ super Yang-Mills theory is consisted only of
a vector multiplet $(A_\mu, \lambda^i_\alpha, \Phi_a), \; i=1, \cdots, 4, \; a=1, \cdots, 6$ which
contains 4-dimensional gauge fields $A_\mu$, four Majorana-Weyl gauginos $\lambda^i_\alpha$
and six adjoint scalar fields $\Phi_a$ in the adjoint representation of gauge group U(N) \cite{n=4sym}.
The action is given by
\begin{eqnarray}\label{n=4-action}
    S &=& \int d^4 x \mathrm{Tr} \left\{- \frac{1}{4} F_{\mu\nu} F^{\mu\nu} - \frac{1}{2} D_\mu \Phi_a
    D^\mu \Phi_a + \frac{g^2}{4}[\Phi_a, \Phi_b]^2 + i \overline{\lambda}_i
    \overline{\sigma}^\mu D_\mu \lambda^i \right. \nonumber \\
    && \qquad \qquad \left. + \frac{i}{2} g \overline{\Sigma}^a_{ij} \lambda^i [ \Phi_a, \lambda^j]
    - \frac{i}{2} g \Sigma^{a,ij} \overline{\lambda}_i [ \Phi^a, \overline{\lambda}_j] \right\},
\end{eqnarray}
where $g$ is a gauge coupling constant and $\overline{\Sigma}^a_{ij}, \; \Sigma^{a,ij}$ are
Clebsch-Gordon coefficients related to the Dirac matrices $\gamma^a$ for $SO(6)_R \cong SU(4)_R$.
A crucial point for a sound progress is that the $\mathcal{N}=4$ super Yang-Mills action (\ref{n=4-action})
and supersymmetry transformations are a dimensional reduction of 10-dimensional $\mathcal{N}=1$
super Yang-Mills theory to four dimensions \cite{n=4sym}:
\begin{eqnarray} \label{n=4susytr}
 &&  \delta A_M = i \overline{\alpha} \Gamma_M \Psi,
 \qquad \delta \Psi = \frac{1}{2} F_{MN} \Gamma^{MN} \alpha,
 \qquad M, N = 0,1, \cdots, 9, \nonumber \\
  && A_M = (A_\mu, \Phi_a), \qquad \Psi = \left(
                                            \begin{array}{c}
                                              P_+ \lambda^i \\
                                              P_- \widetilde{\lambda}_i  \\
                                            \end{array}
                                          \right)
  \quad  \mathrm{with} \;  P_\pm = \frac{1}{2}
(1\pm \gamma_5) \; \mathrm{and} \; \widetilde{\lambda}_i = - C \overline{\lambda}^{iT}, \nonumber\\
&& \Gamma^M =(\gamma^\mu \otimes I_8, \gamma_5 \otimes \gamma^a), \qquad
\Gamma_{11} = \gamma_5 \otimes I_8, \qquad C_{10} = C \otimes \left(
                                                                   \begin{array}{cc}
                                                                     0 & I_4 \\
                                                                     I_4 & 0 \\
                                                                   \end{array}
                                                                 \right),
\end{eqnarray}
where $C_{10}$ and $C$ are the charge conjugation operators for 10 and 4 dimensions, respectively.

Consider a vacuum configuration of the action (\ref{n=4-action}):
\begin{equation}\label{n=4vacuum}
    \langle \Phi_a \rangle_{\mathrm{vac}} = B_{ab} y^b, \quad
    \langle A_\mu \rangle_{\mathrm{vac}} = 0, \quad \langle \lambda^i \rangle_{\mathrm{vac}} = 0.
\end{equation}
Assume that the vacuum expectation value $y^a \in \mathcal{A}_N \; (N \to \infty)$ satisfies
the Moyal-Heisenberg algebra
\begin{equation}\label{n=4moyal}
    [y^a, y^b] = i \theta^{ab} I_{N \times N}
\end{equation}
where the NC parameters $\theta^{ab}= (B^{-1})^{ab}$ are associated with
the Poisson structure $\theta = \frac{1}{2}\theta^{ab} \frac{\partial}{\partial y^a}
\bigwedge \frac{\partial}{\partial y^b} \in \Gamma(\Lambda^2 TM)$ of $M=\mathbb{R}^{6}$.
Of course the commutation relation (\ref{n=4moyal}) is meaningful only when we take
the limit $N \to \infty$. It is obvious that the vacuum configuration (\ref{n=4vacuum}) in this limit
is definitely a solution of the theory and preserves four-dimensional Lorentz symmetry.
Now consider fluctuations of large $N$ matrices around the vacuum (\ref{n=4vacuum}):
\begin{eqnarray} \label{n=4bfluct}
&& D_\mu (x,y) = \partial_\mu - iA_\mu(x,y),  \quad
D_a (x,y) \equiv -i\Phi_a (x,y) = -i\bigl( B_{ab}y^b + A_a(x,y) \bigr), \\
\label{n=4ffluct}
&& \Psi (x, y) = \left(
                                            \begin{array}{c}
                                              P_+ \lambda^i \\
                                              P_- \widetilde{\lambda}_i  \\
                                            \end{array}
                                          \right) (x, y),
\end{eqnarray}
where we assumed that fluctuations also depend on vacuum moduli $y^a$.
This procedure is exactly reverse to the previous matrix representation.
Indeed, if we apply the matrix representation (\ref{matrix-rep}) to fluctuations again,
we can recover the original large $N$ gauge fields.
Therefore let us introduce 10-dimensional coordinates $X^M = (x^\mu, y^a)$ and 10-dimensional
connections defined by
\begin{equation}\label{10d-conn}
    D_M(X) = \partial_M - iA_M (x,y) = (D_\mu, D_a) (x,y)
\end{equation}
whose field strength is given by
\begin{equation}\label{10d-fs}
    F_{MN}(X) = i [D_M, D_N]_\star = \partial_M A_N - \partial_N A_M - i[A_M, A_N]_\star.
\end{equation}
As a consequence of the Moyal-Heisenberg vacuum (\ref{n=4moyal}), according to the map between the NC $\star$-algebra $\mathcal{A}_\theta$ and the matrix algebra $\mathcal{A}_N = \mathrm{End}(\mathcal{H})$,
large $N$ matrices in $\mathcal{N}=4$ vector multiplet on $\mathbb{R}^{3,1}$ are mapped to NC gauge fields
and their superpartners in $\mathcal{N}=1$ vector multiplet on $\mathbb{R}^{3,1} \times \mathbb{R}_{\theta}^6$
where $\mathbb{R}_\theta^{6}$ is a NC space whose coordinate generators $y^a \in \mathcal{A}_\theta$ obey
the commutation relation (\ref{n=4moyal}).

As we remarked before, the $\mathcal{N}=4$ super Yang-Mills action (\ref{n=4-action}) and
the supersymmetry transformations (\ref{n=4susytr}) are obtained by a dimensional reduction
of 10-dimensional $\mathcal{N}=1$ super Yang-Mills theory to four dimensions.
Moreover the orderings in U(N) and NC U(1) gauge theories are compatible with each other as we verified
in eq. (\ref{matrix-comp}). Hence it is straightforward to show that the 4-dimensional $\mathcal{N}=4$
U(N) super Yang-Mills theory (\ref{n=4-action}) can be organized into the 10-dimensional $\mathcal{N}=1$
NC U(1) super Yang-Mills theory with the action
\begin{equation}\label{10dsym-action}
    S = \int d^{10} X \left\{ - \frac{1}{4G_{YM}^2} (F_{MN} - B_{MN})^2 + \frac{i}{2} \overline{\Psi}
    \Gamma^M D_M \Psi \right\}
\end{equation}
where $B$-fields take the same form as eq. (\ref{equiv-ncu1}). The action (\ref{10dsym-action}) is invariant
under $\mathcal{N}=1$ supersymmetry transformations given by
\begin{equation}\label{10dsusytr}
    \delta A_M = i \overline{\alpha} \Gamma_M \Psi, \qquad \delta \Psi = \frac{1}{2} (F_{MN} - B_{MN})
    \Gamma^{MN} \alpha.
\end{equation}
We want to emphasize that the relationship between the 4-dimensional U(N) super Yang-Mills
theory (\ref{n=4-action}) and 10-dimensional NC U(1) super Yang-Mills theory (\ref{10dsym-action})
is not a dimensional reduction but they are exactly equivalent to each other.
Therefore any quantity in lower-dimensional U(N) gauge theory can be transformed into an object
in higher-dimensional NC U(1) gauge theory using the compatible ordering (\ref{matrix-comp}).

For example, a Wilson loop in U(N) gauge theory
\begin{equation}\label{wilson-un}
    W_N = \frac{1}{N} \mathrm{Tr} P \exp \Bigl( i \oint \bigl( A_\mu \dot{x}^\mu + \Phi_a \dot{y}^a \bigr)
    ds \Bigr)
\end{equation}
can be translated into a corresponding NC U(1) Wilson ``line" defined by
\begin{equation}\label{wilson-u1}
    \widehat{W} = \frac{1}{V_6} \int d^6 y  P_\star \exp \Bigl( i \int_{\Gamma} \bigl( B_{ab} \dot{y}^a y^b
    + A_M \dot{x}^M \bigr) ds \Bigr)
\end{equation}
where $V_6$ is a volume of extra 6-dimensional space. The gauge invariance requires
the Wilson loop (\ref{wilson-un}) to be closed. But the Wilson line (\ref{wilson-u1}) is
defined in higher dimensions than (\ref{wilson-un}), so it need not be closed.
It is enough to choose a path $\Gamma$ such that its projection onto $x$-space becomes a closed loop.
(This possibility in $y$-space was already pointed out in \cite{dgo-loop}.)
Actually it perfectly makes sense because, in NC gauge theories, open Wilson lines constitute
a set of gauge invariant operators \cite{jopenw,szabo-f,wgross}. $P$ denotes a path ordering which is taken
only for loop variables $x^M(s)$ satisfying $\dot{x}^2 - \dot{y}^2 = 0$ to preserve supersymmetry
(a minimal surface on the boundary of $AdS_5$) \cite{dgo-loop}. This path ordering with respect to
large $N$ matrices recasts in the path ordering $P_\star$ with respect to star product.
Then the phase factor $B_{ab} \dot{y}^a y^b$ vanishes at leading order because of $\dot{y}^a
= \frac{y^a}{\rho}|\dot{x}|$ with $\rho^2 = \sum_{a=1}^6 y^a y^a$ but there will be a NC
correction $\sim \frac{|\dot{x}|}{\rho}$ at next-to-leading order. So it will be interesting to see
how NC $y$-space affects a singular behavior related to a correction coming from cusps
or intersections of loops.

To recapitulate, one can regard the large $N$ matrices in $\mathcal{N}=4$ supersymmetric gauge theory
as operators in $\mathrm{End}(\mathcal{H})$ acting on a separable Hilbert space $\mathcal{H}$ that
is the Fock space of the Moyal-Heisenberg vacuum (\ref{n=4moyal}). An important point is that
any field $\Phi(x) \in \mathcal{A}_N$ in the limit $N \to \infty$ on four-dimensional
spacetime $\mathbb{R}^{3,1}$ can then be mapped to a NC field $\widehat{\Phi}(x,y)$ defined
on (4+6)-dimensional space $\mathbb{R}^{3,1} \times \mathbb{R}_\theta^{6}$
where $\mathbb{R}_\theta^{6}$ is a NC space defined by the Heisenberg algebra (\ref{n=4moyal}).
In the end one sees that a large $N$ matrix $\Phi(x)$ can be represented by its master
field $\widehat{\Phi}(x,y) \in \mathcal{S}(C^\infty(\mathbb{R}^{3,1}) \otimes \mathcal{A}_\theta)$ which is a higher-dimensional NC U(1) gauge field or its superpartner. Therefore the correspondence between the NC $\star$-algebra $\mathcal{A}_\theta$ and the matrix algebra $\mathcal{A}_N = \mathrm{End}(\mathcal{H})$
leads to the equivalence between 4-dimensional $\mathcal{N}=4$ supersymmetric Yang-Mills theory
with U$(N \to \infty)$ gauge group and 10-dimensional $\mathcal{N}=1$ supersymmetric NC U(1) gauge theory.

Since the large $N$ gauge theory (\ref{n=4-action}) defined on a coherent vacuum (\ref{n=4vacuum}) is
mathematically equivalent to the NC U(1) gauge theory described by the action (\ref{10dsym-action}),
we can try to derive a 10-dimensional gravity dual to the $\mathcal{N}=4$ super Yang-Mills theory
directly from the 10-dimensional NC U(1) gauge theory. First, if we turn off fermions,
i.e. $\Psi = 0$, the equivalence is precisely the case with $m=4$ and $n=3$ in eqs. (\ref{equiv-ncu1})
and (\ref{equiv-u1un}). Thereby the Lorentzian metric on an emergent 10-dimensional
spacetime $\mathcal{M}$ is given by eq. (\ref{d2n-metric}).
First consider a vacuum geometry with $A_M = 0$ and $\Psi = 0$ whose metric is given by
\begin{equation}\label{10vacgeo}
    ds^2 = \lambda^2 (\eta_{\mu\nu} dx^\mu dx^\nu + dy^a dy^a).
\end{equation}
So we see that the vacuum geometry for the $\mathcal{N}=4$ super Yang-Mills theory should be
conformally flat and its conformal class depends on the choice of volume form $\nu = d^4 x \wedge \nu_6$.
There are two interesting cases which are conformally flat:
\begin{eqnarray} \label{vacgeo1}
\nu_6 = d^6 y  \quad  &\Rightarrow&  \quad \lambda^2=1, \quad \mathcal{M} = \mathbb{R}^{9,1}, \\
\label{vacgeo2}
\nu_6 = \frac{d^6 y}{\rho^2}   \quad  &\Rightarrow&  \quad  \lambda^2=\frac{1}{\rho^2}, \;
\mathcal{M} = AdS_5 \times \mathbb{S}^5,
\end{eqnarray}
where $\rho^2 = \sum_{a=1}^6 y^a y^a$. What makes this difference? In order to pose this question,
we need to address the uniqueness of the supersymmetric or BPS vacuum which is consistent with
the isometry of the vacuum geometry (\ref{10vacgeo}), in particular, preserving $SO(6)_R$
Lorentz symmetry.

\begin{figure}
\begin{center}
\includegraphics[width=.6\textwidth]{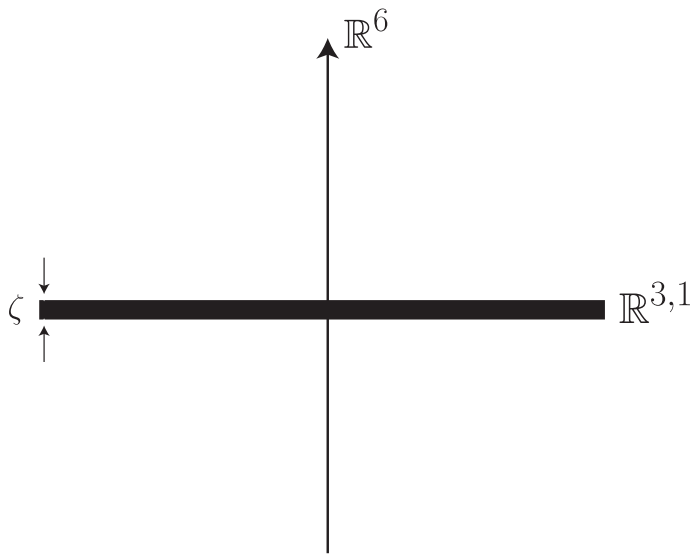}
\end{center}
\caption{D3-brane from NC U(1) instanton}
\label{d3}
\end{figure}

Naturally the supersymmetric or BPS vacuum preserving $SO(6)_R$ Lorentz symmetry is not unique.
The vacuum (\ref{n=4vacuum}) is only one of them.
This vacuum recovers the first case (\ref{vacgeo1})--10-dimensional flat Minkowski spacetime.
Then what is the candidate for the second case (\ref{vacgeo2})?
An educated guess is to consider the stack of instantons at origin in the internal space $\mathbb{R}^6$
so that the directions picked up by instantons are randomly distributed in $SO(6) \cong SU(4)/\mathbb{Z}_2$.
Note that $\mathbb{C}_\theta^3 \cong \mathbb{Z}^3_{\geq 0} \times \mathbb{T}^3$ has a natural
lattice structure defined by the Fock space (\ref{fock-space}) whose lattice spacing is
set by $\zeta = \sqrt{|\theta|}$. We noticed before that the Hermitian NC $U(1)$ instanton
obeying eq. (\ref{hym-eq}) lives in the 8-dimensional vector space $\Lambda_8^2$
in eq. (\ref{2form-dec}) picking up a particular value in the complex
Grassmannian $\mathbb{C}\mathbb{P}^3 = SU(4)/U(3)$ which is the set of complex 1-dimensional
linear subspaces of $\mathbb{C}^4$. Thus we need to consider a small integer lattice
around the origin on which instantons are compactly distributed taking all possible values in $\mathbb{C}\mathbb{P}^3$ like a nucleus containing a lot of nucleons.
Since NC U(1) instantons are BPS states, this superposition
of instantons is probably allowed as a vacuum solution of the $\mathcal{N}=4$ super Yang-Mills theory.
(We noticed a closely related work \cite{g-crystal} which constructs a crystal configuration made
of multi-Taub-NUT solutions which are equivalent to NC U(1) instantons in our case.)
In the classical limit of infinitely many instantons, the stack of instantons at origin
looks spherically symmetric without any preferred direction, so preserves $SO(6)_R$ Lorentz symmetry
as a hydrogen atom does $SO(3)$ rotational symmetry.
As we pointed out in footnote \ref{dp-brane}, these NC U(1) instantons are extended along $\mathbb{R}^{3,1}$
and are supposed to be D3-branes. See Fig. \ref{d3}. Therefore, the configuration of the instanton
lattice corresponds to the stack of $N$ D3-branes in the limit $N \to \infty$.
Considering the fact that the near horizon geometry of the stack of $N$ D3-branes with $N \to \infty$
gives rise to the $AdS_5 \times \mathbb{S}^5$ geometry \cite{dho-fre}, we speculate that the instanton lattice
of infinitely many NC U(1) instantons obeying eq. (\ref{hym-eq}) corresponds to the second case (\ref{vacgeo1})
for the $AdS_5 \times \mathbb{S}^5$ geometry. It will be interesting to understand how our realization
of the $AdS_5 \times \mathbb{S}^5$ geometry is related to the picture in \cite{adsinstanton1,adsinstanton2}
where $AdS_5 \times \mathbb{S}^5$ space emerges as the geometry of the subspace of
multi-instanton collective coordinates which dominates the path integral in the large $N$ limit.

If we include fluctuations generated by 10-dimensional NC U(1) gauge fields $A_M \neq 0$ and
gauginos $\Psi \neq 0$ around a background geometry, either eq. (\ref{vacgeo1}) or
(\ref{vacgeo2}), these excitations will deform the background geometry.
For the bosonic excitations, the deformed geometry will be described by the metric (\ref{d2n-metric}).
For example, the resulting geometry deformed by BPS states is expected to be similar
to the bubbling geometry in Refs. \cite{llm1,lm2}. If we consider the fermionic excitations $\Psi (x,y)$,
we can define fermionic vector fields in the exactly same way as the bosonic case (\ref{gen-vec}):
\begin{equation}\label{fermion-vec}
    \Psi^\star_\alpha (f) \equiv -i [\Psi_\alpha (X), f]_\star = \Psi_\alpha^M (X) \partial_M f + \cdots,
\end{equation}
where $\alpha = 1, \cdots, 16$ is the spinor (Majorana-Weyl) index of $SO(9,1)$. Note that
$\Psi_\alpha^M (X)$ carries both a vector index $M$ and a spinor index $\alpha$ of $SO(9,1)$ and
so we can interpret $\Psi_\alpha^M (X)$ as a 10-dimensional gravitino field.
Combining bosonic and fermionic vector fields together, $(V_A^M, \Psi^M_\alpha) (X)$ form a multiplet
of 10-dimensional $\mathcal{N}=1$ supergravity fields. But this result is not satisfactory
to explain the dual supergravity theory on the $AdS_5 \times \mathbb{S}^5$ background because we
need the $\mathcal{N}=2$ supergravity multiplet for this case.
This failure may not be surprising for the following reason.
We know that the global symmetry group of $\mathcal{N}=4$ super Yang-Mills theory is
$SU(2,2|4)$ which contains the conformal symmetry and conformal supersymmetries \cite{dho-fre}.
From the field theory side, the presence of conformal supersymmetries results
from the fact that the Poincar\'e supersymmetries and the special conformal transformations
do not commute. Since both are symmetries, their commutator must also be a symmetry and
these new generators are responsible for the conformal supersymmetries. From the gravity side,
the conformal symmetry $SO(4,2) = SU(2,2)/\mathbb{Z}_2$ is the isometry of $AdS_5$.
Thus we expect that such a symmetry enhancement for the background geometry (\ref{vacgeo1})
does not happen because the isometry of $\mathbb{R}^{9,1}$ is the usual Poincar\'e symmetry.
So there is no supersymmetry enhancement either. Presumably the 10-dimensional $\mathcal{N}=1$
supergravity multiplet is eligible in this case. Then the question is how to get
the $\mathcal{N}=2$ supergravity multiplet from the 10-dimensional $\mathcal{N}=1$ vector
multiplet. We conjectured before that the $AdS_5 \times \mathbb{S}^5$ geometry arises from
a nontrivial instanton condensate in vacuum which corresponds to a stack of infinitely many D3-branes.
In this case we need to use a corresponding Poisson structure induced by the instanton condensate
to define the supergravity fields, e.g. eq. (\ref{fermion-vec}), which has to explain why
the Poincar\'e symmetry $ISO(3,1)$ is enhanced to the conformal symmetry $SO(4,2)$ in the
instanton background. Unfortunately we do not know the Poisson structure describing the instanton condensate.
It may be highly nontrivial to find it. Thus the question remains open.

\section{Discussion}

Emergence is the essential paradigm for our age (Emergence: The connected lives of ants, brains,
cities, and software, Steven Johnson, 2001). The contemporary physics has revealed growing evidences
that this paradigm can be applied to not only biology and condensed matter systems but also
gravity and spacetime. It might be emphasized again that if gravity is emergent,
then spacetime should be emergent too according to the general theory of relativity.
However the concept of emergent spacetime is very stringent because spacetime plays
a very fundamental role in physics; spacetime plays the background for everything to exist and develop.
Without spacetime, nothing can exist and develop.
A remarkable point in emergent gravity is that the spacetime background is just a condensate in vacuum
that must be allowed for any existence to develop. If true,
everything supported on the spacetime must simultaneously emerge together with emergent spaces
for an internal consistency of the theory.
We argued that matter fields such as leptons and quarks have to emerge too from topological objects in
NC $\star$-algebra $\mathcal{A}_\theta$.

The basic picture of emergent gravity is that gravity and spacetime
are collective manifestations of U(1) gauge fields on NC spacetime.
Thus, in this approach, any spacetime geometry is defined by NC U(1) gauge fields.
It is well-known \cite{nc-top1,nc-top2} that the topology of NC U(1) gauge fields is nontrivial and rich
and NC U(1) instantons \cite{ns-inst,nek-cmp03} represent the pith of their nontrivial topology.
With this perspective we look into these topological solutions applying a deep connection
(see the footnote \ref{deep-connection}) between K-theory, the homotopy groups of classical Lie groups,
and Clifford algebras \cite{k-book1,k-book2}. In particular, the Atiyah-Bott-Shapiro
isomorphism \cite{abs-magic} suggests that these NC U(1) instantons formed along the extra four
or six dimensions can be realized as leptons or quarks.
Moreover NC U(1) instantons are stable excitations over the Moyal-Heisenberg vacuum (\ref{extra-nc2n})
and so originally a part of spacetime geometry, namely, Calabi-Yau $n$-folds,
according to the map (\ref{d2n-vector}). Consequently, we get a remarkable picture, if any, that
leptons and quarks in the Standard model simply arise as stable localized geometries, which are
topological objects in the defining algebra (NC $\star$-algebra) of quantum gravity.
If it is not just a dream, emergent gravity may beautifully embody the emergent quantum mechanics
through a novel unification between spacetime and matter fields \cite{hsy-jpcs12}.

We remark that the overall picture of emergent gravity is very parallel to string theory
as observed in (footnote 3 of) \cite{hsy-jhep09} and (section 6 of) \cite{lee-yang}.
We may grasp an intuitive (though naive) picture for this parallelism.
A Riemannian geometry defined by a pair $(M, g)$ defines an invariant distance in $M$
which may be identified with a geodesic worldline of a particle propagating in the Riemannian manifold.
Instead a symplectic geometry defined by a pair $(M, \Omega)$ measures an area embedded in $M$
which may be identified with a worldsheet minimal surface of a string propagating in the
symplectic manifold. If $M$ is a K\"ahler manifold, i.e., the triple $(g, J, \Omega)$ is compatible,
then there is a different, but closely related point of view.
A (real) geodesic curve in a Riemannian manifold $M$ is a path in $M$. Then the minimal surface
in $M$ from the symplectic perspective is the complex analogue of the real geodesic, which is
called a $J$-holomorphic curve introduced by Gromov \cite{gromovj}.

To be precise, let $(M, J)$ be an almost complex manifold and $(\Sigma, j)$ be a Riemann surface.
A smooth map $u:\Sigma \to M$ is called $J$-holomorphic if the differential $du: T\Sigma \to TM$
is a complex linear map with respect to $j$ and $J$:
\begin{equation}\label{holo-curve}
    du \circ j = J \circ du.
\end{equation}
This statement is equivalent to $\overline{\partial}_J u = 0$ where $\overline{\partial}_J u
:= \frac{1}{2} (du + J \circ du \circ j)$. Then it can be shown \cite{duff-sala,ygoh} that,
when a smooth map $u:\Sigma \to M$ is $J$-holomorphic, we have the identity
\begin{equation}\label{j-area}
    \mathrm{Area}_{gJ}(u) \equiv \int_\Sigma u^* \Omega = \frac{1}{2} \int_\Sigma |du|_J^2.
\end{equation}
This means that any $J$-holomorphic curve minimizes the ``harmonic energy" $S_P (u)
= \frac{1}{2} \int_\Sigma |du|_J^2$ in a fixed homology class which is nothing but the Polyakov action
in string theory. In other words, any $J$-holomorphic curve is a solution of
the worldsheet Polyakov action.
Using the $J$-holomorphic curves, Gromov proved a surprising non-squeezing
theorem \cite{gromovj,duff-sala,ygoh} stating that, if $\phi: B_{2n}(r) \subset \mathbb{C}^n \to (M, \omega)$ is a symplectic embedding of the ball $B_{2n}(r)$ of radius $r$ in $(M, \omega)$  with the standard symplectic form $\omega$ into the cylinder $Z_{2n}(R) = B_2 (R) \times \mathbb{R}^{2n-2}$, then $r \leq R$.

We think that the non-squeezing theorem may be responsible for unexpected (nonlocal) effects
in emergent gravity, e.g., the dark energy. First of all, it may be regarded as a classical manifestation
of the UV/IR mixing in NC field theories \cite{uv-ir}. Actually it was conjectured
in \cite{hsy-jhep09,hsy-jpcs12} that the UV/IR mixing in NC spacetime may be the origin of the dark energy
in our universe incarnated as the mystic energy (\ref{dark-energy}).

We have constructed a global emergent geometry by gluing local data prescribed by NC gauge fields
on local Darboux charts. From the Fedosov quantization approach, it corresponds to solving
the cohomology of quantum Grothendieck connection $\mathfrak{D}$. As we argued in section 6,
the deformation quantization is not a complete quantization but an intermediate stage of
a strict quantization with a Hilbert space. For the former case, the gluing basically means that
two star products defined on nearby overlapping charts are Morita equivalent \cite{jsw-ncl}.
In this case, we can construct Hilbert spaces on local Darboux charts, e.g. eq. (\ref{fock-space}),
on which dynamical NC gauge fields act.
But, for the latter case, it is not obvious how to construct a ``global" Hilbert space.
Even it is not clear whether such a global Hilbert space exists or not.
(This difficulty was also pointed out in \cite{gukov-witten}.) Therefore, the strict quantization
practically asks for the former approach. Then the question is: What is the operation
in the Hilbert space corresponding to the gluing of local data on Darboux charts?
A reasonable answer was put forward by M. Van Raamsdonk \cite{raamsdonk1,raamsdonk2}.
Based on some well-understood examples of gauge/gravity duality, he argued for a picture of quantum gravity
that the emergence of classically connected spacetimes is intimately related to the quantum entanglement
of degrees of freedom in a nonperturbative description of the corresponding quantum system.
This means that if a gravitational description with a large enough geometry emerges,
the quantum system must develop highly entangled low-energy states.
In this picture, quantum spacetimes are roughly analogous to manifolds obtained by gluing
local patches together, with the physics in particular patches of spacetime described
by particular quantum systems. Quantum entanglements between the non-perturbative degrees
of freedom corresponding to different parts of spacetime play a critical role in connecting up
the emergent spacetime. In our case each Darboux chart has its own Hilbert space and on overlaps
they need to be connected each other by unitary transformations obeying
the cocycle condition \cite{jsw-ncl}.
Therefore the Hilbert space representation of Morita equivalent star products may be closely
related to the quantum entanglements building up emergent spacetimes.

\acknowledgments We are very grateful to Joakim Arnlind, Kuerak Chung, Harald Dorn, Hoil Kim,
Sunggeun Lee, Albert Much, Raju Roychowdhury, Alexander Schenkel and Peter Schupp for helpful discussions.
This work was supported by the National Research Foundation of Korea (NRF) grant funded
by the Korea government (MOE) (No. 2011-0010597).
This work was also supported by the National Research Foundation of Korea (NRF) grant funded
by the Korea government (MSIP) through the Center for Quantum Spacetime (CQUeST) of Sogang University
with grant number 2005-0049409.

\newpage

\appendix

\section{Darboux coordinates and NC gauge fields}

In this appendix we will formulate the deformation complex of symplectic structures discussed in section 2
in terms of open string theory. Through this analysis we want to emphasize the local nature of NC gauge fields
by cultivating the Moser flow (\ref{darboux-flow}) to a local description in terms of Darboux coordinates.

Consider a general open string action defined by
\begin{equation}\label{open-action}
    S = \frac{1}{4\pi \alpha'} \int_\Sigma |dX|^2 + \int_\Sigma B + \int_{\partial \Sigma} A
\end{equation}
where $X: \Sigma \to M$ is a map from an open string worldsheet $\Sigma$ to an ambient
spacetime $M$ and $B(\Sigma) = X^* B(M)$ and $A(\partial \Sigma) = X^* A(M)$ are pull-backs of
spacetime fields to the worldsheet $\Sigma$ and the worldsheet boundary $\partial \Sigma$, respectively.
The string action (\ref{open-action}) respects two local gauge symmetries:

(I) Diff($M)$-symmetry: $X \mapsto X' = \phi(X) \in \mathrm{Diff}(M)$,

(II) $\Lambda$-symmetry: $(B, A) \to (B-d\Lambda, A + \Lambda)$, \\
where the gauge parameter $\Lambda$ is a one-form in $M$. Note that the $\Lambda$-symmetry is present
only when $B \neq 0$. The ordinary U(1) gauge symmetry, $A \to A + d\lambda$, is a particular case
with $\Lambda = d\lambda$. When $B=0$, the symmetry is reduced to the U(1) gauge symmetry.
Therefore, in the presence of $B$-fields, the underlying local gauge symmetry is rather enhanced.
Unfortunately the enhanced gauge symmetry due to $B$-fields had not been taken seriously.
Only recently it starts to get some high interests under the names of generalized geometry,
double field theory and higher spin gauge theory.

Let $B$ be a symplectic structure on $M$ and $\theta \equiv B^{-1}$ be a Poisson structure on $M$.
Then the symplectic structure defines a bundle isomorphism $B: TM \to T^*M$
by $X \mapsto \Lambda = - \iota_X B$.
This is the inverse of the anchor map we introduced in eq. (\ref{anchor}). As a result, the $B$-field
transformation, $B \to B + d\iota_X B = (1 + \mathcal{L}_X ) B$, can be understood as a coordinate
transformation generated by the vector field $X \in \Gamma(TM)$.
In other words, the $\Lambda$-symmetry can be considered on par with diffeomorphisms
which basically comes from the Darboux theorem or the Moser lemma in symplectic geometry.
Thus the open string theory on a symplectic manifold $(M, B)$ admits two independent diffeomorphism symmetries.
This enhancement of gauge symmetry already advocates a reason why there must be a radical change of physics
as far as spacetime admits a symplectic structure, namely, a microscopic spacetime becomes NC.

A low energy effective field theory deduced from the open string action (\ref{open-action}) is obtained by integrating out all the massive modes, keeping only massless fields which are slowly varying
at the string scale $l_s^2 = 2 \pi \alpha'$.
From the string action (\ref{open-action}), one can infer that the data of low energy effective field theory
will be specified by the triple $(g, B, A)$ where $g$ denotes a Riemannian metric of
the ambient spacetime $M$. (We are ignoring a possible dilaton coupling in string theory
since it does not play any crucial role in our argument.)
Naturally the low energy effective theory should also respect the above local gauge symmetries
in open string theory. Since the $\Lambda$-symmetry essentially acts as diffeomorphisms and
so two local gauge symmetries should be treated on an equal footing,
one can expect that the low energy effective theory appears with the combination $(g + l_s^2 B, A)$
which signifies an advent of generalized geometry.
The $\Lambda$-symmetry further constrains that the theory has to depend only on
the gauge invariant quantity $\mathcal{F} = B + F$ where $F=dA$.
In the end the low energy effective theory of open strings will be determined by the quantity
$\mathfrak{G} \equiv g + l_s^2 (B + F)$ and the action will be given by the obvious diffeomorphism
invariant measure
\begin{equation}\label{dbi}
    S_{DBI} = \mu_p \int d^{p+1} x \sqrt {\det \mathfrak{G}}.
\end{equation}
The above action is called the Dirac-Born-Infeld (DBI) action.

Now let us apply the chain of symmetry transformations to the quantity $\mathfrak{G}$:
\begin{equation}\label{chain-dbi}
 g + l_s^2 (B + F) \xrightarrow{~F=\mathcal{L}_X B~}  g + l_s^2 (1 + \mathcal{L}_X)B
\xrightarrow{\phi \in \mathrm{Diff}(M)} \phi^*[g + l_s^2 (1 + \mathcal{L}_X)B] = G + \l_s^2 B,
\end{equation}
where $G = \phi^*(g)$. In the last operation of the above chain, the coordinate transformation $\phi \in \mathrm{Diff}(M)$ was chosen such that $\phi^* = (1 + \mathcal{L}_X)^{-1} \approx e^{-\mathcal{L}_X}$.
The Moser lemma (\ref{darboux-flow}) implies that the exponential map $\phi^* \approx e^{-\mathcal{L}_X}$
can be identified with (the leading order of) the Moser flow (\ref{t-flow}).
In terms of local coordinates $\phi: y \mapsto x=x(y)$, the diffeomorphism between $\mathfrak{G}$
and $\mathfrak{G}' \equiv G + \l_s^2 B$ in the map (\ref{chain-dbi}) reads as
\begin{equation}\label{dbi-diff}
    \mathfrak{G}'_{\mu\nu}(y) = \frac{\partial x^a}{\partial y^\mu}
    \frac{\partial x^b}{\partial y^\nu} \mathfrak{G}_{ab} (x)
\end{equation}
where
\begin{equation}\label{dbi-metric}
G_{\mu\nu}(y) = \frac{\partial x^a}{\partial y^\mu}
    \frac{\partial x^b}{\partial y^\nu} g_{ab} (x).
\end{equation}
Consequently we get the equivalence between two different DBI actions
\begin{equation}\label{dbi-equi}
    \int d^{p+1} x \sqrt {\det [g + l_s^2 (B + F)]} =  \int d^{p+1} y \sqrt {\det [G + l_s^2 B]}.
\end{equation}
Note that, though the coordinate transformation to a Darboux frame is defined only locally, the identity (\ref{dbi-equi}) holds globally because both sides are coordinate independent, so local Darboux
charts on the right-hand side can be consistently glued together.
As a result it is possible to obtain a global action for the right-hand side of eq. (\ref{dbi-equi})
by patching the local Darboux charts and the metric (\ref{dbi-metric}) will now be globally defined, i.e.,
\begin{equation}\label{dbi-gmetric}
G_{\mu\nu}(x) = E^a_\mu(x) E^b_\nu(x) \delta_{ab}.
\end{equation}

Let us represent the coordinate transformation $\phi:y \mapsto x=x(y) \in \mathrm{Diff}(M)$
by eq. (\ref{cov-coord}). Note that the dynamical variables on the right-hand side
of eq. (\ref{dbi-equi}) are metric fields $\{ G_{\mu\nu}(y): y \in M \}$ while they on the left-hand side
are U(1) gauge fields $\{F_{\mu\nu}(x): x \in M \}$ in a specific background $(g, B)$.
After substituting the expression (\ref{cov-coord}) of dynamical coordinates into eq. (\ref{dbi-metric}),
one can expand the right-hand side of eq. (\ref{dbi-equi}) around the background $B$-field.
The result is given by \cite{cornalba}
\begin{equation} \label{sw-darboux}
\int d^{p+1} y \sqrt {\det [G + l_s^2 B]} = \int d^{p+1} y \sqrt{\det(l_s^2 B)} \Bigl(1 + \frac{l_s^4}{4}
\widetilde{g}^{ac} \widetilde{g}^{bd} \{C_a, C_b \}_\theta \{C_c, C_d \}_\theta + \cdots \Bigr)
\end{equation}
where $\widetilde{g}^{ab} = \frac{1}{l_s^4} (\theta g \theta)^{ab}$ is a constant open string metric and
$C_a(y) = B_{ab} x^b(y)$ are covariant connections introduced in eq. (\ref{cov-mom}).
As was shown in eq. (\ref{field-moment}), $f_{ab} = \{C_a, C_b \}_\theta + B_{ab}$ are field strengths
of symplectic gauge fields. Therefore we will get NC U(1) gauge theory from the right-hand side
of eq. (\ref{sw-darboux}) after (deformation) quantization. In this respect, the equivalence (\ref{dbi-equi})
of DBI actions represents the SW map between commutative and NC gauge fields.

Some comments are in order to grasp some aspects of emergent gravity.
Note that symplectic or NC gauge fields have been introduced to compensate local deformations of
an underlying symplectic structure by U(1) gauge fields, i.e., the Darboux coordinates in $\phi:y
\mapsto x=x(y) \in \mathrm{Diff}(M)$ obey the relation $\phi^* (B+F) = B$.
This local nature of NC gauge fields is also obvious from the identity (\ref{sw-darboux}) that
they manifest themselves only in a locally inertial frame (in free fall) with the
local metric (\ref{dbi-metric}). If the global metric (\ref{dbi-gmetric}) were used on the left-hand side of eq. (\ref{sw-darboux}), the identification of symplectic or NC gauge fields certainly became ambiguous.
Nevertheless, it may be entertaining to see how the action looks like in terms of the global vector fields in (\ref{dbi-gmetric}). The same calculation as eq. (\ref{sw-darboux}) leads to the result
\begin{equation} \label{complex-darboux}
\int d^{p+1} y \sqrt {\det [G + l_s^2 B]} = \int d^{p+1} y \sqrt{\det(l_s^2 B)} \Bigl(1 + \frac{1}{4}
\mathbb{J}_{ab}\mathbb{J}^{ab} + \cdots \Bigr)
\end{equation}
where
\begin{equation}\label{d-complex}
\mathbb{J}^{ab} \equiv \frac{\theta^{\mu\nu}}{l_s^2} E^a_\mu E^b_\nu = E^a_\mu J^{\mu\nu} E^b_\nu.
\end{equation}
Note that the frame fields in the expression (\ref{d-complex}) are the incarnation of symplectic or
NC gauge fields in Darboux frames. But it may be illusory to find an imprint of
symplectic or NC gauge fields in the expression (\ref{d-complex}). Rather they manifest themselves
as a generic deformation of vacuum complex structure if we intent to interpret $J^{\mu\nu} = \frac{\theta^{\mu\nu}}{l_s^2}$ as a complex structure of $\mathbb{R}^{2n}$, i.e., $J^2 = -1$
and consider $E^a_\mu \approx \delta^a_\mu + h^a_\mu$.
Therefore, with a layman's conviction that the only consistent theory of dynamical metrics is
general relativity (i.e. Einstein gravity) \cite{spin2-weinberg1,spin2-weinberg2},
it should be a sensible idea to derive a spacetime geometry from NC gauge fields.
The problem is how to patch together the local deformations to produce a global spacetime metric
such as eq. (\ref{dbi-gmetric}). Of course the precise procedure is in general intricate.
A useful mathematical device for patching the local information together to obtain a global theory
is to use the theory of jet bundles, which is the subject reviewed in appendix C.

\section{Modular vector fields and Poisson homology}

The aim of this appendix is to explain the property of the modular class of a Poisson manifold and
of its quantization and to introduce a homology theory on Poisson manifolds, using
differential forms, which is to a certain extent dual to the Poisson cohomology (\ref{poisson-coho}).
Some of results will be stated without proofs because a rigorous proof may take us too far away
from our purposes. Instead we will refer to useful references which must fill out the gap
for the rigorous proof.

Let $M$ be a Poisson manifold with Poisson tensor $\theta$ with a trivial canonical class.
Since the bundle $\Lambda^d T^* M$ on $M$ is trivial, a nowhere vanishing regular section of
the canonical bundle always exists, so choose a smooth volume form $\nu$.
But the volume form $\nu$ is defined up to a multiplication by any positive nonvanishing
function $a \in C^\infty(M)$:
\begin{equation}\label{scale-vol}
     \nu \to \widetilde{\nu} = a \nu.
\end{equation}
Take any $f \in C^\infty(M)$. Due to dimensional reasons, the Lie derivative of the volume form $\nu$
in the direction of the Hamiltonian vector $X_f$ must be proportional to itself and thus
there exists a smooth function $\phi_\nu (f) \in C^\infty(M)$ such that
\begin{equation}\label{vol-div}
    \mathcal{L}_{X_f} \nu = \phi_\nu (f) \nu.
\end{equation}
In local coordinates where the volume form is given by $\nu = \upsilon(x) dx^1 \wedge \cdots \wedge dx^d$,
it is easy to calculate the vector field $\phi_\nu$ which is given by
\begin{equation}\label{mod-lvec}
    \phi_\nu = - X_{\log \upsilon} - \partial_\mu \theta^{\mu\nu} \frac{\partial}{\partial x^\nu}.
\end{equation}
From the definition (\ref{vol-div}), it is straightforward to check the following properties:

A: The map $\phi_\nu: f \mapsto \phi_\nu (f)$ is a derivation of $C^\infty(M)$, i.e.,
$\phi_\nu(f \cdot g) = f \phi_\nu (g) + g \phi_\nu (f)$ for $f, g \in C^\infty(M)$,
and thus it is a vector field.

B: The map $\phi_\nu$ is a derivation of $\{-,-\}_\theta$, thus a Poisson vector field, i.e.,
$d_\theta \phi_\nu = 0$.

C: Under the scale transformation (\ref{scale-vol}), the vector field $\phi_\nu$ changes as follows
\begin{equation}\label{vol-scaltr}
    \phi_{\widetilde{\nu}} = \phi_\nu + X_{-\log a}.
\end{equation}

\noindent The above properties can most easily be checked using the local expression (\ref{mod-lvec}).
In particular, it is straightforward to check $d_\theta \phi_\nu = - [\phi_\nu, \theta]_S = 0$
for which it may be necessary to use eq. (\ref{poisson-str}). If $\theta$ is a Poisson tensor,
the vector field $\phi_\nu$ in eq. (\ref{vol-div}) is called the modular vector field of $\theta$
with respect to the volume form $\nu$. These three facts together imply that the modular vector
defines the first Poisson cohomology class $[\phi_\nu] \in H^1_\theta(M)$.
This class is called the Poisson modular class. A Poisson manifold $(M, \theta)$ with $[\phi_\nu] = 0$
will be called {\it unimodular}. It is well known that any symplectic manifold $(M,\omega)$ is unimodular.
(One may use the symplectic volume form $\nu = \frac{\omega^n}{n!}$ to prove that
it gives the zero modular class.)

The Poisson modular class has an interesting interpretation, so-called the infinitesimal KMS condition.
For a compactly supported function $g \in C^\infty(M)$, the following chain of equalities holds:
\begin{eqnarray}\label{poisson-int}
    \int_M \{f, g\}_\theta \nu &=& \int_M \bigl( \mathcal{L}_{X_f}  g \bigr) \nu
    = \int_M \Bigl( \mathcal{L}_{X_f}  ( g \nu )  - g \mathcal{L}_{X_f} \nu \Bigr) \nonumber \\
    &=& \int_M \Bigl( d \bigl( g \iota_{X_f} \nu \bigr)  - g \mathcal{L}_{X_f} \nu \Bigr) \nonumber \\
    &=& - \int_M  g \phi_\nu(f) \nu,
\end{eqnarray}
where, when going from the second line to the third one, we used the Stokes' theorem and
the fact that the function $g$ is compactly supported.
Considering the integral with respect to $\nu$ as a trace $\mathrm{Tr}_\nu$ on the associative
algebra $C^\infty(M)$, the above condition can be written as the form
\begin{equation}\label{poisson-tr}
\mathrm{Tr}_\nu \{f, g\}_\theta = - \mathrm{Tr}_\nu g \phi_\nu(f).
\end{equation}
Therefore the Poisson modular class measures the failure for the trace $\mathrm{Tr}_\nu$ to be also a trace
at the Poisson algebra level. For a unimodular Poisson manifold, it is possible to find a volume form
such that
\begin{equation}\label{unimod-int}
\int_M \{f, g\}_\theta \nu = 0, \qquad \forall f, g \in C^\infty(M)
\end{equation}
with at least one entry compactly supported. The existence of a Poisson trace
is a nontrivial condition (there are many Poisson manifolds having no Poisson trace),
although any symplectic manifold admits a trace. Therefore the modular class of a Poisson manifold is
the obstruction to the existence of a density invariant under the flows of all Hamiltonian vector fields.

The modular vector fields are also related to the canonical homology of a Poisson manifold,
given by the complex in which the chains are differential forms $\Omega^\bullet (M)
= \oplus_{k=0}^d \Omega^k (M)$. We define a homology operator
$\partial_\theta :\Omega^k (M) \to \Omega^{k-1} (M)$ by \cite{p-homology}
\begin{equation}\label{boundary-op}
 \partial_\theta = \iota_\theta \circ d - d \circ \iota_\theta
\end{equation}
where $\iota_\gamma$ is the contraction with a $k$-vector field $\gamma \in \mathcal{V}^k(M)$
with the rule $\iota_{\gamma_1} \iota_{\gamma_2} = \iota_{\gamma_1 \wedge \gamma_2}$
for $\gamma_i \in \mathcal{V}^{k_i}(M)$.
It can be shown that $\partial_\theta^2 = 0$. The corresponding homology of the complex
$(\Omega^\bullet (M), \partial_\theta)$ will be called the Poisson homology of $M$ and denoted by $H^\theta_\bullet (M) = \oplus_{k=0}^d H^\theta_k (M) = \mathrm{Ker} \;
\partial_\theta/\mathrm{Im} \; \partial_\theta$. For example, it is easy to show that
$\{f, g\}_\theta$ is an image of $\partial_\theta$, so the zeroth Poisson homology is
represented by
\begin{equation}\label{0th-homol}
    H^\theta_0 (M) = C^\infty(M)/ \{ C^\infty(M), C^\infty(M) \}_\theta.
\end{equation}
Hence the zeroth Poisson homology can be seen as dual to the space of Poisson traces.
Suppose that $M$ is oriented, so that we can identify densities with differential forms
of top degree. A density $\nu$ is thus a top-dimensional chain for Poisson homology.
Its boundary is given by
\begin{equation}\label{top-homol}
    \partial_\theta \nu = -  d (\iota_\theta \nu) = - \iota_{\phi_{\nu}} \nu
\end{equation}
which one can check by an explicit calculation using the result (\ref{mod-lvec}).
Thus the modular field corresponds to the ($d$ and $\partial_\theta$ exact) $(d-1)$-form
$\partial_\theta \nu = - \iota_{\phi_{\nu}} \nu$.
As a result, a Poisson manifold $(M, \theta)$ is unimodular if and only if there exists a volume
form $\nu$ such that $\partial_\theta \nu = 0$. This means that such a volume form defines
a nontrivial cycle for the higher Poisson homology, so implies $H^\theta_d (M) \neq 0$.

It may be helpful to consider an instructive example \cite{mod-wein}.
Consider a regular Poisson structure on $\mathbb{R}^2 \times \mathbb{S}^1$, with coordinates
$(x, y, t)$, of the form
\begin{equation}\label{reg-poisson}
    \theta = \frac{\partial}{\partial y} \wedge \Bigl( \frac{\partial}{\partial t} + g(x) \frac{\partial}{\partial x} \Bigr)
\end{equation}
where $g(x) = 0$ just at $x=0$. The symplectic leaves for this structure consist of
the cylinder $C$ defined by $x=0$ and a family of planes which spiral around this cylinder.
For $\nu = dt \wedge dx \wedge dy$, we have $\iota_\theta \nu = dx - g(x) dt, \;
d(\iota_\theta \nu) = - g'(x) dx \wedge dt$ and hence $\phi_\nu = g'(x)\frac{\partial}{\partial y}$.
Note that the modular vector field $\phi_\nu$ in this case is coming from the second
part of eq. (\ref{mod-lvec}) with $\theta^{yx} = g(x)$ and $\theta^{yt}=1$, which is not Hamiltonian
unless $g(x)$ is constant, so the modular class of this Poisson structure is nonzero \cite{mod-wein}.

There is another nice formulation for modular vector fields. Let $\nu$ be a volume form on
a $d$-dimensional manifold $M$. Then there is a natural pairing $\nu^\flat : \mathcal{V}^k(M)
\to \Omega^{d-k}(M)$ between a $k$-vector field in $\mathcal{V}^k(M)$ and a $(d-k)$-form
in $\Omega^{d-k}(M)$ via the volume form defined by
\begin{equation}\label{pairing-vf}
\nu^\flat (\Xi) = \iota_\Xi \nu,
\end{equation}
that is, for a given $k$-vector field $\Xi$, there exists any $(d-k)$-vector field $\Pi$ such that
$\langle \nu^\flat (\Xi), \Pi \rangle = \langle \nu, \Xi \wedge \Pi \rangle$.
Thus $\nu^\flat$ corresponds to the Hodge-$*$ operator acting on polyvector fields
in $\mathcal{V}^\bullet (M)$. Define the operator $\delta:  \mathcal{V}^k(M) \to \mathcal{V}^{k-1}(M)$ by
\begin{equation}\label{co-pdiff}
    \delta = (\nu^\flat)^{-1} \circ d \circ \nu^\flat
\end{equation}
which is dual to the codifferential operator $d^\dagger:  \Omega^k(M) \to \Omega^{k-1}(M)$
in the Hodge-de Rham cohomology. Hence $\nu^\flat$ intertwines $\delta$ with $d$, namely, $\nu^\flat
\circ \delta = d \circ \nu^\flat$ which leads to the relation $\nu^\flat \circ \delta^2
= d^2 \circ \nu^\flat$. Since $d^2 = 0$, we also have $\delta^2 = 0$. In a local system of
coordinates $(x^1, \cdots, x^d)$ with $\nu = \upsilon(x) dx^1 \wedge \cdots \wedge dx^d$ and denoting
$\frac{\partial}{\partial x^\mu} = \zeta_\mu$, we have the following
simple formula for the divergence operator:
\begin{equation}\label{div-pop}
    \delta \Xi = \upsilon(x)^{-1} \frac{\partial^2}{\partial x^\mu \partial \zeta_\mu}
    \bigl(\upsilon (x) \Xi \bigr).
\end{equation}
For example, $\delta X = (\nu^\flat)^{-1} \circ \mathcal{L}_X \nu \equiv \mathrm{div} \; X$ of
a vector field $X$ is nothing but the divergence of $X$ with respect to the volume form $\nu$.
Hence we call $\delta$ in (\ref{co-pdiff}) the divergence operator.

Using the result (\ref{div-pop}), it is straightforward to show that the SN bracket (\ref{sn-bracket})
can be expressed by
\begin{equation}\label{snbracket-div}
    [P, Q]_S = (-)^q \delta(P \wedge Q) - (\delta P) \wedge Q - (-)^q P \wedge \delta Q.
\end{equation}
This result immediately leads to the fact that the divergence operator is a graded derivation
of the SN bracket:
\begin{equation}\label{snb-gdiv}
    \delta [P, Q]_S = [P, \delta Q]_S + (-)^{q-1} [\delta P, Q]_S.
\end{equation}
For a Poisson tensor $\theta$, it is then easy to derive from eq. (\ref{snb-gdiv}) and
the definition (\ref{co-pdiff}) the following properties
\begin{eqnarray}\label{snb-gpoi}
    && \delta [\theta, \theta]_S =  - 2 [\delta \theta, \theta]_S = 0, \\
    \label{snb-gvol}
    && \mathcal{L}_{\delta \theta} \nu = d \iota_{\delta \theta} \nu = d \circ \nu^\flat (\delta \theta)
    = d^2 \circ \iota_{\theta} \nu = 0.
\end{eqnarray}
The first property (\ref{snb-gdiv}) implies that the vector field $\delta \theta$
is a Poisson vector field and, according to eq. (\ref{snb-gvol}), it preserves the volume form $\nu$.
Note that the divergence operator transforms under the scale transformation (\ref{scale-vol}) as follows
\begin{equation}\label{div-scale}
    \widetilde{\delta} \Xi = \delta \Xi + [ \Xi, \log a]_S
\end{equation}
where $\widetilde{\delta}$ is the divergence operator with respect to the volume
form $\widetilde{\nu} = a \nu$. In particular, if $\theta$ is a Poisson structure, then the transformation (\ref{div-scale}) is equal to
\begin{equation}\label{dpoi-mod}
   \widetilde{\delta} \theta = \delta \theta + X_{-\log a}.
\end{equation}
Therefore we see that the vector field $\delta \theta$ obeys all the properties (A-C)
for the modular vector field. Indeed the straightforward calculation shows that
$\delta \theta  = \phi_\nu$ in eq. (\ref{mod-lvec}). This proves that the vector field $\delta \theta$
is a modular vector field with respect to $\nu$ and a Poisson manifold $(M, \theta)$ is unimodular
if $H^1_\theta (M) \ni [\delta \theta] = 0$.

For a compact symplectic manifold $(M, \omega)$ of dimension $d=2n$ whose Poisson structure
is given by $\theta = \omega^{-1}$,
there exist natural isomorphisms between the de Rham cohomology $H^k_{dR} (M)$, Poisson cohomology
$H^k_\theta(M)$ and Poisson homology $H^\theta_k (M)$ for $k =0, \cdots, d$ \cite{p-homology}:
\begin{eqnarray}\label{drco-pco}
&& H^k_\theta(M) \cong H^k_{dR} (M), \\
\label{pho-drco}
&& H_k^\theta(M) \cong H^{d-k}_{dR} (M), \\
\label{pho-pco}
&& H_k^\theta(M) \cong H^k_\theta (M).
\end{eqnarray}
Combining (\ref{drco-pco}) with (\ref{pho-drco}), we get the following Poincar\'e duality
between Poisson homology and cohomology
\begin{equation}\label{poin-duality}
 H_k^\theta(M) \cong H^{d-k}_{\theta} (M).
\end{equation}
However the above isomorphisms do not hold for non-symplectic Poisson manifolds
and it is very hard to compute the Poisson (co)homology for them.
But, if a Poisson manifold $(M, \theta)$ is unimodular, it turns out \cite{p-duality1,p-duality2}
that the Poincar\'e duality (\ref{poin-duality}) still holds true.

A trace on a deformed algebra $\mathcal{A}_\theta = C^\infty(M)\bigl[[\hbar]\bigr]$ is by definition
a linear functional $\mu:C_c^\infty(M) \to \hbar^{-n} \mathbb{C} \bigl[[\hbar]\bigr]$ on the compactly
supported functions whose formal extension to $C_c^\infty(M)\bigl[[\hbar]\bigr]$ satisfies
the usual condition $\mu (f \star g) = \mu (g \star f)$. For example, when $(M = \mathbb{R}^{2n}, \omega)$
is a symplectic manifold, the natural trace coming via the Weyl correspondence from the trace
of operators is\footnote{Basically the trace by definition has to preserve the physical dimension of
operators. That is the reason why there is $\hbar^{-n}$ in the trace (\ref{star-trace}).
If the symplectic structure $\omega = \frac{1}{2} B_{\mu\nu} dx^\mu \wedge dx^\nu$ refers to a plain
Euclidean space $\mathbb{R}^{2n}$ rather than a particle phase space,
it is not necessary to include $\hbar^{n}$ in the denominator of the trace (\ref{star-trace})
because $[\omega]$ in this case is dimensionless in itself, i.e., $[B_{\mu\nu}] = (\mathrm{length})^{-2}$.
Nevertheless it may be convenient to keep the deformation parameter $\hbar$ to control the order of deformations.}
\begin{equation}\label{star-trace}
    f \mapsto \mu (f) = (2\pi \hbar)^{-n} \int_M f \frac{\omega^n}{n!}.
\end{equation}
Therefore the trace $\mu$ on a Poisson manifold can exist only when the Poisson manifold is unimodular
obeying (\ref{unimod-int}). A star product is called in \cite{cofs} strongly closed if the functional (\ref{star-trace}) still defines a trace. The existence of a strongly closed star product
on an arbitrary symplectic manifold was shown in \cite{omy2}. Moreover the classification in \cite{nts2}
implies that every star product on a symplectic manifold is equivalent to a strongly closed star product
and the set of traces for a star product on a symplectic manifold forms a 1-dimensional module over
$\mathbb{C} \bigl[[\hbar]\bigr]$, so the trace is essentially unique.
The existence of a strongly closed star product was generalized in \cite{felsho,dolgushev}
to the case of any unimodular Poisson manifolds with an arbitrary volume form.
In particular, it was shown in \cite{felsho} that the divergence of a Poisson bivector field (like
the second term in eq. (\ref{mod-lvec})) is involved with tadpoles (edges with both ends at the same vertex)
in Feynmann diagrams and the anomalous terms vanish for divergence free Poisson bivector fields.
Hence, if we insist on using traces as a NC version of integration, then we are forced to restrict
ourselves to the quantization of unimodular Poisson manifolds. In other words, it is unreasonable
to expect a trace on $\mathcal{A}_\theta$ if we start with a non-unimodular Poisson manifold.

Now our concern is how the modular vector fields can be lifted to a (deformation) quantization.
Note that by definition (the properties A-C below (\ref{mod-lvec}))
the modular vector field is a derivation of $C^\infty (M)$ (Property A) as well as
a Poisson vector field (Property B). Therefore there is no essential obstruction for the lift
of the modular vector fields to derivations of a $\star$-algebra if and only if a Poisson manifold
is unimodular which belongs to a particular class of generic Poisson manifolds.
For example, for a $\star$-product on a symplectic manifold which is always unimodular,
any symplectic vector field $X$ extends to a derivation of NC algebra $\mathcal{A}_\theta
= (C^\infty(M)\bigl[[\hbar]\bigr], \star)$. If $X$ is a Hamiltonian vector field, it can be chosen
as an inner derivation which is precisely the case in eq. (\ref{gmap-derivation}) or
(\ref{gmap-did}). In general any symplectic vector field can be quantized as a derivation of
the quantum algebra $\mathcal{A}_\theta$. See, e.g., Lemma 8.4 in \cite{pxu} for the proof.
For a general Poisson manifold, the quantization problem of the modular class in the formal case
was recently obtained by Dolgushev in \cite{dolgushev}.
In particular, it was shown \cite{dolgushev} that, if $\mathfrak{D}_X$ is a derivation
of $\mathcal{A}_\theta$ constructed from a modular vector field $X$ of Poisson structure $\theta$
via the Kontsevich's formality theorem, then the modular (outer) automorphism of $\mathcal{A}_\theta$
is generated by the exponential map $\exp (\mathfrak{D}_X)$ up to an inner automorphism.

\section{Jet bundles}

In this appendix we briefly review jet bundles which have been often used in this paper.
We refer to \cite{jet-book1,jet-book2,jet-book3} for more detailed exposition.

Suppose that $\pi:E \to M$ is a fiber bundle with fiber $F$.
Introduce local coordinates $x^\mu$ for $M$ and $(x^\mu, z^i)$ for $E$ with coordinates $z^i$
of its standard fiber $F$ and let $\Gamma_p(E)$ denote the set of all local sections
whose domain contains $p \in M$. We say that two sections $\sigma$ and $\sigma'$ of $\pi$ have
a first-order contact at a point $x \in M$ if $\sigma^i(x) = \sigma'^i(x)$ and
$\partial_\mu \sigma^i(x) = \partial_\mu \sigma'^i(x)$. This defines an equivalence relation
on the space of local sections.
They are called the first order jets $j^1_x \sigma$ of sections at $x$.
One can justify that the definition of jets is coordinate independent.
The set of all the 1-jets of local sections of $E \to M$, denoted by $J^1E$, has a natural structure of
a differentiable manifold with respect to the adopted coordinates $(x^\mu, z^i, z^i_\mu)$ such that
\begin{equation}\label{jet-man}
    z^i (j^1_x \sigma) = \sigma^i(x), \qquad z^i_\mu (j^1_x \sigma) = \partial_\mu \sigma^i(x).
\end{equation}
We call $z^i_\mu$ the jet coordinates. They posses the transition functions
\begin{equation}\label{jet-morphism}
z'^i_\mu = \frac{\partial x^\lambda}{\partial x'^\mu}
    (\partial_\lambda + z_\lambda^j \partial_j) z'^i
\end{equation}
with respect to the bundle morphism $z'^i = \phi^i(x,z), \; x'^\mu = \phi^\mu (x)$.
The jet manifold $J^1 E$ admits the natural fibrations
\begin{eqnarray} \label{fibration1}
&& \pi^1: J^1 E \to M \quad \mathrm{by} \quad j^1_x \sigma \mapsto x, \\
\label{fibration2}
&& \pi^1_0: J^1 E \to E \quad \mathrm{by} \quad j^1_x \sigma \mapsto \sigma(x).
\end{eqnarray}
Any section $\sigma$ of $E \to M$ has the jet prolongation to the section of the jet
bundle $J^1 E \to M$ defined by
\begin{equation}\label{prolong}
    (J^1 \sigma)(x) = j^1_x \sigma, \qquad  z^i_\mu \circ J^1 \sigma = \partial_\mu \sigma^i(x).
\end{equation}
An important fact is that there is a one-to-one correspondence between the connections on a fiber
bundle $E \to M$ and the global sections of the affine jet bundle $J^1 E \to E$,
as will be discussed later.

The notion of first jets $j_x^1\sigma$ of sections of a fiber bundle can naturally
be extended to higher order jets. Let $I=(I_1, \cdots, I_n)$ be a multi-index
(an ordered $n$-tuple of integers) and $\partial_I \equiv \frac{\partial^{|I|}}{\partial x^I}
= \prod_{i=1}^n \Bigl( \frac{\partial}{\partial x^{\mu_i}} \Bigl)^{I_i}$ where $|I| = \sum_{i=1}^n I_i$.
Define the local sections $\sigma, \sigma' \in \Gamma_p (E)$ to have the same $k$-jet
at $p \in M$ if
\begin{equation}\label{k-jet}
\frac{\partial^{|I|} \sigma}{\partial x^I} |_p = \frac{\partial^{|I|}\sigma'}{\partial x^I} |_p,
\qquad 0 \leq |I| \leq k.
\end{equation}
A $k$-jet is an equivalence class under this relation and the $k$-jet with representative $\sigma$
is denoted by $j^k_p \sigma$. The holonomic sections $j^k_p \sigma$ are called $k$th-order jet prolongations
of sections $\sigma \in \Gamma_p(E)$. In brief, one can say that sections of $E \to M$
are identified by the $k + 1$ terms of their Taylor series at points of $M$.
The particular choice of coordinates does not matter
for this definition. In this respect, jets may also be seen as a coordinate free version of Taylor expansions.
The $k$-order jet manifold $J^k E$ is then defined by the set of all $k$-jets $j^k_x \sigma$ of
all sections $\sigma$ of $\pi$. Therefore the points of $J^kE$ may be thought of as coordinate free
representations of $k$th-order Taylor expansions of sections of $E$.
The $k$-jet manifold $J^k E$ is endowed with an atlas of the adapted coordinates
\begin{eqnarray}\label{k-jetman1}
   && (x^\mu, z_I^i), \qquad z_I^i \circ J^k \sigma = \partial_{I} \sigma^i(x), \qquad 0 \leq |I| \leq k, \\
\label{k-jetman2}
&& z'^i_{\mu + I} = \frac{\partial x^\lambda}{\partial x'^\mu} d_\lambda z'^i_I,
\end{eqnarray}
where the symbol $d_\mu$ stands for the higher order total derivative defined by
\begin{equation}\label{ho-der}
    d_\mu = \partial_\mu + \sum_{0 \leq |I| \leq k-1} z^i_{\mu + I} \partial_i^{I},
    \qquad d'_\mu = \frac{\partial x^\lambda}{\partial x'^\mu} d_\lambda
\end{equation}
and $\partial^I_i \equiv \frac{\partial}{\partial z^i_I}$.
We call the coordinates in eq. (\ref{k-jetman1}) the natural coordinates on the jet space.

There is a natural projection from $J^2E$  to $J^1E$, the truncation $\pi^2_1$,
characterized by dropping the second-order terms in the Taylor expansion.
In general, one has the natural truncations $\pi^n_m: J^nE \to J^mE$
for all $0 < m < n$ and $\pi^n: J^nE \to M$ by
\begin{equation}\label{n-jet}
\pi^n_m: j^n_x \sigma \mapsto j^m_x \sigma,  \qquad \pi^n: j^n_x \sigma \mapsto x.
\end{equation}
The coordinates (\ref{k-jetman1}) are compatible with the natural surjections $\pi^n_m \; (n>m)$
which form the composite bundle
\begin{equation}\label{chain-jet}
\pi^n: J^nE \xrightarrow{~\pi^n_{n-1}~} J^{n-1}E \xrightarrow{~\pi^{n-1}_{n-2}~} \cdots \xrightarrow{~\pi^1_0~} E
\xrightarrow{~\pi~} M
\end{equation}
with the properties
\begin{equation}\label{comp-jet}
    \pi^k_m \circ \pi^n_k = \pi^n_m, \qquad \pi^k \circ \pi^n_k = \pi^n.
\end{equation}
The composite bundle (\ref{chain-jet}) is constructed by defining $J^{k+1} E$ as the first jet bundle
of $J^k E$ over $M$ and iterating this construction. Then each jet bundle $J^{k+1} E$ becomes
a vector bundle over $J^k E$ and a fiber bundle over $E$.
The inductive limit $\mathcal{E} \equiv J^\infty E$ of the inverse sequence of eq. (\ref{chain-jet})
is defined as a minimal set such that there exist surjections
\begin{equation}\label{inf-jet}
\pi^\infty: \mathcal{E} \to M, \qquad \pi^\infty_0 : \mathcal{E} \to E,
\qquad \pi^\infty_k: \mathcal{E} \to J^k E
\end{equation}
obeying the relations $\pi^\infty_n = \pi^k_n \circ \pi^\infty_k$ for all admissible $k$ and $n < k$.
One can think of elements of $\mathcal{E}$ as being infinite order jets of sections of $\pi: E \to M$
identified by their Taylor series at points of $M$. Therefore a fiber bundle $E$ is a strong deformation
retract of the infinite order jet manifold $\mathcal{E}$. A bundle coordinate atlas $\{\mathcal{U}_E,
(x^\mu, z^i) \}$ of $E \to M$ provides $\mathcal{E}$ with the manifold coordinate atlas
\begin{equation}\label{infty-atlas}
   \{ (\pi^\infty_0)^{-1} (\mathcal{U}_E), (x^\mu, z_I^i) \}_{0 \leq |I|}, \qquad
   z'^i_{\mu + I} = \frac{\partial x^\lambda}{\partial x'^\mu} d_\lambda z'^i_I.
\end{equation}

The tangent vectors to the fibers $F$ form a vector subbundle of $TE$ (because they have good
transformational character) and it is called the vertical vector space denoted by $T^\perp E$.
Note that $\Upsilon$ is tangent to the fiber if and only if $\pi_* \Upsilon = 0$,
hence $T^\perp E = \ker \pi_*$. But, although vectors tangent to $M$ locally complement the vertical
vector space, they do not transform properly on $M$. Thus $TM$ is not a subbundle of $TE$.
A nonlinear connection needs to be introduced to care a selection of complementary vector bundle to
$T^\perp E$ in $TE$. This bundle is usually called the horizontal vector space and denoted by $T^\parallel E$.
The nonlinear connection may be defined via the short exact sequence of vector bundles over $E$:
\begin{equation}\label{exact-seq}
 0 \xrightarrow{~~} T^\perp E  \xrightarrow{~~} TE \xrightarrow{~\pi_*~} \pi^* TM
\xrightarrow{~~} 0
\end{equation}
where $\pi^* TM$ is the pull-back bundle $E \times_M TM$ of $TM$ onto $E$.
A nonlinear connection is a splitting of this short exact sequence: $TE \cong  T^\perp E \oplus T^\parallel E$
where $T^\parallel E \cong \pi^* TM$.

A vector field $X$ on a fiber bundle $\pi: E \to M$ is called projectable if it projects onto a vector field
on $M$, i.e., there exists a vector field $\tau$ on $M$ such that
\begin{equation}\label{pro-vec}
\tau \circ \pi = T \pi \circ X.
\end{equation}
A projectable vector field takes the coordinate form
\begin{equation}\label{pro-x}
    X = X^\mu(x) \partial_\mu + X^i(x,z) \partial_i, \qquad \tau = X^\mu(x) \partial_\mu.
\end{equation}
Its flow generates a local one-parameter group of automorphisms of $E \to M$ over a local one-parameter
group of diffeomorphisms of $M$ whose generator is $\tau$.
A projectable vector field is called vertical if its projection onto $M$ vanishes, i.e., if it lives
in $T^\perp E$. Any projectable vector field $X$ has the following $k$-order jet prolongation to a vector
field on $J^k E$:
\begin{equation}\label{prolong-vec}
    J^k X = X^\mu \partial_\mu + X^i \partial_i + \sum_{1 \leq |I| \leq k} \Bigl( d_I (X^i - z^i_\mu X^\mu)
    + z^i_{\mu + I} X^\mu \Bigr) \partial^I_i
\end{equation}
where we used the compact notation $d_I = d_{\mu_k} \circ \cdots \circ d_{\mu_1}$.
If $X$ is a vertical vector field on $E \to M$, i.e., $X^\mu = 0$, one can see that $J^k X$ is also
a vertical vector field on $J^k E \to M$. Indeed any vector field $\rho_k(X) \equiv J^k X $ on $J^k E$ admits
the canonical decomposition
\begin{eqnarray}\label{dec-provec}
    \rho_k(X) &=& X_H + X_V  \nonumber \\
    &=&  X^\mu d_\mu +  \sum_{|I| \leq k} d_I \Bigl( X^i - X^\mu z^i_{\mu} \Bigr) \partial^I_i
\end{eqnarray}
over $J^{k+1} E$ into the horizontal and vertical parts.
There are also canonical bundle monomorphisms (embeddings) over $J^kE$:
\begin{eqnarray}\label{embedd-h}
    && \eta_k: J^{k+1} E \to T^*M \bigotimes_{J^k E} TJ^k E, \nonumber \\
    && \eta_k = dx^\mu \otimes d_\mu, \\
    \label{embedd-v}
    && \psi_k: J^{k+1} E \to T^*J^k E \bigotimes_{J^k E} T^\perp J^k E, \nonumber \\
    && \psi_k = \sum_{|I| \leq k} \Bigl( dz_I^i - z^i_{\mu + I} dx^\mu \Bigr) \otimes \partial^I_i.
\end{eqnarray}
The one-forms
\begin{equation}\label{contact-form}
    \psi^i_I \equiv dz_I^i - z^i_{\mu + I} dx^\mu
\end{equation}
in eq. (\ref{embedd-v}) are called the local contact forms. A differential one-form $\psi$
on the space $J^k E$ is called a contact form if it is pulled back to the zero form on $M$
by all prolongations. In other words, the one form $\psi \in T^*J^k E$ is a contact form
if and only if, for every open submanifold $U \subset M$ and every $\sigma \in \Gamma_p(E)$,
\begin{equation}\label{contact-pb}
    (j_x^{k+1} \sigma)^* \psi = 0.
\end{equation}
Thus contact forms provide a characterization of the local sections of $\pi^{k+1}$
which are prolongations of sections of $\pi$. The distribution on $J^k E$ generated by the contact forms
is called the Cartan distribution.

It is also possible to consider the limit $k \to \infty$ in eq. (\ref{inf-jet}) for vector fields
and differential forms on a jet bundle $J^k E \to M$.
Let $(x^\mu, z^i_I; \mathcal{U})$ be a standard local chart of $\mathcal{U} \subset \mathcal{E}$ and denote by
$\mathfrak{F}(\mathcal{U}) := C^\infty(\mathcal{E})$ the algebra of functions on $\mathcal{U}$.
Smooth functions on $\mathcal{U}$ may be defined through some finite order $J^n E$:
\begin{equation}\label{infinite-f}
    F: \mathcal{U} \to \mathbb{R}
\end{equation}
by $F = f \circ \pi^\infty_k$ for some smooth $f:J^k E \to \mathbb{R}$.
Let us write $F(x, \mathbf{z}) \in \mathfrak{F}(\mathcal{U})$ for a function on $\mathcal{U}$.
The tangent bundle $T\mathcal{E}$ of $\mathcal{E}$ is the projective limit of $\{(\pi^k)^* TJ^k E \}$.
And the space $\Gamma(T\mathcal{E})$ of the sections of $T\mathcal{E}$ is by definition
the projective limit of $\{ \Gamma\bigl((\pi^n_k)^* TJ^k E \bigr); n \geq k \}$.
$\Gamma(T\mathcal{E})$ acts on the algebra $\mathfrak{F}(\mathcal{U})$ as derivations
in the obvious way and hence carries a natural Lie algebra structure.

The exterior derivative on $\mathfrak{F}(\mathcal{U})$ is defined as usual and the result is given by
\begin{eqnarray}\label{jet-ext-d}
    dF &=& d_\mu F dx^\mu + \partial^I_i F \psi^i_I \nonumber \\
    &\equiv& d_H F + d_V F
\end{eqnarray}
for $F(x, \mathbf{z}) \in \mathfrak{F}(\mathcal{U})$.
In general, the bundle of $p$-forms $\wedge^p T^* \mathcal{E}$
is the injective limit of $\{(\pi^k)^* \wedge^p T^* J^k E \}$ and the space $\Omega^p \mathcal{E}$ of
$p$-forms consists of its sections. Hence a basic differential $p$-form on $\mathcal{E}$ has a local
coordinate expression
\begin{equation}\label{jet-p-form}
 F(x, \mathbf{z}) dx^{\mu_1} \wedge \cdots \wedge dx^{\mu_r} \wedge \psi^{i_1}_{I_1} \wedge \cdots
 \wedge \psi^{i_s}_{I_s}
\end{equation}
with $r+s = p$. A general $p$-form is a finite sum of such terms.
Define the space $\Omega^{r,s} \mathcal{E}$ of forms of type $(r,s)$ to be all linear combinations of
the form (\ref{jet-p-form}). Similarly to eq. (\ref{jet-ext-d}), for $\omega \in \Omega^{r,s} \mathcal{E}$,
there exits a splitting
\begin{equation}\label{p-slitting}
    d \omega = d_H \omega + d_V \omega
\end{equation}
where $d_H \omega \in \Omega^{r+1,s} \mathcal{E}$ and $d_V \omega \in \Omega^{r,s+1} \mathcal{E}$.
In particular, consider a local contact form $\psi_I^i \in \Omega^{0,1} \mathcal{E}$.
Then $d \psi_I^i = - \psi_{\mu + I}^i dx^\mu$ or in succinct form
$\delta \psi^i_I = 0$ and so $d_V \psi^i_I = 0$.
By virtue of $d^2 = 0$ we have $d_H^2 = d_V^2 = d_H d_V + d_V d_H = 0$. Thus, like the Dolbeault differential complex on a complex manifold, there exists a bicomplex $(\Omega^\bullet \mathcal{E}, d_H, d_V)$
of differential forms with the bigrading on $\Omega^\bullet \mathcal{E} = \oplus_{r,s} \Omega^{r,s} \mathcal{E}$.

A connection on a fiber bundle $\pi:E \to M$ is defined as a linear bundle monomorphism
$\Gamma: E \times_M TM \to TE$ over $E$ by
\begin{equation}\label{conn-gamma}
    \Gamma: \dot{x}^\mu \partial_\mu \mapsto \dot{x}^\mu (\partial_\mu + \Gamma^i_\mu \partial_i)
\end{equation}
which splits the exact sequence (\ref{exact-seq}), i.e., $\pi_* \circ \Gamma = \mathrm{Id}_{E \times_M TM}$.
Any connection in a fiber bundle defines a covariant derivative of sections.
If $\sigma: M \to E$ is a section, its covariant derivative is defined by
\begin{eqnarray}\label{conn-sec}
    && \nabla^\Gamma \sigma = D_\Gamma \circ J^1 E: M \to T^* M \times T^\perp E, \nonumber \\
    && \nabla^\Gamma \sigma: (\partial_\mu \sigma^i - \Gamma^i_\mu \circ \sigma) dx^\mu \otimes \partial_i,
\end{eqnarray}
where $D_\Gamma$ is the covariant differential relative to the connection $\Gamma$ defined by
\begin{eqnarray}\label{cov-sec}
    && D_\Gamma: J^1 E \to T^* M \times T^\perp E, \nonumber \\
    && D_\Gamma = (z^i_\mu - \Gamma^i_\mu) dx^\mu \otimes \partial_i.
\end{eqnarray}
A section $\sigma$ is called an integrable section of a connection $\Gamma$ if it belongs to the kernel
of the covariant differential $D_\Gamma$, i.e.,
\begin{equation}\label{int-section}
  \nabla^\Gamma \sigma = 0 \qquad \mathrm{or} \qquad J^1 \sigma = \Gamma \circ \sigma.
\end{equation}
The connection $\Gamma$ can also be seen as a global section $\Gamma: E \to J^1 E$ of the jet bundle
$\pi^1_0: J^1 E \to E$ satisfying
\begin{equation}\label{sec-conn}
    \pi^1_0 \circ \Gamma = \mathrm{Id}_E,
\end{equation}
whose coordinate representation is given by
\begin{equation}\label{coor-conn}
    (x^\mu, z^i, z^i_\mu) \circ \Gamma = (x^\mu, z^i, \Gamma^i_\mu).
\end{equation}
Then the generalization to all higher order jets is obvious.
A $k$-th order connection in a fiber bundle $E \to M$ is a section $\Gamma: E \to J^k E$ which
satisfies
\begin{equation}\label{sec-conn-gen}
    \pi^k_0 \circ \Gamma = \mathrm{Id}_E.
\end{equation}
For an integrable section $\sigma \in \Gamma(E)$, according to eq. (\ref{prolong}),
the connection $\Gamma$ is given by
\begin{equation}\label{int-1sec}
    \Gamma^i_\mu \circ \sigma = \partial_\mu \sigma^i (x)
\end{equation}
or for $\sigma \in \Gamma(J^k E)$, in general, according to eq. (\ref{k-jetman1}),
\begin{equation}\label{int-ksec}
    \Gamma^i_I \circ \sigma = \partial_I \sigma^i (x).
\end{equation}

Now we introduce a flat connection $\nabla^H$ on $\pi^\infty: \mathcal{E} \to M$.
For each $y \in \mathcal{E}$ define a linear subspace $H_y$ of $T_y \mathcal{E}$ by
\begin{equation}\label{flat-h}
H_y = \{ d_H(\partial_\mu) = \partial_\mu + \Gamma^i_{\mu + I}
\partial^I_i \equiv \nabla_\mu \}
\end{equation}
where $\nabla_\mu := d_\mu$ is the alias of the total derivative (\ref{ho-der}) and
\begin{equation}\label{flat-gamma}
 \Gamma^i_{\mu + I} = z^i_{\mu + I}.
\end{equation}
The property (\ref{flat-gamma}) implies that $H_y = \mathrm{Im} \; d_x \sigma_\infty$
for some $\sigma_\infty \in \Gamma(\mathcal{E})$ where $d_x \sigma_\infty : T_x M \to T_y \mathcal{E}$
with $x = \pi^\infty (y)$ is the differential. Hence $H = \bigcup_y H_y$ is a subbundle of $T\mathcal{E}$. Because $d_H^2 = 0$, one can see that $\nabla_\mu$ is a flat connection, i.e.,
$[\nabla_\mu, \nabla_\nu] = 0$. Let $H^\perp = \bigcup_y H^\perp_y$ be the conormal bundle,
where
\begin{equation}\label{conormal}
H^\perp_y = \{ \omega  \in T^*_y \mathcal{E}: \omega|_{H_y} = 0 \}
\end{equation}
for $y \in \mathcal{U} \subset \mathcal{E}$. One can easily check that $\psi^i_I (d_\mu) = 0$ and
$dx^\mu (d_\lambda) = \delta^\mu_\lambda$ for a frame $\{dx^\mu, \psi_I^i \}$ of $T^* \mathcal{U}$.
Therefore, $H^\perp_y$ is spanned by $\{\psi^i_I (y) \}$, i.e., $H^\perp \subset \Omega^{0,1} \mathcal{E}$.
For an integrable (or a flat) section $\sigma_\infty \in \Gamma(\mathcal{E})$,
the connection (\ref{flat-gamma}) can be written as
\begin{equation}\label{flat-section}
\Gamma^i_I \circ \sigma_\infty \equiv \sigma^i_I = z^i_I \circ \sigma_\infty = \partial_I \sigma^i (x)
\end{equation}
which means that $\sigma_\infty^* \psi^i_I = d\sigma^i_I - \sigma^i_{\mu + I} dx^\mu = 0$
or  $\sigma_\infty^* \Gamma(H^\perp) = 0$.

The flat connection $\nabla^H$ lifts $X \in \Gamma(TM)$ up to $\widetilde{X} \in \Gamma(H)
\subset \Gamma(T\mathcal{E})$. Denote this map $\tau: \Gamma(TM) \to \Gamma(T\mathcal{E})$.
Note that $\widetilde{X}$ is uniquely characterized by
\begin{equation}\label{lift-vec}
    \widetilde{X} = d_H (X) = X^\mu (x) \nabla_\mu
\end{equation}
where $X = X^\mu (x) \partial_\mu \in \Gamma(TM)$. Then one can easily check that, for $X, Y \in \Gamma(TM)$,
\begin{equation}\label{lie-homo}
    [\widetilde{X}, \widetilde{Y}] = [X, Y]^\mu (x) \nabla_\mu = [X, Y]^{\thicksim}.
\end{equation}
This means that $\tau: \Gamma(TM) \to \Gamma(T\mathcal{E})$ is a Lie algebra homomorphism.
For example it is an immediate consequence of (\ref{lie-homo}) that $[\nabla_\mu, \nabla_\nu] = 0$.
It should be the case because $d_H^2 = \frac{1}{2} dx^\mu \wedge dx^\nu [d_\mu, d_\nu] = 0$
where $d_\mu := \nabla_\mu$.


\begin{thebibliography}{999}



\bibitem{hsy-ijmp09} H.S. Yang, {\it Emergent gravity from noncommutative spacetime},
\ijmpa{24}{2009}{4473} [\hepth{0611174}].



\bibitem{hsy-jhep09} H.S. Yang, {\it Emergent spacetime and the
origin of gravity}, \jhep{05}{2009}{012} [\arXivid{0809.4728}].



\bibitem{hsy-jpcs12} H.S. Yang, {\it Towards a backround independent
quantum gravity},  {\it J. Phys. Conf. Ser.} {\bf 343} (2012) 012132
[\arXivid{1111.0015}].



\bibitem{big-book} C.W. Misner, K.S. Thorne and J.A. Wheeler, {\it Gravitation},
W.H. Freeman and Company, New York (1973).



\bibitem{hawking-ellis} S.W. Hawking and G.F.R. Ellis, {\it The Large Scale Structure of Space-Time},
Cambridge Univ. Press (1973).



\bibitem{lry2} S. Lee, R. Roychowdhury and H.S. Yang, {\it Test of emergent gravity},
\prd{88}{2013}{086007} [\arXivid{1211.0207}].



\bibitem{japan-matrix} H. Aoki, N. Ishibashi, S. Iso, H. Kawai, Y. Kitazawa and
T. Tada, {\it Noncommutative Yang-Mills in IIB matrix model}, \npb{565}{2000}{176} [\hepth{9908141}].



\bibitem{nc-seiberg} N. Seiberg, {\it A note on background independence in noncommutative gauge theories,
matrix model and tachyon condensation},
\jhep{09}{2000}{003} [\hepth{0008013}].



\bibitem{hsy-epjc09} H.S. Yang, {\it Noncommutative electromagnetism
as a large N gauge theory}, \epjc{64}{2009}{445}
[\arXivid{0704.0929}].




\bibitem{review4} H.S. Yang, {\it Emergent geometry and quantum gravity},
\mpla{25}{2010}{2381} [\arXivid{1007.1795}].



\bibitem{sladler} S.L. Adler, {\it Quantum theory as an emergent phenomenon},
Cambridge Univ. Press, Cambridge (2004).



\bibitem{ams-matter} M.F. Atiyah, N.S. Manton and B.J. Schroers, {\it Geometric models of matter},
{\it Proc. R. Soc. A} {\bf 468} (2012) 1252 [\arXivid{1108.5151}].



\bibitem{sg-book1} V.I. Arnold, {\it Mathematical methods of classical
mechanics}, Springer (1978).


\bibitem{sg-book2} R. Abraham and J.E. Marsden, {\it Foundations of mechanics},
Addison-Wesley (1978).


\bibitem{cornalba} L. Cornalba, {\it D-brane physics and noncommutative Yang-Mills theory},
\atmp{4}{2000}{271} [\hepth{9909081}].


\bibitem{jur-sch} B. Jur\v co and P. Schupp, {\it Noncommutative Yang-Mills
from equivalence of star products}, \epjc{14}{2000}{367}
[\hepth{0001032}].


\bibitem{liu} H. Liu, {\it $*$-Trek II: $*_n$ operations, open Wilson lines and the Seiberg-Witten map},
\npb{614}{2001}{305} [\hepth{0011125}].



\bibitem{ncft-sw} N. Seiberg and E. Witten, {\it String theory and noncommutative geometry},
\jhep{09}{1999}{032} [\hepth{9908142}].



\bibitem{martinec} E.J. Martinec, {\it Evolving notions of geometry in string theory},
{\it Found. Phys.} {\bf 43} (2013) 156.



\bibitem{jsw1} B. Jur\v co, P. Schupp and J. Wess, {\it Noncommutative gauge theory for Poisson manifolds}, \npb{584}{2000}{784} [\hepth{0005005}].



\bibitem{jsw2} B. Jur\v co, P. Schupp and J. Wess, {\it Nonabelian noncommutative gauge theory via
noncommutative extra dimensions}, \npb{604}{2001}{148} [\hepth{0102129}].



\bibitem{jsw-ncl} B. Jur\v co, P. Schupp and J. Wess, {\it Noncommutative line bundle and Morita equivalence}, \lmp{61}{2002}{171} [\hepth{0106110}].



\bibitem{buba-me} H. Bursztyn and S. Waldmann, {\it The characteristic classes of Morita equivalent
star products on symplectic manifolds}, \cmp{228}{2002}{103} [\Math{QA}{0106178}].



\bibitem{kontsevich} M. Kontsevich, {\it Deformation quantization of Poisson manifolds},
\lmp{66}{2003}{157} [\qalg{9709040}].



\bibitem{cft} A.S. Cattaneo, G. Felder and L. Tomassini, {\it From local to glogal deformation
quantization of Poisson manifolds}, \dmj{115}{2002}{329} [\Math{QA}{0012228}].



\bibitem{cftp} A.S. Cattaneo and G. Felder, {\it On the globalization of Kontsevich's star product
and the perturbative sigma model}, \ptps{144}{2001}{38} [\hepth{0111028}].



\bibitem{vdolg-am} V. Dolgushev, {\it Covariant and equivariant formality theorem},
\advm{191}{2005}{147} [\Math{QA}{0307212}].
177 (2005)


\bibitem{d-quant2} B.V. Fedosov, {\it A simple geometrical construction of deformation quantization},
\jdg{40}{1994}{213}.


\bibitem{fedo-book} B.V. Fedosov, {\it Deformation quantization and index theory}, Akademie Verlag,
Berlin (1996).


\bibitem{ads-cft1} J.M. Maldacena, {\it The large N limit of superconformal field theories and supergravity},
\atmp{2}{1998}{231} [\hepth{9711200}].



\bibitem{ads-cft2} S.S. Gubser, I.R. Klebanov and A.M. Polyakov, {\it Gauge theory correlators
from non-critical string theory}, \plb{428}{1998}{105} [\hepth{9802109}].



\bibitem{ads-cft3} E. Witten, {\it Anti de Sitter space and holography}, \atmp{2}{1998}{253}
[\hepth{9802150}].



\bibitem{ads-cft} O. Aharony, S.S. Gubser, J. Maldacena, H. Ooguri and Y. Oz, {\it Large N field theories,
string theory and gravity}, \prep{323}{2000}{183} [\hepth{9905111}].



\bibitem{dho-fre} E. D'Hoker and D.Z. Freedman, {\it Supersymmetric gauge theories and the AdS/CFT
correspondence}, [\hepth{0201253}].



\bibitem{oriti} G.T. Horowitz and J. Polchinski, {\it Gauge/gravity duality}
in {\it Approaches to quantum gravity: Toward a new understanding of space, time and matter},
edited by D. Oriti, Cambridge Univ. Press (2009) [\grqc{0602037}].


\bibitem{blau-theisen} M. Blau and S. Theisen, {\it String theory as a theory of quantum gravity:
A status report}, \grg{41}{2009}{743}.



\bibitem{richard-r} R.J. Szabo, {\it Symmetry, gravity and noncommutativity},
\cqg{23}{2006}{R199} [\{hepth{0606233}].



\bibitem{harold-p} H. Steinacker, {\it Emergent gravity from noncommutative gauge theory},
\jhep{12}{2007}{049} [\arXivid{0708.2426}].



\bibitem{hstein-rev} H. Steinacker, {\it Emergent geometry and gravity from matrix models:
an introduction}, \cqg{27}{2010}{133001} [arXiv:1003.4134].



\bibitem{jun-r} J. Nishimura, {\it The origin of space-time as seen from matrix model simulations},
{\it Prog. Theor. Exp. Phys.} {\bf 2012} (2012) 1A101 [\arXivid{1205.6870}].



\bibitem{epstein} D.B.A. Epstein, {\it Natural tensors on Riemannian manifolds},
\jdg{10}{1975}{631}.


\bibitem{ambient-metric} C. Fefferman and C.R. Graham, {\it The ambient metric}, Princeton Univ. Press (2012).



\bibitem{d-weinstein} A. Weinstein, {\it The local structure of Poisson manifolds},
\jdg{18}{1983}{523}.



\bibitem{vaisman} I. Vaisman, {\it Lectures on the geometry of Poisson manifolds}, Birkh\"auser, Basel (1994).



\bibitem{sg-weinstein} A. Weinstein, {\it Symplectic groupoids and Poisson manifolds},
\bams{16}{1987}{101}.


\bibitem{kara-mas} M.V. Karasev, {\it Analogues of the objects of Lie group theory for nonlinear
Poisson brackets}, {\it  Math. USSR Izv.} {\bf 28} (1987) 497.



\bibitem{raamsdonk1} M. van Raamsdonk, {\it Comments on quantum gravity and entanglement}, [\arXivid{0907.2939}].



\bibitem{raamsdonk2} M. van Raamsdonk, {\it Building up spacetime with quantum entanglement},
\grg{42}{2010}{2323}; \ijmpd{19}{2010}{2429}.



\bibitem{lry1}  S. Lee, R. Roychowdhury and H.S. Yang, {\it Notes on emergent gravity},
\jhep{09}{2012}{030} [\arXivid{1212.3000}].



\bibitem{cofs} A. Connes, M. Flato and D. Stermheimer, {\it Closed star products and cyclic cohomology},
\lmp{24}{1992}{1}.



\bibitem{omy2} H. Omori, Y. Maeda and A. Yoshioka, {\it Existence of a closed star-product},
\lmp{26}{1992}{285}.



\bibitem{mod-wein} A. Weinstein, {\it The modular automorphism group of a Poisson manifold},
\jgp{23}{1997}{379}.



\bibitem{jet-book1} D.J. Saunders, {\it The geometry of jet bundles}, Cambridge Univ. Press,
Cambridge (1989).



\bibitem{jet-book2} I. Kol\'a\v r, P.W. Michor and J. Slov\'ak,
{\it Natural operations in differential geometry}, Springer-Verlag, Berlin (1993).



\bibitem{jet-book3} G. Sardanashvily, {\it Advanced differential geometry for theoreticians:
Fibre bundles, jet manifolds and Lagrangian theory}, LAP, Saarbr\"ucken (2013).



\bibitem{silva-wein} A. Cannas da Silva and A. Weinstein, {\it Geometric models for noncommutative algebras},
AMS (1999).


\bibitem{dufour} J.-P. Dufour and N.T. Zung, {\it Normal forms of Poisson structures},
{\it Geom. Topol. Monogr.} {\bf 17} (2011) 109.



\bibitem{yasi} H.S. Yang and M. Sivakumar, {\it Emergent geometry from quantized spacetime},
\prd{82}{2010}{045004} [\arXivid{0908.2809}].



\bibitem{lee-yang} J. Lee and H.S. Yang, {\it Quantum gravity from noncommutative spacetime},
[\arXivid{1004.0745}].



\bibitem{hsy-mpla06} H.S. Yang, {\it Exact Seiberg-Witten map and induced gravity from
noncommutativity}, \mpla{21}{2006}{2637} [\hepth{0402002}].



\bibitem{ban-yan} R. Banerjee and H.S. Yang, {\it Exact Seiberg-Witten map, induced
gravity and topological invariants in noncommutative field
theories}, \npb{708}{2005}{434} [\hepth{0404064}].



\bibitem{d-quant1} F. Bayen, M. Flato, C. Fronsdal, A. Lichnerowicz and D. Sternheimer,
{\it Deformation theory and quantization. I. Deformations of symplectic structures},
\ap{111}{1978}{61}; {\it II. Physical applications}, \ap{111}{1978}{111}.



\bibitem{cf-sigma} A. Cattaneo and G. Felder, {\it A path integral approach to the Kontsevich
quantization formula}, \cmp{212}{2000}{591} [\Math{QA}{9902090}].



\bibitem{manchon} D. Arnal, D. Manchon and M. Masmoudi, {\it Choix des signes pour la formalit\'e
de M. Kontsevich}, {\it Pac. J. Math.} {\bf 203} (2002) 23 [\Math{QA}{0003003}].



\bibitem{behr-sykora} W. Behr and A. Sykora, {\it Construction of gauge theories on curved noncommutative
spacetime}, \npb{698}{2004}{473} [\hepth{0309145}].



\bibitem{nt-star} R. Nest and B. Tsygan, {\it Algebraic index theorem}, \cmp{172}{1995}{223}.



\bibitem{bc-gutt} M. Bertelson, M. Cahen and S. Gutt, {\it Equivalence of star products}, \cqg{14}{1997}{A93}.



\bibitem{ns-inst} N. Nekrasov and A. Schwarz, {\it Instantons on noncommutative $\mathbb{R}^4$,
and (2; 0) superconformal six dimensional theory}, \cmp{198}{1998}{689} [\hepth{9802068}].



\bibitem{nek-cmp03} N.A. Nekrasov, {\it Noncommutative instantons revisited}, \cmp{241}{2003}{143}
[\hepth{0010017}].



\bibitem{lry3} S. Lee, R. Roychowdhury and H.S. Yang, {\it Topology change of spacetime and resolution
of spacetime singularity in emergent gravity}, \prd{87}{2013}{126002} [\arXivid{1212.3000}].



\bibitem{cheeil} C. Chevalley and S. Eilenberg, {\it Cohomology theory of Lie groups and Lie algebras},
{\it Trans. Amer. Math. Soc.} {\bf 63} (1948) 85.



\bibitem{ahs-1978} M.F. Atiyah, N. Hitchin and I.M. Singer, {\it Self-duality in four-dimensional
Riemannian geometry}, {\it Proc. Roy. Soc. London} {\bf A362} (1978) 425.



\bibitem{oh-yang} J.J. Oh and H.S. Yang, {\it Einstein manifolds as Yang-Mills instantons},
\mpla{28}{2013}{1350097} [\arXivid{1101.5185}].



\bibitem{loy-jhep} J. Lee J.J. Oh and H.S. Yang, {\it An efficient representation of Euclidean gravity I},
\jhep{12}{2011}{025} [\arXivid{1109.6644}].



\bibitem{besse} A.L. Besse, {\it Einstein Manifolds}, Springer-Verlag, Berlin (1987).



\bibitem{kobnom} S. Kobayashi and K. Nomizu, {\it Foundation of differential geometry},
Vol. 1, Interscience, New York (1963).



\bibitem{closed-german} U. M\"uller, C. Schubert and E.M. van de Ven, {\it A closed formula for the Riemann
normal coordinate expansion}, \grg{31}{1999}{1759}.



\bibitem{fedo-man} I. Gelfand, V. Retakh and M. Shubin, {\it Fedosov manifolds},
\advm{136}{1998}{104} [{\tt dg-ga/9707024}].



\bibitem{tamarkin} D.E. Tarmarkin, {\it Topological invariants of connections on symplectic manifolds},
\faa{29}{1995}{258}.



\bibitem{bcm-jhep} F. Bonechi, A.S. Cattaneo and P. Mnev, {\it The Poisson sigma model on closed surfaces},
\jhep{01}{2012}{099} [\arXivid{1110.4850}].



\bibitem{acgutt} M. Ammar, V. Chloup and S. Gutt, {\it Universal star products},
\lmp{84}{2008}{199} [\arXivid{0804.1300}].



\bibitem{pxu} P. Xu, {\it Fedosov $*$-products and quantum momentum maps}, \cmp{197}{1998}{167}
[{\tt q-alg/9608006}].



\bibitem{nc-gravity1} P. Aschieri, C. Blohmann, M. Dimitrijevic, F. Meyer, P. Schupp and J. Wess,
{\it A gravity theory on noncommutative spaces}, \cqg{22}{2005}{3511} [\hepth{0504183}].



\bibitem{nc-gravity2} P. Aschieri, M. Dimitrijevic, F. Meyer and J. Wess, {\it Noncommutative geometry
and gravity}, \cqg{23}{2006}{1883} [\hepth{0510059}].



\bibitem{srsg4} M. Crainic and I. Marcut, {\it On the existence of symplectic realizations},
{\it J. Symplectic Geom.} {\bf 9} (2011) 435 [\arXivid{1009.2085}].



\bibitem{georgi} H. Georgi, {\it Lie Algebras in Particle Physics: From Isospin to Unified Theories},
Advanced Book Program, Addison-Wesley, Reading U.K. (1999).



\bibitem{lie-rep1} C. Chevalley, {\it Theory of Lie group}, Princeton Univ. Press, Princeton (1946).



\bibitem{lie-rep2} J.E. Humphreys, {\it Introduction to Lie algebras and representation theory},
Springer-Verlag, New-York (1972).



\bibitem{srsg5} M. Crainic and R.L. Fernandes,  {\it Integrability of Lie brackets},
\am{157}{2003}{575} [{\tt math.DG/0105033}].


\bibitem{srsg3} M. Crainic and R.L. Fernandes, {\it Integrability of Poisson brackets}, \jdg{66}{2004}{71}
[{\tt math.DG/0210152}].



\bibitem{pxu-cmp} P. Xu, {\it Morita equivalence of Poisson manifolds}, \cmp{142}{1991}{493}.



\bibitem{srsg1} A.S. Cattaneo and G. Felder, {\it Poisson sigma models and symplectic groupoids},
{\it Progr. Math.} {\bf 198} (2001) 61 [{\tt math.SG/0003023}].



\bibitem{srsg2} A.S. Cattaneo, B. Dherin, and G. Felder, {\it Formal symplectic groupoid},
\cmp{253}{2005}{645} [{\tt math.SG/0312380}].



\bibitem{awe-pxu} A. Weinstein and P. Xu, {\it Extensions of symplectic groupoids and quantization},
{\it J. reine angew. Math.} {\bf 417} (1991) 159.



\bibitem{cft-pre} A.S. Cattaneo, G. Felder and L. Tomassini, {\it Fedosov connections on jet bundles
and deformation quantization}, [\Math{QA}{0111290}].



\bibitem{yas-prl} H.S. Yang and M. Salizzoni, {\it Gravitational instantons from gauge theory},
\prl{96}{2006}{201602} [\hepth{0512215}].



\bibitem{sty-plb} M. Salizzoni, A. Torrielli and H.S. Yang, {\it ALE spaces from noncommutative U(1) instantons
via exact Seiberg-Witten map}, \plb{634}{2006}{427} [\hepth{0510249}].



\bibitem{kishimoto} I. Kishimoto, {\it Fuzzy sphere and hyperbolic space from deformation quantization},
\jhep{03}{2001}{025} [\hepth{0103018}].



\bibitem{gh-inst} G. Gibbons and S. Hawking, {\it Gravitational multi-instantons}, \plb{78}{1978}{430}.



\bibitem{real-heaven} C. Boyer and I. Finley, J.D., {\it Killing vectors in selfdual, euclidean Einstein spaces},
\jmp{23}{1982}{1126}.



\bibitem{hsy-epl09} H.S. Yang, {\it Instantons and emergent geometry}, \epl{88}{2009}{31002}
[\hepth{0608013}].



\bibitem{lebrun} C. LeBrun, {\it Explicit self-dual metrics on ${\bf CP}_2 \sharp \cdots \sharp {\bf CP}_2$},
\jdg{34}{1991}{223}.




\bibitem{egu-han1} T. Eguchi and A.J. Hanson, {\it Asymptotically flat self-dual solutions
to Euclidean gravity}, {\it Phys. Lett.} {\bf 74B} (1979) 249.



\bibitem{egu-han2} T. Eguchi and A.J. Hanson, {\it Self-dual solutions to Euclidean gravity},
\ap{120}{1979}{82}.



\bibitem{pra-equi} M. Prasad, {\it Equivalence of Eguchi-Hanson metric to two-center Gibbons-Hawking metric},
\plb{83}{1979}{310}.



\bibitem{joyce} D.D. Joyce, {\it Explicit construction of self-dual 4-manifolds},
\dmj{77}{1995}{519}.



\bibitem{osaka1} Y. Hashimoto, Y. Yasui, S. Miyagi and T. Otsuka, {\it Applications of the Ashtekar
gravity to four-dimensional hyperK\"ahler geometry and Yang-Mills instantons},
\jmp{38}{1997}{5833} [\hepth{9610069}].



\bibitem{osaka2} T. Ootsuka, S. Miyagi, Y. Yasui and S. Zeze, {\it Anti-selfdual Maxwell solutions
on hyperK\"ahler manifold and N = 2 supersymmetric Ashtekar gravity},
\cqg{16}{1999}{1305} [\grqc{9809083}].



\bibitem{book-dasilva} A. Cannas da Silva, {\it Lectures on Symplectic Geometry},
Springer-Verlag, Berlin (2001).



\bibitem{newnir} A. Newlander and L. Nirenberg, {\it Complex analytic coordinates in almost complex
manifolds}, \am{65}{1957}{391}.



\bibitem{comnosym} M. Fern\'andez, M. Gotay and A. Gray, {\it Compact parallelizable four-dimensional
symplectic and complex manifolds}, {\it Proc. Amer. Math. Soc.} {\bf 103} (1988) 1209.



\bibitem{hawking-ebh} S.W. Hawking, {\it Gravitational instantons}, \pla{60}{1977}{81}.



\bibitem{dgms-km} P. Deligne, P. Griffiths, J. Morgan and D. Sullivan, {\it Real homotopy theory of
K\"ahler manifolds}, {\it Invent. Math.} {\bf 29} (1975) 245.



\bibitem{gromov-book} M. Gromov, {\it Partial differential relations}, Springer-Verlag, Berlin (1986).



\bibitem{caval-massey} G.R. Cavalcanti, {\it New aspects of the $dd^c$-lemma}, [{\tt math.DG/0501406}].



\bibitem{gotay-etal} M.J. Gotay, H.B. Grundling and G.M. Tuynman, {\it Obstruction results in
quantization theory}, {\it J. Nonlinear Sci.} {\bf 6} (1996) 469 [{\tt dg-ga/9605001}].



\bibitem{geoq-book1} J. \'Sniatycki, {\it Geometric quantization and quantum mechanics},
Springer-Verlag, Berlin (1980).



\bibitem{geoq-book2} N.M.J. Woodhouse, {\it Geometric quantization}, Clarendon Press, Oxford (1991).



\bibitem{gukov-witten} S. Gukov and E. Witten, {\it Branes and quantization},
\atmp{13}{2009}{1} [\arXivid{0809.0305}].



\bibitem{waldmann-drep} S. Waldmann, {\it States and representations in deformation quantization},
{\it Rev. Math. Phys.} {\bf 17} (2005) 15 [{\tt math.QA/0408217}].



\bibitem{ikkt} N. Ishibashi, H. Kawai, Y. Kitazawa and A. Tsuchiya, {\it A large-N reduced model
as superstring}, \npb{498}{1997}{467} [\hepth{9612115}].



\bibitem{bfss} T. Banks, W. Fischler, S.H. Shenker and L. Susskind, {\it M theory as a matrix model:
A conjecture}, \prd{55}{1997}{5112} [\hepth{9610043}].



\bibitem{mst} L. Motl, {\it Proposals on nonperturbative superstring interactions}, [\hepth{9701025}];
R. Dijkgraaf, E. Verlinde and H. Verlinde, {\it Matrix string theory}, \npb{500}{1997}{43}
[\hepth{9703030}].



\bibitem{highly} J.H. Schwarz, {\it Highly effective actions}, \jhep{01}{2014}{088} [\arXivid{1311.0305}].



\bibitem{highsde1} E. Corrigan, C. Devchand, D.B. Fairlie and J. Nuyts, {\it First-order equations
for gauge fields in spaces of dimension greater than four}, \npb{214}{1983}{452}.



\bibitem{highsde2} R.S. Ward, {\it Completely solvable gauge-field equations in dimension
greater than four}, \npb{236}{1984}{381}.



\bibitem{highsde3} D. Bak, K. Lee and J.-H. Park, {\it BPS equations in six and eight dimensions},
\prd{66}{2002}{025021} [\hepth{0204221}].



\bibitem{yanyun} H.S. Yang and S. Yun, {\it Calabi-Yau manifolds, Hermitian Yang-Mills instantons and
mirror symmetry}, [\arXivid{1107.2095}].



\bibitem{yanyun-unp} H.S. Yang, {\it Calabi-Yau manifolds from NC Hermitian U(1) instantons},
[\arXivid{1411.6115}].



\bibitem{abs-magic} M.F. Atiyah, R. Bott and A. Shapiro, {\it Clifford modules}, {\it Topology} {\bf 3}
(suppl. {\bf 1}) (1964) 3.



\bibitem{geom2} M. Dunajski, {\it Skyrmions from gravitational instantons},
{\it Proc. R. Soc. A} {\bf 469} (2013) 20120576 [\arXivid{1206.0016}];
G. Franchetti and N.S. Manton, {\it Gravitational instantons as models for charged particle systems},
\jhep{03}{2013}{072} [\arXivid{1301.1624}].




\bibitem{b-soibelman} P. Bressler and Y. Soibelman, {\it Mirror symmetry and deformation quantization},
[\hepth{0202128}].



\bibitem{hms} M. Kontsevich, {\it Homological algebra of mirror symmetry}, [{\tt alg-geom/9411018}].




\bibitem{kapustin} A. Kapustin, {\it Topological strings on noncommutative manifolds},
{\it Int. J. Geom. Meth. Mod. Phys.} {\bf 1} (2004) 49 [\hepth{0310057}]; {\it A-branes and
noncommutative geometry}, [\hepth{0502212}].



\bibitem{vpestun} V. Pestun, {\it Topological strings in generalized complex space}, \atmp{11}{2007}{399}
[\hepth{0603145}].


\bibitem{gualtieri} M. Gualtieri, {\it Branes on Poisson varieties}, [\arXivid{0710.2719}].



\bibitem{mkay} M.M. Kay, {\it On the quantization of special K\"ahler manifolds}, [\arXivid{1305.4838}].



\bibitem{isham} J. Butterfield and C.J. Isham, {\it On the emergence of time in guantum gravity},
[\grqc{9901024}].



\bibitem{time-qm} {\it Time in Quantum Mechanics -- Vol. 1}, {\it Lect. Notes Phys.} {\bf 734},
edited by J.G. Muga, R. Sala Mayato and I.L. Egusquiza, Springer, Berlin (2008);
{\it Time in Quantum Mechanics -- Vol. 2}, {\it Lect. Notes Phys.} {\bf 789},
edited by J.G. Muga, A. Ruschhaupt and A. del Campo, Springer, Berlin (2009).



\bibitem{dirac-cq} P.A.M. Dirac, {\it Lectures on quantum mechanics}, Yeshiva Univ. Press,
New York (1964).



\bibitem{m-tduality} W. Taylor, {\it M(atrix) theory: matrix quantum mechanics as a fundamental theory},
\rmp{73}{2001}{419} [\hepth{0101126}].



\bibitem{k-book1} M. Karoubi, {\it K-theory: An introduction}, Springer, Berlin (1978).



\bibitem{k-book2} H.B. Lawson and M.L. Michelsohn, {\it Spin Geometry}, Princeton Univ. Press,
New Jersey (1989).



\bibitem{meinren} E. Meinrenken, {\it Clifford algebras and Lie theory}, Springer, Berlin (2013).



\bibitem{n=4sym} L. Brink, J.H. Schwarz and J. Scherk, {\it Supersymmetric Yang-Mills theories}, \npb{121}{1977}{77}.



\bibitem{dgo-loop} N. Drukker, D.J. Gross and H. Ooguri, {\it Wilson loops and minimal surfaces},
\prd{60}{1999}{125006} [\hepth{9904191}].



\bibitem{jopenw} N. Ishibashi, S. Iso, H. Kawai and Y. Kitazawa, {\it Wilson loops in noncommutative
Yang-Mills}, \npb{573}{2000}{573} [\hepth{9910004}].



\bibitem{szabo-f} J. Ambjorn, Y.M. Makeenko, J. Nishimura and R.J. Szabo, {\it Lattice gauge fields
and discrete noncommutative Yang-Mills theory}, \jhep{05}{2000}{023} [\hepth{0004147}].



\bibitem{wgross} D.J. Gross, A. Hashimoto and N. Itzhaki, {\it Observables of non-commutative gauge
theories}, \atmp{4}{2000}{893} [\hepth{0008075}].



\bibitem{g-crystal} H. Imazato, S. Mizoguchi and M. Yata, {\it Taub-NUT crystal},
\ijmpa{26}{2011}{5143} [\arXivid{1107.3557}].



\bibitem{adsinstanton1} M. Bianchi, M.B. Green, S. Kovacs and G. Rossi, {\it Instantons in
supersymmetric Yang-Mills and D-instantons in IIB superstring theory},
\jhep{08}{1998}{013} [\hepth{9807033}].



\bibitem{adsinstanton2} N. Dorey, T.J. Hollowood, V.V. Khoze, M.P. Mattis and S. Vandoren,
{\it Multi-instanton calculus and the AdS/CFT correspondence in N=4 superconformal field theory},
\npb{552}{1999}{88} [\hepth{9901128}].




\bibitem{llm1} H. Lin, O. Lunin and J. Maldacena, {\it Bubbling AdS space and 1/2 BPS geometries},
\jhep{10}{2004}{025} [\{hepth{0409174}].



\bibitem{lm2} H. Lin and J. Maldacena, {\it Fivebranes from gauge theory},
\prd{74}{2006}{084014} [\hepth{0509235}].


\bibitem{nc-top1} J.A. Harvey, {\it Topology of the gauge group in noncommutative gauge theory},
[\hepth{0105242}].


\bibitem{nc-top2} F. Lizzi, R.J. Szabo and A. Zampini, {\it Geometry of the gauge algebra in
noncommutative Yang-Mills theory}, \jhep{08}{2001}{032} [\hepth{0107115}].



\bibitem{gromovj} M. Gromov, {\it Pseudo-holomorphic curves in symplectic manifolds},
{\it Invent. Math.} {\bf 82} (1985) 307.



\bibitem{duff-sala} D. McDuff and D. Salamon, {\it Introduction to symplectic topology}, 2nd. ed.,
Oxford Univ. Press, Oxford (1998).



\bibitem{ygoh} Y.-G. Oh, {\it Symplectic topology and Floer homology},
available at http://www.math.wisc.edu/$\sim$oh/.



\bibitem{uv-ir} S. Minwalla, M. Van Raamsdonk and N. Seiberg, {\it Noncommutative perturbative dynamics},
\jhep{02}{2000}{020} [\hepth{9912072}].


\bibitem{spin2-weinberg1} S. Weinberg, {\it Photons and gravitons in S-matrix theory:
Derivation of charge conservation and equality of gravitational and inertial mass}, \pr{135}{1964}{B1049}.



\bibitem{spin2-weinberg2} S. Weinberg, {\it Photons and gravitons in perturbation theory:
Derivation of Maxwell's and Einstein's equations}, \pr{138}{1965}{B988}.




\bibitem{p-homology} J.L. Brylinski, {\it A differential complex for Poisson manifolds},
\jdg{28}{1988}{93}.



\bibitem{p-duality1} S. Evens, J.-H. Lu and A. Weinstein, {\it Transverse measures, the modular class
and cohomology pairing for Lie algebroids}, {\it Quart. J. Math. Oxford} {\bf 50} (1999) 417
[{\tt dg-ga/9610008}].



\bibitem{p-duality2} J. Huebschmann, {\it Duality for Lie-Rinehart algebras and the modular class},
{\it J. Reine Angew. Math.} {\bf 510} (1999) 103 [{\tt dg-ga/9702008}].



\bibitem{nts2} R. Nest and B. Tsygan, {\it Algebraic index theorem for families}, \advm{113}{1995}{151}.



\bibitem{felsho} G. Felder and B. Shoikhet, {\it Deformation quantization with traces}, \lmp{53}{2000}{75}
 [{\tt math.QA/0002057}].



\bibitem{dolgushev} V. Dolgushev, {\it The Van den Bergh duality and the modular symmetry of
a Poisson variety}, {\it Selecta Math.} {\bf 14} (2009) 199 [{\tt math.QA/0612288}].



\end{thebibliography}
\end{document}